\documentclass[runningheads,fleqn,epsf]{svmult}
\usepackage{amssymb}

\usepackage{graphicx}
\usepackage{makeidx}
\usepackage{multicol}
\usepackage{subeqnar}
\usepackage{multicol}
\usepackage{cropmark}
\usepackage{physmult}
\usepackage{epsfig}

\begin{document}

\title*{Fluctuation Phenomena in Superconductors}
\author{Anatoly Larkin \and Andrei Varlamov}
\maketitle

\titlerunning{Fluctuation Phenomena} 
\toctitle{Fluctuation Phenomena in
Superconductors}

\authorrunning{A. Larkin \and A. Varlamov}
\vspace{-0.5cm}

\begin{flushright}
\emph{To our friend Lev Aslamazov, in memoriam}
\end{flushright}

\subsection*{\protect\bigskip {\protect\Large Contents}}

\textbf{1.Introduction}

\textbf{2.Ginzburg-Landau formalism. Thermodynamics.}

2.1 Fluctuation contribution to heat capacity

{\small 2.1.1 GL functional.}

{\small 2.1.2} {\small Zero dimensionality: the exact solution.}

{\small 2.1.3} {\small Arbitrary dimensionality: case }$T\geq T_{c}.$

{\small 2.1.4} {\small Arbitrary dimensionality: case }$T<T_{c}.$

2.2 Ginzburg-Levanyuk criterion

2.3 Scaling and renormalization group

2.4 Fluctuation diamagnetism

{\small 2.4.1} {\small Qualitative preliminaries.}

{\small 2.4.2} {\small Zero-dimensional diamagnetic susceptibility.}

{\small 2.4.3} {\small GL treatment of fluctuation magnetization.}

\textbf{3}. \textbf{Fluctuations below critical temperature }(to be
published in the complete version)

\textbf{3}. \textbf{Ginzburg-Landau theory of fluctuations in transport
phenomena}

3.1 Time dependent GL equation

3.2 Paraconductivity

3.3 General expression for paraconductivity

3.4 Fluctuation conductivity of layered superconductor

{\small 3.4.1 In-plane conductivity.}

{\small 3.4.2 Out-of plane conductivity.}

{\small 3.4.3 Analysis of the general expressions.}

3.5 Magnetic field angular dependence of paraconductivity

\textbf{4.} \textbf{Fluctuations near superconductor-insulator transition}

4.1 Quantum phase transition

4.2 3D superconductors

4.3 2D superconductors

{\small 4.3.1 Preliminaries.}

{\small 4.3.2 Boson mechanism of the }$T_{c}${\small \ suppression.}

{\small 4.3.3 Fermion mechanism of }$T_{c}${\small \ suppression.}

\textbf{5. Microscopic derivation of\ the Time-Dependent Ginzburg-Landau
equation }

5.1 Preliminaries

5.2 The Cooper channel of electron-electron interaction

5.3 Superconductor with impurities

{\small 5.3.1 Account for impurities.}

{\small 5.3.2 Propagator.}

5.4. Microscopic theory of fluctuation conductivity of layered superconductor

{\small 5.4.1 Qualitative discussion of different fluctuation contributions}

{\small 5.4.2 Generalities}

{\small 5.4.3 Aslamazov-Larkin contribution}

{\small 5.4.4 Contributions from fluctuations of the density of states}

{\small 5.4.5 Maki-Thompson contribution}

{\small 5.4.6 Discussion}

\textbf{6. Manifestation of fluctuations in various properties}

6.1 The effects of fluctuations on magnetoconductivity

6.2 \ Fluctuations far from $T_{c}$ or in strong magnetic fields

{\small 6.2.1 Fluctuation magnetic susceptibility far from transition.}

{\small 6.2.2 Fluctuation magnetoconductivity far from transition }

{\small 6.2.3 Fluctuations in magnetic fields near }$H_{c2}(0)$

6.3 \ The effect of fluctuations on the Hall conductivity

6.4 Fluctuations in the ultra-clean case

6.5 The effect of fluctuations on the one-electron density of states and on
tunneling measurements

{\small 6.5.1 Density of states }

{\small 6.5.2 The effect of fluctuations on the tunnel current }

6.6 Nonlinear fluctuation effects

6.7 The effect of fluctuation on the optical conductivity

6.8 Thermoelectric power above the superconducting transition

6.9 The effect of fluctuations on NMR characteristics

{\small 6.9.1 Preliminaries.}

{\small 6.9.2 Spin Susceptibility.}

{\small 6.9.3 Relaxation Rate.}

\textbf{7. Conclusions}

\textbf{8. Acknowledgments}

\textbf{9. Bibliography}

\section{Introduction}

A major success of low temperature physics was achieved with the
introduction by Landau of the notion of quasiparticles. According to his
hypothesis, the properties of many body interacting systems at low
temperatures are determined by the spectrum of some low energy, long living
excitations (quasiparticles). Another milestone of many body theory is the
Mean Field Approximation (MFA), which permitted achieving considerable
progress in the theory of phase transitions. Phenomena which cannot be
described by the quasiparticle method or by MFA are usually called
fluctuations. The BCS\ theory of superconductivity is a bright example of
the use of both the quasiparticle description and MFA. The success of the
BCS theory for traditional superconductors was determined by the fact that
fluctuations give small corrections with respect to the MFA results.

During the first half of the century after the discovery of
superconductivity the problem of fluctuation smearing of the superconducting
transition was not even considered. In bulk samples of traditional
superconductors the critical temperature $T_{c}$ sharply divides the
superconducting and the normal phases. It is worth mentioning that such
behavior of the physical characteristics of superconductors is in perfect
agreement both with the Ginzburg-Landau (GL) phenomenological theory (1950) 
\cite{GL50} and the BCS microscopic theory of superconductivity (1957)\cite
{BCS57}.

The characteristics of high temperature and organic superconductors, low
dimensional and amorphous superconducting systems studied today, strongly
differ from those of the traditional superconductors discussed in textbooks.
The transitions turn out to be much more smeared out. The appearance of
superconducting fluctuations above critical temperature leads to precursor
effects of the superconducting phase occurring while the system is still in
the normal phase, sometimes far from $T_{c}$. The conductivity, the heat
capacity, the diamagnetic susceptibility, the sound attenuation, etc. may
increase considerably in the vicinity of the transition temperature.

The first numerical estimation of the fluctuation contribution to the heat
capacity of a superconductor in the vicinity of $T_{c}$ was done by Ginzburg
in 1960 \cite{G60}. In that paper he showed that superconducting
fluctuations increase the heat capacity even above $T_{c}$. In this way
fluctuations change the temperature dependence of the specific heat in the
vicinity of critical temperature, where, in accordance with the
phenomenological GL theory of second order phase transitions, a jump should
take place. The range of temperatures where the fluctuation correction to
the heat capacity of a bulk clean conventional superconductor is relevant
was estimated by Ginzburg \footnote{%
The expression for the width of the strong fluctuation region in terms of
the Landau phenomenological theory of phase transitions was obtained by
Levanyuk \cite{L59}. So in the modern theory of phase transitions the
relative temperature width of fluctuation region is called the
Ginzburg-Levanyuk parameter $Gi_{(D)}$, where $D$ is the effective space
dimensionality.} to be 
\begin{equation}
Gi=\frac{\delta T}{T_{c}}\sim \left( \frac{T_{c}}{E_{F}}\right) ^{4}\sim
10^{-12}\div 10^{-14},  \label{gi}
\end{equation}
where $E_{F}$ is the Fermi energy. The correction occurs in a temperature
range $\delta T$ many orders of magnitude smaller than that accessible in
real experiments.

In the 1950s and 60s the formulation of the microscopic theory of
superconductivity, the theory of type-II superconductors and the search for
high-$T_{c}$ superconductivity attracted the attention of researchers to
dirty systems, superconducting films and filaments. In 1968, in the papers
of L. G. Aslamazov and A. I. Larkin \cite{AL68}, K.Maki \cite{M68} and a
little later in the paper of \ R.S.Thompson \cite{T70} the fundaments of the
microscopic theory of fluctuations in the normal phase of a superconductor
in the vicinity of the critical temperature were formulated. This
microscopic approach confirmed Ginzburg's evaluation \cite{G60} for the
width of the fluctuation region in a bulk clean superconductor. Moreover,
was found that the fluctuation effects increase drastically in thin dirty
superconducting films and whiskers. In the cited papers was demonstrated
that fluctuations affect not only the thermodynamical properties of
superconductor but its dynamics too. Simultaneously the fluctuation smearing
of the resistive transition in bismuth amorphous films was experimentally
found by Glover \cite{G67}, and it was perfectly fitted by the microscopic
theory.

In the BCS theory \cite{BCS57} only the Cooper pairs forming a
Bose-condensate are considered. Fluctuation theory deals with the Cooper
pairs out of the condensate. In some phenomena these fluctuation Cooper
pairs behave similarly to quasiparticles but with one important difference.
While for the well defined quasiparticle the energy has to be much larger
than its inverse life time, for the fluctuation Cooper pairs the ''binding
energy`` $E_{0}$ turns out to be of the same order. The Cooper pair life
time $\tau _{GL}$ is determined by its decay into two free electrons.
Evidently at the transition temperature the Cooper pairs start to condense
and $\tau _{GL}=\infty .$ So it is natural to suppose from dimensional
analysis that $\tau _{GL}\sim \hbar /{k_{B}(T-T_{c}).}$ The microscopic
theory confirms this hypothesis and gives the exact coefficient:

\begin{equation}
\tau _{GL}=\frac{\pi \hbar }{{8k_{B}(T-T_{c})}}.  \label{taugl}
\end{equation}

\smallskip Another important difference of the fluctuation Cooper pairs from
quasiparticles lies in their large size $\xi (T)$. This size is determined
by the distance on which the electrons forming the fluctuation Cooper pair
move away during the pair life-time $\tau _{GL}.$ In the case of an impure
superconductor the electron motion is diffusive with the diffusion
coefficient $\mathcal{D}\sim v_{F}^{2}\tau $ ($\tau $ is the electron
scattering time\footnote{%
Strictly speaking $\tau $ in the most part of future results should be
understood as the electron transport scattering time $\tau _{tr}.$
Nevertheless, as it is well known, in the case of isotropic scattering these
values coincide, so for sake of simplicity we will use hereafter the symbol $%
\tau .$}), and $\xi _{d}(T)=\sqrt{\mathcal{D}\tau _{GL}}\sim v_{F}\sqrt{\tau
\tau _{GL}}.$ In the case of a clean superconductor, where ${k_{B}}T\tau \gg
\hbar ,$ impurity scattering does not affect any more the electron
correlations. In this case the time of electron ballistic motion turns out
to be less than the electron-impurity scattering time $\tau $ and is
determined by the uncertainty principle: $\tau _{bal}$ $\sim $ $\hbar /{k_{B}%
}T.$ Then this time has to be used in this case for determination of the
effective size instead of $\tau $: $\xi _{c}(T)\sim v_{F}\sqrt{\hbar \tau
_{GL}/{k_{B}}T}.$ In both cases the coherence length grows with the approach
to the critical temperature as $(T-T_{c})^{-1/2}$, and we will write it down
in the unique way ($\xi =\xi _{c,d}$):

\begin{equation}
\xi (T)=\frac{\xi }{\sqrt{\epsilon }},\;\;\;\;\epsilon =\frac{T-T_{c}}{T_{c}}%
..  \label{xiGL}
\end{equation}
The microscopic theory in the case of an isotropic Fermi surface gives for $%
\xi $ the precise expression:

\begin{equation}
\xi _{(D)}^{2}=-\frac{v_{F}^{2}\tau ^{2}}{D}\left\{ \psi (\frac{1}{2}+\frac{%
\hbar }{4\pi k_{B}T\tau })-\psi (\frac{1}{2})-\frac{\hbar }{4\pi k_{B}T\tau }%
\psi ^{^{\prime }}(\frac{1}{2})\right\} ,  \label{etad}
\end{equation}
where $\psi (x)$ is the digamma function and $D=3,2,1$ is the space
dimensionality. In the clean ($c$) and dirty ($d$) limits:

\begin{equation}
\xi _{c}=0.133\frac{\hbar v_{F}}{k_{B}T_{c}}\sqrt{\frac{3}{D}}=0.74\xi _{0}%
\sqrt{\frac{3}{D}},  \label{xic}
\end{equation}

\begin{equation}
\xi _{d}=0.36\sqrt{\frac{\hbar v_{F}l}{k_{B}T_{c}}\frac{3}{D}}=0.85\sqrt{\xi
_{0}l}\sqrt{\frac{3}{D}}.  \label{xid}
\end{equation}
Here $l=v_{F}\tau $ is the electron mean free path and $\xi _{0}=\hbar
v_{F}/\pi \Delta (0)$ is the conventional BCS definition of the coherence
length of a clean superconductor at zero temperature. One can see that
\thinspace (\ref{xic}) and \thinspace (\ref{xid}) coincide with the above
estimations \footnote{%
Let us stress some small numerical difference between our Exp.\thinspace (%
\ref{etad}) and the usual definition of the coherence length. We are dealing
near the critical temperature, so the definition \thinspace (\ref{etad}) is
natural and permits us to avoid many numerical coefficients in further
calculations. The cited coherence length $\xi _{0}=\hbar v_{F}/\pi \Delta
(0)=0.18\hbar v_{F}/k_{B}T_{c}$ , as is evident, was introduced for zero
temperature and an isotropic $3D$ superconductor.
\par
It is convenient to determine the coherence length also from the formula for
the upper critical field: $H_{c2}(T)=A(T)\Phi _{0}/2\pi \xi ^{2}(T).$ $%
A(T_{c})=1,$ while its value at $T=0$ depends on the impurities
concentration. For the dirty case the appropriate value was found by K.Maki 
\cite{M1} $A_{d}(0)=0.69,$ for the clean case by L.Gor'kov \cite{G59} $%
A_{c}^{2D}(0)=0.59,$ $A_{c}^{3D}(0)=0.72.$%
\par
{}}.

Finally it is necessary to recognize that fluctuation Cooper pairs can be
really treated as classical objects, but these objects instead of Boltzmann
particles appear as classical fields in the sense of Rayleigh-Jeans. This
means that in the general Bose-Einstein distribution function only small
energies $E(p)$ are involved and the exponent can be expanded:

\begin{equation}
\mathit{n}(p)=\frac{1}{\exp (\frac{E(p)}{k_{B}T})-1}=\frac{k_{B}T}{E(p)}.
\label{ncp}
\end{equation}
This is why the more appropriate tool to study fluctuation phenomena is not
the Boltzmann transport equation but the GL equation for classical fields.
Nevertheless at the qualitative level the treatment of fluctuation Cooper
pairs as particles with the density $n_{(D)}=\int \mathit{n}(p)\frac{d^{D}p}{%
(2\pi \hbar )^{D}}$ often turns out to be useful \footnote{%
This particle density is defined in the ($D)$-dimensional space. This means
that it determines the normal volume density of pairs in $3D$ case, the
density per square unit in $2D$ case and the number of pairs per unit length
in $1D.$ The real three dimensional density $n$ can be defined too: $n=$ $%
d^{D-3}n_{(D)},$ where $d$ is the thickness of the film or wire.}.

Below will be demonstrated both in the framework of the phenomenological
Ginzburg-Landau theory and the microscopic BCS theory that in the vicinity
of the transition

\begin{equation}
E(p)=\alpha k_{B}(T-T_{c})+\frac{\mathbf{p}^{2}}{2m^{\ast }}=\frac{1}{%
2m^{\ast }}\left[ \hbar ^{2}/\xi ^{2}\left( T\right) +\mathbf{p}^{2}\right] .
\end{equation}
Far from transition temperature the dependence $\mathit{n}(p)$ turns out to
be more sophisticated than (\ref{ncp}), nevertheless one can always write it
in the form

\begin{equation}
\mathit{n}(p)=\frac{m^{\ast }k_{B}T}{\hbar ^{2}}\xi ^{2}\left( T\right)
f\left( \frac{\xi (T)p}{\hbar }\right) .
\end{equation}

In classical field theory the notions of the particle distribution function $%
n(p)$ (proportional to $E^{-1}(p)$ in our case) and Cooper pair mass $%
m^{\ast }$ are poorly determined. At the same time the characteristic value
of the Copper pair center of mass momentum can be defined and it turns out
to be of the order of $p_{0}\sim \hbar /\xi (T).$ So for the combination $%
m^{\ast }E(p_{0})$ one can write $m^{\ast }E(p_{0})\sim p_{0}^{2}\sim \hbar
^{2}/\xi ^{2}(T).$ In fact the particles density enters into many physical
values in the combination $n/m^{\ast }.$ As the consequence of the above
observation it can be expressed in terms of the coherence length: 
\begin{equation}
\frac{n_{(D)}}{m^{\ast }}=\frac{k_{B}T}{m^{\ast }E(p_{0})}(\frac{p_{0}}{%
\hbar })^{D}\sim \frac{k_{B}T}{\hbar ^{2}}\xi ^{2-D}(T),  \label{n/m}
\end{equation}
$p_{0}^{D}$ here estimates the result of momentum integration.

For example we can evaluate the fluctuation Cooper pairs contribution to
conductivity by using the Drude formula

\begin{equation}
\sigma =\frac{ne^{2}\tau }{m^{\ast }}\Rightarrow \frac{k_{B}T}{\hbar ^{2}}%
d^{D-3}\xi ^{2-D}(T)(2e)^{2}\tau _{GL}(\epsilon )\sim \epsilon ^{D/2-2}.
\label{sigmaint}
\end{equation}
Analogously a qualitative understanding of the increase in the diamagnetic
susceptibility above the critical temperature may be obtained from the
well-known Langevin expression for the atomic susceptibility\footnote{%
This formula is valid for the dimensionalities $D=2,3,$ when fluctuation
Cooper pair has the possibility to ''rotate'' in the applied magnetic field
and the average square of the rotation radius is $<R^{2}>\sim \xi ^{2}(T).$
''Size'' effects, important for low dimensional samples, will be discussed
later on.}:

\begin{equation}
\chi =-{\frac{e^{2}}{{c^{2}}}}\frac{n}{m^{\ast }}\left\langle
R^{2}\right\rangle \Rightarrow -\frac{4e^{2}}{{c^{2}}}\frac{k_{B}T}{\hbar
^{2}}d^{D-3}\xi ^{4-D}(T)\sim -\epsilon ^{D/2-2}.  \label{chifl32}
\end{equation}

Besides these examples of the direct influence of fluctuations on
superconducting properties, indirect manifestations by means of quantum
interference in the pairing process and of renormalization of the density of
one-electron states in the normal phase of superconductor take place. These
effects, being much more sophisticated, have a purely quantum nature, and in
contrast to the direct Cooper pair contributions require microscopic
consideration. This is why, in developing phenomenological methods through
the first five Sections of this review, we will deal with the direct
fluctuation pair contributions only. The sixth Section is devoted to the
microscopic justification of the time dependent Ginzburg-Landau equation.
The description of the microscopic theory of fluctuations, including
indirect fluctuation effects, discussion of their manifestations in various
physical properties of superconductors will be given in the Sections 7-8.

The first seven Sections are written in detail, so they can serve as a
textbook. On the contrary, in the last (eighth) Section a wide panorama of
fluctuation effects in different physical properties of superconductors is
presented. Thus this Section has more of a handbook character, and the
intermediate calculations often are omitted.

Finally we would like to mention that the number of articles devoted to
superconducting fluctuations published in the last 33 years is of the order
of ten thousands, so our bibliography list does not pretend nor at the
completeness nor at the establishment of the rigorous priorities.

\section{Ginzburg-Landau formalism. Thermodynamics}

\subsection{Fluctuation contribution to heat capacity}

\subsubsection{GL functional.}

The complete description of the thermodynamic properties of a system can be
done through the exact calculation of the partition function\footnote{%
Hereafter $\hbar =k_{B}=c=1.$}:

\begin{equation}
Z=\mathrm{Tr}\left\{ \exp \left( -{\frac{\widehat{\mathcal{H}}}{{T}}}\right)
\right\} .  \label{Z}
\end{equation}

As discussed in the Introduction, in the vicinity of the superconducting
transition, side by side with the fermionic electron states, fluctuation
Cooper pairs of a bosonic nature appear in the system. As already mentioned,
they can be described by means of classical bosonic fields $\Psi (\mathbf{r}%
) $ which can be treated as ''Cooper pair wave functions``. So the
calculation of the trace in \thinspace (\ref{Z}) can be separated into a
summation over the ''fast`` electron degrees of freedom and a further
functional integration carried out over all possible configurations of the
''slow'' Cooper pairs wave functions:

\begin{equation}
Z=\int D\Psi (\mathbf{r})\mathcal{Z}[\Psi (\mathbf{r})],  \label{eb}
\end{equation}
where

\begin{equation}
{\ \mathcal{Z}}[\Psi (\mathbf{r})]=\exp \left( -{\frac{{\ \mathcal{F}}[\Psi (%
\mathbf{r})]}{{T}}}\right)  \label{zphi}
\end{equation}
is the system partition function in a fixed bosonic field $\Psi (\mathbf{r}%
), $ already summed over the electronic degrees of freedom. Here it is
assumed that the classical field dependent part of the Hamiltonian can be
chosen in the spirit of the GL approach\footnote{%
For simplicity the magnetic field is assumed to be zero.}:

\begin{equation}
F[\Psi (r)]=F_{N}+\int dV\left\{ a|\Psi (\mathbf{r})|^{2}+\frac{b}{2}|\Psi (%
\mathbf{r})|^{4}+\frac{1}{4m}|\mathbf{\nabla }\Psi (\mathbf{r})|^{2}\right\}
..  \label{Func}
\end{equation}

Let us discuss the coefficients of this functional. In accordance with the
Landau hypothesis, the coefficient $a$ goes to zero at the transition point
and depends linearly on $T-T_{c}$. Then${\;}a=\alpha T_{c}\epsilon ;$ all
the coefficients ${\alpha ,}b$ and $m$ are supposed to be positive and
temperature independent. Concerning the magnitude of the coefficients it is
necessary to make the following comment. One of these coefficients can
always be chosen arbitrary: this option is related to the arbitrariness of
the Cooper pair wave function normalization. Nevertheless the product of two
of them is fixed by dimensional analysis: $ma\sim \xi ^{-2}(T).$ Another
combination of the coefficients, independent of the wave function
normalization and temperature, is $\alpha ^{2}/b.$ One can see that it has
the dimensionality of the density of states. Since these coefficients were
obtained by a summation over the electronic degrees of freedom, the only
reasonable candidate for this value is the one electron density of states $%
\nu $. The microscopic theory gives the precise coefficients for the above
relations:

\begin{equation}
4m\alpha T_{c}=\xi ^{-2};\;\alpha ^{2}/b=\frac{8\pi ^{2}}{7\zeta (3)}\nu ,
\label{phemicro1}
\end{equation}
where $\zeta (x)$ is the Riemann zeta function, $\zeta (3)=1.202$. One can
notice that the arbitrariness in the normalization of the order parameter
amplitude leads to the unambiguity in the choice of the Cooper mass
introduced in \thinspace (\ref{Func}) as $2m.$ Indeed, this value enters in
\thinspace (\ref{phemicro1}) in the product with $\alpha $ so one of these
parameters has to be fixed. In the case of a clean D-dimensional
superconductor it is natural to suppose that the Copper pair mass is equal
to two free electron masses what results in

\begin{equation}
\;\alpha _{\left( D\right) }=\frac{2D\pi ^{2}}{7\zeta (3)}\frac{T_{c}}{E_{F}}%
.  \label{alpe}
\end{equation}

As the first step in the Landau theory of phase transitions $\Psi \ $ is
supposed to be independent of position. This assumption in the limit of
sufficiently large volume $V$ of the system permits a calculation of the
functional integral in \thinspace (\ref{eb}) by the method of steepest
descent. Its saddle point determines the equilibrium value of the order
parameter

\begin{equation}
|\widetilde{\Psi }|^{2}=\left\{ 
\begin{tabular}{l}
$-{\alpha T}_{c}{\epsilon }/b,\mbox{ }\epsilon <0$ \\ 
$0,\;\epsilon >0$%
\end{tabular}
\right. .  \label{saddle}
\end{equation}
Choosing $\alpha $\ in accordance with \thinspace (\ref{alpe}) one finds
that this value coincides with the superfluid density $n_{s}$ of the
microscopic theory \cite{BCS57}.

The fluctuation part of the free energy related to the transition is
determined by the minimum of the functional \thinspace (\ref{Func}):

\begin{equation}
F=\left( \mathcal{F}[\Psi ]\right) _{\min }=\mathcal{F}[\widetilde{\Psi }%
]=\left\{ 
\begin{tabular}{l}
$F_{N}-\frac{\alpha ^{2}T_{c}^{2}\epsilon ^{2}}{2b}V,\mbox{ }\epsilon <0$ \\ 
$F_{N},\;\epsilon >0$%
\end{tabular}
\right. .  \label{Fmin}
\end{equation}
\smallskip From the second derivative of \thinspace (\ref{Fmin}) one can
find an expression for the jump of the specific heat capacity at the phase
transition point:

\begin{eqnarray}
\Delta C &=&C_{S}-C_{N}=\frac{T_{c}}{V}\left( \frac{\partial S_{S}}{\partial
T}\right) -\frac{T_{c}}{V}\left( \frac{\partial S_{N}}{\partial T}\right) =
\label{skach} \\
&=&-\frac{1}{VT_{c}}\left( \frac{{\ \partial ^{2}}F}{{\ \partial }\epsilon {%
^{2}}}\right) =\frac{\alpha ^{2}}{b}T_{c}=\frac{8\pi ^{2}}{7\zeta (3)}\nu
T_{c}.  \nonumber
\end{eqnarray}

Let us mention that the jump of the heat capacity was obtained because of
the system volume was taken to infinity first, and after this the reduced
temperature $\epsilon $ was set equal to zero.

\subsubsection{Zero dimensionality: the exact solution.}

\begin{figure}[tbp]
\epsfxsize=4in
\centerline {\epsfbox{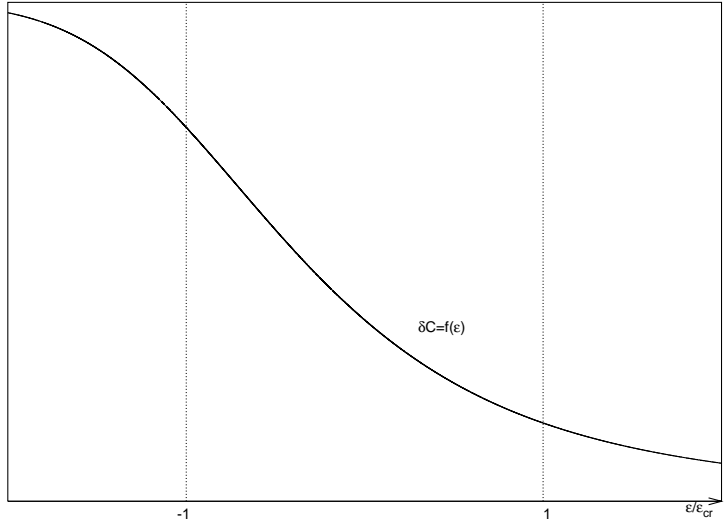}}
\caption{ Temperature dependence of the heat capacity of superconducting
grains in the region of the critical temperature}
\label{0Dhc}
\end{figure}

In a system of finite volume fluctuations smear out the jump in heat
capacity. For a small superconducting sample with the characteristic size $%
d\ll $ $\xi (T)$ the space independent mode $\Psi _{0}=\Psi \sqrt{V}$
defines the main contribution to the free energy:

\begin{eqnarray}
Z_{(0)} &=&\int d^{2}\Psi _{0}\mathcal{F}[\Psi _{0}]=\pi \int d|\Psi
_{0}|^{2}\exp \left( -{\frac{({\ }a|\Psi _{0}|^{2}+\frac{b}{2V}|\Psi
_{0}|^{4})}{{T}}}\right) =  \nonumber \\
&=&\sqrt{\frac{\pi ^{3}VT}{2b}}\exp (x^{2})(1-\mbox{erf}(x))|_{x=a\sqrt{%
\frac{V}{2bT}}}.  \label{stat0}
\end{eqnarray}
By evaluating the second derivative of this exact result \cite{VSchmd68} one
can find the temperature dependence of the superconducting granular heat
capacity (see Fig.~\ref{0Dhc}). It is evident that the smearing of the jump
takes place in the region of temperatures in the vicinity of transition
where $x\sim 1$, i.e.

\[
\epsilon _{cr}=Gi_{(0)}=\frac{\sqrt{7\zeta (3)}}{2\pi }\frac{1}{\sqrt{\nu
T_{c}V}}\approx 13,3\left( \frac{T_{c0}}{E_{F}}\right) \sqrt{\frac{\xi
_{0}^{3}}{V}}, 
\]
where we have supposed that the granule is clean; $T_{c0}$ and $\xi _{0}$
are the critical temperature and the zero temperature coherence length (see
the footnote 3 in Introduction) of the appropriate bulk material. From this
formula one can see that even for granule with the size $d\sim $ $\xi _{0}$
the smearing of the transition still is very narrow.

Far above the critical region, where $Gi_{(0)}\ll \epsilon \ll 1,$ one can
use the asymptotic expression for the $\mbox{erf}(x)$ function and find

\begin{equation}
F_{(0)}=-T\ln Z_{(0)}=-T\ln \frac{\pi }{\alpha \epsilon }.  \label{0Df}
\end{equation}
Calculation of the second derivative gives an expression for the fluctuation
part of the heat capacity in this region:

\begin{equation}
\delta C_{(0)}=1/\epsilon ^{2}.  \label{0Cf}
\end{equation}

The experimental study of the heat capacity of small Sn particles in the
vicinity of transition was done in \cite{TS77}

\subsubsection{Arbitrary dimensionality: case $T\geq T_{c}.$}

It is possible to estimate the fluctuation contribution to the heat capacity
for a specimen of an arbitrary effective dimensionality on the basis of the
following observation. The volume of the specimen may be divided into
regions of size $\xi (T),$ which are weakly correlated with each other. Then
the whole free energy can be estimated as the free energy of one such
zero-dimensional specimen \thinspace (\ref{0Df}), multiplied by their number 
$N_{(D)}=V\xi ^{-D}(T):$

\begin{equation}
F_{(D)}=-TV\xi ^{-D}(T)\ln \frac{\pi }{\alpha \epsilon }.  \label{Feva}
\end{equation}
This formula gives the correct temperature dependence for the free energy
for even dimensionalities. A more accurate treatment removes the $\ln
\epsilon $ dependence from it in the case of the odd dimensions.

Let us begin with the calculation of the fluctuation contribution to the
heat capacity in the normal phase of a superconductor. We restrict ourselves
to the region of temperatures beyond the immediate vicinity of transition,
where this correction is still small. In this region one can omit the fourth
order term in $\Psi (\mathbf{r})$ with respect to the quadratic one and
write down the GL functional, expanding the order parameter in a Fourier
series:

\begin{equation}
F[\Psi _{\mathbf{k}}]=F_{N}+\sum_{\mathbf{k}}[a+\frac{\mathbf{k}^{2}}{4m}%
]|\Psi _{\mathbf{k}}|^{2}=\alpha T_{c}\sum_{\mathbf{k}}\left( \epsilon +\xi
^{2}\mathbf{k}^{2}\right) |\Psi _{\mathbf{k}}|^{2}.  \label{fglfourier}
\end{equation}
Here $\Psi _{\mathbf{k}}={\frac{1}{\sqrt{V}}}\int \Psi (\mathbf{r})e^{-i%
\mathbf{kr}}dV$ and the summation is carried out over the vectors of the
reciprocal space. Now we see that the free energy functional appears as a
sum of energies of the independent modes $\mathbf{k}$. The functional
integral for the partition function \thinspace (\ref{zphi}) can be separated
to a product of Gaussian type integrals over these modes:

\begin{equation}
Z=\prod_{\mathbf{k}}\int d^{2}\Psi _{\mathbf{k}}\exp \left\{ -{\alpha }%
(\epsilon +\frac{\mathbf{k}^{2}}{4m\alpha T_{c}}{\ })|\Psi _{\mathbf{k}%
}|^{2}\right\} .  \label{partit}
\end{equation}
Carrying out these integrals, one gets the fluctuation contribution to the
free energy:

\begin{equation}
F(\epsilon >0)=-T\ln Z=-T\sum_{\mathbf{k}}\ln \frac{{\ \pi }}{{\ }\alpha {(}%
\epsilon +\frac{\mathbf{k}^{2}}{4m\alpha T_{c}}{)}}.  \label{ffe}
\end{equation}
The appropriate correction to the specific heat capacity of a superconductor
at temperatures above the critical temperature may thus be calculated. We
are interested in the most singular term in ${\epsilon }^{-1}$, so the
differentiation over the temperature can be again replaced by that over $%
\epsilon :$

\begin{equation}
\delta C_{+}=-{\frac{1}{VT}}_{c}\left( {\frac{{\ \partial ^{2}F}}{{\
\partial \epsilon ^{2}}}}\right) =\frac{1}{V}\sum_{\mathbf{k}}\frac{1}{%
(\epsilon +\frac{\mathbf{k}^{2}}{4m\alpha T_{c}})^{2}}{.}  \label{C+}
\end{equation}

The result of the summation over $\mathbf{k}$ strongly depends on the linear
sizes of the sample, i.e. on its effective dimensionality. As it is clear
from (\ref{C+}), the scale with which one has to compare these sizes is
determined by the value $(4m\alpha T_{c}\epsilon )^{-\frac{1}{2}}$ which, as
was already mentioned above, coincides with the effective size of Cooper
pair $\xi (T)$. Thus, if all dimensions of the sample considerably exceed $%
\xi (T),$ one can integrate over $(2\pi )^{-3}{%
L_{x}L_{y}L_{z}dk_{x}dk_{y}dk_{z}}$ instead of summing over $%
n_{x},n_{y},n_{z}$. In the case of arbitrary dimensionality the fluctuation
correction to the heat capacity turns out to be

\begin{equation}
\delta C_{+}=\frac{V_{D}}{V}\int \frac{1}{{(\epsilon +\frac{\mathbf{k}^{2}}{%
4m\alpha T_{c}})^{2}}}{\frac{{d^{D}}\mathbf{k}}{({2\pi })^{D}}}=\vartheta
_{D}\frac{V_{D}}{V}{\frac{(4m\alpha T_{c})^{\frac{D}{2}}}{{\ \epsilon ^{2-{%
\frac{D}{2}}}}}},  \label{fhc}
\end{equation}
where $V_{D}=V,S,L,1$ for $D=3,2,1,0.$ For the coefficients $\vartheta _{D}$
it is convenient to write an expression valid for an arbitrary
dimensionality $D,$ including fractional ones. For a space of fractional
dimensionality we just mention that the momentum integration in spherical
coordinates is carried out according to the rule: $\int d^{D}\mathbf{k/}%
\left( 2\pi \right) ^{D}\mathbf{=}\mu _{D}\int k^{D-1}dk,$ where

\begin{equation}
\mu _{D}=\frac{D}{2^{D}\pi ^{D/2}\Gamma (D/2+1)}  \label{mu}
\end{equation}
and $\Gamma (x)$ is a gamma-function. The coefficient in (\ref{fhc}) can be
expressed in terms of the gamma-function too:

\begin{equation}
\vartheta _{D}=\frac{\Gamma (2-D/2)}{2^{D}\pi ^{D/2}}  \label{thetad}
\end{equation}
yielding $\vartheta _{1}=1/4,\vartheta _{2}=1/4\pi $ and $\vartheta
_{3}=1/8\pi .$

In the case of small particles with characteristic sizes $d\lesssim \xi
(\epsilon )$ the appropriate fluctuation contribution to the free energy and
the specific heat capacity coincides with the asymptotics of the exact
results (\ref{0Df}) and (\ref{0Cf}). From the formula given above it is easy
to see that the role of fluctuations increases when the effective
dimensionality of the sample or the electron mean free path decrease.

\subsubsection{Arbitrary dimensionality: case $T<T_{c}.$}

The general expressions (\ref{eb}) and (\ref{Func}) allow one to find the
fluctuation contribution to heat capacity below $T_{c}$. For this purpose
let us restrict ourselves to the region of temperatures not very close to $%
T_{c}$ from below, where fluctuations are sufficiently weak. In this case
the order parameter can be written as the sum of the equilibrium $\widetilde{%
\Psi }$ (see (\ref{saddle})) and fluctuation $\psi \mathbf{(r})$ parts:

\begin{equation}
\Psi \mathbf{(r})=\widetilde{\Psi }+\psi \mathbf{(r}).  \label{psisum}
\end{equation}
Keeping in (\ref{Func}) the terms up to the second order in $\psi \mathbf{(r}%
)$ and up to the fourth order in $\widetilde{\Psi }$, one can find

\begin{eqnarray}
{\ \mathcal{Z}}[\widetilde{\Psi }] &=&\exp (-\frac{a\widetilde{\Psi }^{2}+b/2%
\widetilde{\Psi }^{4}}{T})\prod_{\mathbf{k}}\int d\mbox{Re}\psi _{\mathbf{k}%
}d\mbox{Im}\psi _{\mathbf{k}}\times  \label{zbelow} \\
&&\times \exp \left\{ -{\frac{1}{T}}[(3b\widetilde{\Psi }^{2}+a+\frac{%
\mathbf{k}^{2}}{4m})\mbox{Re}^{2}\psi _{\mathbf{k}}+(b\widetilde{\Psi }%
^{2}+a+\frac{\mathbf{k}^{2}}{4m})\mbox{Im}^{2}\psi _{\mathbf{k}}]\right\} . 
\nonumber
\end{eqnarray}
Carrying out the integral over the real and imaginary parts of the order
parameter one can find an expression for the fluctuation part of the free
energy:

\begin{equation}
F=-\frac{T}{2}\sum_{\mathbf{k}}\left\{ \ln \frac{{\ \pi }T_{c}}{{\ }3b%
\widetilde{\Psi }^{2}{+}a+\frac{\mathbf{k}^{2}}{4m}}{+}\ln \frac{\pi T_{c}{\ 
}}{{\ }b\widetilde{\Psi }^{2}+a+\frac{\mathbf{k}^{2}}{4m}}\right\} {.}
\label{fflb}
\end{equation}
Let us discuss this result. It is valid both above and below ${T}_{c}.$ The
two terms in it correspond to the contributions of the modulus and phase
fluctuations of the order parameter. Above ${T}_{c}$ $\ \widetilde{\Psi }%
\equiv 0$ and these contributions are equal: phase and modulus fluctuations
in the absence of $\widetilde{\Psi }$ represent just two equivalent degrees
of freedom of the scalar complex order parameter. Below ${T}_{c},$ the
symmetry of the system decreases (see (\ref{saddle})). The order parameter
modulus fluctuations remain of the same diffusive type as above ${T}_{c},$
while the character of the phase fluctuations, in accordance with the
Goldstone theorem, changes dramatically.

Substitution of (\ref{saddle}) to (\ref{fflb}) results in the disappearance
of the temperature dependence of the phase fluctuation contribution and,
calculating the second derivative, one sees that only the fluctuations of
the order parameter modulus contribute to the heat capacity. As a result the
heat capacity, calculated below $T_{c},$ turns out to be proportional to
that found above:

\[
\delta C_{-}=2^{{\frac D2}-2}\delta C_{+}. 
\]

Hence, in the framework of the theory proposed we found that the heat
capacity of the superconductor tends to infinity at the transition
temperature. Strictly speaking, the restrictions of the above approach do
not permit us to discuss seriously this divergence at the critical point
itself. The calculations in principle are valid only in that region of
temperatures where the fluctuation correction is small. We will discuss in
the next Section the quantitative criteria for the applicability of this
perturbation theory.

\subsection{Ginzburg-Levanyuk criterion}

The fluctuation corrections to the heat capacity obtained above allow us to
answer quantitatively the question: where are the limits of applicability of
the GL theory?

This theory is valid not too near to the transition temperature, where the
fluctuation correction is still small in comparison with the heat capacity
jump.\ Let us define as the Ginzburg-Levanyuk number $Gi_{(D)}$ \cite
{L59,G60} the value of the reduced temperature at which the fluctuation
correction (\ref{fhc}) equals the value of $\Delta C$ (\ref{skach})\footnote{%
One can see that some arbitrariness occurs in this definition. For instance
the $Gi$ number could be defined as the reduced temperature at which the AL
correction to conductivity is equal to the normal value of conductivity (as
it was done in \cite{L99,LO01}). Such definition results in the change of
the numerical factor in $Gi$ number:
\par
$Gi_{(2,\sigma )}=1.44Gi_{(2,h.c.)}.$%
\par
{}}: 
\begin{equation}
Gi_{(D)}=\frac{1}{\alpha }\left[ \frac{V_{D}}{V}\vartheta _{D}b(4m)^{\frac{D%
}{2}}T_{c}^{\frac{D}{2}-1}\right] ^{\frac{2}{4-D}}.  \label{gid}
\end{equation}
Substituting into this formula the microscopic values of the GL theory
parameters (\ref{phemicro1}) one can find

\begin{equation}
Gi_{(D)}=\left[ \frac{7\zeta (3)\vartheta _{D}}{8\pi ^{2}}\left( \frac{V_{D}%
}{V}\right) \frac{1}{\nu _{D}T_{c}\xi ^{D}}\right] ^{\frac{2}{4-D}}.
\label{gimicro}
\end{equation}
Since $\nu _{D}T_{c}\sim \nu _{D}v_{F}/\xi _{c}\sim p_{F}^{D-1}\xi _{c}^{-1}$
$\sim \frak{a}^{1-D}\xi _{c}^{-1}$ one can convert this formula to the form

\[
Gi_{(D)}\sim \left[ \frac{7\zeta (3)\vartheta _{D}}{8\pi ^{2}}\left( \frac{%
V_{D}}{V}\right) \frac{\xi _{c}\frak{a}^{D-1}}{\xi ^{D}}\right] ^{\frac{2}{%
4-D}}, 
\]
where $\frak{a}$ \ is the interatomic distance. It is worth mentioning that
in bulk conventional superconductors, due to the large value of the
coherence length $(\xi _{c}\sim 10^{-6}\div 10^{-4}cm),$ which drastically
exceeds the interatomic distance $(\frak{a}\sim 10^{-8}cm)$, the fluctuation
correction to the heat capacity is extremely small. However, the fluctuation
effect increases for small effective sample dimensionality and small
electron mean free path. For instance, the fluctuation heat capacity of a
superconducting granular system is readily accessible for experimental study.

Using the microscopic expression for the coherence length (\ref{etad}), the
Ginzburg-Levanyuk number (\ref{gimicro}) can be evaluated for different
cases of clean (c) and dirty (d) superconductors of various dimensionalities
and geometries (film, wire, whisker and granule are supposed to have $3D$
electronic spectrum):

\begin{tabular}{|c|c|c|c|}
\hline
$Gi_{(3)}$ & $Gi_{(2)}$ & $Gi_{(1)}$ & $Gi_{(0)}$ \\ \hline
\begin{tabular}{l}
$80\left( \frac{T_{c}}{E_{F}}\right) ^{4},\;\left( c\right) $ \\ 
$\frac{1.6}{\left( p_{F}l\right) ^{3}}\left( \frac{T_{c}}{E_{F}}\right)
,\;\left( d\right) $%
\end{tabular}
& 
\begin{tabular}{l}
$\left( \frac{T_{c}}{E_{F}}\right) ,\;\left( c\right) $ \\ 
$\frac{0.27}{p_{F}l},\;\;\left( d\right) $ \\ 
$\frac{1.3}{p_{F}^{2}ld},\;\left( d\right) \mbox{ film}$%
\end{tabular}
& 
\begin{tabular}{l}
$0.5,\;\left( c\right) \mbox{ }$ \\ 
$1.3\left( p_{F}^{2}S\right) ^{-2/3}\left( T_{c}\tau \right)
^{-1/3},\;\left( d\right) $ wire \\ 
$2.3\left( p_{F}^{2}S\right) ^{-2/3},\;\left( c\right) $ whisker
\end{tabular}
& $
\begin{tabular}{l}
$\frac{\sqrt{7\zeta (3)}}{2\pi }\frac{1}{\sqrt{\nu T_{c}V}}\approx $ \\ 
$13.3\frac{T_{c0}}{E_{F}}\sqrt{\frac{\xi _{0}^{3}}{V}}$%
\end{tabular}
$ \\ \hline
\end{tabular}

\begin{center}
Table 1.
\end{center}

One can see that for the three dimensional clean case the result coincides
with the original Ginzburg evaluation and demonstrates the negligibility of
the superconducting fluctuation effects in clean bulk materials.

\subsection{Scaling and renormalization group}

In the above study of the fluctuation contribution to heat capacity we have
restricted ourselves to the temperature range out of the direct vicinity of
the critical temperature: $|\epsilon |\gtrsim Gi_{(D)}.$ As
we have seen the fluctuations in this region turn out to be weak and
neglecting their interaction was justified. In this Section we will discuss
the fluctuations in the immediate vicinity of the critical temperature ( 
$|\epsilon |\lesssim Gi_{(D)}$) where this interaction
turns out to be of great importance.

We will start with the scaling hypothesis, i.e. with the belief that in the
immediate vicinity of the transition the only relevant length scale is $\xi
(T)$. The temperature dependencies of all other physical quantities can be
expressed through $\xi (T)$. This means, for instance, that the formula for
the fluctuation part of the free energy (\ref{Feva}) with the logarithm
omitted is still valid in the region of critical fluctuations\footnote{%
The logarithm in (\ref{Feva}) is essential for the case $D=2.$ This case
will be discussed later.}

\begin{equation}
F_{(D)}\sim -\xi ^{-D}(\epsilon ),  \label{frcr}
\end{equation}
the coherence length is a power function of the reduced temperature: $\xi
(\epsilon )\sim \epsilon ^{-\nu }$. The corresponding formula for the
fluctuation heat capacity can be rewritten as

\begin{equation}
\delta C\sim -{\frac{{\ \partial ^{2}F}}{{\ \partial \epsilon ^{2}}}}\sim
\epsilon ^{D\nu -2}.  \label{cscale}
\end{equation}

As was demonstrated in the Introduction, the GL functional approach, where
the temperature dependence of $\xi (T)$ is determined only by the diffusion
of the electrons forming Cooper pair, $\xi (\epsilon )\sim \epsilon ^{-1/2}$
and $\delta C\sim \epsilon ^{-1/2}.$ These results are valid for the GL
region ( $|\epsilon |\gtrsim Gi)$ only, where the interaction between
fluctuations can be neglected. In the immediate vicinity of the transition
(so-called critical region), where $|\epsilon |\lesssim Gi,$ the interaction
of fluctuations becomes essential. Here fluctuation Cooper pairs affect the
coherence length themselves, changing the temperature dependencies of $\xi
(\epsilon )$ and $\delta C(\epsilon ).$ In order to find the heat capacity
temperature dependence in the critical region one would have to calculate
the functional integral with the fourth order term, accounting for the
fluctuation interaction, as was done for $0D$ case. For the $3D$ case up to
now it is only known how to calculate a Gaussian type functional integral.
This was done above when, for the GL region, we omitted the fourth order
term in the free energy functional (\ref{Func}).

\begin{figure}[tbp]
\epsfxsize=10cm
\centerline {\epsfbox{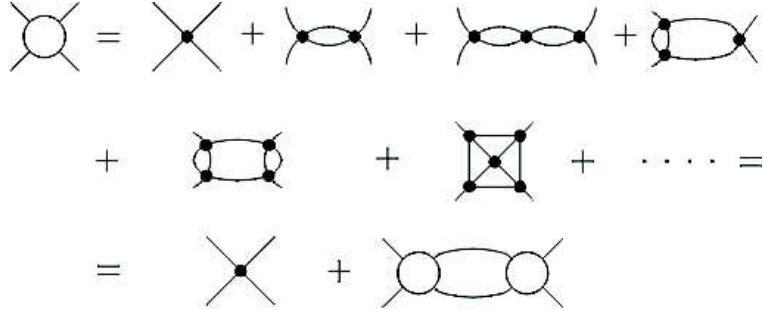}}
\caption{Examples of diagrams for the fluctuation contribution to $b$.}
\label{RG}
\end{figure}

The first evident step in order to include in consideration of the critical
region would be to develop a perturbation series in $b$ \footnote{%
Let us mention that this series is an asymptotic one, i.e. it does not
converge even for small $b$. One can easily see this for small negative $b,$
when the integral for the partition function evidently diverges. This is
also confirmed by the exact $0D$ solution (\ref{stat0}).}. Any term in this
series has the form of a Gaussian integral and can be represented by a
diagram, where the solid lines correspond to the correlators $\left\langle
\Psi (r)\Psi ^{\ast }(r^{^{\prime }})\right\rangle $. The ''interactions '' $%
b$ are represented by the points where four correlator lines intersect (see
Fig. \ref{RG}).

This series can be written as

\[
C\sim \sum_{n=0}c_{n}\left( \frac{{\ }Gi_{(D)}}{{\ \epsilon }}\right) ^{%
\frac{4-D}{2}n}. 
\]
For $\epsilon \gtrsim Gi$ it is enough to keep\ only the first two terms to
reproduce the perturbational result obtained above. For $\epsilon \lesssim
Gi $ all terms have to be summed. It turns out that the coefficients $c_{n}$
can be calculated for the space dimensionality $D\rightarrow 4$ only. In
this case the complex diagrams from Fig.~\ref{RG} (like the diagram similar
to an envelope) are small by the parameter $\varepsilon =4-D$ and in order
to calculate $c_{n}$ it is sufficient to sum the relatively simple
''parquet'' type diagrammatic series. Such a summation results in the
substitution of the ''bare'' vertex $b$ by some effective interaction $%
\widetilde{b}$ which diminishes and tends to zero when the temperature
approaches the transition point. Such\ a method was originally worked out in
quantum field theory \cite{LaKh54,DSTm57,S56}. For the problem of a phase
transition such a summation was first accomplished in \cite{LKh}.

Instead of a direct summation of the diagrams it is more convenient and
physically obvious to use the method of the renormalization group\textbf{. }%
In the case of quantum field theory this method was known long ago \cite{Gel
Man,SP51}. For phase transition theory it was proposed in \cite{LKh,Di
Castro} but the most simple and evident formulation was presented by Wilson 
\cite{Wilson}. The idea of the renormalization group method consists in
separating the functional integration over ''fast'' ($\psi _{|{k|>}\Lambda }$%
) and ''slow''($\psi _{|{k|<}\Lambda }$) fluctuation modes. If the cut off $%
\Lambda $ is large enough the fast mode contribution is small and the
integration over them is Gaussian. After the first integration over fast
modes the functional obtained depends on the slow ones only. They can, in
their turn, be divided on slow ($|k|<\Lambda _{1}$) and fast ($\Lambda
_{1}<|k|<\Lambda $), and the procedure can be repeated. Moving step by step
ahead in this way one can calculate the complete partition function.

As an example of the first step of \ renormalization, the partition function
calculation below $T_{c}$ can be recalled. There we separated the order
parameter into the space-independent part $\widetilde{\Psi }$ (''slow''
mode) and the fluctuation part $\psi ({r})$ (''fast'' mode) which was
believed to be small in magnitude. Being in the GL region it was enough to
average over the fast variables just once, while in the critical region the
renormalization procedure requires subsequent approximations.

The cornerstone of the method consists in the fact that in the critical
region at any subsequent step the free energy functional has the same form.
For $D$ close to $4$ this form coincides with the initial free energy GL
functional but with the coefficients $a_{\Lambda }$ and $b_{\Lambda }$
depending on $\Lambda .$ We will perform these calculations by the method of
mathematical induction. Let us suppose that after the $(n-1)-$st step the
free energy functional has the form:

\begin{eqnarray}
\mathcal{F}[\widetilde{\Psi }_{\Lambda _{n-1}}] &=&F_{N,\Lambda _{n-1}}+\int
dV\left\{ a_{\Lambda _{n-1}}|\widetilde{\Psi }_{\Lambda _{n-1}}|^{2}+\right.
\label{finv} \\
&&+\left. \frac{b_{\Lambda _{n-1}}}{2}|\widetilde{\Psi }_{\Lambda
_{n-1}}|^{4}+\frac{1}{4m}|\mathbf{\nabla }\widetilde{\Psi }_{\Lambda
_{n-1}}|^{2}\right\} .  \nonumber
\end{eqnarray}
Writing $\widetilde{\Psi }_{\Lambda _{n-1}}$ in the form $\widetilde{\Psi }%
_{\Lambda _{n-1}}=\widetilde{\Psi }_{\Lambda _{n}}+\psi _{\Lambda _{n}}$ and
choosing $\Lambda _{n}$ close enough to $\Lambda _{n-1}$ it is possible to
make $\psi _{\Lambda _{n}}$ so small, that one can restrict the functional
to the quadratic terms in $\psi _{\Lambda _{n}}$ only and perform the
Gaussian integration in complete analogy with (\ref{zbelow}). The important
property of the spaces with dimensionalities close to $4$ is the possibility
to choose $\Lambda _{n}\ll \Lambda _{n-1}$ and still to have $\psi _{\Lambda
_{n}}\ll \widetilde{\Psi }_{\Lambda _{n}}.$ In this case $\widetilde{\Psi }%
_{\Lambda _{n}}$can be taken as coordinate independent, and one can use the
result directly following from (\ref{zbelow}):

\begin{eqnarray}
\mathcal{F}[\widetilde{\Psi }_{\Lambda _{n}}] &=&F_{N,\Lambda _{n-1}}+\int
dV\left\{ a_{\Lambda _{n}}|\widetilde{\Psi }_{\Lambda _{n}}|^{2}+\frac{%
b_{\Lambda _{n}}}{2}|\widetilde{\Psi }_{\Lambda _{n}}|^{4}+\frac{1}{4m}|%
\mathbf{\nabla }\widetilde{\Psi }_{\Lambda _{n}}|^{2}\right\}  \nonumber \\
&&-\frac{T}{2}\sum_{\Lambda _{n}<|\mathbf{k|<\Lambda }_{n-1}}\left\{ \ln {%
\frac{{\ \pi }T_{c}}{{\ (3b_{\Lambda _{n}}|\widetilde{\Psi }_{\Lambda
_{n}}|^{2}+a_{\Lambda _{n}}+\frac{\mathbf{k}^{2}}{4m})}}+}\right.  \nonumber
\\
&&\left. \ln {\frac{\pi T_{c}{\ }}{b_{\Lambda _{n}}|\widetilde{\Psi }%
_{\Lambda _{n}}|^{2}+a_{\Lambda _{n}}+\frac{\mathbf{k}^{2}}{4m}}}\right\} .
\label{induc}
\end{eqnarray}
Expanding the last term in (\ref{induc}) in a series in ${\widetilde{\Psi }%
_{\Lambda _{n}}}$ one can get for $\mathcal{F}[\widetilde{\Psi }_{\Lambda
_{n}}]$ the same expression (\ref{finv}) with the substitution of ${\Lambda }%
_{n-1}\rightarrow {\Lambda }_{n}.$ From (\ref{induc}) follows that

\[
F_{n,\Lambda _{n}}=F_{n,\Lambda _{n-1}}-T\sum_{\Lambda _{n}<|\mathbf{%
k|<\Lambda }_{n-1}}\ln {\frac{{\ \pi }T_{c}}{{\ (a_{\Lambda _{n}}+\frac{%
\mathbf{k}^{2}}{4m})}},} 
\]

\begin{equation}
a_{\Lambda _{n}}=a_{\Lambda _{n-1}}+2T\sum_{\Lambda _{n}<|\mathbf{k|<\Lambda 
}_{n-1}}{\frac{{b_{\Lambda _{n}}}}{{\ (a_{\Lambda _{n}}+\frac{\mathbf{k}^{2}%
}{4m})}}}  \label{recur}
\end{equation}

\[
b_{\Lambda _{n}}=b_{\Lambda _{n-1}}-5T\sum_{\Lambda _{n}<|\mathbf{k|<\Lambda 
}_{n-1}}{\frac{{b_{\Lambda _{n}}^{2}}}{{\ (a_{\Lambda _{n}}+\frac{\mathbf{k}%
^{2}}{4m}})^{2}}.} 
\]
Passing to a continuous variable $\Lambda _{n}\rightarrow \Lambda $ one can
rewrite these recursion equations as the set of differential equations:

\begin{equation}
\frac{\partial F(\Lambda )}{\partial \Lambda }=-T\mu _{D}\Lambda ^{D-1}\ln {%
\frac{\pi T_{c}}{{\ (a(\Lambda )+\frac{\Lambda ^{2}}{4m}})},}  \label{freeen}
\end{equation}

\begin{equation}
\left\{ 
\begin{array}{c}
\frac{\partial a(\Lambda )}{\partial \Lambda }=-2T\mu _{D}{\frac{b(\Lambda
)\Lambda ^{D-1}}{{\ (a(\Lambda )+\frac{\Lambda ^{2}}{4m}})}} \\ 
\frac{\partial b(\Lambda )}{\partial \Lambda }=5T\mu _{D}{\frac{%
b^{2}(\Lambda )\Lambda ^{D-1}}{{\ (a(\Lambda )+\frac{\Lambda ^{2}}{4m}})^{2}}%
}
\end{array}
\right. .  \label{sysdif}
\end{equation}
These renormalization group equations are evidently valid for small enough $%
\Lambda $ only, where the transition from discrete to continuous variables
is justified. This means at least

\begin{equation}
\Lambda ^{2}/4m\ll T_{c0}Gi_{(D)}.  \label{remote}
\end{equation}
in order to move away from the first approximation.

Let us recall that in the framework of the Landau theory of phase
transitions the coefficient $a(T_{c0})=0$ at the transition point and this
can be considered as the MFA definition of the critical temperature $T_{c0}.$
The same statement for the function $a=a(\Lambda )$ in the framework of the
renormalization group method can be written as the $a(T_{c0},\Lambda \sim
\xi ^{-1})=0.$ With the decrease of $\Lambda $ the effect of critical
fluctuations more and more is taken into account and the renormalized value
of the critical temperature decreases, being defined by the equation: $%
a(T_{c}(\Lambda ),\Lambda )=0.$ Finally, after the application of the
complete renormalization procedure, one can define the real critical
temperature $T_{c},$ shifted down with respect to $T_{c0}$ due to the effect
of fluctuations, from the equation:

\begin{equation}
a(T_{c},\Lambda =0)=0.  \label{critdef}
\end{equation}
It is easy to find this shift in the first approximation. Indeed, let us
integrate the first equation of (\ref{sysdif}) over $\Lambda $ in limits $%
[0,\xi ^{-1}].$ The main contribution to the integral will be determined by
the region where ${a(\Lambda )\ll \frac{\Lambda ^{2}}{4m}.}$ \ Being far
from the critical point one can assume that the coefficient $b=const$ and
then

\[
a(\xi ^{-1})=\alpha \delta T_{c}=\int_{a(0)}^{a(\xi ^{-1})}da=-8mT\mu
_{D}b\int_{0}^{1/\xi }\Lambda ^{D-3}d\Lambda . 
\]
For the $3D$ case this gives the shift of the critical temperature $\delta
T_{c}$ due to fluctuations \footnote{%
In the same way we can analyze the shift of the critical temperature in $2D$%
\ case too and obtain:
\par
\begin{equation}
\frac{\delta T_{c}^{(2)}}{T_{c}}=-2Gi_{(2)}\ln \frac{1}{4Gi_{(2)}}.
\label{2shift}
\end{equation}
\ 
\par
As we will show below both $3D$ and $2D$ results for $\delta T_{c}$ coinside
with those obtained by the analysis of the effect of fluctuations on
superconducting density in the perturbation approach.}

\begin{equation}
\frac{\delta T_{c}^{(3)}}{T_{c}}\sim -\frac{2mb}{\pi \alpha \xi }=-\frac{b}{%
2\pi T_{c}\alpha ^{2}}\frac{1}{\xi ^{3}}=-\frac{7\zeta (3)}{16\pi ^{3}\nu
T_{c}\xi ^{3}}=-\frac{8}{\pi }\sqrt{Gi_{(3)}}.  \label{rgshift}
\end{equation}

Let us come back to study of properties of the system of equations (\ref
{sysdif}). One can find its partial solution at $T$ $=T_{c}$ in the form:

\begin{eqnarray}
a(T_{c},\Lambda ) &=&\frac{4-D}{4m[5+(4-D)]}\Lambda ^{2}  \label{difsol} \\
b(T_{c},\Lambda ) &=&\frac{5}{16m^{2}T\mu _{D}}\frac{4-D}{[5+(4-D)]^{2}}%
\Lambda ^{4-D}.  \nonumber
\end{eqnarray}
These power solutions are correct in the domain of validity of the system (%
\ref{sysdif}) itself, i.e. for small enough $\Lambda $\ defined by the
condition (\ref{remote}). Nevertheless, in a space of dimensionality close
to $4$ ($D=4-\varepsilon ,\varepsilon \ll 1$) it is possible to extend their
validity up to the GL region and to observe their crossover to the GL
results:$\ a(T_{c})=0;$ $b(T_{c})=b_{0}=const$. \ Indeed, in this case, due
to proportionality of $a(\Lambda )$ to $\varepsilon \rightarrow 0,$ \ one
can omit it in the denominators of the system (\ref{sysdif}) and to write
down the solution for $b(\Lambda )$ in the form

\begin{equation}
b^{-1}(T_{c},\Lambda )=b_{0}^{-1}+\frac{80m^{2}}{(4-D)}T\mu _{D}(\Lambda
^{D-4}-\xi ^{4-D}).  \label{bresh}
\end{equation}
We have chosen the constant of integration as $b_{0}^{-1}-$ $\frac{80m^{2}}{%
(4-D)}T\mu _{D}\xi ^{4-D}$ in order to match the renormalization group and
GL solutions at the value of $\Lambda =$ $\Lambda _{\max }\sim \xi ^{-1}.$

Now let us pass to study of the function $a(T,\Lambda )$ for the same
interesting case of space dimensionality $D\rightarrow 4$ for temperatures
slightly different ( but still close enough) from $T_{c},$ where one can
write 
\[
a(T,\Lambda )=a(T_{c},\Lambda )+\alpha (T_{c},\Lambda )T_{c}\epsilon . 
\]
The first term on the right hand side is determined by Eq.(\ref{difsol}). In
order to determine $\alpha (T_{c},\Lambda )$ let us expand the first
equation in (\ref{sysdif}) in terms of $\epsilon $

\begin{equation}
\frac{\partial \alpha (T_{c},\Lambda )}{\partial \Lambda }=2T\mu _{D}\frac{%
b(T_{c},\Lambda )\Lambda ^{D-1}}{\left( a(T_{c},\Lambda )+{\frac{\Lambda ^{2}%
}{4m}}\right) ^{2}}\alpha (T_{c},\Lambda ).  \label{alphacri}
\end{equation}
For

\begin{equation}
\Lambda ^{2}/4m\gtrsim \alpha (T_{c},\Lambda )T_{c}\epsilon
\label{lambdacond}
\end{equation}
we can again use the solution (\ref{bresh}) for $b(T_{c},\Lambda )$ and omit 
$a(T_{c},\Lambda )$ in the denominator of (\ref{alphacri}). The constant of
integration, appearing in the process of solution of (\ref{alphacri}), is
chosen in accordance with the condition that for $\Lambda =$ $\Lambda _{\max
}\sim \xi ^{-1}$ we match $\alpha (T_{c},\Lambda )=\alpha (T_{c0},\xi
^{-1})=\alpha _{0}$ with the GL theory:

\begin{equation}
\alpha (T_{c},\Lambda )=\alpha _{0}\left[ 1+\frac{80m^{2}b_{0}}{(4-D)}T\mu
_{D}(\Lambda ^{D-4}-\xi ^{4-D})\right] ^{-2/5}.  \label{alp}
\end{equation}

The condition (\ref{lambdacond}) can be written as $\Lambda \gtrsim \xi
^{-1}(T),$ where $\xi (T)$ is the generalized coherence length, determined
by the equation:

\[
\xi ^{-2}(T)=4m\alpha \left( T_{c},\xi ^{-1}(T)\right) T_{c}\epsilon . 
\]
Such a definition is valid at any temperature. For example, far enough from
the critical point, in the GL region, $\alpha \left( T_{c},\xi
^{-1}(T)\right) =\alpha _{0}$ and one reproduces the result (\ref{xiGL}).
Vice versa, in the critical region the main contribution on the right hand
side of the Eq. (\ref{alp}) results from the second term containing $\Lambda
^{D-4}$ so, putting $\xi ^{-1}(T)=\Lambda ,$ one can rewrite the
self-consistent equation for $\xi (T)$ and get

\begin{equation}
\xi (T)=(4m)^{(1-D)/2}\frac{4-D}{20b_{0}T\mu _{D}\sqrt{T_{c}\alpha _{0}}}%
\epsilon ^{-\nu },  \label{xieff}
\end{equation}
where $2\nu =[1-(4-D)/5]^{-1}.$ As was already mentioned, strictly speaking
this result was carried out for $\ \varepsilon =4-D\ll 1$, so it is
confident up to the first in $\varepsilon $ expansion only: $\nu
=1/2+\varepsilon /10.$ Nevertheless extending it to $\varepsilon =1$ ($D=3)$
one can obtain $\nu _{3}=3/5.$

Let us pass to the calculation of the critical exponent of the heat capacity
in the immediate vicinity of the transition. For this purpose one can
calculate the second derivative of equation (\ref{freeen}) with respect to $%
\epsilon $:

\begin{equation}
\frac{\partial C(\Lambda )}{\partial \Lambda }=T^{2}\mu _{D}\frac{\Lambda
^{D-1}\alpha ^{2}(\Lambda )}{(\alpha {(\Lambda )T}_{c}\epsilon {+\frac{%
\Lambda ^{2}}{4m}})^{2}}  \label{cdif}
\end{equation}
The heat capacity renormalized by fluctuations has the value $C(\Lambda =0)$
which is the result of integration over all fluctuation degrees of freedom.
Carrying out the integration of (\ref{cdif}) over all $\Lambda $ $\lesssim $ 
$\xi ^{-1}$ one can divide the domain of integration on the right hand side
in two: $\Lambda $ $\lesssim $ $\xi ^{-1}(T)$ and $\xi ^{-1}(T)\lesssim
\Lambda $ $\lesssim $ $\xi ^{-1}.$ In the calculation of the integral over
the region $\xi >\Lambda \gtrsim \xi ^{-1}(T)$ the inequality $\alpha {%
(\Lambda )T}_{c}\epsilon {\ll \frac{\Lambda ^{2}}{4m}}$ holds, and the
function $\alpha {(\Lambda )}$ can be omitted in the denominator. In the
numerator of (\ref{cdif}) one can use for $\alpha {(\Lambda )}$ the solution
(\ref{alp}). In the region $\Lambda \lesssim \xi ^{-1}(T)$ one has to use
the partial solution (\ref{difsol}) for $\alpha {(\Lambda )}$ and can find
that the contribution of this domain has the same singularity as that from
the region $\Lambda \gtrsim \xi ^{-1}(T),$ but with a coefficient
proportional to $(4-D)^{2}=\varepsilon ^{2}$, hence negligible in our
approximation. The result is:

\begin{equation}
C(\Lambda =0)=\alpha _{0}^{2}[(4mT)^{2}\mu _{D}]^{\frac{1}{5}}\left[ \frac{4%
}{5}\frac{(4-D)}{b_{0}}\right] ^{4/5}\frac{5}{4-D}\xi ^{\frac{4-D}{5}}(T). 
\nonumber
\end{equation}
Substituting the expression for $\xi (T)$ one can finally find

\begin{equation}
C=2^{12/5}\alpha _{0}^{2}[5\mu _{D}\frac{m^{2}T^{2}}{b_{0}^{4}(4-D)}]^{\frac{%
1}{5}}\epsilon ^{-\alpha },  \label{ccrit}
\end{equation}
confirming the validity of the scaling hypothesis and the relation (\ref
{cscale}). The critical exponent in (\ref{ccrit}) is

\[
\alpha =\frac{(4-D)}{10[1-(4-D)/5]}\approx \varepsilon /10. 
\]

One can see that generally speaking the critical exponents $\nu $ and $%
\alpha $ appear in the form of series in powers of $\varepsilon .$ More
cumbersome calculations permit finding the next approximations for them in $%
\varepsilon =4-D.$ Nevertheless it is worth mentioning that even the first
approximation, giving $\nu _{3}=3/5$ and $\alpha _{3}=1/10$ for $\varepsilon
=1,$ is already weakly affected by the following steps of the expansion in
powers of $\varepsilon $ \cite{PokPat}.

One can notice that the exercise performed in this Section has more academic
than practical character. Indeed, the results obtained turn out applicable
to the analysis of the critical region of a $3D$ superconductor only if $Gi$
is so small that the theoretical predictions are hardly experimentally
observable. \ Nevertheless we demonstrated the RG method which helps to see
the complete picture of the fluctuations manifestation in the vicinity of
the\ critical temperature.

\subsection{ Fluctuation diamagnetism}

\subsubsection{Qualitative preliminaries.}

In this Section we discuss the effect of fluctuations on the magnetization
and the susceptibility of a superconductor above the transition temperature.
Being the precursor effect for the Meissner diamagnetism, the fluctuation
induced magnetic susceptibility has to be a small correction with respect to
the diamagnetism of a superconductor but it can be comparable to or even
exceed the value of the normal metal diamagnetic or paramagnetic
susceptibility and can be easily measured experimentally. As was already
mentioned in the Introduction the temperature dependence of the fluctuation
induced diamagnetic susceptibility can be qualitatively analyzed on the
basis of the Langevin formula, but some precautions in the case of low
dimensional samples have to be made.

As regards the $3D$ case we would like just to mention here that Exp.(\ref
{chifl32}), presented in terms of $\xi (T),$ has a wider region of
applicability than the GL one. Namely, the scaling arguments are valid for
diamagnetic susceptibility too and one can write the general relation

\begin{equation}
\chi _{(3)}\sim -e^{2}T\xi (T)\sim -\chi _{P}\epsilon ^{-1/2}\left\{ 
\begin{tabular}{l}
$1,\;\epsilon \gtrsim Gi$ \\ 
$\left( \frac{\epsilon }{Gi}\right) ^{1/2-\nu },\;\epsilon \lesssim Gi$%
\end{tabular}
\right. ,  \label{3dkhi}
\end{equation}
which is valid in the region of critical fluctuations in the immediate
vicinity of the transition temperature too. Here, in order to define the
scale of fluctuation effects, we have introduced the Pauli paramagnetic
susceptibility $\chi _{P}=e^{2}v_{F}/4\pi ^{2}.$ Moreover, the Langevin
formula permits us to extend the estimation of the fluctuation diamagnetic
effect to the other side beyond the GL region: to high temperatures $T\gg
T_{c}$. The coherence length far from the transition becomes a slow function
of temperature. In a clean superconductor, far from $T_{c},$ $\xi (T)\sim
v_{F}/T,$ so one can write

\begin{equation}
\chi _{(3c)}(T\gg T_{c})\sim -e^{2}T\xi (T)\sim -\chi _{P}  \label{3dfl}
\end{equation}
and see that the fluctuation diamagnetism turns out to be of the order of
the Pauli paramagnetism even far from the transition. More precise
microscopic calculations of $\chi _{(3)}(T\gg T_{c})$ lead to the appearance
of $\ln ^{2}(T/T_{c})$ in the denominator of (\ref{3dfl}).

In the $2D$ case Exp.(\ref{chifl32}) is applicable for the estimation of $%
\chi _{(2)}$ in the case when the magnetic field is applied perpendicular to
the plane, permitting $2D$\ rotations of fluctuation Cooper pairs in it:

\begin{equation}
\chi _{(2c)}(T)\sim e^{2}\frac{n}{m}<R^{2}>\sim e^{2}T\xi ^{2}(T)\sim -\chi
_{P}\frac{E_{F}}{T-T_{c}}.  \label{khi2}
\end{equation}
This result is valid for a wide range of temperatures and can exceed the
Pauli paramagnetism by factor $\frac{E_{F}}{T}$ even far from the critical
point (we consider the clean case here).

For a thin film ($d\ll \xi (T))$ perpendicular to the magnetic field the
fluctuation Cooper pairs behave like effective $2D$ rotators, and the
formula (\ref{chifl32}) still can be used, though one has to take into
account that the susceptibility in this case is calculated per unit square
of the film. So for the realistic case (from the experimental point of view)
of the dirty film, one has just use in (\ref{khi2})\ the expression (\ref
{xid}) for the coherence length:

\begin{equation}
\chi _{(2d)}\sim \frac{e^{2}T}{d}\xi ^{2}(T)\sim -\chi _{P}\left( \frac{l}{d}%
\right) \frac{T_{c}}{T-T_{c}}.  \label{filmper}
\end{equation}

Let us discuss now the important case of a layered superconductor (for
example, a high temperature superconductor). It is usually supposed that the
electrons move freely in conducting planes separated by a distance $s$.
Their motion in the perpendicular direction has a tunneling character, with
effective energy\textit{\ }$\mathit{J}$. The related velocity and coherence
length can be estimated as $v_{z}=\partial E(\mathbf{p})/\partial p_{\perp
}\sim \mathit{J}/p_{\perp }\sim s\mathit{J}$ and $\xi _{z,(c)}\sim s\mathit{J%
}/T$ for clean case. In dirty case the anisotropy can be taken into account
in the spirit of \ formula (\ref{xid}) yielding $\xi _{z,(d)}\sim \sqrt{%
\mathit{D}_{\perp }/T}\sim s\mathit{J}\sqrt{\tau /T}.$

We start from the case of a weak magnetic field applied perpendicular to
layers. The effective area of a rotating fluctuation pair is $\xi
_{x}(\epsilon )\xi _{y}(\epsilon ).$\ The density of Cooper pairs in the
conducting layers (\ref{n/m}) has to be modified for the anisotropic case.
Its isotropic $3D$\ value is proportional to $1/\xi (\epsilon ),$\ that now
has to be read as $\sim 1/\sqrt{\xi _{x}(\epsilon )\xi _{y}(\epsilon )}.$\ \
The anisotropy of the electron motion leads to a concentration of
fluctuation Cooper pairs in\ the conducting layers and hence, to an
effective increase of the Cooper pairs density of $\sqrt{\xi _{x}(\epsilon
)\xi _{y}(\epsilon )}/\xi _{z}(\epsilon )$\ times its isotropic value. This
increase is saturated when $\xi _{z}(\epsilon )$\ reaches the interlayer
distance $s$, so finally the anisotropy factor appears in the form $\sqrt{%
\xi _{x}(\epsilon )\xi _{y}(\epsilon )}/\max \{s,\xi _{z}(\epsilon )\}$\ and
the square root in its numerator is removed in the Langevin formula (\ref
{chifl32}), rewritten for this case

\begin{equation}
\chi _{(layer,\perp )}(\epsilon ,H\rightarrow 0)\sim -e^{2}T\frac{\xi
_{x}(\epsilon )\xi _{y}(\epsilon )}{\max \{s,\xi _{z}(\epsilon )\}}.
\label{khipe}
\end{equation}
The existence of a crossover between the $2D$ and $3D$ temperature regimes
in this formula is evident: as the temperature tends to $T_{c}$ the
diamagnetic susceptibility temperature dependence changes from $1/\epsilon $
to $1/\sqrt{\epsilon }$ . This happens when$\ $the reduced temperature
reaches its crossover value $\epsilon _{cr}=r$ $\left( \xi _{z}(\epsilon
_{cr})\sim s\right) $. The anisotropy parameter

\begin{equation}
r=\frac{4\xi _{z}^{2}(0)}{s^{2}}=\frac{\mathit{J}^{2}}{T}\left\{ 
\begin{array}{c}
\frac{\pi \ \tau }{4},\;T\tau \ll 1 \\ 
\frac{7\zeta (3)}{8\pi ^{2}T},\;T\tau \gg 1
\end{array}
\right.  \label{r1}
\end{equation}
plays an important role in the theory of layered superconductors \footnote{%
We use here a definition of $r$ following from microscopic theory (see
Section 6).}.

It is interesting to note that this intrinsic crossover, related to the
spectrum anisotropy, has an opposite character to the geometric crossover
which happens in thick enough films when $\xi (T)\ $reaches $d$ . In the
latter case the characteristic $3D$ $\;1/\sqrt{\epsilon }\;-$ dependence
taking place far enough from $T_{c}$ (where $\xi (T)\ll d)$, is changed to
the\ $2D\;\;1/\epsilon $ law (see (\ref{filmper})) in the immediate vicinity
of transition ( where $\xi (T)\gg d$) \cite{VY91}. It is worth mentioning
that in a strongly anisotropic layered superconductor the
fluctuation-induced susceptibility may considerably exceed the normal metal
dia- and paramagnetic effects even relatively far from $T_{c}$ \cite
{TT72,PBS77}.

Let us consider a magnetic field applied along the layers. First it is
necessary to mention that the fluctuation diamagnetic effect disappears in
the limit $\mathit{J}\sim \xi _{z}\rightarrow 0.$ Indeed, for the formation
of a circulating current it is necessary to tunnel twice, so

\[
\chi _{(layer,\parallel )}\sim -e^{2}T\frac{\ \xi _{z}^{2}}{\max \{s,\xi
_{z}(T)\}}\sim -\chi _{P}(\frac{s\mathit{J}}{v_{F}})\frac{\mathit{J}/T}{%
\sqrt{\epsilon }\max \{\sqrt{\epsilon },\mathit{J}/T\}}. 
\]

In the general case of an anisotropic superconductor, choosing the $z$ axis
along the direction of magnetic field $H$, the following extrapolation of
the results obtained may be written

\begin{equation}
\chi \sim -e^{2}T\frac{\xi _{x}^{2}(\epsilon )\xi _{y}^{2}(\epsilon )}{\max
\{\frak{a},\xi _{x}(\epsilon )\}\max \{\frak{b},\xi _{y}(\epsilon )\}\max \{%
\frak{s},\xi _{z}(\epsilon )\}}.  \label{anis}
\end{equation}
This general formula is useful for the analysis of the fluctuation
diamagnetism of anisotropic superconductors or samples of some specific
shape: granular, quasi-$1D,$ quasi-$2D,$ and $3D.$ It is also applicable to
the case of a thin film ($d\ll \xi _{z}(\epsilon ))$ placed perpendicular to
the magnetic field: it is enough to replace $\xi _{z}(\epsilon )$ by $d$ in (%
\ref{anis}). Nevertheless the formula (\ref{anis}) cannot be applied to the
cases of thin films in parallel fields, wires and granules. In those cases
the Langevin formula (\ref{chifl32}) can still be used with the replacement
of $\left\langle R^{2}\right\rangle \rightarrow d^{2}:$

\[
\chi _{(D)}\sim -\chi _{P}\left( \frac{T}{v_{F}}\right) \xi
^{2-D}d^{D-1}\sim \epsilon ^{D/2-1}. 
\]
The magnetic field dependence of the fluctuation part of \ free energy in
these cases is reduced only to account for the quadratic shift of the
critical temperature versus magnetic field.

For $3D$ systems or in the case of a film in a perpendicular magnetic field
the critical temperature depends on $H$ linearly, while the magnetic field
dependent part of the free energy for $H\ll H_{c2}^{\ast }(-\epsilon )$ (the
line $H_{c2}^{\ast }(-\epsilon )$ is mirror-symmetric to the $%
H_{c2}(\epsilon )$\ with respect to $y$-axis passing through $T=T_{c}$) is
proportional to $H^{2}$. This is why the magnetic susceptibility is
determined by Eq.(\ref{anis}) for weak enough magnetic fields $H\ll \Phi
_{0}/[\xi _{x}(\epsilon )\xi _{y}(\epsilon )]=H_{c2}(\epsilon )\ll $ $%
H_{c2}(0)$ only. In the vicinity of $T_{c}$ these fields are small enough.

\subsubsection{Zero-dimensional diamagnetic susceptibility.}

For quantitative analysis of the fluctuation diamagnetism we start by
writing down the GL functional for the free energy (see Exp.(\ref{Func})) in
the presence of the magnetic field

\begin{eqnarray}
\mathcal{F}[\Psi (\mathbf{r})] &=&F_{n}+\int dV\left\{ a|\Psi (\mathbf{r}%
)|^{2}+\frac{b}{2}|\Psi (\mathbf{r})|^{4}+\frac{1}{4m}|\left( -i\mathbf{%
\nabla -}2e\mathbf{A}\right) \Psi (\mathbf{r})|^{2}+\right.  \nonumber \\
&&\left. +\frac{\mathbf{B}^{2}}{8\pi }-\frac{\mathbf{H}\cdot \mathbf{B}}{%
4\pi }\right\} .  \label{GLMF}
\end{eqnarray}
where $\mathbf{A}$ is vector potential. As long as fluctuation effects are
comparatively small, the average magnetic field in the metal $\mathbf{B}$
may be assumed to be equal to the external field $\mathbf{H.}$ Thus we omit
the last two terms in (\ref{GLMF}) (see later on).

The fluctuation contribution to the diamagnetic susceptibility in the
simplest case of a ''zero-dimensional'' superconductor (spherical
superconducting granule of diameter $d\ll \xi (\epsilon )$) was considered
by V.Shmidt \cite{VSchmd68}. In this case the order parameter does not
depend on the space variables and the free energy can be calculated exactly
for all temperatures including the critical region in the same way as was
done for the case of the heat capacity in\ the absence of a magnetic field.
Formally the effect of a magnetic field in this case is reduced to the
renormalization of the coefficient $a,$ or, in other words, to the
suppression of the critical temperature. This is why one can use the same
formula (\ref{stat0}) for the partition function with the critical
temperature $T_{c}$ shifted by magnetic field as\footnote{%
Let us stress the difference between the $H^{2}$ shift of the critical
temperature for a zero-dimensional granule and the linear shift in the case
of bulk material.}:

\begin{equation}
T_{c}(H)=T_{c}(0)(1-\frac{4\pi ^{2}\xi ^{2}}{\Phi _{0}^{2}}<\mathbf{A}^{2}>).
\label{tchqv}
\end{equation}
Here $\Phi _{0}=\frac{\pi }{e}$ is the magnetic flux quantum and $<\mathbf{%
.....}>$\ means the averaging over the sample volume.

Such a trivial dependence of the properties of $0D$ samples on magnetic
field immediately allows one to understand its effect on the heat capacity
of a granular sample. Indeed, with the growth of the field the temperature
dependence of the heat capacity presented in Fig.~\ref{0Dhc} just moves in
the direction of lower temperatures.

In the GL region $Gi_{(0)}$ $\lesssim \epsilon $ one can write the
asymptotic expression (\ref{0Df}) for the free energy:

\[
F_{(0)}(\epsilon ,H)=-T\ln \frac{\pi }{\alpha (\epsilon +\frac{4\pi ^{2}\xi
^{2}}{\Phi _{0}^{2}}<\mathbf{A}^{2}>)}. 
\]
In the case of a spherical particle the relation $<\mathbf{A}^{2}>=\frac{1}{%
10}H^{2}d^{2}$ can be used in full analogy with the calculation of the
moment of inertia of a solid sphere. In this way an expression for the $0D$
fluctuation magnetization valid for all fields $H\ll H_{c2}(0)$ can be found:

\begin{equation}
M_{(0)}(\epsilon ,H)=-\frac{\partial F_{(0)}(\epsilon ,H)}{\partial H}=-T%
\frac{\frac{2\pi ^{2}\xi ^{2}}{5\Phi _{0}^{2}}d^{2}}{(\epsilon +\frac{\pi
^{2}\xi ^{2}}{5\Phi _{0}^{2}}H^{2}d^{2})}H.  \label{M0ex}
\end{equation}
One can see that the fluctuation magnetization turns out to be negative and
linear up to some crossover field, which can be called the temperature
dependent upper critical field of the granule $H_{c2(0)}(\epsilon )\sim 
\frac{\Phi _{0}}{d\xi (\epsilon )}=\frac{\xi }{d}H_{c2}(0)\sqrt{\epsilon }$
at which it reaches a minimum. At higher fields $H_{c2(0)}(\epsilon
)\lesssim H\ll H_{c2}(0)$ the fluctuation magnetization of the $0D$\ granule
decreases as $1/H.$ In the weak field region $H\ll H_{c2(0)}(\epsilon )$ the
diamagnetic susceptibility is:

\[
\chi _{(0)}(\epsilon ,H)=-\frac{12\pi T\xi _{0}^{2}}{5\Phi _{0}^{2}d}\frac{1%
}{\epsilon }\approx -2\cdot 10^{2}\chi _{P}\left( \frac{\xi }{d}\right) 
\frac{1}{\epsilon } 
\]
which coincides with our previous estimate in its temperature dependence but
the numerical factor found is very large. Let us underline that the
temperature dependence of the $0D$ fluctuation diamagnetic susceptibility
turns out to be less singular than the $0D$ heat capacity correction: $%
\epsilon ^{-1}$ instead of $\epsilon ^{-2}$.

The expression for the fluctuation part of free energy (\ref{ffe}) is also
applicable to the cases of a wire or a film placed in a parallel field: as
was already mentioned above all its dependence on magnetic field is
manifested by the shift of the critical temperature (\ref{tchqv}). In the
case of the wire in a parallel field one has to choose the gauge of the
vector-potential $\mathbf{A=}\frac{1}{2}\mathbf{H\times r}$ yielding $%
\left\langle \mathbf{A}^{2}\right\rangle _{(wire,\parallel )}=\frac{%
H^{2}d^{2}}{32}$ (the calculation of this average is analogous to that of
the moment of inertia of a solid sphere). For a wire in a perpendicular
field, or a film in a parallel field, the gauge has to be chosen in the form 
$\mathbf{A=}(0,Hx,0)$ (to avoid the appearance of currents perpendicular to
surface). One can find $\left\langle \mathbf{A}^{2}\right\rangle
_{(wire,\perp )}=\frac{H^{2}d^{2}}{16}$ for a wire and $\left\langle \mathbf{%
A}^{2}\right\rangle _{(film,\parallel )}=\frac{H^{2}d^{2}}{12}$ for a film.

Calculating the second derivative of Eq.(\ref{ffe}) with the appropriate
magnetic field dependencies of the critical temperature one can find the
following expressions for the diamagnetic susceptibility:

\begin{equation}
\chi _{(D)}(\epsilon )=-2\pi \frac{\xi T}{v_{F}}\chi _{P}\left\{ 
\begin{tabular}{l}
$\frac{1}{\sqrt{\epsilon }},\;$wire in parallel field \\ 
$\frac{2}{\sqrt{\epsilon }},\;\;$wire in perpendicular field \\ 
$\frac{d}{3\xi }\ln \frac{1}{\epsilon },\;\;$film in parallel field
\end{tabular}
\right. .  \label{diageo}
\end{equation}

\subsubsection{GL treatment of fluctuation magnetization.}

Let us analyze quantitatively, on the basis of the GL functional, the
temperature and field dependencies of the fluctuation magnetization. We will
carry on the discussion for a layered superconductor. As was already
mentioned this system has a great practical importance because of its direct
applicability to high temperature superconductors, where the fluctuation
effects are very noticeable.\ Moreover, the general results obtained will
allow us to analyze $3D$ and $2D$ situations as limiting cases. The effects
of a magnetic field are more pronounced for perpendicular orientation, so
let us consider first this case.

The generalization of the GL functional for a layered superconductor
(Lawrence-Doniach (LD) functional \cite{LD70}) in a perpendicular magnetic
field can be written as 
\begin{eqnarray}
\mathcal{F}_{LD}\left[ \Psi \right] &=&{\sum_{l}}\int d^{2}r\left( a\left|
\Psi _{l}\right| ^{2}+\frac{b}{2}\left| \Psi _{l}\right| ^{4}+\frac{1}{4m}%
\left| \left( \mathbf{\nabla }_{\parallel }-2ie\mathbf{A}_{\parallel
}\right) |\Psi _{l}\right| ^{2}\right.  \nonumber \\
&&\left. +\mathcal{J}\left| \Psi _{l+1}-\Psi _{l}\right| ^{2}\right) ,
\label{LDF}
\end{eqnarray}
where $\Psi _{l}$\ is the order parameter of the $l-$th superconducting
layer and the phenomenological constant $\mathcal{J}$ is proportional to the
Josephson coupling between adjacent planes. The gauge with $A_{z}=0$ is
chosen in (\ref{LDF}). In the immediate vicinity of $T_{c}$\ the LD
functional is reduced to the GL one with the effective mass $M=(4\mathcal{J}%
s^{2})^{-1}$ along $c$-direction,\ where $s$\ is the inter-layer spacing.
One can relate the value of $\mathcal{J}$\emph{\ }to the coherence length
along the $c$-direction: $\mathcal{J}=2\alpha T_{c}\xi _{z}^{2}/s^{2}.$
Since we are dealing with the GL region the fourth order term in (\ref{LDF})
can be omitted.

As it is well known the Landau representation is the most appropriate for
problems related with the motion of a charged particle in a uniform magnetic
field. The fluctuation Cooper pair wave function $\phi _{nk_{z}}(\mathbf{r})$
can be written as the product of a plane wave propagating along the magnetic
field direction and\ a Landau state wave function. Let us expand the order
parameter $\Psi _{l}(\mathbf{r)}$ on the basis of these eigenfunctions:

\begin{equation}
\Psi _{l}(\mathbf{r})=\sum_{\mathbf{n},k_{z}}\Psi _{n,k_{z}}\phi _{nk_{z}}(%
\mathbf{r})\exp (ik_{z}l),  \label{psiland}
\end{equation}
where $\mathbf{n}$ is the quantum number related with the degenerate Landau
state and $k_{z}$ is the momentum component along the direction of the
magnetic field. Substituting this expansion into (\ref{LDF}) one can find
the LD free energy as a functional of the $\Psi _{n,k_{z}}$ coefficients:

\begin{equation}
\mathcal{F}_{LD}\left[ \Psi _{n,k_{z}}\right] =\sum_{\mathbf{n}%
,k_{z}}\left\{ \alpha T_{c}\epsilon +{\ }\frac{H}{2m\Phi _{0}}{\left( n+{%
\frac{1}{2}}\right) +}\mathcal{J}\left( {1-\cos (k}_{z}{s)}\right) \right\}
|\Psi _{n,k_{z}}|^{2}.  \label{GLFEmf}
\end{equation}
In complete analogy with the case of an isotropic spectrum the functional
integral over the order parameter configurations $\Psi _{n,k_{z}}$ in the
partition function can be reduced to a product of ordinary Gaussian
integrals, and the fluctuation part of the free energy in a magnetic field
takes the form:

\begin{equation}
F(\epsilon ,H)=-{\frac{2\pi {S}H}{\Phi _{0}}T}\sum_{n,k_{z}}\ln \frac{{\ \pi
T}}{{\ }\alpha T_{c}\epsilon +\ \frac{H}{2m\Phi _{0}}\left( n+{\frac{1}{2}}%
\right) +J\left( {1-\cos (k}_{z}{s)}\right) }.  \label{frla}
\end{equation}
Here the summation over the degenerate states of each Landau level was
performed (${S}$ is the sample cross-section) and results in appearance of
the number of particle states $(2\pi {HS}/\Phi _{0})$ with the definite
quantum numbers $n$ and $k_{z}.$ The summation over $n$ has to be performed
through all occupied states, i.e. the upper limit of the sum is $N\sim
2m\Phi _{0}E_{F}/H.$

In the limit of weak fields one can carry out the summation over the Landau
states by means of the Euler-Maclaurin's transformation

\[
\sum_{n=0}^{N}f(n)=\int_{-1/2}^{N+1/2}f(n)-\frac{1}{24}\left[ f^{^{\prime
}}(N+1/2)-f^{^{\prime }}(-1/2)\right] 
\]
and obtain

\begin{equation}
F(\epsilon ,H)=F(\epsilon ,0)+{\frac{\pi {ST}H^{2}}{24m\Phi _{0}^{2}}}%
\int_{-\pi /s}^{\pi /s}\frac{\mathcal{N}sdk_{z}}{2\pi }\left\{ \frac{{\ 1}}{{%
\ }\alpha T_{c}\epsilon {+}\mathcal{J}\left( {1-\cos (k_{z}s)}\right) }%
\right\} .  \label{magenergy}
\end{equation}
Here $\mathcal{N}$ is the total number of layers. After the momentum
integration one gets:

\[
F(\epsilon ,H)=F(\epsilon ,0)+{\frac{\pi {V}H^{2}}{24m\alpha s\Phi _{0}^{2}}}%
\frac{{\ 1}}{{\ }\sqrt{\epsilon {(\epsilon +r)}}} 
\]
with the anisotropy parameter defined as\footnote{%
Let us stress the difference between $\mathit{J}$ and$\ \mathcal{J}$ in the
two definitions (\ref{r1}) and (\ref{r2}) of the anisotropy parameter $r.$
The first one was introduced as the electron tunneling matrix element, while
the second one enters in the LD functional as the characteristic Josephson
energy for the order parameter. Later on, in the framework of the
microscopic theory, it will be demostrated that, in accordance with our
qualitative definition, $r$ $\sim $ $\mathit{J}^{2},\;$while $\mathcal{J}$
turns out to be proportional to $\mathit{J}^{2}$ too. In the dirty case it
depends on the relaxation time of the electron scattering on impurities: $%
\mathcal{J}\sim \alpha \mathit{J}^{2}\max \{\tau ,1/T\}.$ Hence both
definitions (\ref{r1}), appearing in the qualitative consideration, and (\ref
{r2}), following from the LD model, are consistent.}

\begin{equation}
r=\frac{2\mathcal{J}}{\alpha T}=\frac{4\xi _{z}^{2}(0)}{s^{2}}.  \label{r2}
\end{equation}
The magnetic susceptibility in a weak field turns out \cite{Y72,AL73} to be 
\begin{equation}
\chi _{(layer,\perp )}=-{\frac{e^{2}T}{3\pi s}}\frac{{\ \xi }_{xy}^{2}}{{\ }%
\sqrt{\epsilon {(\epsilon +r)}}}.  \label{khiLD}
\end{equation}
These results confirm the qualitative estimation (\ref{khipe}) additionally
providing the exact value of the numerical coefficient and the temperature
dependence in the crossover region. In the limit $r\gg \epsilon $ Exp.(\ref
{khiLD}) transforms into the diamagnetic susceptibility of the $3D$
anisotropic superconductor \cite{ASchmid}.

For a film of thickness $d$ the integral over $k_{z}$ in Exp.(\ref{magenergy}%
) has to be replaced by a summation over the discrete $k_{z}$ and when $\xi
_{z}(T)\ \gg d$ only the term with $k_{z}=0$ has to be taken into account:

\begin{equation}
\chi _{(film,\perp )}=-{\frac{e^{2}T}{3\pi d}}\frac{{\ \xi }_{xy}^{2}}{{\ }%
\epsilon }.  \label{filmperp}
\end{equation}
Note that these formulas predict a nontrivial increase of diamagnetic
susceptibility for clean metals \cite{AL73}. The usual statement that
fluctuations are most important in dirty superconductors with a short
electronic mean free path does not hold in the particular case of
susceptibility because here $\xi $ turns out to be in the numerator of the
fluctuation correction.

Now we will demonstrate that, besides the crossovers in its temperature
dependence, the fluctuation induced magnetization is a nonlinear function of
magnetic field too, and these nonlinearities, different for various
dimensionalities, take place at relatively low fields.\ This, strong in
comparison with the expected scale of $H_{c2}(0),$ manifestation of the
nonlinear regime in fluctuation magnetization and hence, field dependent
fluctuation susceptibility, was the subject of the intensive debates in
early seventies \cite{ASchmid,SchH,Pr,PAW69,AKE,G1,G2,G3,LS72,FG75} (see
also the old but excellent review of W.J. Skocpol and M. Tinkham \cite{ST75}%
) and after the discovery of HTS \cite{TK88,QA88,LKJ89,CRLRV00} ( see also
very recent detailed essay of T.Mishonov and E.Penev \cite{MP00} with
references there). We will mainly follow here the paper of Buzdin et al. 
\cite{Buz96}, dealing with the fluctuation magnetization of a layered
superconductor, which permits observing in a unique way all variety of the
crossover phenomena in temperature and magnetic field.

Let us go back to the general expression (\ref{frla}) and evaluate it
without taking the magnetic field to small. The difficulty in dealing with
it consists in the divergence of the sum over Landau levels $n$. This
divergence can be regularized (see \cite{MP00,M90}), but let us observe that
in order to calculate the magnetization we must know the magnetic field
dependent part of the free energy only. So a very convenient method to
bypass the divergence problem \cite{K94} is to calculate the difference $%
F(H)-F(0),$ turning the sum over Landau states in $F(\epsilon ,0)$ into an
integral and then, in its turn, turning this integral into a sum of \
integrals over the unit length intervals $x\in \lbrack n-1/2,n+1/2].$ Then

\begin{eqnarray*}
F(\epsilon ,0) &=&-\lim_{H\rightarrow 0}{\frac{2\pi {V}H}{\Phi _{0}}T}%
\int_{-\pi /s}^{\pi /s}\frac{dk_{z}}{2\pi }\sum_{n=0}^{\infty
}\int_{-1/2}^{1/2}dx\times \\
&&\times \ln {{\frac{{\ \pi T}}{{\ }\alpha T_{c}\epsilon +{\ }\frac{H}{%
2m\Phi _{0}}{\left( x+n+{\frac{1}{2}}\right) +}\mathcal{J}\left( {1-\cos (k}%
_{z}{s)}\right) }.}}
\end{eqnarray*}
and by introducing the dimensionless variable \footnote{%
Let us remind that the exact definition of $H_{c2}(0)$ contains the
numerical coefficient $A(0)$ (see footnote 3).}

\begin{equation}
h=\frac{H}{H_{c2}(0)},\;H_{c2}(0)=2m\alpha T_{c}/e=\Phi _{0}/2\pi \xi
_{xy}^{2},  \label{dimlh}
\end{equation}
one can write

\begin{eqnarray}
&&F(\epsilon ,H)-F(\epsilon ,0)=  \label{deltaF} \\
&=&-\frac{TV}{2\pi s\xi _{xy}^{2}}h\int_{-\pi }^{\pi }dz{\sum_{n=0}}%
\int_{-1/2}^{1/2}dx\ln \frac{(2n+1+2x)h+r/2(1-\cos z)+\epsilon }{%
(2n+1)h+r/2(1-\cos z)+\epsilon }.  \nonumber
\end{eqnarray}
Performing the integrations over $z$ and $x$ in (\ref{deltaF}) and
differentiating with respect to $h$ we finally obtain a very convenient
general expression for the fluctuation magnetization in a layered
superconductor:

\begin{eqnarray}
M(\epsilon ,H)=\frac{T}{\Phi _{0}s} &&\sum_{n=0}^{\infty }\left\{ n\ln \frac{%
\varphi (R_{n}+1)}{\varphi (R_{n})}+\right.  \nonumber \\
&&\left. \ln \frac{\varphi (R_{n}+1)}{\varphi (R_{n}+1/2)}-\frac{n+1/2}{%
\sqrt{(R_{n}+1/2)^{2}-\rho ^{2}}}\right\}  \label{magn}
\end{eqnarray}
with $\varphi (x)=x+\sqrt{x^{2}-\rho ^{2}},R_{n}=n+\epsilon /2h+\rho $ and $%
\rho =r/2h.$ The sum in (\ref{magn}) converges as $1/n^{2}$\ and it provides
a volume magnetization expression that can be compared with experiment.

Let us comment on the different crossovers in the $M(\epsilon ,H)$ field
dependence analyzing the general formula (\ref{magn}). Let us fix the
temperature $\epsilon \ll r.\ $In this case the $c$-axis\ coherence length
exceeds the interlayer distance ($\xi _{z}\gg s$ ) and in the absence of a
magnetic field the fluctuation Cooper pairs motion has a $3D$ character.
Supposing the magnetic field to be not too high ($h\ll r)$\ we may perform
an expansion\ in (\ref{magn}) in$1/\rho $ and obtain 
\begin{eqnarray}
M_{(3)}(\epsilon &\ll &r,h\ll r)=\frac{T\sqrt{2}h^{1/2}}{\Phi _{0}\xi _{z}}%
\times  \nonumber \\
&&{\sum_{n}}\left\{ \left( n+1\right) \left( n+1+\frac{\epsilon }{2h}\right)
^{1/2}-\right.  \nonumber \\
&&\left. -n\left( n+\frac{\epsilon }{2h}\right) ^{1/2}-\frac{3}{2}\frac{(n+1+%
\frac{2\epsilon }{3h})}{\sqrt{n+\frac{1}{2}+\frac{\epsilon }{2h}}}\right\} .
\label{lowfield}
\end{eqnarray}

For weak fields $(h\ll \epsilon )$ the magnetization grows linearly with
magnetic field, justifying our preliminary qualitative results.
Nevertheless, this linear growth is changed to the nonlinear $3D$ high field
regime $M\sim \sqrt{H}$ already in the region of a relatively small fields $%
H_{c2}(\epsilon )\lesssim H\;(\epsilon \lesssim h).\ $ \ The further
increase of magnetic field leads to the next crossover in the magnetization
field dependence at $h\sim r$ . However the Exp.(\ref{lowfield}) was
obtained in the assumption $h\ll r$ as an expansion over $1/\rho ,$\ and it
does not work any more. Handling with\ the Hurvitz zeta-function the
summation\ in (\ref{magn}) for $3D$\ case can be carried out for an
arbitrary field \cite{AKE}:

\begin{eqnarray}
M_{(3)}(\epsilon &\ll &r,h)=3\frac{T}{\Phi _{0}s}\left( \frac{2}{r}\right)
^{1/2}\sqrt{h}\times \\
&&\left[ \zeta \left( -\frac{1}{2},\frac{1}{2}+\frac{\epsilon }{2h}\right)
-\zeta \left( \frac{1}{2},\frac{1}{2}+\frac{\epsilon }{2h}\right) \frac{%
\epsilon }{6h}\right] .  \nonumber
\end{eqnarray}
One can see from this formula that for large fields the magnetization
saturates at the value $M_{\infty }$ \cite{KLB73}: 
\begin{equation}
M(h\gg r)\rightarrow M_{\infty }=-\frac{\ln 2}{2}\frac{T}{\Phi _{0}s}=-0.346%
\frac{T}{\Phi _{0}s},  \label{satur}
\end{equation}
that is a typical for $2D$ superconductors. Therefore at $h\sim r$\ we have
a $3D$ $\rightarrow $ $2D$ crossover in $M(H)$\ behavior in spite of the
fact that all sizes of fluctuation Cooper pair exceed considerably the
lattice parameters. The effective ''bidimensionalization'' of the
fluctuations is related to the effect of a strong magnetic field which
''freezes out'' the rotations along its direction. Let us stress that this
crossover occurs in the region of already strongly non-linear dependence of $%
M(H)$ and therefore for a rather strong magnetic field from the experimental
point of view in HTS.

Fixing the temperature $\epsilon \gg r$ in the formula (\ref{magn}) one can
find the general formula for $2D$ fluctuation regime \cite{MP00}:

\begin{eqnarray}
M_{(2)}(\epsilon &\gg &r,h)=\frac{T}{\Phi _{0}s}\left\{ \ln \Gamma \left( 
\frac{1}{2}+\frac{\epsilon }{2h}\right) -\frac{1}{2}\ln \left( 2\pi \right)
\right.  \nonumber \\
&&\left. -\frac{\epsilon }{2h}\left[ \psi \left( \frac{1}{2}+\frac{\epsilon 
}{2h}\right) -1\right] \right\} ,  \label{mishon}
\end{eqnarray}
where $\Gamma \left( x\right) $ is the Euler gamma-function and $\psi \left(
x\right) =d\ln \Gamma (x)/dx$ is already cited in Introduction
digamma-function. Using (\ref{mishon}) one can directly pass from the linear
regime in a weak magnetic field corresponding to (\ref{khi2}) to the
saturation of magnetization (\ref{satur}) in strong fields.

Near the line of the upper critical field $(h_{c2}(\epsilon )=-\epsilon )$\
the contribution of the term with $n=0$\ in the sum (\ref{magn}) becomes
most important and for the magnetization the expression

\[
-M(h)\sim \frac{h}{\sqrt{(h-h_{c2})(h-h_{c2}+r)}} 
\]
can be obtained. It contains the already familiar for us ''$0D$'' regime $%
(r\ll h-h_{c2}\ll 1),$\ where the magnetization decreases as $-M(h)\sim 
\frac{1}{h-h_{c2}}$ (compare with (\ref{M0ex})), while for $h-h_{c2}<<r$\
the regime becomes ''$1D$'' and the magnetization decreases slower, as $%
-M(h)\sim \frac{1}{\sqrt{h-h_{c2}}}$\ . \ 

Such an analogy is observed in the next orders in $Gi$ too. In the Ref. \cite
{HFL91} the analogy was demonstrated \ for the example of the first eleven
terms for the $2D$ case and nine for the $3D$ case. Summation of the series
of high order fluctuation contributions to the heat capacity by the
Pade-Borel method resulted in its temperature dependence similar to the $0D$
and $1D$ cases without magnetic field. Nevertheless a considerable
difference has not be forgotten: in the $0D$ and $1D$ cases no phase
transition takes place while in the $2D$ and $3D$ cases in a magnetic field
a phase transition of first order to the Abrikosov vortex lattice state
occurs.

\begin{figure}[tbp]
\centerline {\epsfig {file=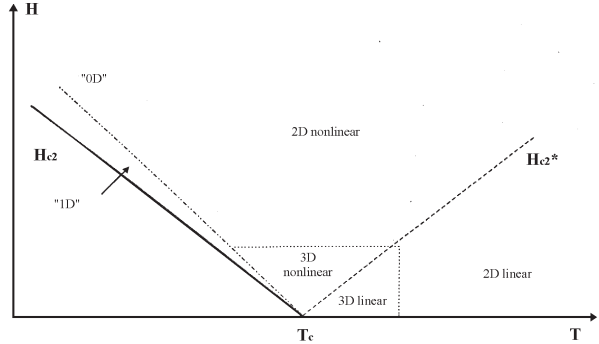,width=8cm}}
\caption{\textit{\ }Schematic representation of the different regimes for
fluctuation magnetization in the (H,T) diagram. The line $H_{c2}^{\ast }(T)$%
\ is mirror-symmetric to the $H_{c2}(T)$\ line with respect to a $y$-axis
passing through $T=T_{c}.$\ This line defines the crossover between linear
and non-linear behavior of the fluctuation magnetization above $T_{c}.$}
\label{BD}
\end{figure}
\begin{figure}[tbp]
\centerline {\includegraphics[width=6cm]{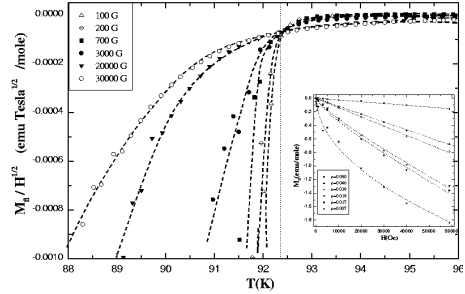}}
\caption{\textit{\ }Fluctuation magnetization of a YBaCO123 \ normalized on $%
\protect\sqrt{H}$ as the function of temperature in accordance with the
described theory$\ $shows the crossing of the isofield curves at $%
T=T_{c}(0)=92.3K$. The best fit obtained for anisotropy parameter r=0.09. In
the inset the magnetization curves as the function of magnetic field are
reported.}
\label{Rigamonti}
\end{figure}


\ In conclusion, the fluctuation magnetization of a layered\ superconductor
in the vicinity of the transition temperature turns out to be a complicated
function of temperature and magnetic field, and it evidently cannot be
factorized on these variables. The fit of the experimental data is very
sensitive to the anisotropy parameter $r$\ and allows determination of the
latter with a rather high precision \cite{Bar95,JGT98} . In Fig. \ref
{Rigamonti} the successful application of the described approach to fit the
experimental data on YBa$_{2}$Cu$_{3}$O$_{7}$ is shown \cite{LRT01}.

\section[Ginzburg-Landau formalism. Transport]{ Ginzburg-Landau theory of
fluctuations in transport phenomena}

The appearance of fluctuating Cooper pairs above $T_{c}$ leads to the
opening of a ''new channel'' for charge transfer. In the Introduction the
fluctuation Cooper pairs were treated as carriers with charge $2e$ while
their lifetime $\tau _{GL}$ was chosen to play the role of the scattering
time in the Drude formula. Such a qualitative consideration results in the
Aslamazov-Larkin (AL) pair contribution to conductivity (\ref{sigmaint})
(so-called paraconductivity\footnote{%
This term may have different origins. First of all, evidently,
paraconductivity is analogous to paramagnetism and means excess
conductivity. Another possible origin is an incorrect onomatopoeic
translation from the Russian ``paroprovodimost'~'' that means pair
conductivity}).

Below we will present the generalization of the phenomenological GL
functional approach to transport phenomena. Dealing with the fluctuation
order parameter, it is possible to describe correctly the paraconductivity
type fluctuation contributions to the normal resistance and
magnetoconductivity, Hall effect, thermoelectric power and thermal
conductivity at the edge of transition. Unfortunately the indirect
fluctuation contributions are beyond the possibilities of the description by
Time-Dependent GL approach and they will be calculated in the framework of
the microscopic theory (see Sections 6-8).

\subsection{Time dependent GL equation}

In previous Sections we have demonstrated how the GL functional formalism
allows one to account for fluctuation corrections to thermodynamical
quantities. Let us discuss the effect of fluctuations on the transport
properties of a superconductor above the critical temperature.

In order to find the value of paraconductivity, some time-dependent
generalization of the GL equations is required. Indeed, the conductivity
characterizes the response of the system to the applied electric field. It
can be defined as $\mathbf{E}=-{\partial }\mathbf{A}/{\partial t}$ but, in
contrast to the previous Section, $\mathbf{A}$ has to be regarded as being
time dependent. The general non-stationary BCS equations are very
complicated, even in the limit of slow time and space variations of the
field and the order parameter. For our purposes it will be sufficient,
following \cite{Sch66,CK67,AT96,WA68,DCY69,UD91,GE68}, to write a model
equation in the vicinity of $T_{c}$, which in general correctly reflects the
qualitative aspects of the order parameter dynamics and in some cases is
exact.

Let us keep in mind the GL functional formalism introduced above. If a
deviation from equilibrium is assumed, then it is no more possible to derive
the GL equations starting from the condition that the variational derivative
of the free energy is zero. At the same time, in the absence of equilibrium $%
\Psi $ begins to depend on time. For small deviations from the equilibrium
it is natural to assume that in the process of order parameter relaxation
its time derivative $\partial \Psi /\partial t$ is proportional to the
variational derivative of the free energy $\delta \mathcal{F}/\delta \Psi
^{\ast },$ which is equal to zero at the equilibrium. But this is not all:
side by side with the normal relaxation of the order parameter the effect of
thermodynamical fluctuations on it has to be taken into account. This can be
done by the introduction the Langevin forces $\zeta (\mathbf{r},t)$ in the
right-hand side of equation describing the order parameter dynamics.\
Finally, gauge invariance requires that $\partial \Psi /\partial t$ should
be included in the equation in the combination $\partial \Psi /\partial
t+2ie\varphi \Psi ,$ where $\varphi $ is the scalar potential of the
electric field. By concluding all these speculations one can write the model
time-dependent GL equation (TDGL) in the form

\begin{equation}
-\gamma _{GL}\left( \frac{\partial }{\partial t}+2ie\varphi \right) \Psi =%
\frac{\delta \mathcal{F}}{\delta \Psi ^{\ast }}+\zeta (\mathbf{r},t).
\label{DeltaF}
\end{equation}
with the GL functional $\mathcal{F}$ determined by (\ref{Func}),(\ref{GLMF}%
),(\ref{LDF}). The dimensionless coefficient $\gamma _{GL}$ in the
left-hand-side of the equation can be related to pair life-time $\tau _{GL}$
(\ref{taugl}): $\gamma _{GL}=\alpha T_{c}\epsilon \tau _{GL}=\pi \alpha /8$
by the substitution in (\ref{DeltaF}) of the first term of (\ref{Func}) only%
\footnote{%
It will be shown below that taking into account the electron-hole asymmetry
leads to the appearance of an imaginary part of $\gamma _{GL}$ proportional
to the derivative $\partial \ln (\rho v^{2}\tau )/\partial E|_{E_{F}}\sim 
\mathcal{O}(1/E_{F}).$ This is important for such phenomena as fluctuation
Hall effect or fluctuation thermopower and, having in mind writing the most
general formula, we will suppose $\gamma _{GL}=\pi \alpha /8+i\mbox{Im}%
\gamma _{GL},$ where necessary.}.

Neglecting the fourth order term in the GL functional, equation (\ref{DeltaF}%
) \ can be rewritten in operator form as

\begin{equation}
\lbrack \widehat{L}^{-1}-2ie\gamma _{GL}\varphi (r,t)]\Psi (\mathbf{r}%
,t)=\zeta (\mathbf{r},t)  \label{GLop}
\end{equation}
with the TDGL operator $\widehat{L}$ and Hamiltonian $\widehat{\;\mathcal{H}}
$ defined as

\begin{equation}
\widehat{L}=\left[ \gamma _{GL}\frac{\partial }{\partial t}+\widehat{%
\mathcal{H}}\right] ^{-1},\widehat{\;\mathcal{H}}=\alpha T_{c}\left[
\epsilon -\widehat{\xi }^{2}(\widehat{\mathbf{\nabla }}-2ie\mathbf{A})^{2}%
\right] .  \label{hamiso}
\end{equation}
We have introduced here the formal operator of the coherence length $%
\widehat{\xi }$ to have the possibility to deal with an arbitrary type of
spectrum. For example, in the most interesting case for our applications to
layered superconductors, the action of this operator is defined by the Exp.(%
\ref{LDF}).

In the absence of an electric field one can write the formal solution of
equation (\ref{GLop}) as

\begin{equation}
\Psi ^{(0)}(\mathbf{r},t)=\widehat{L}\zeta (\mathbf{r},t).  \label{formal}
\end{equation}
The Langevin forces introduced above must satisfy the
fluctuation-dissipation theorem, which means that the correlator $<\Psi _{%
\mathbf{p}}^{(0)\ast }(t^{\prime })\Psi _{\mathbf{p}}^{(0)}(t)>$ at
coinciding moments of time has to be the same as $<|\Psi _{\mathbf{p}%
}|^{2}>, $ obtained by averaging over fluctuations in thermal equilibrium. This requirement is fulfilled if the Langevin forces 
$\zeta (\mathbf{r},t)$ and $\zeta ^{\ast }(\mathbf{r},t)$ are correlated by
the Gaussian white-noise law

\begin{equation}
<\zeta ^{\ast }(\mathbf{r},t)\zeta (\mathbf{r}^{^{\prime }},t^{^{\prime
}})>=2T\mbox{Re}\gamma _{GL}\delta (\mathbf{r}-\mathbf{r}^{^{\prime
}})\delta (t-t^{^{\prime }}).  \label{lan}
\end{equation}
To show it let us restrict ourselves for sake of simplicity to the case of $%
A=0$ and calculate the correlator 
\begin{eqnarray}
&<&\Psi ^{\ast }(\mathbf{r},t)\Psi (\mathbf{r}^{\prime },t)>=<\zeta ^{\ast }(%
\mathbf{r},t)\widehat{\widetilde{L^{\ast }}}\widehat{L}\zeta (\mathbf{r}%
^{^{\prime }},t)>=  \label{lstarl} \\
&=&2T\mbox{Re}\gamma _{GL}\int \frac{d\mathbf{p}}{(2\pi )^{D}}e^{i\mathbf{%
p(r-r}^{^{\prime }}\mathbf{)}}\int_{-\infty }^{\infty }\frac{d\Omega }{2\pi }%
L^{\ast }(\mathbf{p},\Omega )L(\mathbf{p},\Omega ).  \nonumber
\end{eqnarray}
$L(\mathbf{p},\Omega )$ can be found by making the Fourier transform in (\ref
{hamiso}):

\begin{equation}
L(\mathbf{p},\Omega )=(-i\gamma _{GL}\Omega +\varepsilon _{\mathbf{p}})^{-1},
\label{propagator}
\end{equation}
resulting in

\begin{equation}
\left\langle \Psi ^{\ast }(\mathbf{r},t)\Psi (\mathbf{r}^{\prime
},t)\right\rangle _{\mathbf{p}}=2T\mbox{Re}\gamma _{GL}\int_{-\infty
}^{\infty }\frac{d\Omega }{2\pi }\frac{1}{\gamma _{GL}^{2}\Omega
^{2}+\varepsilon _{\mathbf{p}}^{2}}=\left\langle |\Psi _{\mathbf{p}%
}|^{2}\right\rangle ,
\end{equation}
where

\begin{equation}
\varepsilon _{\mathbf{p}}=\alpha T_{c}(\epsilon +\widehat{\xi }^{2}\mathbf{p}%
^{2}).  \label{epsp}
\end{equation}
is the Cooper pair energy spectrum.

\subsection{Paraconductivity}

Let us try to clear up the reason why the simple Drude formula works so well
for complex phenomenon like paraconductivity. For this purpose let us try to
derive the Boltzmann master equation for the fluctuation Cooper pair
distribution function

\begin{equation}
n_{\mathbf{p}}(t)=\int \left\langle \Psi (\mathbf{r},t)\Psi ^{\ast }(\mathbf{%
r}^{\prime },t)\right\rangle \exp (-i\mathbf{p(r-r}^{\prime }))d(\mathbf{r-r}%
^{\prime }).  \label{denCP}
\end{equation}
Let us recall that in the state of thermal equilibrium $n_{\mathbf{p}%
}^{(0)}=<|\Psi _{\mathbf{p}}|^{2}>=T/\varepsilon _{\mathbf{p}}.$

In order to determine the electric field dependence of $n_{\mathbf{p}}$ let
us write its time derivative using \thinspace (\ref{DeltaF})

\begin{eqnarray}
\frac{\partial n_{\mathbf{p}}(t)}{\partial t} &=&\int d(\mathbf{r-r}^{\prime
})e^{-i\mathbf{p(r-r}^{\prime })}\left[ <\frac{\partial \Psi (\mathbf{r},t)}{%
\partial t}\Psi ^{\ast }(\mathbf{r}^{\prime },t)>+<\Psi (\mathbf{r}^{\prime
},t)\frac{\partial \Psi ^{\ast }(\mathbf{r},t)}{\partial t}>\right] = 
\nonumber \\
&=&\int d(\mathbf{r-r}^{\prime })e^{-i\mathbf{p(r-r}^{\prime })}\left[ 2%
\frac{e}{i}[\varphi (\mathbf{r})-\varphi (\mathbf{r}^{\prime })]<\Psi (%
\mathbf{r},t)\Psi ^{\ast }(\mathbf{r}^{\prime },t)>\right. +  \label{boltz1}
\\
&&\left. +\frac{2}{\gamma _{GL}}\mbox{Re}\left\langle \Psi (\mathbf{r},t)%
\frac{\delta \mathcal{F}}{\delta \Psi }(\mathbf{r}^{\prime },t)\right\rangle
+\frac{2}{\gamma _{GL}}\mbox{Re}\left\langle \zeta (\mathbf{r},t)\Psi ^{\ast
}(\mathbf{r}^{\prime },t)\right\rangle \right] ,  \nonumber
\end{eqnarray}
where $\mathcal{F}$ is determined by (\ref{GLMF}). Expressing the scalar
potential by the electric field $E$ one can transform the first term of the
last integral into $-2e\mathbf{E}\frac{\partial n_{\mathbf{p}}}{\partial 
\mathbf{p}}.$ The term with the variational derivative can be evaluated by
means of (\ref{fglfourier}) and expressed in the form $-\frac{2}{\gamma _{GL}%
}\varepsilon _{\mathbf{p}}n_{\mathbf{p}}.$ More cumbersome is the evaluation
of the last term, containing the Langevin force. To the first approximation
it is possible to use here the order parameter $\Psi ^{(0)}(\mathbf{r},t)$
(see (\ref{formal})) unperturbed by the electric field as the convolution $%
\widehat{L}\zeta (\mathbf{r},t).$ In this way, using (\ref{lan}) and (\ref
{lstarl}) we calculate the\ last average in (\ref{boltz1})

\begin{eqnarray*}
\frac{2}{\gamma _{GL}}\int d(\mathbf{r-r}^{\prime })e^{-i\mathbf{p(r-r}%
^{\prime })}\mbox{Re}\left\langle \zeta ^{\ast }(\mathbf{r},t)\widehat{L}%
\zeta (\mathbf{r},t)\right\rangle &=& \\
4T\int \frac{d\mathbf{p}}{(2\pi )^{D}}e^{i\mathbf{p(r-r}^{^{\prime }}\mathbf{%
)}}\mbox{Re}\int_{-\infty }^{\infty }\frac{d\Omega }{2\pi }L(\mathbf{p}%
,\Omega ) &=&\frac{2T}{\gamma _{GL}}=\frac{2}{\gamma _{GL}}\varepsilon _{%
\mathbf{p}}n_{\mathbf{p}}^{(0)}
\end{eqnarray*}
and obtain the transport equation

\begin{equation}
\frac{\partial n_{\mathbf{p}}}{\partial t}+2e\mathbf{E}\frac{\partial n_{%
\mathbf{p}}}{\partial \mathbf{p}}=-\frac{2}{\gamma _{GL}}\varepsilon _{%
\mathbf{p}}\left( n_{\mathbf{p}}-n_{\mathbf{p}}^{(0)}\right) =-\frac{2}{\tau
_{\mathbf{p}}}\left( n_{\mathbf{p}}-n_{\mathbf{p}}^{(0)}\right) .
\label{Boln}
\end{equation}
In the absence of a magnetic field $\varepsilon _{\mathbf{p}}$ was
determined by (\ref{epsp}) and the momentum dependent lifetime,
corresponding to the Ginzburg-Landau one, can be introduced:

\[
\tau _{\mathbf{p}}=\gamma _{GL}/\varepsilon _{\mathbf{p}}=\frac{\tau
_{GL}(\epsilon )}{1+\xi ^{2}(\epsilon )\mathbf{p}^{2}}. 
\]
Let us stress the appearance of the coefficient $2$ on the right hand side
of the equation (\ref{Boln}). This means that the real Cooper pair lifetime,
characterizing its density decay, is $\tau _{\mathbf{p}}/2$ .

The effect of a weak electric field on the fluctuation Cooper pair
distribution function in the linear approximation is determined by

\begin{equation}
n_{\mathbf{p}}^{(1)}=-\frac{e\mathbf{E}\gamma _{GL}}{\varepsilon _{\mathbf{p}%
}}\frac{\partial n_{\mathbf{p}}^{(0)}}{\partial \mathbf{p}}=\ \frac{eT\gamma
_{GL}}{\varepsilon _{\mathbf{p}}^{3}}\mathbf{E}\cdot \frac{\partial
\varepsilon _{\mathbf{p}}}{\partial \mathbf{p}}.  \label{Boln1}
\end{equation}
Substituting this formula into the expression for the electric current side by side with the Cooper pair velocity \ $\mathbf{v}_{%
\mathbf{p}}=\partial \varepsilon _{\mathbf{p}}/\partial \mathbf{p}$ one can
find

\begin{equation}
\mathbf{j}^{\alpha }=\sum_{\mathbf{p}}\left( 2e\mathbf{v}^{\alpha }n\mathbf{%
_{p}}\right) \mathbf{=}\sigma ^{\alpha \beta }E^{\beta },  \nonumber
\end{equation}
where the paraconductivity tensor components are:

\begin{equation}
\sigma _{(D)}^{\alpha \beta }=\frac{\pi }{4}e^{2}\alpha T\sum_{\mathbf{p}}%
\frac{\mathbf{v}_{\mathbf{p}}^{\alpha }\mathbf{v}_{\mathbf{p}}^{\beta }}{%
\varepsilon _{\mathbf{p}}^{3}}.  \label{paraten}
\end{equation}
In the case of isotropic spectrum

\begin{equation}
\sigma _{(D)}^{\alpha \beta }=\frac{\pi }{4}e^{2}\alpha T\sum_{\mathbf{p}}%
\frac{\mathbf{v}_{\mathbf{p}}^{\alpha }\mathbf{v}_{\mathbf{p}}^{\beta }}{%
\varepsilon _{\mathbf{p}}^{3}}=\left\{ 
\begin{array}{ll}
\displaystyle{\frac{e^{2}}{32\xi }}\frac{1}{\sqrt{\epsilon }} & \mathrm{%
3D\,\;case}, \\ 
\displaystyle{\frac{e^{2}}{16d}\frac{1}{\epsilon }} & \mathrm{%
2D\;film,thickness:}d\ll \xi , \\ 
\displaystyle{\frac{\pi e^{2}\xi }{16S}\frac{1}{\epsilon ^{3/2}}} & \mathrm{%
1D\;wire,cross\,-section:}S\ll \xi ^{2}.
\end{array}
\right.  \label{ALmicro}
\end{equation}

One can compare this result with that carried out in the Introduction from
qualitative consideration, based on the Drude formula. Those simple
speculations reflect correctly the physics of the phenomenon but were
carried out with the assumption of the momentum independence of the
relaxation time $\tau _{\mathbf{p}},$ taken as $\tau _{\mathbf{0}}=\frac{\pi 
}{8(T-T_{c})}.$ As we have just seen, in reality $\tau _{\mathbf{p}}$
decreases rapidly with increase of the momentum, \ the excess $"2"$ appeared
in (\ref{Boln}) because of the wave nature of fluctuation Cooper pairs;
accounting for this circumstance results in \thinspace the precise
coefficients of (\ref{ALmicro}), different from (\ref{sigmaint}).

An especially simple form the paraconductivity results in the $2D$ case,
where, calculated per unit square, it depends on the reduced temperature
only:

\begin{equation}
\sigma _{\square }(T)=\frac{e^{2}}{16\hbar }\frac{T}{T-T_{c}}.
\label{para2D}
\end{equation}
The coefficient in this formula turns out to be a universal constant and is
given by the value $\hbar /e^{2}=4.1k\Omega .$ For electronic spectra of
other dimensionalities this universality is lost, and the paraconductivity
comes to depend on the electron mean free path.

Let us compare $\sigma _{\square }(T)$ with the normal electron Drude part $%
\sigma _{n}=n_{e}e^{2}\tau /m$ by writing the total conductivity

\begin{equation}
\sigma =\frac{e^{2}}{\hbar }(\frac{p_{F}l}{2\pi \hbar }+\frac{1}{16\epsilon }%
).  \label{drudeal}
\end{equation}
One sees that at $\epsilon _{cr}=0.4/(p_{F}l)\sim Gi_{(2c)}$ the fluctuation
correction reaches the value of the normal conductivity. Let us recall that
the same order of magnitude for the $2D$ Ginzburg-Levanyuk number was
obtained above from the heat capacity study. We will discuss the region\ of
applicability of (\ref{drudeal}) below in Section 8.6.

It is worth mentioning that the results derived here for paraconductivity
are valid with the assumption of weak fluctuations: for the temperature
range $\epsilon \lesssim Gi_{(D)}$ they are not anymore applicable.
Nevertheless, one can see that for not very dirty films, with $%
p_{F}^{2}ld\gg 1,$ a wide region of temperatures $Gi_{(2d)}\ll \epsilon \ll
1 $ exists where the temperature dependence of conductivity is determined by
fluctuations and in this region the localization effects are negligible.

The transport equation (\ref{Boln}) was originally derived many years ago by
L.G.Aslamazov and A.I.Larkin \cite{Asl70}. Recently T.Mishonov et al. \cite
{ToAnIn01} rederived Eq.(\ref{Boln}) and solved it for $n_{\mathbf{p}}$ in
the case of an arbitrary electric field.

\subsection{General expression for paraconductivity}

Unfortunately the applicability of the master equation derived is restricted
to weak magnetic fields $(H\ll H_{c2}(\epsilon ))$. For stronger fields $%
H_{c2}(\epsilon )\lesssim H\ll H_{c2}(0)$ the simple evaluation of averages
in \thinspace (\ref{boltz1}) turns out to be incorrect, the density matrix
has to be introduced and the master equation loses its attractive
simplicity. At the same time, as we already know, namely these fields,
quantizing the fluctuation Cooper pair motion, present special interest.
This is why in order to include in the scheme the magnetic field and
frequency dependencies of the paraconductivity, we return to the analysis of
the general TDGL equation \thinspace (\ref{DeltaF}) without the objective to
reduce it to a Boltzmann type transport equation.

Let us solve it in the case when the applied electric field can be
considered as a perturbation. The method will much \ resemble an exercise
from a course of quantum mechanics. To carry out the necessary generality
side by side with a formal simplicity of expressions we will introduce some
kind of subscript $\{i\}$ which includes the complete set of quantum numbers
and time. By a repeated subscript a summation over a discrete and
integration over continuous variables (time in particular) will be supposed.

We will look for the response of the order parameter to a weak electric
field applied in the form

\begin{equation}
\Psi _{k_{z}}(\mathbf{r,}t)=\Psi _{\{i\}}^{(0)}+\Psi _{\{i\}}^{(1)},
\label{e}
\end{equation}
where $\Psi _{\{i\}}^{(0)}$ is determined by (\ref{formal}). Substituting
this expression into (\ref{GLop}) and restricting our consideration to
linear terms in the electric field we can write

\[
(\widehat{L}^{-1})_{\{ik\}}\Psi _{\{k\}}^{(1)}=2ie\gamma _{GL}\varphi
_{\{il\}}\Psi _{\{l\}}^{(0)} 
\]
with the solution in the form

\begin{equation}
\Psi _{\{i\}}^{(1)}=2ie\gamma _{GL}\widehat{L}_{\{ik\}}\varphi _{\{kl\}}%
\widehat{L}_{\{lm\}}\zeta _{\{m\}}.  \nonumber
\end{equation}
Let us substitute the order parameter (\ref{e}) in the quantum mechanical
expression for current

\begin{equation}
\mathbf{j}=2e\mbox{Re}\left[ \Psi _{\{i\}}^{(0)\ast }\widehat{\mathbf{v}}%
_{\{ik\}}\Psi _{\{k\}}^{(1)}+\Psi _{\{i\}}^{(1)\ast }\widehat{\mathbf{v}}%
_{\{ik\}}\Psi _{\{k\}}^{(0)}\right] ,  \label{cu}
\end{equation}
where $\widehat{\mathbf{v}}_{\{ik\}}$ is the velocity operator which can be
expressed by means of the commutator of $\mathbf{r}$ with Hamiltonian
\thinspace (\ref{hamiso}):

\begin{equation}
\widehat{\mathbf{v}}_{\{ik\}}=i\{\widehat{\mathcal{H}},\mathbf{r}\}_{\{ik\}}.
\label{veloc}
\end{equation}
The second term of \thinspace (\ref{cu}) can be written by means of a
transposed velocity operator (which is Hermitian) as the complex conjugated
value of the first one:

\begin{equation}
\Psi _{\{i\}}^{(1)\ast }\widehat{\mathbf{v}}_{\{ik\}}\Psi
_{\{k\}}^{(0)}=(\Psi _{\{k\}}^{(0)\ast }\widetilde{\widehat{\mathbf{v}}}%
_{\{ik\}}\Psi _{\{i\}}^{(1)})^{\ast },  \label{e1}
\end{equation}
which results in 
\begin{eqnarray}
\mathbf{j} &=&2\mbox{Re}\{\Psi _{\{i\}}^{(0)\ast }(2e\widehat{\mathbf{v}}%
_{\{ik\}})\Psi _{\{k\}}^{(1)}\}=  \label{e1j} \\
&=&-8e^{2}\mbox{Im}\{\gamma _{GL}\widehat{L}_{\{ki\}}^{\ast }\widehat{%
\mathbf{v}}_{\{il\}}\widehat{L}_{\{lm\}}\varphi _{\{mn\}}\widehat{L}%
_{\{np\}}\zeta _{\{k\}}^{\ast }\zeta _{\{p\}}\}.  \nonumber
\end{eqnarray}
Let us average now (\ref{e1j}) over the Langevin forces moving the operator $%
\widehat{L}_{\{ki\}}^{\ast }$ from the beginning to the end of the trace and
using (\ref{lan}). One finds

\begin{equation}
\mathbf{j}=-16Te^{2}\mbox{Re}(\gamma _{GL})\mbox{Im}\{\gamma _{GL}\widehat{%
\mathbf{v}}_{\{il\}}\widehat{L}_{\{lm\}}\varphi _{\{mn\}}\widehat{L}_{\{np\}}%
\widehat{L}_{\{pi\}}^{\ast }\}.  \label{e1j2}
\end{equation}
Now we choose the representation where the $\widehat{L}_{\{lm\}}$ operator
is diagonal (it is evidently given by the eigenfunctions of the Hamiltonian (%
\ref{hamiso}) ):

\begin{equation}
L_{\{m\}}(\Omega )=\frac{1}{\varepsilon _{\{m\}}-i\Omega \gamma _{GL}},
\label{lmo}
\end{equation}
where $\varepsilon _{\{m\}}$ are the appropriate energy eigenvalues. Then we
assume that the electric field is coordinate independent but is a
monochromatic periodic function of time: 
\begin{equation}
\varphi (r\mathbf{,}t)=-E^{\beta }r^{\beta }\exp (-i\omega t).  \label{e3}
\end{equation}
In doing the Fourier transform in \thinspace (\ref{e1j2}) one has to
remember that the time dependence of the matrix elements $\varphi _{\{mn\}}$
results in a shift of the frequency variable of integration $\Omega
\rightarrow \Omega -\omega $ in both L-operators placed after $\varphi
_{\{mn\}}$ or, what is the same, to a shift of the argument of the previous $%
\widehat{L}_{\{lm\}}$ for $\omega :$

\begin{eqnarray}
\mathbf{j}_{\omega }^{\alpha } &=&16Te^{2}\mbox{Re}(\gamma _{GL})\times
\label{e11} \\
&&\times \int \frac{d\Omega }{2\pi }\Re \{\gamma _{GL}\widehat{\mathbf{v}}%
_{\{il\}}^{\alpha }\widehat{L}_{\{l\}}(\Omega +\omega )[-ir_{\{li\}}^{\beta }%
\widehat{]L}_{\{i\}}(\Omega )\widehat{L}_{\{i\}}^{\ast }(\Omega )\}\mathbf{E}%
^{\beta },  \nonumber
\end{eqnarray}
where $\Re f(\omega )\equiv \lbrack f(\omega )+f^{\ast }(-\omega )]/2.$

Let us express the matrix element $\mathbf{r}_{\{li\}}$ by means of $%
\widehat{\mathbf{v}}_{\{li\}}$ using the commutation relation \thinspace (%
\ref{veloc}). One can see that in the representation chosen

\begin{equation}
\widehat{\mathbf{r}}_{\{li\}}^{\beta }=i\frac{\widehat{\mathbf{v}}%
_{\{li\}}^{\beta }}{\varepsilon _{\{i\}}-\varepsilon _{\{l\}}}  \label{coord}
\end{equation}
and, carrying out the frequency integration in (\ref{e11}), finally write
for the fluctuation conductivity tensor $(\mathbf{j}_{\omega }^{\alpha
}=\sigma ^{\alpha \beta }(\omega )\mathbf{E}^{\beta }):$

\begin{eqnarray}
\sigma ^{\alpha \beta }(\epsilon ,H,\omega ) &=&8e^{2}T\mbox{Re}(\gamma
_{GL})\sum_{\{i,l\}=0}^{\infty }  \label{condgeneral} \\
&&\Re \left[ \gamma _{GL}\frac{\widehat{\mathbf{v}}_{\{il\}}^{\alpha }%
\widehat{\mathbf{v}}_{\{li\}}^{\beta }}{\varepsilon _{\{i\}}(\gamma
_{GL}\varepsilon _{\{i\}}+\gamma _{GL}^{\ast }\varepsilon _{\{l\}}-i|\gamma
_{GL}|^{2}\omega )(\varepsilon _{\{l\}}-\varepsilon _{\{i\}})}\right] . 
\nonumber
\end{eqnarray}
This is the most general expression which describes the d.c.,
galvanomagnetic and high frequency paraconductivity contribution. In the
case when we are interested in diagonal effects only, where it is enough to
accept $\gamma _{GL}$ as real $(\gamma _{GL}=\mbox{Re}\gamma _{GL}=\pi
\alpha /8)$ omitting its small imaginary part, the last expression can be
simplified by means of symmetrization of the summation variables):

\begin{equation}
\sigma ^{\alpha \alpha }(\epsilon ,H,\omega )=\frac{\pi }{2}\alpha
e^{2}T\sum_{\{i,l\}=0}^{\infty }\Re \left[ \frac{\widehat{\mathbf{v}}%
_{\{il\}}^{\alpha }\widehat{\mathbf{v}}_{\{li\}}^{\alpha }}{\varepsilon
_{\{i\}}\varepsilon _{\{l\}}(\varepsilon _{\{i\}}+\varepsilon
_{\{l\}}-i\gamma _{GL}\omega )}\right] .  \label{sym}
\end{equation}

Let us demonstrate the calculation of the d.c. paraconductivity in the
simplest case of a metal with an isotropic spectrum. In this case we choose
a plane wave representation. By using $\varepsilon _{\mathbf{p}}$ defined by
(\ref{epsp}) one has

\begin{equation}
\widehat{\mathbf{v}}_{\{\mathbf{pp}^{^{\prime }}\}}=\mathbf{v}_{\mathbf{p}%
}\delta _{\mathbf{pp}^{^{\prime }}},\;\mathbf{v}_{\mathbf{p}}=\frac{\partial
\varepsilon _{\mathbf{p}}}{\partial \mathbf{p}}=2\alpha T_{c}\xi ^{2}\mathbf{%
p}.  \label{vis}
\end{equation}
We do not need to keep the imaginary part of $\gamma _{GL}$, which is
necessary to calculate particle-hole asymmetric effects only. Then the
fluctuation conductivity calculated from (\ref{sym}) coincides exactly with (%
\ref{paraten}).

\subsection{Fluctuation conductivity of layered superconductor}

Let us return to the discussion of our general formula \thinspace (\ref{sym}%
) for the fluctuation conductivity tensor. A magnetic field directed along
the c-axis still permits separation of variables even in the case of a
layered superconductor. The Hamiltonian in this case can be written as in
\thinspace (\ref{GLFEmf}),\thinspace (\ref{hamiso}):

\begin{equation}
\widehat{\mathcal{H}}=\alpha T_{c}\left( \epsilon -\xi _{xy}^{2}(\mathbf{%
\nabla }_{xy}-2ie\mathbf{A}_{xy})^{2}-\frac{r}{2}(1-\cos (k_{z}s)\right) .
\label{hamlay}
\end{equation}
It it is convenient to work in the Landau representation, where the
summation over $\{i\}$ is reduced to one over the ladder of the Landau
levels $i=0,1,2.$. (each is degenerate with a density $2eH$ per unit square)
and integration over the c-axis momentum in the limits of the Brillouine
zone. The eigenvalues of the Hamiltonian \thinspace (\ref{hamlay}) can be
written in the form

\begin{equation}
\varepsilon _{n}=\alpha T_{c}[\epsilon +\frac{r}{2}(1-\cos
(k_{z}s))+h(2n+1)]=\varepsilon _{k_{z}}+\alpha T_{c}h(2n+1),
\label{eigenmag}
\end{equation}
where $h=\frac{eH}{2m\alpha T_{c}}$ was already defined by Eq.\thinspace (%
\ref{dimlh}). For the velocity operators one can write

\begin{equation}
\widehat{\mathbf{v}}^{x,y}=\frac{1}{2m}(-i\mathbf{\nabla -}2ie\mathbf{A)}%
^{x,y};\;\widehat{\mathbf{v}}^{z}=-\frac{\alpha rs}{2}T_{c}\sin (k_{z}s).
\end{equation}

\subsubsection{In-plane conductivity.}

Let us start from the calculation of the in-plane components. The
calculation of the velocity operator matrix elements requires some special
consideration. First of all let us stress that the required matrix elements
have to be calculated for the eigenstates of a quantum oscillator whose
motion is equivalent to the motion of a charged particle in a magnetic
field. The commutation relation for the oscillator's velocity components is
well known (see \cite{LLqm}):

\begin{equation}
\lbrack \widehat{\mathbf{v}}^{x},\widehat{\mathbf{v}}^{y}]=i\frac{eH}{2m^{2}}%
=\frac{i\alpha T_{c}}{m}h.  \label{commut}
\end{equation}
In order to calculate the necessary matrix elements let us present the
velocity operator components in the form of boson-type creation and
annihilation operators $\widehat{a}^{+},\widehat{a}$ with commutation
relation $[\widehat{a},\widehat{a}^{+}]=1$ :

\[
\widehat{\mathbf{v}}^{x,y}=\sqrt{\frac{\alpha T_{c}h}{2m}}\left( 
\begin{tabular}{c}
$\widehat{a}^{+}+\widehat{a}$ \\ 
$i\widehat{a}^{+}-i\widehat{a}$%
\end{tabular}
\right) 
\]
One can check that the correct commutation relation (\ref{commut}) \ is
fulfilled. Taking into account that

\[
<l|\widehat{a}|n>=<n|\widehat{a}^{+}|l>=\sqrt{n}\delta _{n,l+1} 
\]
it is seen that the only non-zero matrix elements of the velocity operator
are

\[
<l|\widehat{\mathbf{v}}^{x,y}|n>=\sqrt{\frac{\alpha T_{c}h}{2m}}\left( 
\begin{tabular}{c}
$\sqrt{l}\delta _{l,n+1}+\sqrt{n}\delta _{n,l+1}$ \\ 
$i\sqrt{l}\delta _{l,n+1}-i\sqrt{n}\delta _{n,l+1}$%
\end{tabular}
\right) . 
\]
Using these relations the necessary product of matrix elements can be
calculated:

\begin{equation}
<l|\widehat{\mathbf{v}}^{x}|n><n|\widehat{\mathbf{v}}^{x}|l>=\frac{\alpha
T_{c}h}{2m}(l\delta _{l,n+1}+n\delta _{n,l+1}).  \nonumber
\end{equation}
Its substitution into (\ref{sym}) and accounting of the degeneracy of the
Landau levels $2eH=4m\alpha T_{c}h$ gives for the diagonal component of the
in-plane conductivity tensor

\begin{eqnarray}
\sigma ^{xx}(\epsilon ,H,\omega ) &=&\frac{\pi \alpha ^{2}T_{c}^{2}e^{2}}{4m}%
h\sum_{\{n\}=0}^{\infty }\sum_{\left\{ l\right\} =0}\Re \lbrack \frac{%
(l\delta _{l,n+1}+n\delta _{n,l+1})}{\varepsilon _{l}\varepsilon
_{n}(\varepsilon _{l}+\varepsilon _{n}-i\gamma _{GL}\omega )}] \\
&=&\pi e^{2}\left( \alpha T_{c}\right) ^{3}h^{2}\int_{-\frac{\pi }{s}}^{%
\frac{\pi }{s}}\frac{dk_{z}}{2\pi }\sum_{n=0}^{\infty }\Re \lbrack \frac{n+1%
}{\varepsilon _{n+1}\varepsilon _{n}(\varepsilon _{n+1}+\varepsilon
_{n}-i\gamma _{GL}\omega )}].  \nonumber
\end{eqnarray}
Expanding the denominator into simple fractions we reduce the problem to the
calculation of the $c$-axis momentum integral, which can be carried out in
the general case by use of the identity:

\begin{equation}
\int_{0}^{2\pi }\frac{dx}{2\pi }\frac{1}{\cos x-z}=-\frac{1}{\sqrt{z^{2}-1}}
\label{identkz}
\end{equation}
valid for any complex parameter $z\neq 1$ with the proper choice of the
square root branch. Using it we write the general expression for the
in-plane component of the fluctuation conductivity tensor

\begin{eqnarray}
\sigma ^{xx}(\varepsilon ,h,\omega ) &=&\frac{e^{2}h}{16s}\sum_{n=0}^{\infty
}(n+1)\left\{ \frac{1}{h-i\widetilde{\omega }}\frac{1}{\sqrt{[\epsilon
+h(2n+1)][r+\epsilon +h(2n+1)]}}+\right.  \nonumber \\
&&\frac{1}{h+i\widetilde{\omega }}\frac{1}{\sqrt{[\epsilon
+h(2n+3)][r+\epsilon +h(2n+3)]}}-  \label{sigmah} \\
&&\left. \frac{2h}{h^{2}+\widetilde{\omega }^{2}}\frac{1}{\sqrt{[\epsilon
+h(2n+2)-i\widetilde{\omega }][r+\epsilon +h(2n+2)-i\widetilde{\omega }]}}%
\right\} .  \nonumber
\end{eqnarray}
where $\widetilde{\omega }=\frac{\pi \omega }{16T_{c}}.$

\subsubsection{Out-of plane conductivity.}

The situation with the out-of plane component of paraconductivity turns out
to be even simpler because of the diagonal structure of the $\widehat{%
\mathbf{v}}_{\{in\}}^{z}=-\frac{\alpha rs}{2}T_{c}\sin (k_{z}s)\times \delta
_{in}\times \delta (k_{z}-k_{z^{^{\prime }}}).$ Taking into account that the
Landau state degeneracy $2eH=h\xi _{xy}^{-2}$ we write

\begin{eqnarray*}
\sigma ^{zz}(\epsilon ,H,\omega ) &=&\frac{1}{2}\pi \alpha
e^{2}T\sum_{\{i,l\}=0}^{\infty }\Re \left[ \frac{\widehat{\mathbf{v}}%
_{\{il\}}^{\alpha }\widehat{\mathbf{v}}_{\{li\}}^{\alpha }}{\varepsilon
_{\{i\}}\varepsilon _{\{l\}}(\varepsilon _{\{i\}}+\varepsilon
_{\{l\}}-i\alpha T_{c}\widetilde{\omega })}\right] = \\
&=&\frac{\pi e^{2}\left( \alpha T_{c}\right) ^{3}}{16}\left( \frac{sr}{\xi
_{xy}}\right) ^{2}h\sum_{n=0}^{\infty }\int_{-\frac{\pi }{s}}^{\frac{\pi }{s}%
}\frac{dk_{z}}{2\pi }\Re \lbrack \frac{\sin ^{2}(k_{z}s)}{\varepsilon
_{n}^{2}(k_{z})[\varepsilon _{n}(k_{z})-i\alpha T_{c}\widetilde{\omega }]}].
\end{eqnarray*}
The following transformations are similar to the calculation of the in-plane
component: we expand the integrand into simple fractions and perform the $%
k_{z}-$integration by means of the identity (\ref{identkz}). The final
expression can be written as

\begin{eqnarray}
\sigma ^{zz}(\epsilon ,H,\omega ) &=&\frac{\pi e^{2}}{32s}\left( \frac{sr}{%
\xi _{xy}}\right) ^{2}h\sum_{n=0}^{\infty }\left( -\frac{\partial }{\partial
\lambda }\right) \times  \label{sigmaz} \\
&&\Re \left( \frac{1}{\lambda +i\widetilde{\omega }}\right) \left\{ \frac{1}{%
\sqrt{(\epsilon +h(2n+1)+\lambda )(\epsilon +h(2n+1)+\lambda +r)}}-\right. 
\nonumber \\
&&\left. \frac{1}{\sqrt{(\epsilon +h(2n+1)-i\widetilde{\omega })(\epsilon
+h(2n+1)-i\widetilde{\omega }+r)}}\right\} \left| _{\lambda =0}\right. . 
\nonumber
\end{eqnarray}

\subsubsection{Analysis of the general expressions.}

In principle the expressions derived above give an exact solution for the
a.c.$(\omega \ll T)$ paraconductivity tensor of a layered superconductor in
a perpendicular magnetic field $H\ll $ $H_{c2}$ $(h\ll 1)$ in the vicinity
of the critical temperature $(\epsilon \ll 1).$ The interplay of the
parameters $r,\epsilon ,\omega $ $h$ entering into (\ref{sigmah})-(\ref
{sigmaz}), as we have seen in the example of fluctuation magnetization,
yields a variety of crossover phenomena.

1. The simplest and most important results which can be derived are the
components of the $d.c.$ paraconductivity ($\omega =0$) of layered
superconductor in the absence of magnetic field. Keeping $\omega =0$ and
setting $h\rightarrow 0$ one can change the summations over Landau levels
into integration and find

\[
\sigma ^{xx}(\varepsilon ,h\rightarrow 0,\omega =0)=\frac{e^{2}}{16s}\frac{1%
}{\sqrt{[\epsilon (r+\epsilon )]}}, 
\]

\[
\sigma ^{zz}(\epsilon ,h\rightarrow 0,\omega =0)=\frac{e^{2}s}{32\xi
_{xy}^{2}}\left( {\frac{{\epsilon +r/2}}{{\ [\epsilon (\epsilon +r)]^{1/2}}}}%
-1\right) . 
\]

2. The Aslamazov-Larkin contribution to the magnetoconductivity can be
studied by putting $\omega =0$ and keeping magnetic field as arbitrary. We
will not go into details and just report the results (following \cite{BV98}
with some revision of the coefficient in $3D$ case)

\bigskip

\begin{tabular}{|l|l|l|l|}
\hline
& $h\ll \epsilon $ & $
\begin{array}{c}
\epsilon \ll h\ll r \\ 
(3D)
\end{array}
\quad $ & $
\begin{array}{c}
\max \{\epsilon ,r\}\ll h \\ 
(2D)
\end{array}
\quad $ \\ \hline
$\sigma ^{xx}$ & $\sigma ^{xx}(\epsilon ,h=0)-\frac{{e^{2}}}{2^{8}{s}}\frac{{%
[8\epsilon (\epsilon +r)+3r^{2}]}}{{[\epsilon (\epsilon +r)]^{5/2}}}h{^{2}}$
& $\frac{{e^{2}}}{{4s}}\frac{1}{\sqrt{2hr}}$ & $\frac{{e^{2}}}{{8s}}\frac{1}{%
h}$ \\ \hline
$\sigma ^{zz}$ & $\sigma ^{zz}(\epsilon ,h=0)-\frac{{e^{2}s}}{2^{8}\xi
_{xy}^{2}}\frac{{r^{2}(\epsilon +r/2)}}{{[\epsilon (\epsilon +r)]^{5/2}}}%
h^{2}$ & $\frac{3.24{e^{2}s}}{\xi _{xy}^{2}}\sqrt{\frac{r}{h}}$ & $\frac{%
7\zeta (3){e^{2}s}}{2^{9}\xi _{xy}^{2}}\frac{r^{2}}{h^{2}}$ \\ \hline
\end{tabular}

\begin{center}
Table 2.
\end{center}

Here it is worth making an important comment. The proportionality of the
fluctuation magnetoconductivity to $h^{2}$ is valid when using the
parametrization $\epsilon =(T-T_{c0})/T_{c0}$ only. As it well known, a weak
field shifts the critical temperature linearly, which often makes it
attractive to analyze the experimental data by choosing as the reduced
temperature parameter $\epsilon _{h}=(T-T_{c}(H))/T_{c}(H).$ In this
parametrization one can get a term in the magnetoconductivity linear in $h$
, which previously was cancelled out by the magnetic field renormalization
of the critical temperature. So it is important to recognize that the effect
of a weak magnetic field on\ the fluctuation conductivity cannot be reduced
to a simple replacement of $T_{c0}$ by $T_{c}(H)$ in the appropriate formula
without the field. Vice versa, this effect is exactly compensated by the
change in the functional dependence of the paraconductivity in magnetic
field, and finally it contains the negative quadratic contribution only.

3. Letting the magnetic field go to zero and considering nonzero frequency
of the electromagnetic field one can find general expressions for the
components of the $a.c$ paraconductivity tensor. They are cumbersome enough
and in the complete form can be found, for instance, in \cite{VBML99}. We
recall here the simplified asymptotics for the $\mbox{Re}\sigma $ in the $2D$
regime only:

\begin{eqnarray}
\mbox{Re}\sigma _{(2D)}^{xx}(r &\ll &\epsilon ,\widetilde{\omega })=
\label{sigomx} \\
&=&\frac{e^{2}}{16s}\frac{1}{\epsilon }\left[ \frac{2\epsilon }{\widetilde{%
\omega }}\arctan \frac{\widetilde{\omega }}{\epsilon }-\left( \frac{\epsilon 
}{\widetilde{\omega }}\right) ^{2}\ln \left[ 1+\left( \frac{\widetilde{%
\omega }}{\epsilon }\right) ^{2}\right] \right]  \nonumber
\end{eqnarray}

\begin{equation}
\mbox{Re}\sigma _{(2D)}^{zz}(r\ll \epsilon ,\widetilde{\omega })=\frac{e^{2}%
}{2^{8}s}\left( \frac{sr}{\xi _{xy}}\right) ^{2}\left( \frac{1}{\widetilde{%
\omega }}\right) ^{2}{\ln }[{1+}\left( \frac{\widetilde{\omega }}{\epsilon }%
\right) ^{2}].  \label{sigomz}
\end{equation}

The general formulas (\ref{sigmah})-(\ref{sigmaz}) allow one to study the
different crossovers in the $a.c.$ conductivity of layered superconductor in
the presence of magnetic field of various intensity. We leave this exercise
for the reader having some practical interest in the problem.

\subsection{Magnetic field angular dependence of paraconductivity}

We have seen above that in the case of a geometry with a magnetic field
directed along the c-axis many sophisticated fluctuation features of layered
superconductors can be studied in the most general form. Nevertheless even
the attempt to explore the d.c. conductivity in a longitudinal magnetic
field (directed in ab plane) \cite{E79} or, moreover, with the field
directed at some arbitrary angle $\theta $ with the c-axis leads to the
appearance of the a vector potential component in the argument of $\cos
(k_{z}s)$ and the problem requires a nontrivial calculation of the matrix
elements over the Mathieu functions.

We already learned that at temperatures very near to the critical one ($%
\epsilon \ll r)$ the $3D$ fluctuation regime takes place. Here the size of
the Cooper pairs along the c-axis is so large that the peculiarities of the
layered structure do not play any more role. This means that only small
values of $k_{z}$ are important in the $k_{z}$-integrations, where the $\cos
(k_{z}s)$ in (\ref{GLFEmf}) can be expanded and the LD functional is reduced
to its traditional GL\ form with an anisotropic effective mass tensor:

\begin{eqnarray}
\mathcal{F}[\Psi ] &=&\int d^{3}\mathbf{r}\left\{ a|\Psi |^{2}+\frac{B}{2}%
|\Psi |^{4}+\sum_{\mu =1}^{3}\frac{1}{4m_{\mu }}\left| \left( \frac{1}{i}%
\frac{d}{dx_{\mu }}-2eA_{\mu }\right) \Psi \right| ^{2}+\right.  \nonumber \\
&&\left. +\frac{B^{2}}{8\pi }-\frac{\mathbf{H}\cdot \mathbf{B}}{4\pi }%
\right\} .  \label{GLanimas}
\end{eqnarray}
We will demonstrate below that in this case a scaling approach provides a
direct access to the most general results by rescaling the anisotropic
problem to the corresponding isotropic one on the initial level of the GL
approach \cite{BGL92}.

Let us suppose that the external field $\mathbf{H}$ is chosen to lie in the $%
y-z$ plane and makes angle $\theta $ with the$\ z-$ axis. For sake of
simplicity and because the oxide superconductors are within high accuracy
uniaxial materials, we choose $m_{x}=m_{y}=m^{\ast },$ while $%
m_{z}^{-1}=2\alpha s^{2}r$ (compare with (\ref{LDF})). The effective
anisotropy parameter $\mathcal{\gamma }_{a}^{2}=m^{\ast }/m_{z}=2\alpha
s^{2}rm^{\ast }<1$ is introduced. In (\ref{GLanimas}) the anisotropy enters
only in the gauge-invariant gradient term, so the simple rescaling of the
coordinate axes: $x=\widetilde{x},y=\widetilde{y},z=\mathcal{\gamma }_{a}%
\widetilde{z}$ together with the scaling of the vector potential: $\mathbf{%
A=(}\widetilde{A}_{x},\widetilde{A}_{y},\widetilde{A}_{z}/\mathcal{\gamma }%
_{a}\mathbf{)}$ will render this term isotropic. The magnetic field
evidently is rescaled to $\mathbf{B=(}\widetilde{B}_{x}/\mathcal{\gamma }%
_{a},\widetilde{B}_{y}/\mathcal{\gamma }_{a},\widetilde{B}_{z}\mathbf{)}$
and the last three terms in (\ref{GLanimas}), describing the magnetic-field
energy, are transformed to

\begin{eqnarray*}
\delta \mathcal{F}[\Psi ] &=&\frac{\mathcal{\gamma }_{a}}{8\pi }\int d^{3}%
\widetilde{\mathbf{r}}\left[ \frac{1}{4m}\sum_{\mu =1}^{3}\left| \left( 
\frac{\hbar }{i}\frac{d}{d\widetilde{x}_{\mu }}-\frac{2e}{c}\widetilde{A}%
_{\mu }\right) \Psi \right| ^{2}+(\frac{\widetilde{\mathbf{B}}_{xy}^{2}}{%
\mathcal{\gamma }_{a}}+\widetilde{B}_{z}^{2})-\right. \\
&&\left. -2(\frac{\widetilde{\mathbf{B}}_{xy}\cdot \mathbf{H}_{xy}}{\mathcal{%
\gamma }_{a}}+\widetilde{B}_{z}H_{z})\right] .
\end{eqnarray*}
In short, we have removed the anisotropy from the gradient term but
reintroduced it into the magnetic energy term. In general it is not possible
to make both terms isotropic in the Gibbs energy simultaneously. However,
depending on the physical question addressed, we can neglect fluctuations in
the magnetic field, as was mostly done above.

Let us demonstrate how the method works for the example of the d.c.
fluctuation conductivity tensor which was calculated above for a magnetic
field directed along the c-axis. We restrict our consideration to the $3D$
region $(\epsilon \ll r).$ One can write the scaling relations between the
electric field and current components before and after the scaling
transformation by means of a conductivity tensor and the anisotropy
parameter:

\begin{eqnarray}
j_{x,y} &=&\widetilde{j}_{x,y}\ \qquad j_{z}\sim ev_{z}\sim \mathcal{\gamma }%
_{a}\widetilde{j}_{z}  \label{surel} \\
E_{x,y} &=&\widetilde{E}_{x,y}\ \qquad E_{z}\sim \frac{\partial \varphi }{%
\partial z}\sim \frac{1}{\mathcal{\gamma }_{a}}\widetilde{E}_{z}.  \nonumber
\end{eqnarray}
Now let us rewrite the relations between the current and electric field
vectors before and after the scale transformation

\begin{eqnarray*}
j_{\alpha } &=&\sigma _{\alpha \beta }E_{\beta } \\
\widetilde{j}_{\alpha } &=&\widetilde{\sigma }_{\alpha \beta }\widetilde{E}%
_{\beta }.
\end{eqnarray*}
Comparing them with (\ref{surel}) and introducing the operator of the direct
scaling transformation $T_{\alpha \beta }$

\[
T_{\alpha \beta }=\left( \mbox{%
\begin{tabular}{ccc}
$1$ & $0$ & $0$ \\ 
$0$ & $1$ & $0$ \\ 
$0$ & $0$ & $\mathcal{\gamma }_{a}$%
\end{tabular}
}\right) \mbox{,} 
\]
one can write $j_{\alpha }=T_{\alpha \mu }\widetilde{j}_{\mu },\;E_{\alpha
}=(T^{-1})_{\alpha \mu }\widetilde{E}_{\mu }$ and express the conductivity
tensor as

\[
\sigma _{\alpha \beta }=T_{\alpha \mu }\widetilde{\sigma }_{\mu \rho
}T_{\rho \beta }. 
\]

Now let us work in\ the already isotropic coordinate frame. We suppose that
initially the magnetic field was directed along the c-axis and now we rotate
it in the X-Z plane by the angle $\widetilde{\theta }$ with respect to the
initial direction. The conductivity tensor will be transformed by the usual
matrix law:

\[
\widetilde{\sigma }_{\alpha \beta }(\widetilde{\theta })=R_{\alpha \mu }^{T}%
\widetilde{\sigma }_{\mu \rho }(0)R_{\rho \beta }=R_{\alpha \mu
}^{T}(T^{-1})_{\mu \varsigma }\sigma _{\varsigma \eta }(0)(T^{-1})_{\eta
\delta }R_{\delta \beta } 
\]
and

\[
\sigma _{\alpha \beta }(\widetilde{\theta })=T_{\alpha \gamma }\widetilde{%
\sigma }_{\gamma \delta }(\widetilde{\theta })T_{\delta \beta }=T_{\alpha
\gamma }R_{\gamma \mu }^{T}(T^{-1})_{\mu \varsigma }\sigma _{\varsigma \eta
}(0)(T^{-1})_{\eta \delta }R_{\delta \kappa }T_{\kappa \beta }, 
\]
where

\[
R_{\alpha \beta }=\left( 
\begin{tabular}{ccc}
$\cos \widetilde{\theta }$ & $0$ & $-\sin \widetilde{\theta }$ \\ 
$0$ & $1$ & $0$ \\ 
$\sin \widetilde{\theta }$ & $0$ & $\cos \widetilde{\theta }$%
\end{tabular}
\right) . 
\]
Finally the fluctuation conductivity tensor $\sigma _{\alpha \beta }(\theta
) $ in the initial tetragonal system with the magnetic field directed at the
angle $\theta $ with respect to the c-axis can be expressed by means of the
effective transformation operator $M_{\alpha \beta }$ :

\[
\sigma _{\alpha \beta }(\theta )=M_{\alpha \varsigma }^{T}(\widetilde{\theta 
})\sigma _{\varsigma \eta }(0,\widetilde{H})M_{\eta \beta }(\widetilde{%
\theta }), 
\]
with

\[
M_{\alpha \beta }(\widetilde{\theta })=(T^{-1})_{\alpha \delta }R_{\delta
\kappa }T_{\kappa \beta }=\left( 
\begin{tabular}{ccc}
$\cos \widetilde{\theta }$ & $0$ & $-\frac{1}{\gamma }\sin \widetilde{\theta 
}$ \\ 
$0$ & $1$ & $0$ \\ 
$\mathcal{\gamma }_{a}\sin \widetilde{\theta }$ & $0$ & $\cos \widetilde{%
\theta }$%
\end{tabular}
\right) . 
\]
The angle $\widetilde{\theta }$ can be expressed via the renormalized
magnitude of the magnetic field $\widetilde{H}=\sqrt{H_{c}^{2}+\mathcal{%
\gamma }_{a}^{2}H_{x}^{2}}:$

\[
\cos \widetilde{\theta }=\frac{\widetilde{H}_{z}}{\widetilde{H}}=\frac{\cos
\theta }{\sqrt{\cos ^{2}\theta +\mathcal{\gamma }_{a}^{2}\sin ^{2}\theta }}%
;\sin \widetilde{\theta }=\frac{\mathcal{\gamma }_{a}\sin \theta }{\sqrt{%
\cos ^{2}\theta +\mathcal{\gamma }_{a}^{2}\sin ^{2}\theta }}. 
\]

In the case of the paraconductivity of a layered superconductor with the
magnetic field applied at an arbitrary angle $\theta $ the answer can be
written in the general form by means of the three diagonal components of
conductivity $\sigma _{ii}(0,\widetilde{H})$ in the perpendicular field $%
\widetilde{H}$:

\begin{eqnarray*}
\sigma _{\alpha \beta }(\theta ) &=&\frac{1}{\cos ^{2}\theta +\mathcal{%
\gamma }_{a}^{2}\sin ^{2}\theta }\times \\
&& \\
&&\times \left( 
\begin{tabular}{ccc}
$
\begin{array}{c}
\sigma _{xx}\cos \theta + \\ 
+\mathcal{\gamma }_{a}^{4}\sigma _{zz}\sin ^{2}\theta
\end{array}
$ & $0$ & $
\begin{array}{c}
\sigma _{zz}^{2}\mathcal{\gamma }_{a}\sin ^{2}\theta \cos \theta - \\ 
-\sigma _{xx}\sin \theta
\end{array}
$ \\ 
$0$ & $\sigma _{yy}(\cos ^{2}\theta +\mathcal{\gamma }_{a}^{2}\sin
^{2}\theta )$ & $0$ \\ 
$\sigma _{zz}^{2}\mathcal{\gamma }_{a}\sin \theta \cos \theta $ & $0$ & $%
\sigma _{zz}\cos ^{2}\theta $%
\end{tabular}
\right)
\end{eqnarray*}

In the simplest case of\ a longitudinal field $\theta =90^{0}:$

\begin{equation}
\sigma _{\alpha \beta }(90^{0},H)=\left( 
\begin{tabular}{ccc}
$\mathcal{\gamma }_{a}^{2}\sigma _{zz}(0,\mathcal{\gamma }_{a}H)$ & $0$ & $-%
\mathcal{\gamma }_{a}^{-2}\sigma _{zz}(0,\mathcal{\gamma }_{a}H)$ \\ 
$0$ & $\sigma _{yy}(0,\mathcal{\gamma }_{a}H)$ & $0$ \\ 
$0$ & $0$ & $0$%
\end{tabular}
\right) .
\end{equation}

\section[Fluctuations near S-I transition]{Fluctuations near
superconductor-insulator transition}

\subsection{Quantum phase transition}

It is usually supposed that the temperature of the superconducting
transition does not depend on the concentration of non-magnetic impurities
(Anderson's theorem \cite{a59,ag58}). Nevertheless when the degree of
disorder is very high Anderson localization takes place, and it would be
difficult to expect that under conditions of strong electron localization
superconductivity can exist, even if there is inter-electron attraction.
This means that at $T=0$ the phase transition takes place with a change of
the disorder strength or carrier concentration. Such a transition is called
a quantum phase transition since at zero temperature the classical
fluctuations are absent. Indeed, one can see from (\ref{ncp}) that in the
limit $T\rightarrow 0$ the thermal fluctuation Cooper pairs vanish.

In the metallic phase of a disordered system the conductivity is mostly
determined by the weakly decaying fermionic excitations, their dynamics
yielding the familiar Drude formula (the method which accounts for the
fermionic excitations will be referred to as the Fermi approach later on).
Inside the critical region the charge transfer due to fluctuation Cooper
pairs turns out to be more important. In some approximation, the pairs may
be considered as Bose particles. Therefore the approach dealing with the
fluctuation pairs will be called below the Bose approach.

Let us suppose that at temperature $T=0$ the superconducting state occurs in
a weakly disordered system. In principle two scenarios of the development of
the situation are possible with an increase of the disorder strength: the
system at some critical disorder strength can go from the superconducting
state to the metallic state or to the insulating state. The first scenario
is natural and takes place in the following cases: if the effective constant
of the inter-electron interaction changes its sign with the growth of the
disorder; if the effective concentration of magnetic impurities increases
together with the disorder growth; if the pairing symmetry of
superconducting state is nontrivial it can be destroyed even by the weak
disorder level. We will study here the second scenario where the
superconductor becomes an insulator with disorder increase. This means that
at some disorder degree range, higher than the localization edge when the
normal phase does not exist\ any more at finite temperatures,
superconductivity can still survive. From the first glance this statement
seems strange: what does superconductivity mean if the electrons are already
localized? And if it really can take place beyond the metallic phase, at
what value of disorder strength and in which way does the superconductivity
finally disappears?

One has to have in mind that localization is a quantum phenomenon in its
nature and with the approach to the localization edge the coherence length
of localization $\ell $ grows. From the insulator side of the transition
vicinity this means the existence of large scale regions where delocalized
electrons exist. If the energy level spacing in such regions does not exceed
the value of superconducting gap Cooper pairs still can be formed by the
delocalized electrons of this region.

The problem can be reformulated in other, already familiar, way: how does
the critical temperature of the superconducting transition decrease with the
increase of the disorder strength? In the previous Sections we have already
tried to solve it by discussing the critical temperature fluctuation shift.
We have seen that the fluctuation shift of the critical temperature is
proportional to $\sqrt{Gi_{(3)}}$ for a $3D$ superconductor and to $%
Gi_{(2)}\ln \{1/Gi_{(2)}\}$ for $2D.$ This means that the critical
temperature is not changed noticeably as long as the Ginzburg-Levanyuk
number remains small. So one can expect the complete suppression of
superconductivity when $Gi\sim 1$ only. For further consideration it is
convenient to separate the $3D$ and $2D$ cases because the physical pictures
of the superconductor-insulator transition for them are quite different.

\subsection{3D superconductors}

As one can see from Table 1 in the $3D$ case the Ginzburg-Levanyuk number
remains small at $p_{F}l\sim 1:$ $Gi_{(3d)}$ $\approx $ $\left( \frac{T_{c}}{%
E_{F}}\right) \ll 1.${} Nevertheless, approaching the edge of localization,
the width of the fluctuation region increases \cite{kk}. In the framework of
the self-consistent theory of localization \cite{VW} such growth of the
width of the fluctuation region was found in paper \cite{BVS}.

Instead of the cited self-consistent theory let us make some more general
assumptions concerning the character of the metal-insulator (M-I) transition
in the absence of superconductivity \cite{L99}. We suppose that in the case
of very strong disorder and not very strong Coulomb interaction the M-I
transition is of second order. The role of ''temperature'' for this
transition is played by the ''disorder strength'' which is characterized by
the dimensionless value $g=\frac{p_{F}l.}{2\pi }$ With its decrease the
conductivity of the metallic phase decreases and at some critical value $%
g_{c}$ tends to zero as 
\begin{equation}
\sigma =e^{2}p_{F}(g-g_{c})^{\varkappa }.  \label{17}
\end{equation}
This is the critical point of the Anderson (M-I) transition. We assume that
the thermodynamic density of states remains constant at the transition point.

The electron motion in metallic phase far enough from the M-I transition has
a diffusion character and the conductivity can be related to the diffusion
coefficient $\mathcal{D}=p_{F}l/3m$ by the Einstein relation: $\sigma =\nu
e^{2}\mathcal{D}\mathit{.}$ One can say that diffusion like ''excitations''
with the spectrum $\omega (q)=iDq^{2}$ propagate in the system. At the point
of the M-I transition normal diffusion terminates and conductivity, together
with $\mathcal{D}\mathit{,}$ turns zero. In accordance with scaling ideas,
the diffusion coefficient can be assumed here to be a power function of $q$: 
$\mathcal{D}(q)\sim q^{z-2},$ with the dynamical critical exponent $z>2$.
The anomalous diffusion excitation spectrum in this case would take the form 
$\omega \sim q^{z}$.

In the insulating phase ($g<g_{c})$ some local, anomalous diffusion,
confined to regions of the scale $\ell ,$ is still possible. It cannot
provide charge transfer through out all the system, so $\mathcal{D}(q=0)=0,$
but for small distances ($q\gtrsim \ell ^{-1}$) anomalous diffusion takes
place. Analogously, in the metallic phase ($g>g_{c})$ the diffusion
coefficient in the vicinity of the transition has an anomalous dependence on 
$q$ for $q\gtrsim \ell ^{-1}$ and weakly depends on it for $q\lesssim \ell
^{-1}.$ So one can conclude that the diffusion coefficient for $q\gtrsim
\ell ^{-1}$ from both sides of the transition has the same $q$-dependence as
for all $q$ in the transition point. It can be written in the form

\begin{equation}
\mathcal{D}(q)=\frac{g}{3m}\left[ \frac{\varphi (q\ell )}{p_{F}\ell }\right]
^{z-2},\;\;\varphi (x)=\left\{ 
\begin{tabular}{l}
$x,\;x\gg 1$ \\ 
$1,\;x\ll 1,g>g_{c}$ \\ 
$0,\;x\ll 1,g<g_{c}$%
\end{tabular}
\right. ,  \label{anomdif}
\end{equation}
where the dimensionless localization length $\ell $, characterizing the
spatial scale near the transition, grows with the approach to the transition
point like

\begin{equation}
\ell (g)=\frac{1}{p_{F}}(g-g_{c})^{-\frac{\varkappa }{z-2}}.
\end{equation}
The critical exponent in this formula is found from the Einstein relation in
the vicinity of the M-I transition.

At finite temperatures, instead of the critical point $g_{c},$a crossover
from metallic to insulating behavior of $\sigma (g)$ takes place. The width
of the crossover region is $\widetilde{g}-g_{c},$where $\widetilde{g}$ is
determined from the relation $\mathcal{D}\ell ^{-2}(\widetilde{g})\sim
E_{F}[p_{F}\ell (\widetilde{g})]^{-z}\sim T$ (we have used the second
asymptotic of (\ref{anomdif})). In this region the diffusion coefficient is

\begin{equation}
\mathcal{D}(T)\sim T\ell ^{2}\sim \frac{T}{p_{F}^{2}}\left( \frac{E_{F}}{T}%
\right) ^{\frac{2}{z}}  \label{dcross}
\end{equation}
and it depends weakly on the $g-g_{c}.$ Beyond this region the picture of
the transition remains the same as at $T=0.$

Let us consider now what happens to superconductivity in the vicinity of the
localization transition. In the mean field approximation (BCS) the
thermodynamic properties of a superconductor do not depend on the character
of the diffusion of excitations. This should be contrasted with the
fluctuation theory, where such a dependence clearly exists. We will show
that the type of superconducting transition depends on the dynamical
exponent $z$. If $z>3$, the transition to superconductivity occurs on the
metallic side of the localization transition (we will refer to such a
transition as S-N transition). If $z<3$, the transition to superconductivity
occurs from the insulating state directly (S-I transition).

Let us study how the superconducting fluctuations affect the transition
under discussion. In spirit of the GL approach fluctuation phenomena in the
vicinity of the transition can be described in the framework of the GL
functional (\ref{fglfourier}). The coherence length in the metallic region,
far enough from the Anderson transition, was reported in Introduction to be
equal $\xi ^{2}=\xi _{c}l=0.42\mathcal{D}/T$ . In the vicinity of the M-I
transition we still believe in the diffusive character of the electron
motion resulting in the pair formation. The only difference from the
previous consideration is the anomalous character of the quasiparticle
diffusion. So in order to describe the superconducting fluctuations
simultaneously near superconducting (in temperature) and Anderson (in $g$)
transitions let us use the GL functional (\ref{fglfourier}) with the $k$%
-dependent diffusion coefficient (\ref{anomdif}).

The value of $Gi$ can be estimated from the expression for the fluctuation
contribution to heat capacity (\ref{fhc}) taken at $\epsilon \sim Gi,$ where
the fluctuation correction reaches the value of the heat capacity jump: 
\begin{equation}
1\sim \frac{T}{\nu }\int \frac{d^{3}q}{\left( TGi+\mathcal{D}(q)q^{2}\right)
^{2}},  \label{Giloc}
\end{equation}
with $T\simeq T_{c}$. Let us approach the M-I transition from the metallic
side. If we are far enough from transition, $Gi$ is small and the integral
in (\ref{Giloc}) is determined by the region of small momenta $\mathcal{D}%
(q)q^{2}\lesssim TGi$ :

\begin{equation}
Gi\sim \frac{T}{\nu ^{2}\mathcal{D}^{3}(q=0)}.
\end{equation}
Two scenarios are possible: $Gi$ becomes of the order of $1$ in the metallic
phase, or it remains small up to the crossover region, where finally reaches
its saturation value. In the first case we can use the second asymptotic of (%
\ref{anomdif}) for $\mathit{D}(q)$ and find:

\begin{equation}
Gi\sim \frac{T_{c}}{E_{F}}(p_{F}\ell )^{3z-6}.
\end{equation}
One can see that $Gi$ becomes of the order of $1$ at $p_{F}\ell _{M}\sim
\left( E_{F}/T\right) ^{\frac{1}{3z-6}}.$ Comparing this value with $%
p_{F}\ell (\widetilde{g})\sim \left( E_{F}/T\right) ^{1/z}$ at the limit of
the crossover region we see that for $z>3$ the first scenario is realized.
Concluding the first scenario discussion we see that the superconducting
critical temperature goes to zero at $\ell =\ell _{M},$ still in the
metallic phase, so at $T=0$ a superconductor-normal phase (S-N) type quantum
phase transition takes place.

The second scenario takes place for $z<3$ when $Gi$ remains small even at
the edge of crossover region, reaching there the value

\begin{equation}
Gi\sim \left( \frac{T_{c}}{E_{F}}\right) ^{\frac{2(3-z)}{z}}\ll 1.
\end{equation}
In the crossover region the diffusion coefficient, and hence $Gi$, almost do
not vary. This is why the temperature of superconducting transition remains
almost frozen with further increase of disorder driving the system through
the Anderson transition. The abrupt growth of $Gi$ and decrease of $T_{c}$
take place when the system finally goes from the crossover to the insulating
region. In the insulator phase the diffusion coefficient $\mathit{D}%
(q\lesssim l^{-1})=0$ and from (\ref{Giloc}) one can find for $Gi:$

\begin{equation}
Gi\sim \sqrt{\frac{E_{F}}{T_{c}}}\frac{1}{(p_{F}\ell )^{3/2}}
\end{equation}
Comparing this result with the Table 1 it is easy to see that it coincides
with the Ginzburg-Levanyuk number for a zero-dimensional granule of size $%
\ell \ll \xi (T).$ Hence we see that in the second scenario the Ginzburg
number reaches $1$ and, respectively, $T_{c}\rightarrow 0$ at $p_{F}\ell
_{I}\sim \left( \frac{E_{F}}{T_{c}}\right) ^{1/3},$ which is far enough from
the M-I transition point. This is why in this case one can speak about the
realization at $T=0$ of a superconductor-insulator (S-I) type quantum phase
transition. The scale $\ell _{I}$ determines the size of the ''conducting''
domains in the insulating phase, where the level spacing reaches the order
of the superconducting gap. It is evident that in the domain of scale $\ell
\lesssim $ $\ell _{I}$ superconductivity cannot be realized.

In the vicinity of a quantum phase transition one can expect the appearance
of non monotonic dependencies of the resistance on temperature and magnetic
field. Indeed, starting from the zero resistance superconducting phase and
increasing temperature from $T=0,$ the system passes through the
localization region, where the resistance is high, to high temperatures
where some hopping charge transfer will decrease the resistance again. The
analogous speculations are applicable to the magnetic field effect: first
the magnetic field ''kills'' superconductivity and increases the resistance,
then it destroys localization and decreases it.

\begin{figure}[tbp]
\epsfxsize=4in
\centerline{\epsfbox{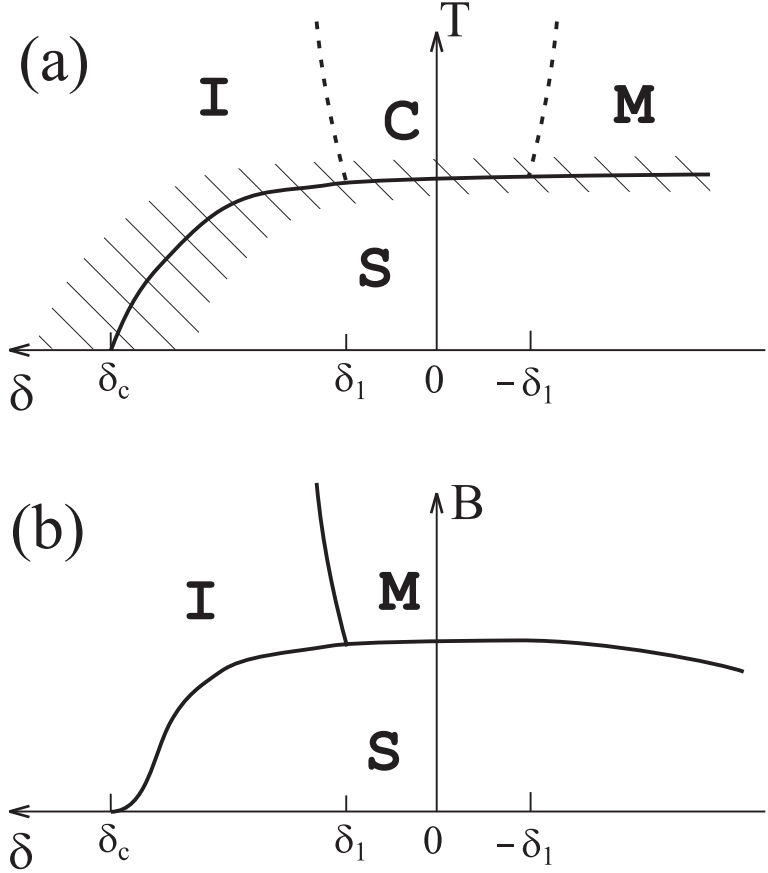}}
\caption{Phase diagram in the temperature -- disorder plane for a
three-dimensional superconductor.}
\label{PD3}
\end{figure}

The phase diagram in the $(T,g)$ plane has the form sketched in Fig.~\ref
{PD3}. For $g_{I}=g_{c}-(T_{c}/E_{F})^{3(z-2)/\varkappa }$, an S-I
transition takes place at $T=0$. Increasing the temperature from $T=0$ in
the region $0\lesssim g\lesssim g_{I}$ we remain in the insulating phase
with exponential dependence of resistance on temperature. For $g_{I}\lesssim
g\lesssim \widetilde{g}_{-}$ at low temperatures $0\leq $ $T<T_{c}(g)$ the
system stays in the superconducting state which goes to the insulating phase
at higher temperatures. In the vicinity of the Anderson transition ($%
\widetilde{g}_{-}\lesssim g\lesssim \widetilde{g}_{+})$ the superconducting
state goes with growth of the temperature to some crossover metal-insulator
state which is characterized by a power decrease of the resistivity with the
increase of temperature. Finally at $g_{c}\lesssim g$ the superconducting
phase becomes of the BCS type and at $T=T_{c}$ it goes to a metallic phase.

The phase diagram in the magnetic field -- disorder plane is similar to that
in the $(T,g)$ plane with the only difference that at $T=0$ there is no
crossover region, instead a phase transition takes place.

\subsection{ 2D superconductors}

\subsubsection{Preliminaries.}

As was demonstrated in Section 4, according to the conventional theory of
paraconductivity, the sheet conductivity in the vicinity of the
superconducting transition is given by a sum of the electron residual
conductivity $ge^{2}$ (Fermi part) and the conductivity of the Cooper pair
fluctuations (Bose part) (see ~(\ref{drudeal})). This expression is valid in
the Ginzburg Landau region when the second term is a small correction to the
first one. The width of the critical region can be determined from the
requirement of equality of both contributions in (\ref{drudeal})\footnote{%
It is worth mentioning that this definition of the Ginzburg Levanyuk number (%
$\widetilde{Gi}_{(2d)}=\frac{\pi }{8p_{F}l}$) agrees with that defined from
the heat capacity fluctuations ($Gi_{(2d)}=\frac{0.3}{p_{F}l}$).}:

\begin{equation}
\epsilon _{cr}=\frac{1}{16g}=1.3Gi_{(2d)}.
\end{equation}
In accordance with general scaling ideas one can believe that inside the
fluctuation region the conductivity should obey the form: 
\begin{equation}
\sigma (T)=ge^{2}f\left( \frac{\epsilon }{Gi_{(2d)}}\right) .  \label{4si}
\end{equation}
Concerning the scaling function $f(x)$, we know its asymptotes in the mean
field region $(x\gg 1)$ and just above the BKT transition \cite{B73,TK74}: 
\begin{equation}
f(x)=\left\{ 
\begin{array}{ll}
1+x^{-1}, & x\gg 1 \\ 
\exp \left[ -b(x-x_{\mathrm{BKT}})^{-1/2})\right] , & x\rightarrow x_{%
\mathrm{BKT}}=-4
\end{array}
\right. .  \label{5si}
\end{equation}
The BKT transition temperature $T_{c}^{\mathrm{BKT}}$ is determined by 
\begin{equation}
n_{s2}(T_{c})=\frac{4mT_{c}}{\pi }.  \label{BTK}
\end{equation}
and one can find its value by comparing the superfluid density $%
n_{s}$~from (\ref{BTK}) with that found in the BCS scheme: 
\begin{equation}
T_{c}^{\mathrm{BKT}}=T_{c0}(1-4Gi).  \label{7}
\end{equation}
Here we assumed that the Ginzburg parameter is small, so that the BKT
transition temperature does not deviate much from the mean-field BCS
transition temperature $T_{c0}$.

\subsubsection{Boson mechanism of the $T_{c}$ suppression.}

The classical and quantum fluctuations reduce $n_{s}$ and therefore,
suppress $T_{c}^{\mathrm{BKT}}$. At some $g=g_{c}\sim 1$, the superfluid
density $n_{s}$, and simultaneously $T_{c}^{\mathrm{BKT}},$ go to zero. In
the vicinity of this critical concentration of impurities $T_{c}^{\mathrm{BKT%
}}\ll T_{c0}$. Thus a wide new window of intermediate temperatures $T_{c}^{%
\mathrm{BKT}}\ll T\ll T_{c0}$ opens up. In this window, according the
dynamical quantum scaling conjecture~\cite{M}, one finds 
\begin{equation}
\sigma =e^{2}\varphi \left( \frac{T}{T_{c}^{BKT}}\right) .  \label{8}
\end{equation}
At $T-T_{c}^{\mathrm{BKT}}\ll T_{c}^{\mathrm{BKT}}$ the
Berezinski-Kosterlitz-Thouless law (\ref{4si})-(\ref{5si}) should hold, so $%
\varphi (x)=f(x)$ and is exponentially small. In the intermediate region $%
T_{c}^{\mathrm{BKT}}\ll T\ll T_{c}^{\mathrm{BCS}}$ the duality hypothesis
gives $\varphi (x)=\pi /2.$ Let us derive this relation.

We will start from the assumption that in the region $T_{c}^{\mathrm{BKT}%
}\ll T\ll T_{c0}$ the conductivity is a universal function of temperature
which does not depend on the pair interaction type. Being in the framework
of the classical approach, let us suppose that in a weak electric field
pairs move with the velocity $\mathbf{v}=\mathbf{F}/\eta \mathbf{,}$ where $%
\mathbf{F}=2e\mathbf{E}$ is the force acting on the pairs. The current
density $\mathbf{j}=2en\mathbf{v=}\sigma \mathbf{E,}$ (here $n$ is the pair
density), so one can relate the conductivity with the effective viscosity $%
\eta $: $\sigma \mathbf{=}4e^{2}n/\eta \mathbf{.}$

Let us recall that we are dealing with a quantum fluid, so another,
superconducting, view on the problem of its motion near the quantum phase
transition exists. One can say that with the increase of $Gi$ the role of
quantum fluctuations grows too and fluctuation vortices carrying the
magnetic flux quantum $\Phi _{0}=\pi /e$ are generated. With electric
current flow in the system the Lorentz (Magnus) force acts on a vortex: $%
F=j\Phi _{0}.$ The electric field is equal to the rate of magnetic flux
transfer, i.e. to the density of the vortex current: $\mathbf{E=}\Phi
_{0}n_{v}\mathbf{v}_{v}\mathbf{=}\Phi _{0}n_{v}\mathbf{F/\eta _{v},}$ where $%
n_{v}$ is the density and $\eta _{v}$ is the viscosity of the vortex liquid.
As a result $\mathbf{E=}\Phi _{0}^{2}n_{v}\mathbf{j/\eta _{v}=j/}\sigma 
\mathbf{.}$ So one can conclude that for vortices the velocity is
proportional to the voltage, and the force is proportional to the current.
For Cooper pairs (bosons) the situation is exactly the opposite.

The duality hypothesis consists in the assumption that at the critical point
the pair and the vortex liquid density flows are equal: $n_{v}\mathbf{v}_{v}%
\mathbf{=}$ $n\mathbf{v.}$ Comparing these quantities, expressed in terms of
the conductivity from the above relations, one can find a universal value
for the conductivity at the critical point

\begin{equation}
\sigma =\frac{2e}{\Phi _{0}}=\frac{2e^{2}}{\pi }.
\end{equation}

One can restrict oneself to a less strong duality hypothesis, supposing the
product $n\eta =CT^{\delta }$ with a universal $\delta $ exponent both for
the pair and the vortex liquids, while the constant $C$ for them is
different. In this case, based on duality, is possible to demonstrate that $%
\delta =0$ and the conductivity is temperature independent up to $T_{c0}$
but its value is not universal any more and can vary from one sample to
another.

To conclude, let us emphasize that in the framework of the boson scenario of
superconductivity suppression, the BCS critical temperature is changed
insignificantly, while the ``real'' superconducting transition temperature $%
T_{c}^{\mathrm{BKT}}\rightarrow 0$.

\subsubsection{ Fermion mechanism of $T_{c}$ suppression.}

Apart from the above fluctuation (boson) mechanism of the suppression of the
critical temperature in the $2D$ case, there exists another, fermionic
mechanism. The suppressed electron diffusion results in a poor dynamical
screening of the Coulomb repulsion which, in turn, leads to the
renormalization of the inter-electron interaction in the Cooper channel. and
hence to the dependence of the critical temperature on the value of the
high-temperature sheet resistivity of the film. As long as the correction to
the non-renormalized BCS transition temperature $T_{c0}$ is still small, one
finds~\cite{O73,fu,VD86}: 
\begin{equation}
T_{c}=T_{c0}\left( 1-\frac{1}{12\pi ^{2}g}\ln ^{3}\frac{1}{T_{c0}\tau }%
\right) .  \label{10}
\end{equation}
At small enough $T_{c0}$ this mechanism of critical temperature suppression
turns out to be the principal one. The suppression of $T_{c}$ down to zero
in this case may happen in principle even at $g\gg 1.$ A renormalization
group analysis gives \cite{F87} the corresponding critical value of
conductance 
\begin{equation}
g_{c}=\left( \frac{1}{2\pi }\ln \frac{1}{T_{c0}\tau }\right) ^{2}.
\label{11}
\end{equation}

Here we should recall that the typical experimental~\cite{gm} values of $%
g_{c}$ are in the region $g_{c}\sim 1-2$, and do not differ dramatically
from the predictions of the boson duality assumption $g_{c}=\frac{2}{\pi }$.
If one attempts to explain the suppression of $T_{c}$ within the fermion
mechanism, one should assume that $\ln \frac{1}{T_{c0}\tau }>5.$ Then,
according to Eq.~(\ref{11}), $g_{c}>2/\pi $ and the boson mechanism is not
important. On the contrary, if $\ln \frac{1}{T_{c0}\tau }<4,$ then Eq.~(\ref
{10}) gives a small correction for $T_{c}$ even for $g_{c}=2/\pi $ and the
fermion mechanism becomes unimportant. The smallness of the critical
temperature $T_{c}$ compared to the Fermi energy is the cornerstone of the
BCS theory of superconductivity and it is apparently satisfied even in high-$%
T_{c}$ materials. Nevertheless it is necessary to use the theoretically
large logarithmic parameter with care, if one needs $\ln \frac{1}{T_{c0}\tau 
}$ to be as large as $4$.

\section[Microscopic derivation of the TDGL equation]{Microscopic derivation
of\ the Time-Dependent Ginzburg-Landau equation}

\subsection{Preliminaries}

We have seen above how the phenomenological approach based on the GL
functional allows one to describe fluctuation Cooper pairs (Bose particles)
near the superconducting transition and to account for their contribution to
different thermodynamical and transport characteristics of the system. Now
we pass to the discussion of the microscopic description of fluctuation
phenomena in superconductors. The development of the microscopic approach is
necessary for the following reasons:

1. This description permits microscopic determination of the values of the
phenomenological parameters of the GL theory.

2. This method is more powerful than the phenomenological GL approach and
permits treatment of fluctuation effects quantitatively even far from the
transition point and for magnetic fields strong as $H_{c2},$ taking into
account the contributions of dynamical and short wavelength fluctuations.

3. The electron energy relaxation times in metals are relatively large $%
(\tau _{\varepsilon }\gg \hbar /T)$ which causes the electron low frequency
dynamics to be sensitive to the nearness to the superconducting transition.
This is why the temperature dependence of fluctuation corrections can be
determined generally speaking not only by the Cooper pair motion but also by
changes in the single-electron properties.

4. There are some fluctuation phenomena in which the direct Cooper pair
contribution is considerably suppressed or even absent altogether. Among
them we can mention the nuclear magnetic relaxation rate, tunnel
conductivity, c-axis transport in strongly anisotropic layered metals,
thermoelectric power and heat conductivity where the fluctuation pairing
manifests itself by means of the indirect influence on the properties of the
single-particle states of electron system.

Formally in the above consideration averaging over the superconducting order
parameter has been accomplished by means of a functional integration over
all its possible bosonic field configurations. In this description we have
dealt with the fluctuation Cooper pair related effects only and the method
of the functional integration turned out to be simple and effective for
their description. In the following Sections we will develop the
diagrammatic method of Matsubara temperature Green functions which is more
adequate for the description of the properties of a Fermi system of
interacting electrons.

\subsection{The Cooper channel of electron-electron interaction}

Let us start the microscopic description of fluctuation phenomena in\ a
superconductor from the electron Hamiltonian. We will choose it in the
simple BCS form\footnote{%
We suppose that reader is familiar with the BCS formulation of the theory of
superconductivity (see for example, \cite{AGD}).}:

\begin{equation}
\mathcal{H=}\sum_{\mathbf{p,}\sigma }E\mathbf{(p)}\widetilde{\psi }_{\mathbf{%
p,}\sigma }^{+}\widetilde{\psi }_{\mathbf{p,}\sigma }+\mathit{g}\sum_{%
\mathbf{p,p}^{\prime },\mathbf{q},\sigma ,\sigma ^{\prime }}\widetilde{\psi }%
_{\mathbf{p+q,}\sigma }^{+}\widetilde{\psi }_{-\mathbf{p,-}\sigma }^{+}%
\widetilde{\psi }_{-\mathbf{p}^{\prime }\mathbf{,-}\sigma ^{\prime }}%
\widetilde{\psi }_{\mathbf{p}^{\prime }+\mathbf{q,}\sigma ^{\prime }}.
\label{BCSham}
\end{equation}
The momentum conservation law side by side with singlet pairing are already
taken into account in the interaction term. Here $E\mathbf{(p)}$ is the
quasiparticle spectrum of the normal metal; $\mathit{g}$ is the negative
constant of electron-electron attraction which is supposed to be momentum
independent and different from zero in a narrow domain of momentum space in
the vicinity of the Fermi surface where

\[
p_{F}-\frac{\omega _{D}}{v_{F}}<|\mathbf{p}|,|\mathbf{p}^{\prime }|<p_{F}+%
\frac{\omega _{D}}{v_{F}}. 
\]
$\widetilde{\psi }_{\mathbf{p,}\sigma }^{+}$ and $\widetilde{\psi }_{\mathbf{%
p,}\sigma }$ are the creation and annihilation field operators in the
Heisenberg representation, so the first term is just the kinetic energy of
the non-interacting Fermi gas. The interaction term is chosen in the
traditional form characteristic for the electron-phonon mechanism of
superconductivity\footnote{%
Fluctuations in the framework of more realistic Eliashberg \cite{E60} model
of superconductivity were studied by B.Narozhny \cite{N94}. He demonstrated
that \ the strong coupling does not change drastically the results of the
weak coupling approximation. The critical exponents turn out to be exactly
the same as in the framework of the GL theory, which provides an adequate
description of paraconductivity in strong coupling superconductors. The
robustness of the critical exponents and their dependence in GL region on
the space dimensionallity only was stressed in \cite{BVS} in relation to the
discussion of the paraconductivity at the edge of the
superconductor-insulator transition.}.

For the description of the properties of an interacting electron system with
the Hamiltonian (\ref{BCSham}) we will use the formalism of the Matsubara
temperature diagrammatic technique. The state of a non-interacting
quasiparticle is described by its Green function

\begin{equation}
G(\mathbf{p},\varepsilon _{n})=\frac{1}{i\varepsilon _{n}-\xi (\mathbf{p})},
\label{1elec}
\end{equation}
where $\varepsilon _{n}=(2n+1)\pi T$ is a fermion Matsubara frequency and $%
\xi (\mathbf{p})=E\mathbf{(p)-}E_{F}$ is the quasiparticle energy measured
from the Fermi level.

\begin{figure}[tbp]
\centerline {\epsfig {file=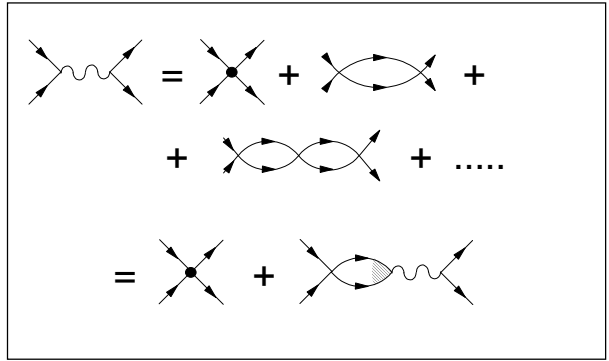,width=8cm}}
\caption{The Dyson equation for the fluctuation propagator (wavy line) in
the ladder approximation. Solid lines represent one-electron Green
functions, bold points correspond to the model electron-electron
interaction. }
\label{FluPro}
\end{figure}

As it is well known the effective electron-electron attraction leads to a
reconstruction of the ground state of the electron system which formally
manifests itself by the appearance at the critical temperature of a pole in
the two particle Green function

\[
\mathcal{L}(p,p^{\prime },q)=<T_{\tau }[\widetilde{\psi }_{p+q\mathbf{,}%
\sigma }\widetilde{\psi }_{-p\mathbf{,-}\sigma }\widetilde{\psi }_{p^{\prime
}+q\mathbf{,}\sigma ^{\prime }}^{+}\widetilde{\psi }_{-p^{\prime }\mathbf{,-}%
\sigma ^{\prime }}^{+}]>, 
\]
where $T_{\tau }$ is the time ordering operator and $4D$ vector notations
are used \cite{AGD}. As it well known the two particle Green function can be
expressed in terms of the\textbf{\ }vertex part \cite{AGD}. In the case
under consideration it is the vertex part of the electron-electron
interaction in the Cooper channel $L(\mathbf{q},\Omega _{k}),$ which will be
called below the fluctuation propagator\textbf{. }The Dyson equation for $L(%
\mathbf{q},\Omega _{k}),$ accounting for the e-e attraction in the ladder
approximation, is represented graphically in Fig.\ref{FluPro} . It can be
written down analytically as

\begin{equation}
L^{-1}(\mathbf{q},\Omega _{k})=\mathit{g}^{-1}-\Pi (\mathbf{q},\Omega _{k}),
\label{dyspro}
\end{equation}
where the polarization operator\textbf{\ }$\Pi (\mathbf{q},\Omega _{k})$ is
defined as a loop of two single-particle Green functions:

\begin{equation}
\Pi (\mathbf{q},\Omega _{k})=T\sum_{\varepsilon _{n}}\int {\frac{{d^{3}}%
\mathbf{p}}{{(2\pi )^{3}}}}G(\mathbf{p+q},\varepsilon _{n+k})G(-\mathbf{p}%
,\varepsilon _{-n}).  \label{genprop}
\end{equation}

Let us emphasize, that the two quantities introduced above, $\mathcal{L}%
\left( p,p^{\prime },q\right) $ and $L(q),$ are closely connected with each
other. The former being integrated over momenta $p$ and $p^{\prime }$
becomes an average of the product of two order parameters: 
\begin{equation}
\int dpdp^{\prime }\mathcal{L}\left( p,p^{\prime },q\right) ={\frac{1}{g^{2}}%
}\,\left\langle \Delta _{q}\Delta _{q}^{\ast }\right\rangle ,  \label{dd}
\end{equation}
where $\Delta _{q}$ is the superconducting gap proportional to the
condensate wave function $\Psi $. Thus, this quantity represents the
coefficient in the linear term in the GL equation. In terms of the
polarization operator introduced above it can be written as 
\[
\int dpdp^{\prime }\mathcal{L}\left( p,p^{\prime },q\right) \propto {\frac{%
\Pi }{1-g\Pi }.} 
\]
Comparing this equation with Eq.(\ref{dyspro}) for the fluctuation
propagator we see that the corresponding expressions are very similar. After
analytical continuation to the real frequencies the fluctuation propagator $%
L(q,i\Omega )$ coincides with the quantity defined by Eq.(\ref{dd}) (up to a
constant).

One can calculate the propagator (\ref{dyspro}) using the one-electron Green
functions of the normal metal (\ref{1elec}). For sake of convenience of
future calculations let us define the correlator of two one-electron Green
functions

\begin{eqnarray}
\mathcal{P}(\mathbf{q},\varepsilon _{1},\varepsilon _{2}) &=&\int \frac{d^{3}%
\mathbf{p}}{\left( 2\pi \right) ^{3}}G(\mathbf{p+q},\varepsilon _{1})G(%
\mathbf{-p},\varepsilon _{2})=  \label{a2} \\
&=&2\pi \nu \Theta (-\varepsilon _{1}\varepsilon _{2})\left\langle \frac{1}{%
|\varepsilon _{1}-\varepsilon _{2}|+i\Delta \xi (\mathbf{q,p})|_{_{\epsilon 
\mathbf{(p)=}E_{F}}}}\right\rangle _{F.S.},  \nonumber
\end{eqnarray}
where $\Theta (-\varepsilon _{1}\varepsilon _{2})$ is Heavyside step
function, $\nu $ is the one-electron density of states, $<>_{F.S.}=\int 
\frac{d\Omega _{\mathbf{p}}}{4\pi }$ means the averaging over the Fermi
surface,

\begin{equation}
\Delta \xi (\mathbf{q,p})|_{_{\epsilon \mathbf{(p)=}E_{F}}}=[\xi (\mathbf{q+p%
})-\xi (-\mathbf{p})]|_{\epsilon \mathbf{(p)=}E_{F}}\approx (\mathbf{v}_{%
\mathbf{p}}\mathbf{q)}_{\xi (\mathbf{p})=0}.  \nonumber
\end{equation}
The last approximation is valid not too far from the Fermi surface, i.e.
when $(\mathbf{v}_{\mathbf{p}}\mathbf{q)}_{\xi (\mathbf{p})=0}\ll E_{F}.$

It is impossible to carry out the angular averaging in (\ref{a2}) for a
general anisotropic spectrum. Nevertheless in the following calculations of
fluctuation effects in the vicinity of critical temperature only small
momenta $\mathbf{v}_{\mathbf{p}}\mathbf{q}\ll T$ will be involved in the
integrations, so we can restrict our consideration here to this region,
where one can expand the integrand in powers of $\mathbf{v}_{\mathbf{p}}%
\mathbf{q}$. Indeed, the presence of $\Theta (-\varepsilon _{1}\varepsilon
_{2})$ leaves the difference of the two fermionic frequencies in (\ref{a2})
to be of the order of the temperature which permits this expansion. The
first term in $\mathbf{v}_{\mathbf{p}}\mathbf{q}$ will evidently be averaged
out, so with quadratic accuracy one can find:

\begin{equation}
\mathcal{P}(\mathbf{q},\varepsilon _{1},\varepsilon _{2})=2\pi \nu \frac{%
\Theta (-\varepsilon _{1}\varepsilon _{2})}{|\varepsilon _{1}-\varepsilon
_{2}|}\left( 1-2\frac{\langle (\mathbf{v}_{\mathbf{p}}\mathbf{q})^{2}\rangle
_{F.S.}}{|\varepsilon _{1}-\varepsilon _{2}|^{2}}\right) .  \label{pclean}
\end{equation}

Now one can calculate the polarization operator

\begin{eqnarray}
{\Pi }(\mathbf{q},\Omega _{k}) &=&T\sum_{\varepsilon _{n}}\mathcal{P}(%
\mathbf{q},\varepsilon _{n+k},\varepsilon _{-n})=  \label{b4} \\
&=&\nu \left[ \sum_{n\geq 0}\frac{1}{n+1/2+\frac{|\Omega _{k}|}{4\pi T}}{\ }%
-2\frac{\langle (\mathbf{v}_{\mathbf{p}}\mathbf{q})^{2}\rangle _{F.S.}}{%
(4\pi T)^{2}}\sum_{n=0}^{\infty }\frac{1}{\left( n+1/2+\frac{|\Omega _{k}|}{%
4\pi T}\right) ^{3}}\right] .  \nonumber
\end{eqnarray}
The calculation of the sums in (\ref{b4}) can be carried out in terms of the
logarithmic derivatives of the $\Gamma $-function $\psi ^{(n)}(x).$ It worth
mentioning that the first sum is well known in the BCS\ theory, one can
recognize in it the so-called ''Cooper logarithm''; its logarithmic
divergence at the upper limit $(\psi (x\gg 1)\approx \ln x)$ is cut off by
the Debye energy ($N_{\max }=\frac{\omega _{D}}{2\pi T})$ and one gets$:$%
\begin{eqnarray}
\frac{1}{\nu }\Pi (\mathbf{q},\Omega _{k}) &=&\psi \left( \frac{1}{2}+\frac{%
|\Omega _{k}|}{4\pi T}+\frac{\omega _{D}}{2\pi T}\right) -\psi \left( \frac{1%
}{2}+\frac{|\Omega _{k}|}{4\pi T}\right) -  \label{b6} \\
&&-\frac{\langle (\mathbf{v}_{\mathbf{p}}\mathbf{q})^{2}\rangle _{F.S.}}{%
(4\pi T)^{2}}\psi ^{^{\prime \prime }}\left( \frac{1}{2}+\frac{|\Omega _{k}|%
}{4\pi T}\right) .  \nonumber
\end{eqnarray}
The critical temperature in the BCS theory is determined as the temperature $%
T_{c}$ at which the pole of $L(0,0,T_{c})$ occurs 
\[
L^{-1}(\mathbf{q=}0,\Omega _{k}=0,T_{c})=\mathit{g}^{-1}-\Pi (0,0,T_{c})=0, 
\]
\begin{equation}
T_{c}=\frac{2\mathit{\gamma }_{E}}{\pi }\omega _{D}\exp \left( -\frac{1}{\nu 
\mathit{g}}\right) ,  \label{gvst}
\end{equation}
where\ $\mathit{\gamma }_{E}=1.78$\ \ is the Euler constant. Introducing the
reduced temperature $\epsilon =\ln (\frac{T}{T_{c}})\ $ one can write the
propagator as 
\begin{equation}
L^{-1}(\mathbf{q},\Omega _{k})=-\nu \left[ \epsilon +\psi (\frac{1}{2}+\frac{%
|\Omega _{k}|}{4\pi T})-\psi (\frac{1}{2})-\frac{\langle (\mathbf{v}_{%
\mathbf{p}}\mathbf{q})^{2}\rangle _{F.S.}}{(4\pi T)^{2}}\psi ^{^{\prime
\prime }}\left( \frac{1}{2}+\frac{|\Omega _{k}|}{4\pi T}\right) \right] .
\label{propa}
\end{equation}
We found (\ref{propa}) for bosonic imaginary Matsubara frequencies $i\Omega
_{k}=2\pi iTk.$ These frequencies are necessary for the calculation of
fluctuation contributions to any thermodynamical characteristics of the
system.

In the vicinity of the transition point one can restrict oneself in
summations of the expressions with $L(\mathbf{q},\Omega _{k})$ over
Matsubara frequencies to the so-called static approximation, taking into
account the term with $\Omega _{k}=0$ only, which turns out to be the most
singular term in $\epsilon \ll 1$ . This approximation physically means that
the product of Heisenberg field operators $\widetilde{\psi }_{p\mathbf{,}%
\sigma }\widetilde{\psi }_{-p\mathbf{,-}\sigma }$ appears here like a
classical field $\Psi ,$ which in the phenomenological approach describes
the Cooper pair wave function and in the vicinity of critical temperature is
proportional to the fluctuation order parameter. Having in mind namely this
GL region of temperatures we restricted ourselves above by the assumption of
small momenta $\mathbf{v}_{\mathbf{p}}\mathbf{q}\ll T$. In these conditions
the static propagator reduces to

\begin{equation}
L(\mathbf{q},0)=-\frac{1}{\nu }\frac{1}{\epsilon +\xi ^{2}\mathbf{q}^{2}}.
\end{equation}
\ With an accuracy of a numerical factor and the total sign this correlator
coincides with the expression for $\left\langle |\Psi _{%
\mathbf{q}}|^{2}\right\rangle $. By this expression we also have finally
obtained the microscopic value of the coherence length $\xi $ for a clean
superconductor with an isotropic $D$-dimensional Fermi surface which was
often mentioned previously (compare with (\ref{xic}))

\begin{equation}
\xi _{(D)}^{2}=\frac{7\zeta (3)\mathbf{v}_{\mathbf{F}}^{2}}{16D\pi ^{2}T^{2}}%
..
\end{equation}

In order to describe the fluctuation contributions to transport phenomena
one has to start from the analytical continuation of the propagator (\ref
{propa}) from the discrete set of $\Omega _{k}\geq 0$ to the whole upper
half-plane of imaginary frequencies. The analytical properties of $\psi
^{(n)}(x)-$functions (which have poles at $x=0,-1,-2...$) permit one to
obtain the retarded propagator $L^{R}(\mathbf{q},-i\Omega )$ by simple
substitution $i\Omega _{k}\rightarrow \Omega $ . For small $\Omega \ll T$
the $\psi -$functions can be expanded in $-i\Omega /4\pi T$ and the
propagator acquires the simple pole form :

\begin{equation}
L^{R}(\mathbf{q},\Omega )=-\frac{1}{\nu }\frac{1}{-\frac{i\pi }{8T}\Omega
+\epsilon +\xi ^{2}\mathbf{q}^{2}}=\frac{8T}{\pi \nu }\frac{1}{i\Omega
-\left( \tau _{GL}^{-1}+\frac{8T}{\pi }\xi ^{2}\mathbf{q}^{2}\right) }.
\label{anafp}
\end{equation}
This expression provides us with the microscopic value of the GL relaxation
time $\tau _{GL}=\frac{\pi }{8(T-T_{c})},$ widely used above in the
phenomenological theory. Moreover, comparison of the microscopically derived
(\ref{anafp}) with the phenomenological expressions (\ref{hamiso}), (\ref
{propagator}) and (\ref{lmo}) shows that $\alpha T_{c}=\nu $ and $\gamma
_{GL}=\pi \nu /8T_{c}.$

In evaluating $L(\mathbf{q},\Omega _{k})$ we neglected the effect of
fluctuations on the one-electron Green functions. This is correct when
fluctuations are small, i.e. not too near to the transition temperature. The
exact criterion of this approximation will be discussed in the following.

\subsection{Superconductor with impurities}

\subsubsection{Account for impurities.}

In order to study fluctuations in real systems like superconducting alloys
or high temperature superconductors one has to perform an impurity average\
in the graphical equation for the fluctuation propagator (see Fig. \ref
{FluPro}).\ This procedure can be done in the framework of the
Abrikosov-Gorkov approach \cite{AGD}, which we shortly recall below.

\begin{figure}[tbp]
\centerline{\epsfig {file=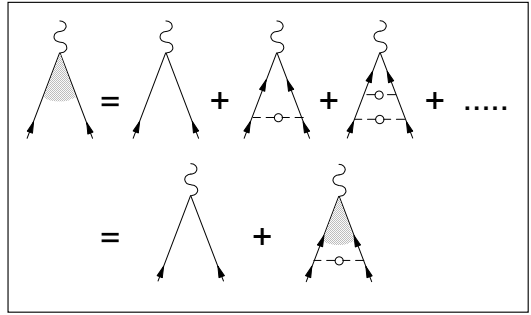,width=10cm}}
\caption{The equation for the vertex part $\protect\lambda (q,\protect\omega
_{1},\protect\omega _{2})$ in the ladder approximation. Solid lines
correspond to bare one-electron Green functions and dashed lines to the
impurity random potential correlators. }
\label{cooperon}
\end{figure}

Let us start from the equation for the electron Green function in the
potential of impurities $U(\mathbf{r})$:

\begin{equation}
\left( E-U(\mathbf{r})-\widehat{H}\right) G_{E}(\mathbf{r,r}^{\prime
})=\delta (\mathbf{r-r}^{\prime })
\end{equation}
If we solve this equation using the perturbation theory for the impurity
potential and average the solution, then the average product of two Green
functions, can be presented as series, each term of which is associated with
a graph drawn according the rules of diagrammatic technique (see Fig. \ref
{cooperon}). In this technique solid lines correspond to bare Green
functions and dashed lines to random potential correlators. We assume that
the impurity system random potential $U(r)$\ is distributed according to the
Gauss $\delta -$correlated law. Then all the correlators can be represented
as the products of pair correlators

\begin{equation}
\left\langle U(r)\right\rangle =0,\;\left\langle U(r)U(r^{\prime
})\right\rangle =\left\langle U^{2}\right\rangle \delta (r-r^{\prime }),
\label{impcor}
\end{equation}
where the angle brackets denote averaging over the impurity configuration.
Equation \ref{impcor} corresponds to the Born approximation for the electron
interaction with short range impurities, and $\left\langle
U^{2}\right\rangle =C_{imp}\left( \int V(\mathbf{r})d\mathbf{r}\right) ^{2}$
where $C_{imp}$ is the impurity concentration and $V(\mathbf{r})$ is the
potential of the single impurity.

In conductors (far enough from the metal-insulator transition) the mean free
path is much greater than the electron wavelength $l\gg \lambda =2\pi /p_{F}$
(which in practice means the mean free path up to tens of interatomic
distances). As is well known \cite{AGD} for the electron spectra with
dimensionality $D>1$ the angular integration in momentum space reduces
considerably the contribution of the diagrams with intersecting impurity
lines what permits to omit them to the leading approximation in $\left(
p_{F}l\right) ^{-1}$. In this approximation the one-electron Green function
keeps the same form as the bare one (\ref{1elec}) with the only substitution 
\begin{equation}
\varepsilon _{n}\Rightarrow \widetilde{\varepsilon }_{n}=\varepsilon _{n}+%
\frac{1}{2\tau }sign(\varepsilon _{n}),  \label{epstild}
\end{equation}
where $1/\tau =2\pi \nu \left\langle U^{2}\right\rangle $ is the frequency
of elastic collisions.

Another effect of the coherent scattering on the same impurity by both
electrons forming a Cooper pair is the renormalization of the vertex part $%
\lambda (\mathbf{q},\varepsilon _{1},\varepsilon _{2})$ in the
particle-particle channel. Let us demonstrate the details of its
calculation. The renormalized vertex $\lambda (\mathbf{q},\varepsilon
_{1},\varepsilon _{2})$ is determined by a graphical equation of the ladder
type (see Fig.~\ref{cooperon} ). Here after the averaging over the impurity
configurations the value $\left\langle U^{2}\right\rangle =\frac{1}{2\pi \nu
\tau }$ is associated with the dashed line. In the momentum representation
this, generally speaking, integral equation is reduced to the algebraic one

\begin{equation}
\lambda ^{-1}(\mathbf{q},\varepsilon _{1},\varepsilon _{2})=1-\frac{1}{2\pi
\nu \tau }\mathcal{P}(\mathbf{q},\widetilde{\varepsilon }_{1},\widetilde{%
\varepsilon }_{2}),  \label{lapi}
\end{equation}
where $\mathcal{P}(\mathbf{q},\widetilde{\varepsilon }_{1},\widetilde{%
\varepsilon }_{2})$ was defined above by (\ref{a2}).

Now one has to perform a formal averaging of the general expression (\ref{a2}%
) over the Fermi surface ($\langle {\ ...}\rangle _{F.S.}$). Restricting
consideration to small momenta

\begin{equation}
\Delta \xi (\mathbf{q,p})|_{|\mathbf{p}|=p_{F}}\ll |\widetilde{\varepsilon }%
_{1}-\widetilde{\varepsilon }_{2}|.  \label{a4}
\end{equation}
the calculation of $\lambda (\mathbf{q},\omega _{1},\omega _{2})$ for the
practically important case of an arbitrary spectrum can be done analogously
to (\ref{pclean}). Indeed, expanding the denominator of (\ref{a2}) one can
find 
\begin{equation}
\lambda (\mathbf{q},\omega _{1},\omega _{2})=\frac{|\widetilde{\varepsilon }%
_{1}-\widetilde{\varepsilon }_{2}|}{|\varepsilon _{1}-\varepsilon _{2}|+%
\frac{\langle (\Delta \xi (\mathbf{q,p})|_{|\mathbf{p}|=p_{F}})^{2}\rangle
_{F.S.}}{\tau |\tilde{\omega}_{1}-\tilde{\omega}_{2}|^{2}}\Theta
(-\varepsilon _{1}\varepsilon _{2})}.  \label{lamb}
\end{equation}
It is easy to see that assumed restriction on momenta is not too severe and
is almost always satisfied in calculations of \ fluctuation effects at
temperatures near $T_{c}.$ In this region of temperatures the effective
propagator momenta are determined by $|\mathbf{q}|_{eff}\sim \lbrack \xi
^{GL}(T)]^{-1}=\xi ^{-1}\sqrt{\epsilon }\ll \xi ^{-1},$ while the Green
function $\mathbf{q}$-dependence becomes important for much larger momenta $%
q\sim \min \{\xi ^{-1},l^{-1}\},$ which is equivalent to the limit of the
condition (\ref{a4}).

\begin{figure}[tbp]
\centerline {\includegraphics[width=12cm]{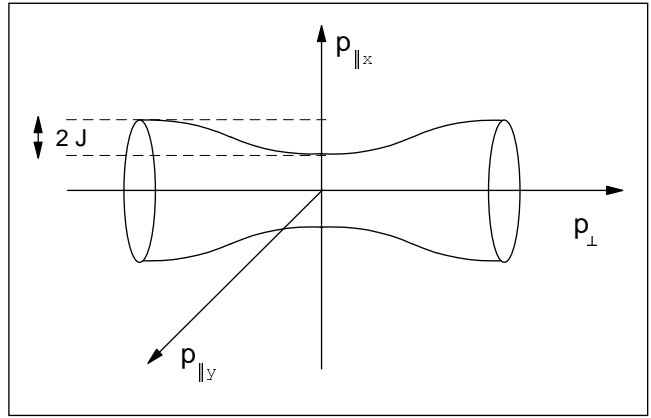}}
\caption{The Fermi surface in the form of a corrugated cylinder}
\label{fcorrug}
\end{figure}

The average in (\ref{lamb}) can be calculated for some particular types of
spectra. For example in the cases of $2D$ and $3D$ isotropic spectra it is
expressed in terms of the diffusion coefficient $\mathcal{D}_{(D)}$ : 
\begin{equation}
\langle (\Delta \xi (\mathbf{q,p})|_{|\vec{p}|=p_{F}})^{2}\rangle
_{F.S.(D)}=\tau ^{-1}\mathcal{D}_{(D)}q^{2}=\frac{v_{F}^{2}q^{2}}{D}.
\label{a3}
\end{equation}
Another important example is already familiar case of quasi-two-dimensional
electron motion in a layered metal: 
\begin{equation}
\xi (\mathbf{p})=E(\mathbf{p}_{\parallel })+J\cos (p_{z}s)-E_{F},  \label{d1}
\end{equation}
where $E(\mathbf{p}_{\parallel })=\mathbf{p}_{\parallel }^{2}/(2m)$, $%
\mathbf{p}\equiv (\mathbf{p}_{\parallel },p_{z})$, $\mathbf{p}_{\parallel
}\equiv (p_{x},p_{y})$, $J$ is the effective nearest-neighbor interlayer
hopping energy for quasiparticles. We note that $J$ characterizes the width
of the band in the $c$-axis direction taken in the strong-coupling
approximation and can be identified with the effective energy of electron
tunneling between planes\textit{\ }(see (\ref{r1}) and footnote 14). The
Fermi surface, defined by the condition $\xi (\mathbf{p})=0,$ is a
corrugated cylinder (see Fig. \ref{fcorrug}). In this case the average (\ref
{a3}) is written in a more sophisticated form: 
\begin{equation}
\langle (\Delta \xi (\mathbf{q,p})|_{|\vec{p}|=p_{F}})^{2}\rangle _{F.S.}={\ 
\frac{1}{2}}(v_{F}^{2}\mathbf{q}^{2}+4J^{2}\sin ^{2}(q_{z}s/2))=\tau ^{-1}%
\widehat{\mathcal{D}}q^{2},  \label{xi2lay}
\end{equation}
where we have introduced the definition of the generalized diffusion
operator $\widehat{\mathcal{D}}$ in order to deal with an arbitrary\
anisotropic spectrum.

\subsubsection{Propagator.}

In Section 4, in the process of the microscopic derivation of the TDGL
equation, the fluctuation propagator was introduced. This object is of first
importance for the microscopic fluctuation theory and it has to be
generalized for the case of an impure metal with an anisotropic electron
spectrum. This is easy to do using the averaging procedure presented in the
previous Section. Formally it is enough to use in equation (\ref{dyspro})
the polarization operator $\Pi (\mathbf{q},\Omega _{k})$ averaged over
impurity positions, which can be expressed in terms of $\mathcal{P}(\mathbf{q%
},\widetilde{\varepsilon }_{n+k},\widetilde{\varepsilon }_{-n})$ introduced
above:

\begin{eqnarray}
\Pi (\mathbf{q},\Omega _{k}) &=&T\sum_{\omega _{n}}\lambda (q,\varepsilon
_{n+k},\varepsilon _{-n})\mathcal{P}(\mathbf{q},\widetilde{\omega }_{n+k},%
\widetilde{\omega }_{-n})=  \label{imppol} \\
&=&T\sum_{\omega _{n}}\frac{1}{\left[ \mathcal{P}(\mathbf{q},\widetilde{%
\varepsilon }_{n+k},\widetilde{\varepsilon }_{-n})\right] ^{-1}-\frac{1}{%
2\pi \nu \tau }}.  \nonumber
\end{eqnarray}
For relatively small $\mathbf{q}$ ($\Delta \xi (\mathbf{q,p})|_{|E\mathbf{(p)%
}|=E_{F}}\ll |\widetilde{\varepsilon }_{n+k}-\widetilde{\varepsilon }%
_{-n}|\sim \max \{T,\tau ^{-1}\})$ and $\Omega \ll T$ one can find an
expression for the fluctuation propagator, which can be useful in studies of
fluctuation effects near $T_{c}$\ ($\epsilon \ll 1$)\ for the dirty and
intermediate but not very clean case ($T\tau \ll 1/\sqrt{\epsilon }).$
Expanding (\ref{lamb}) in powers of $\left( \Delta \xi (\mathbf{q,p})|_{|E%
\mathbf{(p)}|=E_{F}}/|2\widetilde{\varepsilon }_{n}+\Omega _{k}|\right) ^{2}$
it is possible write $L^{R}(q,\Omega )$ in a form almost completely
coinciding with Exp.(\ref{anafp}):

\begin{equation}
L^{R}(\mathbf{q},\Omega )=-\frac{1}{\nu }\frac{1}{\epsilon -i\frac{\pi
\Omega }{8T}+\xi ^{2}(T\tau )\mathbf{q}^{2}}.  \label{dirtypro}
\end{equation}
Let us stress that the phenomenological coefficient $\gamma _{GL}$ turns out
to be equal to the same value $\frac{\pi \nu }{8T}$ as in clean case, and
hence does not depend on the impurity concentration. The only difference in
comparison with the clean case is in appearance of a dependence of the
natural effective coherence length on the elastic relaxation time. In the
isotropic D-dimensional case it can be written as 
\begin{eqnarray}
\xi _{(D)}^{2}(T\tau ) &=&\left( 4m\alpha T\right) ^{-1}=\eta _{(D)}=
\label{xigen} \\
&&-\frac{\tau ^{2}v_{F}^{2}}{D}\left[ \psi (\frac{1}{2}+\frac{1}{4\pi T\tau }%
)-\psi (\frac{1}{2})-\frac{1}{4\pi T\tau }\psi ^{^{\prime }}(\frac{1}{2})%
\right]  \nonumber
\end{eqnarray}
(we introduced here the parameter $\eta _{(D)}$ frequently used in the
microscopic theory)\footnote{%
Let us recall that its square determines the product of the GL parameter $%
\alpha $ and the Cooper pair mass entering in the GL functional. In clean
case we supposed the letter equal to two free electron masses and defined $%
\alpha $ in accordance with (\ref{alpe}). As we just have seen in the case
of the impure superconductor $\xi $\ depends on impurity concentration and
this dependence, in principle, can be attributed both to $\alpha $ or $m.$ \
For our further purposes it is convenient to leave $\alpha $ in the same
form (\ref{alpe}) as in the case of a clean superconductor. The Cooper pair
mass in this case becomes dependent on the electron mean free pass what
physically can be attributed to the diffusion motion of the electrons
forming the pair.}.

The generalization of (\ref{dirtypro}) \ for the case of a layered
electronic spectrum is evident:

\begin{equation}
L^{R}(q\mathbf{,}\Omega )=-\frac{1}{\nu }\frac{1}{\epsilon -i\frac{\pi
\Omega }{8T}+\eta _{(2)}\mathbf{q}_{\Vert }^{2}+r\sin ^{2}(q_{z}s/2)}.
\label{layerpro}
\end{equation}

One has to remember that the Exp. (\ref{dirtypro}) was derived in the
assumption of small momenta $\Delta \xi (\mathbf{q,p})|_{|E\mathbf{(p)}%
|=E_{F}}\ll |\widetilde{\varepsilon }_{n+k}-\widetilde{\varepsilon }%
_{-n}|\sim \max \{T,\tau ^{-1}\},$ so the range of its applicability is
restricted to the GL region of temperatures $\epsilon =\ln (\frac{T}{T_{c}}%
)\ll 1,$ where the integrands of diagrammatic expressions have singularities
at small momenta of the Cooper pair center of mass.

Finally let us express the Ginzburg-Levanyuk parameter for the important $2D$
case in terms of the microscopic parameter $\eta _{\left( 2\right) }$. In
accordance with (\ref{gimicro}) and the definition (\ref{xigen}): 
\begin{equation}
{Gi}_{\left( 2\right) }\left( T\tau \right) =\frac{7\zeta (3)}{16\pi ^{2}}\ 
\frac{1}{mT_{c}\eta _{\left( 2\right) }\left( T\tau \right) }.
\label{Gi2micro}
\end{equation}
One can see that this general definition in the limiting cases of a clean
and dirty metal results in the same values ${Gi}_{\left( 2c\right) }$ and ${%
Gi}_{\left( 2d\right) }$ as was reported in Table 1.

\subsection{Microscopic theory of fluctuation conductivity of layered
superconductor}

\subsubsection{Qualitative discussion of different fluctuation contributions}

In Section 4 the direct fluctuation effect on conductivity, related with the
charge transfer by means of fluctuation Cooper pairs, was discussed in
details. Nevertheless, below\ in this Section we return to its discussion
and will demonstrate its calculation by means of the microscopic theory.
This will be done in purpose to prepare the basis for studies of the
Aslamazov-Larkin contribution in the variety of physical values like
magnetoconductivity near the upper critical field, conductivity far from
transition point, fluctuation conductivity in ultra-clean limit, Hall
conductivity etc.

Microscopic approach permits also to calculate the cited above indirect
fluctuation effects like so called DOS and MT contributions. We will start
now from their qualitative discussion.

The important consequence of the presence of fluctuating Cooper pairs above $%
T_{c}$ is the decrease of the one-electron density of states at the Fermi
level. Indeed, if some electrons are involved in pairing they can not
simultaneously participate in charge transfer and heat capacity as
single-particle excitations. Nevertheless, the total number of the
electronic states can not be changed by the Cooper interaction, and only a
redistribution of the levels along the energy axis is possible \cite
{ARW70,CCRV90}. In this sense one can speak about the opening of a
fluctuation pseudo-gap at the Fermi level. The decrease of the one-electron
density of states at the Fermi level leads to a reduction of the normal
state conductivity. This, indirect, fluctuation correction to the
conductivity is called the \textit{density of states (DOS)} contribution and
it appears side by side with the \textit{paraconductivity (or
Aslamazov-Larkin contribution)}. It has the opposite (negative) sign and
turns out to be much less singular in $(T-T_{c})^{-1}$ in comparison with
the AL contribution, so that in the vicinity of $T_{c}$ it was usually
omitted. However, in many cases \cite{AL73,ILVY93,KG93,BDKLV93,AV80,ARV83},
when for some special reasons the main, most singular, corrections are
suppressed, the DOS correction becomes of major importance. Such a situation
takes place in many cases of actual interest (quasiparticle current in
tunnel structures, c-axis transport in strongly anisotropic high temperature
superconductors, NMR relaxation rate, thermoelectric power).

The correction to the normal state conductivity above the transition
temperature related with the fluctuation DOS renormalization for the dirty
superconductor can be evaluated qualitatively. Indeed, the fact that some
electrons ($\Delta \mathcal{N}_{e}$ per unit volume) participate in
fluctuation Cooper pairing means that the effective number of carriers
taking part in one-electron charge transfer diminishes leading to a decrease
of conductivity (we deal here with the longitudinal component): 
\begin{equation}
\delta \sigma _{xx}^{DOS}=-\frac{\Delta \mathcal{N}_{e}e^{2}\tau }{m}=-\frac{%
2n_{s}e^{2}\tau }{m},  \label{Drudos}
\end{equation}
where $n_{s}$ is the superfluid density coinciding with the Cooper pairs
concentration. The latter can be identified with the average value of the
square of the order parameter modulus already calculated as the correlator

\begin{equation}
\left\langle \Psi ^{\ast }(0)\Psi (\mathbf{r})\right\rangle (T>T_{c0})=\sum_{%
\mathbf{k}}\left\langle |\Psi _{k}|^{2}\right\rangle \exp \left( i\mathbf{kr}%
\right) =\frac{1}{4\pi \alpha \xi ^{2}}\mathbf{K}_{0}\left( \frac{r}{\xi
(\epsilon )}\right) ,
\end{equation}
with $r\sim \xi $. For the $2D$ case, which is of the most interest to us,
one finds:

\begin{equation}
n_{s}=\frac{1}{4\pi \alpha \xi ^{2}}\frac{1}{s}\mathbf{K}_{0}(\sqrt{\epsilon 
})=\frac{7\zeta \left( 3\right) }{\pi ^{4}v_{F}^{2}}\frac{E_{F}}{s\tau }\ln {%
\ \frac{1}{\epsilon },}
\end{equation}
where we have used the explicit expression (\ref{alpe}) for $\alpha $ and $%
\xi $. As we will see the corresponding expression for the fluctuation DOS
correction to conductivity (\ref{Drudos}) coincides with the accuracy of 2\
with the microscopic expression (\ref{DOSq2d}) which will be carried out
below.

The third, purely quantum, fluctuation contribution is generated by the
coherent scattering of the electrons forming a Cooper pair on the same
elastic impurities. This is the so called \textit{anomalous Maki-Thompson
(MT)} contribution \cite{M68,T70} which can be treated as the result of
Andreev scattering of the electron by fluctuation Cooper pairs. This
contribution often turns out to be important in conductivity and other
transport phenomena. Its temperature singularity near $T_{c}$ is similar to
that of the paraconductivity, although being extremely sensitive to electron
phase-breaking processes and to the type of orbital symmetry of pairing it
can be suppressed. Let us evaluate it.

The physical origin of the Maki-Thompson correction consists in the fact
that the Cooper interaction of electrons with the almost opposite momenta
changes the mean free path (diffusion coefficient) of electrons. As we have
already seen in the previous Section the amplitude of this interaction
increases drastically when $T\rightarrow T_{c}:$

\[
g_{eff}=\frac{g}{1-g\ln \frac{\omega _{D}}{2\pi T}}=\frac{1}{\ln \frac{T}{%
T_{c}}}\approx \frac{T}{T-T_{c}}=\frac{1}{\epsilon }. 
\]

What is the reason of this growth? One can say that the electrons scatter
one at another in resonant way with the virtual Cooper pairs formation. Or
it is possible to imagine that the electrons undergo the Andreev scattering
at fluctuation Cooper pairs binding in the Cooper pair themselves. The
probability of such induced pair irradiation (let us remind that Cooper
pairs are Bose particles) is proportional to their number in the final
state, i.e. $\mathit{n}(p)$ (\ref{ncp}). For small momenta $\mathit{n}%
(p)\sim 1/\epsilon .$

One can ask why such interaction does not manifest itself considerably far
from the transition point? The matter of fact that so intensively interacts
just small number of electrons with the total momentum $q\lesssim \xi
^{-1}(T).$ In accordance with the Heisenberg principle the minimal distance
between such electrons is of the order of $\sim \xi (T).$ From the other
hand such electrons in purpose to interact have to approximate one another
up to the distance of the Fermi length $\lambda _{F}\sim 1/p_{F}.$ The
probability of such event may be estimated in the spirit of the
self-intersection trajectories contribution evaluation in the weak
localization theory \cite{LK82,Ab88}.

In the process of diffusion motion the distance between two electrons
increases with the time growth in accordance with the Einstein law: $%
R(t)\sim \left( Dt\right) ^{1/2}.$ Hence the scattering probability

\[
W\sim \int_{t_{\min }}^{t_{\max }}\ \frac{\lambda _{F}^{D-1}}{R^{D}(t)}%
v_{F}\ dt. 
\]
The lower limit of the integral can be estimated from the condition $%
R(t_{\min })$ $\sim \xi (T)$ (only such electrons interact in the resonant
way). The upper limit is determined by the phase breaking time $\tau
_{\varphi }$ since for larger time intervals the phase coherence, necessary
for the pair formation, is broken. In result the relative correction to
conductivity \ due to such processes is equal to the product of the
scattering probability\ on the effective interaction constant: $\delta
\sigma ^{MT}/\sigma =W$ $g_{eff}.$ In the $2D$ case

\[
\delta \sigma ^{MT}\sim \frac{e^{2}}{8\epsilon }\ln \frac{D\tau _{\varphi }}{%
\xi ^{2}(T)}. 
\]
This result will be confirmed below in the frameworks of the microscopic
consideration.

\subsubsection{Generalities}

Let us pass to the microscopic calculation of the fluctuation conductivity
of the layered superconductor. We begin by discussing the quasiparticle
normal state energy spectrum. While models with several conducting layers
per unit cell and with either intralayer or interlayer pairing have been
considered \cite{KL91}, it has been shown \cite{LK93} that all of these
models give rise to a Josephson pair potential that is periodic in $k_{z}$,
the wave-vector component parallel to the $c$-axis, with period $s$, the $c$%
-axis repeat distance. While such models differ in their superconducting
densities of states, they all give rise to qualitatively similar fluctuation
propagators, which differ only in the precise definitions of the parameters
and in the precise form of the Josephson coupling potential. Ignoring the
rather unimportant differences between such models in the Gaussian
fluctuation regime above $T_{c}(H)$, we therefore consider the simplest
model of a layered superconductor, in which there is one layer per unit
cell, with intralayer singlet $s$-wave pairing. These assumptions lead to
the simple spectrum (\ref{d1}) and hence to a Fermi surface having the form
of a corrugated cylinder (see Fig.\ref{fcorrug}).

Some remarks regarding the normal-state quasiparticle momentum relaxation
time are necessary. In the ''old'' layered superconductors the materials
were generally assumed to be in the dirty limit (like $TaS_{2}$(pyridine)$%
_{1/2}$). In the high-$T_{c}$ cuprates, however, both single crystals and
epitaxial thin films are nominally in the ''intermediate'' regime, with $%
l/\xi _{xy}\approx 2-5$. In addition, the situation in the cuprates is
complicated by the presence of phonons for $T\simeq T_{c}\simeq 100K$, the
nearly localized magnetic moments on the Cu$^{2+}$ sites, and by other
unspecified inelastic processes. In this Section we assume simple elastic
intralayer scattering and restrict our consideration to the local limit in
the fluctuation Cooper pair motion. This means that we consider the case of
not too clean superconductors, keeping the impurity concentration $n_{i}$ \
and reduced temperature such that the resulting mean-free path satisfies the
requirement $l<\xi _{xy}(T)=\frac{\xi _{xy}}{\sqrt{\varepsilon }}$ and the
impurity vertex \ can be taken in the local form \ (\ref{lamb}) with $%
\langle (\Delta \xi (\mathbf{q,p})|)^{2}\rangle _{F.S.}$ determined by (\ref
{xi2lay}). The phase-breaking time $\ \tau _{\varphi }$ is supposed to be
much larger than $\tau $.

The most general relation between the current density $\mathbf{j(r,}t\mathbf{%
)}$\ and vector-potential $\mathbf{A(r}^{\prime },t^{\prime }\mathbf{)}$ \
is given through the so-called electromagnetic response operator $Q_{\alpha
\beta }(\mathbf{r,r}^{\prime },t,t^{\prime })$ \cite{AGD}:

\[
\mathbf{j(r,}t\mathbf{)}=\int \ Q_{\alpha \beta }(\mathbf{r,r}^{\prime
},t,t^{\prime })\mathbf{A(r}^{\prime },t^{\prime }\mathbf{)}\ \mathbf{dr}%
^{\prime }\ dt, 
\]
Assuming space and time homogeneity, one can take the Fourier transform of
this relation and compare it with the definition of the conductivity tensor $%
j_{\alpha }=\sigma _{\alpha \beta }E_{\beta }.$ This permits us to express
the conductivity tensor in terms of the retarded electromagnetic response
operator

\begin{equation}
\sigma _{\alpha \beta }(\omega )=-{\frac{1}{{i\omega }}}[Q_{\alpha \beta
}]^{R}(\omega )\ .  \label{sigmaQ}
\end{equation}

\begin{figure}[tbp]
\epsfxsize=4in
\centerline{\epsfbox{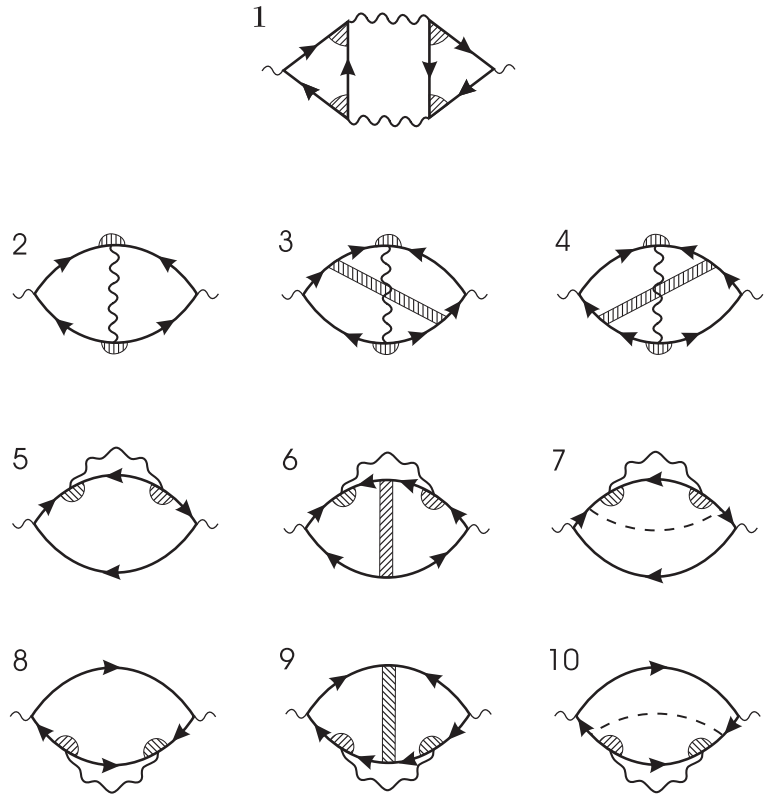}}
\caption{Feynman diagrams for the leading-order contributions to the
fluctuation conductivity. Wavy lines are fluctuation propagators, thin solid
lines with arrows are impurity-averaged normal-state Green's functions,
shaded semicircles are vertex corrections arising from impurities, dashed
lines with central crosses are additional impurity renormalisations and
shaded rectangles are impurity ladders. Diagram 1 represents the
Aslamazov-Larkin term; diagrams 2--4 represent the Maki-Thompson type
contributions; diagrams 5--10 arise from corrections to the normal state
density of states.}
\label{conddia}
\end{figure}

The electromagnetic response operator $Q_{\alpha \beta }(\omega _{\nu })$,
defined on Matsubara frequencies $\omega _{\nu }=(2\nu +1)\pi T,$ \ can be
presented as the correlator of \ two exact one-electron Green functions \cite
{AGD} averaged over impurities and accounting for interactions, in our case
the particle-particle interactions in the Cooper channel. {\ }The
appropriate diagrams corresponding to the first order of perturbation theory
in the fluctuation amplitude are shown in Fig. \ (\ref{conddia}).

With each electromagnetic field component $A_{\alpha }$ we associate the
external vertex $ev_{\alpha }(p)=e{\frac{{\partial \xi (p)}}{{\partial
p_{\alpha }}}}$. For the longitudinal conductivity tensor elements (parallel
to the layers, for which $\alpha =x,y$), the resulting vertex is simply $%
ep_{\alpha }/m$. For the $c$-axis conductivity, the vertex is given by 
\begin{equation}
ev_{z}(p)=e{\frac{{\partial \xi (p)}}{{\partial p_{z}}}}=-eJs\sin (p_{z}s).
\label{vz}
\end{equation}
Each solid line in the diagrams represents a \ one-electron Green function
averaged over impurities (\ref{1elec}), a wavy line represents a fluctuation
propagator $L(\mathbf{q},\Omega _{k})$ (\ref{dirtypro}), three-leg vertices
were defined by the Exp.(\ref{lamb}). The four-leg impurity vertex\textbf{,}%
\ appearing in diagrams 3-4, 9-10 of the Fig. \ref{conddia}, is called the
Cooperon in the weak localization theory (see, for example,\cite{AA85})\ It
is easy to see that it differs from the above three-leg vertex only by the
additional factor $\left( 2\pi \nu \tau \right) ^{-1}$. We do not
renormalize the current vertices: it is well known (see \cite{AGD}) that
this renormalization only leads to the substitution of the scattering time $%
\tau $ by the transport one $\tau _{tr}$. We integrate over the internal
Cooper pair momentum $\mathbf{q}$ and electron momentum $\mathbf{p}$ and sum
over the internal fermionic and bosonic\ Matsubara frequencies, with
momentum and energy conservation at each internal vertex (fluctuation
propagator endpoint) in the analytical expressions for the diagrams
presented in Fig. (\ref{conddia}).

After these necessary introductory remarks and definitions we pass to the
microscopic calculation of the different fluctuation contributions.

\subsubsection{Aslamazov-Larkin contribution}

We first examine the AL paraconductivity (diagram 1 of Fig.(\ref{conddia})).
Actually this contribution was already studied in the Section 4 in the
framework of the TDGL equation but, in order to demonstrate how the method
works, we will carry out here the appropriate calculations in the
microscopic approach, as was originally done by Aslamazov and Larkin \cite
{AL68}.

The AL contribution to the electromagnetic response operator tensor has the
form: 
\begin{eqnarray}
Q_{\alpha \beta }^{AL}(\omega _{\nu }) &=&2e^{2}T\sum_{\Omega _{k}}\int {%
\frac{{d^{3}}\mathbf{q}}{{(2\pi )^{3}}}}B_{\alpha }(\mathbf{q},\Omega
_{k},\omega _{\nu })L(\mathbf{q},\Omega _{k})\times  \label{QAL} \\
&&\times B_{\beta }(\mathbf{q},\Omega _{k},\omega _{\nu })L(\mathbf{q}%
,\Omega _{k}+\omega _{\nu }),  \nonumber
\end{eqnarray}
where the three Green function block is given by 
\begin{eqnarray}
B_{\alpha }(\mathbf{q},\Omega _{k},\omega _{\nu }) &=&T\sum_{\varepsilon
_{n}}\lambda (\mathbf{q},\varepsilon _{n+\nu },\Omega _{k}-\varepsilon
_{n})\lambda (\mathbf{q},\varepsilon _{n},\Omega _{k}-\varepsilon _{n})\times
\label{block} \\
&&\times \int {\frac{{d^{3}}\mathbf{p}}{{\ (2\pi )^{3}}}}v_{\alpha }(\mathbf{%
p})G(\mathbf{p},\varepsilon _{n+\nu })G(\mathbf{p},\varepsilon _{n})G(%
\mathbf{q}-\mathbf{p},\Omega _{k}-\varepsilon _{n}).  \nonumber
\end{eqnarray}
Expanding $G(\mathbf{q-p},\Omega _{k}-\varepsilon _{n})$ over $\mathbf{q}$\
one find that the angular integration over the Fermi surface kills the first
term and leaves nonzero the second term of the expansion only. Then the $\xi
-$integration is performed by means of the Cauchy theorem. The further
summation over the fermionic frequency is cumbersome, so we will show it for
the example of the simplest case of a dirty superconductor with $T\tau \ll
1. $ In this case the main sources of the $\varepsilon _{n}-$dependence in (%
\ref{block}) are the $\lambda $-vertices and that originating from the Green
functions can be neglected by the parameter $T\tau \ll 1$ (indeed, one can
see that $\varepsilon _{n}\sim T$ are important in vertices, while in Green
functions $\varepsilon _{n}\gtrsim \tau ^{-1}$ only). The remaining
summation in (\ref{block}) is performed in the same way as was done in (\ref
{dirtypro}) and results in:

\begin{eqnarray}
B_{\alpha }(\mathbf{q},\Omega _{k},\omega _{\nu }) &=&\nu \frac{\eta _{(2)}}{%
v_{F}^{2}}\left\langle v_{\alpha }q_{\beta }v_{\beta }\right\rangle
_{FS}\times  \label{bl1} \\
&&\times \frac{8T}{\pi \omega _{\nu }}\left[ \psi (\frac{1}{2}+\frac{|\Omega
_{k}|+\omega _{\nu }+\widehat{D}q^{2}}{4\pi T})-\psi (\frac{1}{2}+\frac{%
|\Omega _{k}|+\widehat{D}q^{2}}{4\pi T})\right. +  \nonumber \\
&&+\left. \psi (\frac{1}{2}+\frac{|\Omega _{k+\nu }|+\omega _{\nu }+\widehat{%
D}q^{2}}{4\pi T})-\psi (\frac{1}{2}+\frac{||\Omega _{k+\nu }|+\widehat{D}%
q^{2}}{4\pi T})\right] .  \nonumber
\end{eqnarray}

\begin{figure}[tbp]
\epsfxsize=10cm
\centerline {\epsfbox{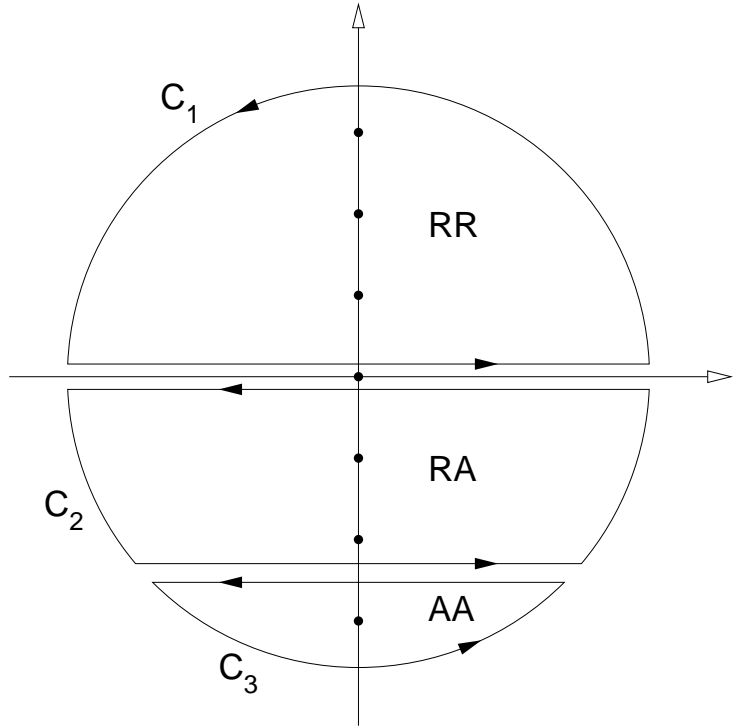}}
\caption{{}The contour of integration in the plane of complex frequencies.}
\label{figAL}
\end{figure}

Now let us return to the general expression for $Q_{\alpha \beta
}^{AL}(\omega _{\nu })$ and transform the $\Omega _{k}-$ summation into a
contour integral, \ using the identity \cite{E61}

\[
T\sum_{\Omega _{k}}f(\Omega _{k})=\frac{1}{4\pi i}\oint_{C}dz\coth \frac{z}{%
2T}f(-iz), 
\]
where $z=i\Omega _{k}$ is a variable in the plane of complex frequency and
the contour $C$ encloses all bosonic Matsubara frequencies over which the
summation is carried out. In our case the contour $C$ can be chosen as a
circle with radius tending to infinity (see Fig. \ref{figAL}):

\begin{eqnarray}
Q_{\alpha \beta }^{AL}(\omega _{\nu }) &=&\frac{e^{2}}{2\pi i}\int {\frac{{%
d^{3}}\mathbf{q}}{{(2\pi )^{3}}}}\oint_{C}dz\coth \frac{z}{2T}B_{\alpha }(%
\mathbf{q},-iz+\omega _{\nu },-iz)\times  \label{analicon} \\
&&\times L(\mathbf{q},-iz)B_{\beta }(\mathbf{q},-iz+\omega _{\nu },-iz)L(%
\mathbf{q},-iz+\omega _{\nu }).  \nonumber
\end{eqnarray}

One can see that the integrand function in (\ref{analicon}) has breaks of
analyticity at the lines $\mbox{Im}z=0$ and $\mbox{Im}z=-i\omega _{\nu }.$
Indeed, the fluctuation propagator $L(\mathbf{q},\Omega _{k})$ and Green
function blocks $B_{\alpha }(\mathbf{q},\Omega _{k},\omega _{\nu })$ were
defined on the bosonic Matsubara frequencies only, while now we have to use
them as functions of the continuous variable $z$. As it is well known from
the properties of Green functions in the complex plane $z\ ,$ two analytical
functions, related with $L(\mathbf{q},\Omega _{k}),$ can be introduced. The
first one, $L^{R}(\mathbf{q},-iz)$ (retarded), is analytic in the upper
half-plane ($\mbox{Im}z$ $>0$), while the second one, $L^{A}(\mathbf{q},-iz)$
(advanced), has no singularities in the lower half-plane ($\mbox{Im}z$ $<0).$
As we have seen above the same lines separate the domains of the analyticity
of the Green function blocks, so the functions $B^{RR},B^{RA},B^{AA}$
analytic in each domain can be introduced (with the appropriate choices of
the $|\Omega _{k+\nu }|$ and $|\Omega _{k}|$ signs in the arguments of the $%
\psi -$\ functions, see (\ref{bl1})). This means that\ by cutting the $z$%
-plane along the lines $\mbox{Im}z=0$ and $\mbox{Im}z$ $=-i\omega _{\nu }$
we can reduce the calculation of the contour integral in (\ref{analicon}) to
the sum of \ three integrals along the contours $C_{1},C_{2},C_{3}$ which
enclose domains of \ well defined analyticity of the integrand function. The
integral along the large circle evidently vanishes and the contour integral
is reduced to four integrals along the cuts of the plane in Fig.\ref{figAL}:

\begin{eqnarray*}
I(\mathbf{q},\omega _{\nu }) &=&\oint_{C_{1}+C_{2}+C_{3}}dz\coth \frac{z}{2T}%
B_{\alpha }(\mathbf{q},-iz)L(\mathbf{q},-iz)B_{\beta }(\mathbf{q},-iz)L(%
\mathbf{q},-iz+\omega _{\nu })= \\
&=&\int_{-\infty }^{\infty }dz\coth \frac{z}{2T}L^{R}(\mathbf{q},-iz+\omega
_{\nu })\left[ B_{\alpha }^{RR}B_{\beta }^{RR}L^{R}(\mathbf{q},-iz)-\right.
\\
&&\left. -B_{\alpha }^{RA}B_{\beta }^{RA}L^{A}(\mathbf{q},-iz)\right]
+\int_{-\infty -i\omega _{\nu }}^{\infty -i\omega _{\nu }}dz\coth \frac{z}{2T%
}L^{A}(\mathbf{q},-iz)\times \\
&&\left[ B_{\alpha }^{RA}B_{\beta }^{RA}L^{R}(\mathbf{q},-iz+\omega _{\nu
})-B_{\alpha }^{AA}B_{\beta }^{AA}L^{A}(\mathbf{q},-iz+\omega _{\nu })\right]
..
\end{eqnarray*}
Now one can shift the variable in the last integral to $z=z^{\prime
}-i\omega _{\nu },\;$ take into account that $i\omega _{\nu }$\ is the
period of $\coth \frac{z}{2T}$ \ and get an expression analytic in $i\omega
_{\nu }\rightarrow \omega .$

In the vicinity of $T_{c}$, due to the pole structure of the fluctuation
propagators in (\ref{QAL}), the leading contribution to the electromagnetic
response operator $Q_{\alpha \beta }^{AL(R)}$ arises from them rather than
from the frequency dependence of the vertices $B_{\alpha }$, so we can
neglect the $\Omega _{k}$- and $\omega _{\nu }$-dependencies of the Green
functions blocks and use the expression for\ $B_{\alpha }(\mathbf{q},0,0)$
valid for small $\mathbf{q}_{ab}\mathbf{\ }$only: 
\begin{equation}
B_{\alpha }(\mathbf{q})=-2\nu \frac{\eta _{(2)}}{v_{F}^{2}}\left\{ 
\begin{array}{c}
v_{F}^{2}q_{\alpha },\;\mbox{ }\alpha =x,y \\ 
sJ^{2}\sin q_{z}s,\;\mbox{ }\alpha =z
\end{array}
\right. .  \nonumber
\end{equation}
Detailed calculations demonstrate that this result can be generalized to an
arbitrary impurity concentration just by using the expression (\ref{xigen})
for\ $\eta _{(2)}$. Finally:

\begin{eqnarray*}
Q_{\alpha \beta }^{AL(R)}(\omega )=\frac{2e^{2}}{\pi }\int {\frac{{d^{3}q}}{{%
(2\pi )^{3}}}}B_{\alpha }(\mathbf{q})B_{\beta }(\mathbf{q}) &&\int_{-\infty
}^{\infty }dz\coth \left( \frac{z}{2T}\right) \left[ L^{R}(\mathbf{q}%
,-iz-i\omega )\right. \\
&&\left. +L^{A}(\mathbf{q},-iz+i\omega )\right] \mbox{Im}L^{R}(\mathbf{q}%
,-iz).
\end{eqnarray*}

Being interested here in the d.c. conductivity one can expand the integrand
function in $\omega .$ It is possible to show that the zeroth order term is
cancelled by the same type contributions from all other diagrams (this
cancellation confirms the absence of anomalous diamagnetism above the
critical temperature). The remaining integral can be integrated by parts and
then carried out taking into account that the contribution most singular in $%
\epsilon $\ comes from the region $z\sim \epsilon \ll T:$ 
\begin{eqnarray}
\sigma _{xx}^{AL} &=&\frac{e^{2}}{2\pi T}\int {\frac{{d^{3}}\mathbf{q}}{{%
(2\pi )^{3}}}}B_{x,z}^{2}(\mathbf{q},0,0)\int_{-\infty }^{\infty }\frac{dz}{%
\sinh ^{2}\frac{z}{2T}}\left[ \mbox{Im}L^{R}(\mathbf{q},-iz)\right] ^{2}= 
\nonumber \\
&=&{\frac{{\pi ^{2}e^{2}\eta _{(2)}^{2}}}{{s}}}\int {\frac{{d^{2}\mathbf{q}}%
}{{(2\pi )^{2}}}}{\frac{\mathbf{q}^{2}}{\left[ {(\eta _{(2)}\mathbf{q}%
^{2}+\epsilon )(\eta _{(2)}\mathbf{q}^{2}+\epsilon +r)}\right] {^{3/2}}}} 
\nonumber \\
&=&\frac{e^{2}}{16s}\frac{1}{[\epsilon (\epsilon +r)]^{1/2}}\rightarrow {\ 
\frac{{e^{2}}}{{16s}}}\left\{ 
\begin{array}{c}
1/\sqrt{\epsilon r},\;\mbox{ }\epsilon \ll r \\ 
1/\epsilon ,\;\mbox{ }\epsilon \gg r
\end{array}
\right. .  \label{d12}
\end{eqnarray}
where the Lawrence-Doniach anisotropy parameter $r$ \cite{LD70} was already
defined by (\ref{r1}).

In the same way one can evaluate the AL contribution to the transverse
fluctuation conductivity \cite{K74,ILVY93,Buz92a}: 
\begin{eqnarray}
\sigma _{zz}^{AL} &=&{\frac{{\pi e^{2}sr^{2}}}{{32}}}\int {\frac{{d^{2}%
\mathbf{q}}}{{\ (2\pi )^{2}}}}{\frac{1}{\left[ {(\eta _{(2)}\mathbf{q}%
^{2}+\epsilon )(\eta _{(2)}\mathbf{q}^{2}+\epsilon +r)}\right] {^{3/2}}}=}
\label{d12_2} \\
&=&{\frac{{e^{2}s}}{{32}\eta _{(2)}}}\left( {\frac{{\epsilon +r/2}}{{\
[\epsilon (\epsilon +r)]^{1/2}}}}-1\right) \rightarrow {\frac{{e^{2}s}}{{\ 64%
}\eta _{(2)}}}\left\{ 
\begin{array}{c}
\sqrt{r/\epsilon },\;\mbox{for }\epsilon \ll r \\ 
\left( r/2\epsilon \right) ^{2},\;\mbox{for }\epsilon \gg r
\end{array}
\right. .  \nonumber
\end{eqnarray}
Note, that contrary to the case of in-plane conductivity, the critical
exponent for $\sigma _{zz}$ above the Lawrence-Doniach crossover temperature 
$T_{LD}$ (for which $\epsilon (T_{LD})=r$) is $2$ instead of $1$, so the
crossover occurs from the 0D to 3D regimes. This is related with the
tunneling (so from the band structure point of view effectively zero
dimensional) character of electron motion along the c-axis.

\subsubsection{Contributions from fluctuations of the density of states}

In original paper of Aslamazov and Larkin \cite{AL68} the most singular AL
contribution to conductivity, heat capacity and other properties of a
superconductor above the critical temperature was considered. The diagrams
of the type 5-6 were pictured and correctly evaluated as less singular in $%
\epsilon .$ Nevertheless the specific form of the AL contribution to the
transverse conductivity of a layered superconductor, which may be
considerably suppressed for small interlayer transparency, suggested to
re-examine the contributions from diagrams 5-10 of Fig.\ref{conddia} which
are indeed less divergent in $\epsilon $, but turn out to be of lower order
in the transmittance and of the opposite sign with respect to the AL one 
\cite{ILVY93,KG93}. These, so-called DOS, diagrams describe the changes in
the normal Drude-type conductivity due to fluctuation renormalization of the
normal quasiparticles density of states above the transition temperature
(see Section 8.5). In the dirty limit, the calculation of contributions to
the longitudinal fluctuation conductivity $\sigma _{xx}$ from such diagrams
was discussed in \cite{AM78,ARV83}. Contrary to the case of the AL
contribution, the in-plane and out-of-plane components of the DOS
contribution differ only in the square of the ratio of effective Fermi
velocities in the parallel and perpendicular directions. This allows us to
calculate both components simultaneously. The contribution to the
fluctuation conductivity due to diagram 5 is 
\begin{eqnarray*}
Q_{\alpha \beta }^{5}(\omega _{\nu }) &=&2e^{2}T\sum_{\Omega _{k}}\int {%
\frac{{d^{3}}\mathbf{q}}{{(2\pi )^{3}}}}L(\mathbf{q},\Omega
_{k})T\sum_{\varepsilon _{n}}\lambda ^{2}(\mathbf{q},\varepsilon _{n},\Omega
_{k}-\varepsilon _{n})\times \\
&&\int {\frac{{d^{3}}\mathbf{p}}{{(2\pi )^{3}}}}v_{\alpha }(\mathbf{p}%
)v_{\beta }(\mathbf{p})G^{2}(\mathbf{p},\varepsilon _{n})G(\mathbf{q-p}%
,\Omega _{k}-\varepsilon _{n})G(\mathbf{p},\varepsilon _{n+\nu
}),~~~~~~~~~~~~
\end{eqnarray*}
and diagram 6 gives an identical contribution. Evaluation of the
integrations over the in-plane momenta $\mathbf{p}$ and the summation over
the internal frequencies $\varepsilon _{n}$ are straightforward. Treatment
of the other internal frequencies $\Omega _{k}$ is less obvious, but in
order to obtain the leading singular behavior in the vicinity of transition
it suffices to set $\Omega _{k}=0$ \cite{ARV83}. After integration over $%
q_{z}$, we have \cite{ILVY93,BDKLV93}: 
\begin{eqnarray}
\sigma _{\alpha \beta }^{5+6} &=&-{\frac{\pi {e^{2}}}{{2s}}}A_{\alpha \beta }%
{\kappa _{1}\eta }_{\left( 2\right) }\int_{|\mathbf{q}|\leq \xi ^{-1}}{\frac{%
{d^{2}\mathbf{q}}}{{(2\pi )^{2}}}}{\frac{1}{\left[ {(\epsilon +\eta _{(2)}%
\mathbf{q}^{2})(\epsilon +r+\eta _{(2)}\mathbf{q}^{2})}\right] {^{1/2}}}} 
\nonumber \\
&\approx &-{\frac{{e^{2}\kappa _{1}}}{{8s}}}A_{\alpha \beta }\ln \left( 
\frac{2}{{\epsilon ^{1/2}+(\epsilon +r)^{1/2}}}\right) ,  \label{d15}
\end{eqnarray}
where $A_{xx}=A_{yy}=1,A_{zz}=(sJ/v_{F})^{2},A_{\alpha \neq \beta }=0$ and 
\[
\kappa _{1}={\frac{{2(v_{F}\tau )^{2}}}{{\pi ^{2}\eta }_{\left( 2\right) }}}%
\left[ \psi ^{^{\prime }}\left( {\frac{1}{2}}+{\frac{1}{{4\pi T\tau }}}%
\right) -{\frac{3}{{4\pi T\tau }}}\psi ^{^{\prime \prime }}\left( {\frac{1}{2%
}}\right) \right] . 
\]
In order to cut off the ultra-violet divergence in $q$ we have introduced
here a cut off parameter $q_{\mathrm{max}}=\xi ^{-1}=$ ${\eta _{(2)}^{-1/2}}$
in complete agreement with Section 3. Let us stress that in the framework of
the phenomenological GL theory we attributed this cut off to the breakdown
of the GL approach at momenta as large as $q\sim \xi ^{-1}.$ The microscopic
approach developed here permits to see how this cut off appears: the
divergent shortwave-length contribution arising from GL-like fluctuation
propagators is automatically restricted by the $q$-dependencies of the
impurity vertices and Green functions, which appear at the scale $q\sim
l^{-1}.$

In a similar manner, the equal contributions from diagrams 7 and 8 sum to 
\begin{eqnarray}
\sigma _{\alpha \beta }^{7+8} &=&-{\frac{\pi {e^{2}}}{{2s}}}A_{\alpha \beta
}\kappa {_{2}\eta _{(2)}}\int_{|\mathbf{q}|\leq q_{\mathrm{max}}}{\frac{{%
d^{2}\mathbf{q}}}{{(2\pi )^{2}}}}{\frac{1}{\left[ (\epsilon +\eta _{(2)}{%
\mathbf{q}^{2})(}\epsilon {+r+\eta _{(2)}\mathbf{q}^{2})}\right] {^{1/2}}}} 
\nonumber  \label{d17} \\
&\approx &-{\frac{{e^{2}}\kappa {_{2}}}{{8s}}}A_{\alpha \beta }\ln \left( 
\frac{2}{\epsilon {^{1/2}+(}\epsilon {+r)^{1/2}}}\right) ,
\end{eqnarray}
\[
\kappa _{2}={\frac{\left( v_{F}{\tau }\right) ^{2}}{{2\pi ^{3}T\tau }}}\psi
^{^{\prime \prime }}\left( {\frac{1}{2}}\right) . 
\]
Comparing (\ref{d15}) and (\ref{d17}), we see that in the clean limit, the
main contributions from the DOS fluctuations arise from diagrams 5 and 6. In
the dirty limit, diagrams 7 and 8 are also important, having -1/3 the value
of diagrams 5 and 6, for both $\sigma _{xx}$ and $\sigma _{zz}$. Diagrams 9
and 10 are not singular in $\epsilon <<1$ at all and can be neglected. The
total DOS contribution to the in-plane and $c$-axis conductivity is
therefore 
\begin{equation}
\sigma _{\alpha \beta }^{DOS}=-\frac{e^{2}}{2s}\kappa (T\tau )A_{\alpha
\beta }\ln \left( \frac{2}{{\epsilon ^{1/2}+(\epsilon +r)^{1/2}}}\right) ,
\label{DOSq2d}
\end{equation}
where 
\begin{eqnarray}
\kappa (T\tau ) &=&{\kappa _{1}+\kappa _{2}}={\frac{{-\psi ^{^{\prime }}}%
\left( {{\frac{1}{2}}+{\ \frac{1}{{4\pi \tau T}}}}\right) {+{\frac{1}{{2\pi
\tau T}}}\psi ^{^{\prime \prime }}}\left( {{\frac{1}{2}}}\right) }{{\pi ^{2}}%
\left[ {\psi \left( {{\frac{1}{2}}+{\ \frac{1}{{4\pi \tau T}}}}\right) -\psi 
}\left( {{{\frac{1}{2}}}}\right) {-{\frac{1}{{4\pi \tau T}}}\psi ^{^{\prime
}}}\left( {{\frac{1}{2}}}\right) \right] }}  \nonumber \\
&\rightarrow &\left\{ 
\begin{array}{c}
56\zeta (3)/\pi ^{4}\approx 0.691,\;\mbox{ }T\tau \ll 1 \\ 
8\pi ^{2}\left( T\tau \right) ^{2}/\left[ 7\zeta (3)\right] \approx
9.384\left( T\tau \right) ^{2},\;\mbox{ }1\ll T\tau \ll 1/\sqrt{\epsilon }
\end{array}
\right.  \label{DOSint}
\end{eqnarray}
is a function of $\tau T$ only. \ As it will be shown below at the upper
limit $T\tau \sim 1/\sqrt{\epsilon }$ the DOS contribution reaches the value
of the other fluctuation contributions and in the limit of $T\tau
\rightarrow \infty $ exactly eliminates the Maki-Thompson one.

\subsubsection{Maki-Thompson contribution}

We now consider another quantum correction to fluctuation conductivity which
is called the Maki-Thompson (MT) contribution (diagram 2 of Fig. \ref
{conddia}). It was firstly discussed by Maki \cite{M68} in a paper which
appeared almost simultaneously with the paper of Aslamazov and Larkin \cite
{AL68}. Both these articles gave rise to the microscopic theory of
fluctuations in superconductor. Maki found that, in spite of the seeming
weaker singularity of diagram 2 with respect to the AL one (it contains one
propagator only, while the AL\ one contains two of them) it can contribute
to conductivity comparably or even stronger than AL one.

Since the moment of its discovery the MT contribution became the subject of
intense controversy. In its original paper Maki found that in $3D$ case this
fluctuation correction is four times larger than the AL one. In $2D\ $case
the result was\textbf{\ }striking\textbf{:} the MT contribution simply
diverged. This paradox was, at least at the level of recipe, resolved by
Thompson \cite{T70}: he proposed to cut off the infra-red divergence in the
Cooper pair center of mass momentum integration by\ the introduction of the
finite length $l_{s}$ of inelastic scatterings of electrons on paramagnetic
impurities. In the further papers of Patton \cite{Pat71}, Keller and
Korenman \cite{KK72} it was cleared up that the presence of paramagnetic
impurities or other external phase-breaking sources is not necessary: the
fluctuation Cooper pairing of two electrons results in a change of the
quasiparticle phase itself and the corresponding phase-breaking time $\tau
_{\varphi }$ appears as a natural cut off parameter of the MT divergence in
the strictly 2D case. The minimal quasi-two-dimensionality of the electron
spectrum, as we will show below, automatically results in a cut off of the
MT divergence.

Although the MT contribution to\ the in-plane conductivity is expected to be
important in the case of low pair-breaking, experiments on high-temperature
superconductors have shown that the excess in-plane conductivity can usually
be explained in terms of the fluctuation paraconductivity alone. Two
possible explanations can be found for this fact. The first one is that the
pair-breaking in these materials is not weak. The second is related with the 
$d-$wave symmetry of pairing which kills the anomalous Maki-Thompson process 
\cite{Yip,CLRV96}. We will consider below the case of s-pairing, where the
Maki-Thompson process is well pronounced.

The appearance of the anomalously large MT contribution is nontrivial and
worth being discussed. We consider the scattering lifetime $\tau $ and the
pair-breaking lifetime $\tau _{\varphi }$ to be arbitrary, but satisfying $%
\tau _{\varphi }>\tau $. In accordance with diagram 2 of\ Fig.(\ref{conddia}%
) the analytical expression for the MT contribution to the electromagnetic
response tensor can be written as 
\begin{equation}
Q_{\alpha \beta }^{MT}(\omega _{\nu })=2e^{2}T\sum_{\Omega _{k}}\int {\frac{{%
d^{3}}\mathbf{q}}{{(2\pi )^{3}}}}L(\mathbf{q},\Omega _{k})I_{\alpha \beta }(%
\mathbf{q},\Omega _{k},\omega _{\nu }),  \label{d21}
\end{equation}
where 
\begin{equation}
I_{\alpha \beta }(\mathbf{q},\Omega _{k},\omega _{\nu })=T\sum_{\varepsilon
_{n}}\lambda (\mathbf{q},\varepsilon _{n+\nu },\Omega _{k-n-\nu })\lambda (%
\mathbf{q},\varepsilon _{n},\Omega _{k-n})\times  \label{d22}
\end{equation}
\[
\times \int {\frac{{d^{3}}\mathbf{p}}{{(2\pi )^{3}}}}v_{\alpha }(\mathbf{p}%
)v_{\beta }(\mathbf{q-p})G(\mathbf{p},\varepsilon _{n+\nu })G(\mathbf{p}%
,\varepsilon _{n})G(\mathbf{q-p},\Omega _{k-n-\nu })G(\mathbf{q-p},\Omega
_{k-n})~. 
\]

In the vicinity of $T_{c}$, it is possible to restrict consideration to the
static limit of the MT diagram, simply by setting $\Omega _{k}=0$ in (\ref
{d21}). Although dynamic effects can be important for the longitudinal
fluctuation conductivity well above $T_{LD}$, the static approximation is
correct very close to $T_{c}$, as shown in \cite{AV80,RVV91}. \ The main $q-$%
dependence in (\ref{d21}) arises from the propagator and vertices $\lambda .$
This is why we can assume $q=0$ in Green functions and to calculate the
electron momentum integral passing, as usual, to a $\xi (\mathbf{p})$
integration:

\begin{eqnarray}
I_{\alpha \beta }(q,0,\omega _{\nu }) &=&\pi \nu \left\langle v_{\alpha
}(p)v_{\beta }(q-p)\right\rangle _{FS}\times  \label{d22a} \\
&&\times T\sum_{\varepsilon _{n}}\frac{1}{\left( |2\varepsilon _{n+\nu }|+%
\widehat{D}q^{2}\right) }\frac{1}{\left( |2\varepsilon _{n}|+\widehat{D}%
q^{2}\right) }\frac{1}{|\widetilde{\varepsilon }_{n+\nu }|+|\widetilde{%
\varepsilon }_{n}|}~.  \nonumber
\end{eqnarray}
In evaluating the sum over the Matsubara frequencies $\varepsilon _{n}$ in (%
\ref{d22a}) it is useful to split it into the two parts. In the first $%
\varepsilon _{n}$ belongs to the domains $]-\infty ,-\omega _{\nu }[$ and $%
[0,\infty \lbrack $, which finally give two equal contributions. This gives
rise to the \textit{regular} part of the MT diagram. The second, \textit{%
anomalous}, part of the MT diagram arises from the summation over $%
\varepsilon _{n}$ in the domain $[-\omega _{\nu },0[$. In this interval, the
further analytic continuation over $\omega _{\nu }$ leads to the appearance
of an additional diffusive pole:

\begin{eqnarray*}
I_{\alpha \beta }(q,0,\omega _{\nu }) &=&I_{\alpha \beta }^{(reg)}(q,\omega
_{\nu })+I_{\alpha \beta }^{(an)}(q,\omega _{\nu })=\nu \left\langle
v_{\alpha }(p)v_{\beta }(q-p)\right\rangle _{FS} \\
&&\times \left[ 2\pi T\sum_{n=0}^{\infty }\frac{1}{\left( 2\varepsilon
_{n+\nu }+\widehat{D}q^{2}\right) }\frac{1}{\left( 2\varepsilon _{n}+%
\widehat{D}q^{2}\right) }\frac{1}{2\varepsilon _{n}+\omega _{\nu }+\tau ^{-1}%
}~+\right. \\
&&\left. +\frac{\pi T}{\omega _{\nu }+\tau ^{-1}}\sum_{n=-\nu }^{-1}\frac{1}{%
\left( 2\varepsilon _{n+\nu }+\widehat{D}q^{2}\right) }\frac{1}{\left(
-2\varepsilon _{n}+\widehat{D}q^{2}\right) }\right] .
\end{eqnarray*}
\newline
The limits of summation in the first sum do not depend on $\omega _{\nu },$
so it is an analytic function of this argument and can be continued to the
upper half-plane of the complex frequency by the simple substitution $\omega
_{\nu }\rightarrow -i\omega .$ Then, tending $\omega \rightarrow 0,$ one can
expand the sum over powers of $\omega $ and perform the summation in terms
of digamma-function:

\begin{eqnarray}
I_{\alpha \beta }^{(reg)}(q,\omega _{\nu }) &=&\nu \left\langle v_{\alpha }(%
\mathbf{p})v_{\beta }(\mathbf{q-p})\right\rangle _{FS}\times  \nonumber \\
&&\left\{ const+\frac{i\omega }{2}\frac{\partial }{\partial \left( \widehat{D%
}q^{2}\right) }\left( \frac{\partial }{\partial \left( \widehat{D}%
q^{2}\right) }+\frac{\partial }{\partial \left( \tau ^{-1}\right) }\right)
\cdot \right. \\
&&\left. \frac{1}{\tau ^{-1}-\widehat{D}q^{2}}\left[ \psi \left( \frac{1}{2}+%
\frac{\omega _{\nu }+\widehat{D}q^{2}}{4\pi T}\right) -\psi \left( \frac{1}{2%
}+\frac{\widehat{D}q^{2}}{4\pi T}\right) \right] \right\} .  \nonumber
\end{eqnarray}
The values of characteristic momenta $q\ll l^{-1}$\ are determined by the
domain of convergibility of the final integral of the propagator $L(\mathbf{q%
},0)$ in (\ref{d21}) (analogously to (\ref{d17}) and one can neglect $%
\widehat{D}q^{2}$ with respect to $\tau ^{-1}.$ In result 
\begin{eqnarray}
I_{\alpha \beta }^{(reg)R}(q,\omega &\rightarrow &0)=\nu \left\langle
v_{\alpha }(\mathbf{p})v_{\beta }(\mathbf{q-p})\right\rangle _{FS}\times
\label{regfi} \\
&&\left\{ const+\frac{i\omega \tau ^{2}}{4}\left[ \psi ^{\prime }\left( 
\frac{1}{2}+\frac{1}{4\pi \tau T}\right) -\psi ^{\prime }\left( \frac{1}{2}%
\right) -\frac{\psi ^{\prime \prime }\left( \frac{1}{2}\right) }{4\pi T\tau }%
\right] \right\} .  \nonumber
\end{eqnarray}
The appearance of the constant in $Q_{\alpha \beta }(\omega _{\nu })$ was
already discussed in the case of the AL contribution and, as was mentioned
there, it is cancelled with the similar contributions of the other diagrams 
\cite{AV80} and we will not consider it any more.

Now let us pass to the calculation of $I_{\alpha \beta }^{(an)}(q,\omega
_{\nu }).$ Expanding the summing function in simple fractions one can
express the result of summation in terms of digamma-functions

\begin{eqnarray}
I_{\alpha \beta }^{(an)}(q,\omega _{\nu }) &=&\frac{1}{4}\frac{\nu
\left\langle v_{\alpha }(\mathbf{p})v_{\beta }(\mathbf{q-p})\right\rangle
_{FS}}{\omega _{\nu }+\tau ^{-1}}\frac{1}{\omega _{\nu }+\widehat{D}q^{2}}%
\times  \label{angen} \\
&&\left[ \psi \left( \frac{1}{2}+\frac{2\omega _{\nu }+\widehat{D}q^{2}}{%
4\pi T}\right) -\psi \left( \frac{1}{2}+\frac{\widehat{D}q^{2}}{4\pi T}%
\right) \right] .  \nonumber
\end{eqnarray}
Doing the analytical continuation $i\omega _{\nu }\rightarrow \omega
\rightarrow 0$ and taking into account that in the further $q$-integration
of $I_{\alpha \beta }^{(an)R}(q,\omega \rightarrow 0),$ due to the singular
at small $q$ propagator, the important range is $\widehat{D}q^{2}\ll T$, one
can find

\begin{equation}
I_{\alpha \beta }^{(an)R}(q,\omega \rightarrow 0)=-\frac{i\pi \omega \tau }{%
16T}\frac{\nu \left\langle v_{\alpha }(\mathbf{p})v_{\beta }(\mathbf{q-p}%
)\right\rangle _{FS}}{-i\omega +\widehat{D}q^{2}}.  \label{anfinal}
\end{equation}
\ Because of the considerable difference in the angular averaging of the
different tensor components we discuss the MT contribution to the in-plane
and out of plane conductivities separately.

Taking into account that $\left\langle v_{x}(\mathbf{p})v_{x}(\mathbf{q-p}%
)\right\rangle _{FS}=-v{_{F}^{2}/2}$ one can find that the calculation of
the regular part of MT diagram to the in-plane conductivity is completely
similar to the corresponding DOS contribution and here we list the final
result\ \cite{BDKLV93} only: 
\[
\sigma _{xx}^{MT(reg)}=-\frac{e^{2}}{2s}\tilde{\kappa}\ln \left( \frac{2}{{%
\epsilon ^{1/2}+(\epsilon +r)^{1/2}}}\right) , 
\]
where 
\begin{eqnarray}
{\tilde{\kappa}}(T\tau ) &=&{\frac{{-\psi ^{^{\prime }}\left( {{\frac{1}{2}}+%
{\frac{1}{{4\pi \tau T}}}}\right) +\psi ^{^{\prime }}}\left( {{\frac{1}{2}}}%
\right) {+{\frac{1}{{4\pi T\tau }}}\psi ^{^{\prime \prime }}}\left( {{\frac{1%
}{2}}}\right) }{{\pi ^{2}}\left[ {\psi \left( {{\frac{1}{2}}+{\frac{1}{{4\pi
\tau T}}}}\right) -\psi \left( {{\frac{1}{2}}}\right) -{\frac{1}{{4\pi \tau T%
}}}\psi ^{^{\prime }}\left( {{\frac{1}{2}}}\right) }\right] }}  \nonumber
\label{d25} \\
&\rightarrow &\left\{ 
\begin{array}{c}
28\zeta (3)/\pi ^{4}\approx 0.346,\;\mbox{for }T\tau \ll 1 \\ 
\pi ^{2}/\left[ 14\zeta (3)\right] \approx 0586,\;\mbox{for }1\ll T\tau \ll
1/\sqrt{\epsilon }
\end{array}
\right.
\end{eqnarray}
is a function only of $\tau T$. We note that this regular MT term is
negative, as is the overall DOS contribution.

For the anomalous part of the in-plane MT contribution we have: 
\begin{eqnarray}
\sigma _{xx}^{MT(an)} &=&8e^{2}\eta _{(2)}T\int {\frac{{d^{3}q}}{{(2\pi )^{3}%
}}}\frac{1}{[1/\tau _{\varphi }+\hat{D}q^{2}][\epsilon +\eta _{(2)}\mathbf{q}%
^{2}+{\frac{r}{2}}(1-\cos q_{z}s)]}  \nonumber \\
&=&\frac{e^{2}}{4s(\epsilon -\gamma _{\varphi })}\ln \left( \frac{\epsilon
^{1/2}+(\epsilon +r)^{1/2}}{\gamma _{\varphi }^{1/2}+(\gamma _{\varphi
}+r)^{1/2}}\right) ,  \label{d25_1}
\end{eqnarray}
where, in accordance with\ \cite{T70}, the infra-red divergence for the
purely 2D case $(r=0)$ is cut off at $Dq^{2}\sim 1/\tau _{\varphi }$%
\footnote{%
The detailed study of the phase-breaking time, its energy dependence and the
effect on the MT contribution was done in \cite{Rei92}}. The dimensionless
parameter 
\[
\gamma _{\varphi }={\frac{{2\eta }}{v{_{F}^{2}\tau \tau _{\varphi }}}}%
\rightarrow {\frac{{\pi }}{{\ 8T\tau _{\varphi }}}}\left\{ 
\begin{array}{c}
1,\;\mbox{ }T\tau \ll 1 \\ 
7\zeta (3)/\left( 2\pi ^{3}T\tau \right) ,\;\mbox{ }1\ll T\tau \ll 1/\sqrt{%
\epsilon }
\end{array}
\right. 
\]
is introduced for simplicity. If $r\neq 0$ the MT contribution turns out to
be finite even with {$\tau _{\varphi }$}${=\infty .}$ Comparison of the
expressions (\ref{d12}) and (\ref{d25_1}) indicates that in the weak
pair-breaking limit, the MT diagram makes an important contribution to the
longitudinal fluctuation conductivity: it is four times larger than the AL
contribution in the 3D regime, and even logarithmically exceeds it in the 2D
regime above $T_{LD}$. For finite pair-breaking, however, the MT
contribution is greatly reduced in magnitude.

We now consider the calculation of the MT contribution to the transverse
conductivity. The explicit expressions for $v_{z}(p)$ and $v_{z}(q-p)$ (see
Exp. (\ref{vz})), result in $\left\langle v_{x}(p)v_{x}(q-p)\right\rangle
_{FS}$ $=\frac{1}{2}J^{2}s^{2}\cos q_{z}s$. We take the limit $J\tau <<1$ in
evaluating the remaining integrals, which may then be performed exactly.

\bigskip The regular part of the MT contribution to the transverse
conductivity is 
\begin{eqnarray*}
\sigma _{zz}^{MT(\mathrm{reg})} &=&-{\frac{{e^{2}s^{2}\pi r\tilde{\kappa}}%
(T\tau )}{{4}}}\int {\ \frac{{d^{3}q}}{{(2\pi )^{3}}}}{\frac{{\cos q_{z}s}}{{%
\epsilon +\eta _{(2)}\mathbf{q}^{2}+{\ \frac{r}{2}}(1-\cos q_{z}s)}}} \\
&=&-{\frac{{e^{2}sr\tilde{\kappa}}(T\tau )}{{16\eta _{(2)}}}}\left( {\frac{{%
\ (\epsilon +r)^{1/2}-\epsilon ^{1/2}}}{{r^{1/2}}}}\right) ^{2}
\end{eqnarray*}
This term is smaller in magnitude than is the DOS one, and therefore makes a
relatively small contribution to the overall fluctuation conductivity. In
the 3D regime below $T_{LD}$, it is proportional to $J^{2}$, and in the 2D
regime above $T_{LD}$, it is proportional to $J^{4}$.

For the anomalous part of the MT diagram one can find 
\begin{eqnarray}
\sigma _{zz}^{MT(\mathrm{an)}} &=&{\frac{{\pi e^{2}J^{2}s^{2}\tau }}{{4}}}%
\int {\frac{{d^{3}q}}{{(2\pi )^{3}}}}{\frac{{\cos q_{z}s}}{{[1/\tau
_{\varphi }+\hat{\mathbf{D}}q^{2}][\epsilon +\eta _{(2)}\mathbf{q}^{2}+{%
\frac{r}{2}}(1-\cos q_{z}s)]}}}  \nonumber \\
&=&{\frac{{\pi e^{2}s}}{{4(\epsilon -\gamma _{\varphi })}}}\int {\frac{{d^{2}%
\mathbf{q}}}{{\ (2\pi )^{2}}}}\left[ \frac{{\gamma _{\varphi }+\eta _{(2)}%
\mathbf{q}^{2}+r/2}}{\left[ {(\gamma _{\varphi }+\eta _{(2)}\mathbf{q}%
^{2})(\gamma _{\varphi }+\eta _{(2)}\mathbf{q}^{2}+r)}\right] {^{1/2}}}%
-\right.  \nonumber \\
&&\left. -{\frac{{\epsilon +\eta _{(2)}\mathbf{q}^{2}+r/2}}{\left[ {%
(\epsilon +\eta _{(2)}\mathbf{q}^{2})(\epsilon +\eta _{(2)}\mathbf{q}^{2}+r)}%
\right] {^{1/2}}}}\right]  \nonumber \\
&=&{\frac{{e^{2}s}}{{16}\eta _{(2)}}}\left( {\frac{{\gamma _{\varphi
}+r+\epsilon }}{{\ [\epsilon (\epsilon +r)]^{1/2}+[\gamma _{\varphi }(\gamma
_{\varphi }+r)]^{1/2}}}}-1\right) .\ \ \ \ \ \ \ \ \ \ \ \ \ \ \ \ \ \ 
\label{d26}
\end{eqnarray}
In examining the limiting cases of (\ref{d26}), it is useful to consider the
cases of weak ($\gamma _{\varphi }<<r$, $\Longleftrightarrow $ $J^{2}\tau
\tau _{\varphi }>>1/2$) and strong ($\gamma _{\varphi }>>r$,$%
\Longleftrightarrow $ $J^{2}\tau \tau _{\varphi }<<1/2$) pair-breaking
separately\footnote{%
{}
\par
Physically the value $J^{2}\tau $. characterizes the effective interlayer
tunneling rate \ \cite{BDKLV93,V94}. When $1/\tau _{\phi }<<J^{2}\tau
<<1/\tau $, the quasiparticles scatter many times before tunneling to the
neighboring layers, and the pairs live long enough for them to tunnel
coherently. When $J^{2}\tau <<1/\tau _{\phi }$, the pairs decay before both
paired quasiparticles tunnel.}. For weak pair-breaking, we have 
\[
\sigma _{zz}^{MT(\mathrm{an})}\rightarrow {\frac{{e^{2}s}}{{16}\eta _{(2)}}}%
\left\{ 
\begin{array}{c}
\sqrt{r/\gamma _{\varphi }},\;\mbox{ }\epsilon \ll \gamma _{\varphi }\ll r
\\ 
\sqrt{r/\epsilon },\;\mbox{ }\gamma _{\varphi }\ll \epsilon \ll r \\ 
r/\left( 2\epsilon \right) ,\;\mbox{ }\gamma _{\varphi }\ll r\ll \epsilon
\end{array}
\;\right. . 
\]
In this case, there is the usual 3D to 2D dimensional crossover in the
anomalous MT contribution at $T_{LD}$. There is an additional crossover at $%
T_{\varphi }$(where $T_{c}<T_{\varphi }<T_{LD}$), characterized by $\epsilon
(T_{\varphi })=\gamma _{\varphi }$, below which the anomalous MT term
saturates. Below $T_{LD}$, the MT contribution is proportional to $J$, but
in the 2D regime above $T_{LD}$, it is proportional to $J^{2}$.

For strong pair-breaking 
\[
\sigma _{zz}^{MT(\mathrm{an})}\rightarrow {\frac{{e^{2}s}}{{32}\eta _{(2)}}}%
\left\{ 
\begin{array}{c}
r/\gamma _{\varphi },\;\mbox{ }\epsilon \ll r\ll \gamma _{\varphi } \\ 
r^{2}/\left( 4\gamma _{\varphi }\epsilon \right) ,\;\mbox{ }r\ll \min
\{\gamma _{\varphi },\epsilon \}
\end{array}
\right. . 
\]
In this case, the 3D regime (below $T_{LD}$) is not singular, and the
anomalous MT contribution is proportional to $J^{2}$, rather than $J$ for
weak pair-breaking. In the 2D regime, it is proportional to $J^{4}$ for
strong pair-breaking, as opposed to $J^{2}$ for weak pair-breaking. In
addition, the overall magnitude of the anomalous MT contribution with strong
pair-breaking is greatly reduced from that for weak pair-breaking.

Let us now compare the regular and anomalous MT contributions. Since these
contributions are opposite in sign, it is important to determine which will
dominate. For the in-plane resistivity, the situation is straightforward:
the anomalous part always dominates over the regular and the latter can be
neglected. The case of $c-$axis resistivity requires more discussion. Since
we expect $\tau _{\varphi }\geq \tau $, strong pair-breaking is likely in
the dirty limit. When the pair-breaking is weak, the anomalous term is
always of lower order in $J$ than the regular term, so the regular term can
be neglected. This is true for both the clean and dirty limits. The most
important regime for the regular MT term is the dirty limit with strong
pair-breaking. In this case, when $\tau _{\varphi }T\sim 1$, the regular and
anomalous terms are comparable in magnitude. In short, it is usually a good
approximation to neglect the regular term, except in the dirty limit with
relatively strong pair-breaking and only for the out-of-plane conductivity.

Finally let us mention that the contributions from the two other diagrams of
the MT type (diagrams 3 and 4 of Fig. \ref{conddia}) in the vicinity of
critical temperature can be omitted: one can check that they have an
additional square of the Cooper pair center of mass momentum $q$ in the
integrand of $q$-integration with respect to diagram 2 and hence turn out to
be less singular in $\epsilon $.

\subsubsection{Discussion}

Although the in-plane and out-of plane components of the fluctuation
conductivity tensor of a layered superconductor contain the same fluctuation
contributions, their temperature behavior may be qualitatively different. In
fact, for $\sigma _{xx}^{fl}$, the negative contributions are considerably
less than the positive ones in the entire experimentally accessible\
temperature range above the transition, and it is a positive monotonic
function of the temperature. Moreover, for HTS compounds, where the
pair-breaking is strong and the MT contribution is in the overdamped regime,
it is almost always enough to take into account only the paraconductivity to
fit experimental data. Some examples of the experimental findings for
in-plane fluctuation conductivity of HTS\ materials on can see in \cite
{AL88,FTP87,HS90,AAR88,PMDGC89,KGKGK89,BNV89,KW90}.

\begin{figure}[tbp]
\epsfxsize=6cm
\centerline {\epsfbox{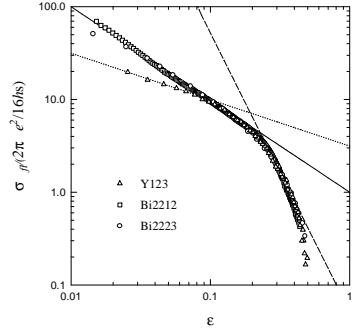}}
\caption{{} The normalised excess conductivity for samples of YBCO-123
(triangles), BSSCO-2212(squares) and BSSCO-2223 (circles) plotted against $%
\protect\epsilon =\ln T/T_{c}$ on a ln-ln plot as described in \protect\cite
{FPFV97}. The dotted and solid lines are the AL theory in 3D and 2D
respectively. The dashed line is the extended theory of \protect\cite{RVV91}%
.. }
\label{fluplane}
\end{figure}

In Fig. \ref{fluplane} \ the fluctuation part of in-plane conductivity$\
\sigma _{xx}^{fl}$ is plotted as a function of $\epsilon =\ln T/T_{c}$ on a
double logarithmic scale for three HTS samples (the solid line represents
the $2D$ AL behavior $\left( 1/\epsilon \right) $, the dotted line
represents the $3D$ one: $3.2/\sqrt{\epsilon }$) \cite{FPFV97}. One can see
that paraconductivity of the less anisotropic YBCO compound asymptotically
tends to the 3D behavior ($1/\epsilon ^{1/2}$) for $\epsilon <0.1$, showing
the LD crossover at $\epsilon \approx 0.07$; \ the curve for more
anisotropic 2223 phase of BSCCO starts to bend for $\epsilon <0.03$ while
the most anisotropic 2212 phase of BSCCO shows a $2D$ behavior in the whole
temperature range investigated. All three compounds show a universal $2D$
temperature behavior above the LD crossover up to the limits of the GL
region. \ It is interesting that around $\epsilon \approx 0.24$ all the
curves bend down and follow the same asymptotic $1/\epsilon ^{3}$ behavior
(dashed line). Finally at the value $\epsilon \approx 0.45$ all the curves
fall down indicating the end of the observable fluctuation regime.

Reggiani et al. \cite{RVV91} extended the $2D$ AL theory to the high
temperature region by taking into account the short wavelength fluctuations.
The following universal formula for $2D$ paraconductivity of a clean 2D
superconductor as a function of the generalized reduced temperature $%
\epsilon =\ln T/T_{c}$ \ was obtained\footnote{%
In Section 7.2 we will demonstrate how such a dependence $\left( 1/\ln
^{3}(T/T_{\mathrm{c}})\right) $ appears by accounting for short wavelength
fluctuations for the $2D$ fluctuation susceptibility.}: 
\[
\sigma _{xx}^{fl}=\frac{e^{2}}{16s}f(\epsilon ) 
\]
with $f(\epsilon )=\epsilon ^{-1},\;\epsilon \ll 1$ and $f(\epsilon
)=\epsilon ^{-3},\;\epsilon \gtrsim 1.$

\begin{figure}[tbp]
\epsfxsize=8cm
\centerline {\epsfbox{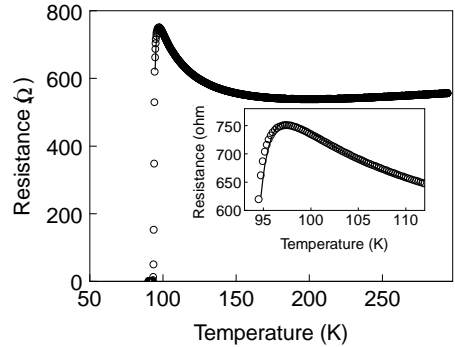}}
\caption{{}\ Fit of the temperature dependence of the transverse resistance
of an underdoped BSCCO c-axis oriented film with the results of the
fluctuation theory \protect\cite{BMV93}. The inset shows the details of the
fit in the temperature range between $T_{c}$ and 110K.}
\label{flucaxis}
\end{figure}

In the case of the out-of-plane conductivity the situation is quite
different. Both positive contributions (AL and anomalous MT) are suppressed
by the interlayer transparency, leading to a competition between positive
and negative terms. This can lead to a maximum in the c-axis fluctuation
resistivity which occurs in the 2D regime (in the case discussed $J\tau
<<1,r\kappa <<1$ and $\gamma _{\varphi }\kappa >1)$ : 
\[
\epsilon _{m}/r\approx \frac{1}{{(8r}\kappa {)^{1/2}}}-\frac{1}{{8\kappa }}%
\left[ \tilde{\kappa}-{\frac{1}{{2\gamma _{\varphi }}}}\right] . 
\]
This nontrivial effect of fluctuations on the transverse resistance of a
layered superconductor allows a successful fit to the data observed on
optimally and overdoped HTS samples (see, for instance, Fig.\ref{flucaxis})
where the growth of the resistance still can be treated as the correction.

The fluctuation mechanism of the growth of the transverse resistance can be
easily understood in a qualitative manner. Indeed to modify the in-plane
result (\ref{para2D}) for the case of c-axis paraconductivity one has to
take into account the hopping character of the electronic motion in this
direction. If the probability of one-electron interlayer hopping is $%
\mathcal{P}_{1}$, then the probability of coherent hopping for two electrons
during the fluctuation Cooper pair lifetime $\tau _{GL}$ is the conditional
probability of these two events: $\mathcal{P}_{2}=\mathcal{P}_{1}(\mathcal{P}%
_{1}\tau _{GL}).$ The transverse paraconductivity may thus be estimated as $%
\sigma _{\perp }^{AL}\sim \mathcal{P}_{2}\sigma _{\parallel }^{AL}\sim 
\mathcal{P}_{1}^{2}\frac{1}{\epsilon ^{2}},$ in complete accordance with (%
\ref{d12_2}). We see that the temperature singularity of $\sigma _{\perp
}^{AL}$ turns out to be stronger than that in $\sigma _{\parallel }^{AL},$
however for a strongly anisotropic layered superconductor $\sigma _{\perp
}^{AL}$ is considerably suppressed by the square of the small probability of
inter-plane electron hopping which enters in the pre-factor. It is this
suppression which leads to the necessity of taking into account the DOS
contribution to the transverse conductivity. The latter is less singular in
temperature but, in contrast to the paraconductivity, manifests itself in
the first, not the second, order in the interlayer transparency $\sigma
_{\perp }^{DOS}\sim -\mathcal{P}_{1}\ln {\ \frac{1}{\varepsilon }.}$ The DOS
fluctuation correction to the one-electron transverse conductivity is
negative and, being proportional to the first order of $\mathcal{P}_{1}$,
can completely change the traditional picture of fluctuations just rounding
the resistivity temperature dependence around transition. The shape of the
temperature dependence of the transverse resistance mainly is determined by
competition of the opposite sign contributions: the paraconductivity and MT
term, which are strongly temperature dependent but are suppressed by the
square of the barrier transparency and the DOS contribution which has a
weaker temperature dependence but depends only linearly on the barrier
transparency.

\section[Manifestation of fluctuations ...]{Manifestation of fluctuations in
various properties}

In this Section we will demonstrate the applications of the microscopic
theory of fluctuations. The limited volume does not permit us to deliver
here the systematic review of the modern theory and we restrict ourselves
only by presentation of the several representative recent studies. The
details one can find in the articles cited in subtitles.

It is necessary to underline that the comparison of the results of
fluctuation theory with the experimental findings on HTS materials has to be
considered sooner in qualitative than quantitative context. Indeed, as is
clear now, the superconductivity in the most of HTS compounds has the
nontrivial symmetry. Moreover, as was discussed in the previous Section,
these compounds are rather clean than dirty. Both these complications can be
taken into account (see for example \cite{Yip,BDKLV93}) but this was not
done in the majority of the cited papers.

\subsection{ The effects of fluctuations on magnetoconductivity \protect\cite
{BDKLV93,BMV95,BV98,Wahl00}}

The experimental investigations of the fluctuation magnetoconductivity are
of special interest first because this physical value weakly depends on the
normal state properties of superconductor and second due to its special
sensitivity to temperature and magnetic field. The role of AL contribution
for both the in-plane and out-of-plane magnetoconductivities was studied
above in the framework of the phenomenological approach. The microscopic
calculations of the other fluctuation corrections to the in-plane
magnetoconductivity conductivity show that the MT contribution has the same
positive sign and temperature singularity as the AL one. In the case of weak
pair-breaking it can even considerably exceed the latter. The negative DOS
contribution, like in the case of the zero-field conductivity, turns out to
be considerably less singular and many authors (see e.g. Refs. \cite
{SH91,MIKY89,SIKYH92,HARKM93,LLM95,LHKS95,HRJH95,L95,WM96,L95B},\cite
{VRQ97,SLP95}) successfully explained the in-plane magnetoresistance data in
HTS using the AL and MT contributions only \cite{HL88,AHL89,BM90,BMT91}.

Turning to the out-of-plane magnetoconductivity of a layered superconductor
one can find a quite different situation. Both the AL and MT contributions
turn out to be here of the second order in the interlayer transparency and
this circumstance makes the less singular DOS contribution, which remains
however of first order in transparency, to be competitive with the main
terms \cite{BMV95}. The large number of microscopic characteristics involved
in this competition, like the Fermi velocity, interlayer transparency,
phase-breaking and elastic relaxation times, gives rise to the possibility
of occurrence of different scenarios for various compounds. The c-axis
magnetoresistance of a set of HTS materials shows a very characteristic
behavior above $T_{c0}$. In contrast to the ab-plane magnetoresistance which
is positive at all temperatures, the magnetoresistance along the c-axis has
been found in many HTS compounds (BSSCO \cite{YMH95,NTHKT94,HASH,HLW96},
LSSCO \cite{KMT96}, YBCO \cite{AHE} and TlBCCO \cite{WAHL98}) to have a
negative sign not too close to $T_{c0}$ and turn positive at lower
temperatures. We will show how this behavior find its explanation within the
fluctuation theory \cite{BDKLV93}.

We consider here the effect of a magnetic field parallel to the $c$-axis. In
this case both quasiparticles and Cooper pairs move along Landau orbits
within the layers. The $c$-axis dispersion remains unchanged from the
zero-field form. In the chosen geometry one can generalize the zero-field
results reported in the previous Section to finite field strengths simply by
the replacement of the two-dimensional integration over $\mathbf{q}$ by a
summation over the Landau levels 
\[
\int {\frac{{d^{2}\mathbf{q}}}{{(2\pi )^{2}}}}\rightarrow {\frac{H}{\Phi _{0}%
}}\sum_{n}={\frac{h}{{2\pi \eta }_{(2)}}}\sum_{n} 
\]
(let us remind that ${\eta }_{(2)}=\xi _{xy}^{2}$). So the general
expressions for all fluctuation corrections to the c-axis conductivity in a
magnetic field can be simply written in the form \cite{BDKLV93}:

\begin{equation}
\sigma _{zz}^{AL}=\frac{e^{2}sr^{2}h}{64{\xi }_{xy}^{2}}\sum_{n=0}^{\infty }%
\frac{1}{\left\{ [\epsilon +h(2n+1)][r+\epsilon +h(2n+1)]\right\} ^{3/2}}
\label{ALzz}
\end{equation}
\begin{equation}
\sigma _{zz}^{DOS}=-\frac{e^{2}sr\kappa h}{8{\xi }_{xy}^{2}}\sum_{n=0}^{1/h}%
\frac{1}{\left\{ [\epsilon +h(2n+1)][r+\epsilon +h(2n+1)]\right\} ^{1/2}}
\label{doszz}
\end{equation}
\begin{equation}
\sigma _{zz}^{MT(reg)}=-\frac{e^{2}s\tilde{\kappa}h}{4{\xi }_{xy}^{2}}%
\sum_{n=0}^{\infty }\left( \frac{\epsilon +h(2n+1)+r/2}{\left\{ [\epsilon
+h(2n+1)][r+\epsilon +h(2n+1)]\right\} ^{1/2}}-1\right)  \label{MTrg}
\end{equation}
\[
\sigma _{zz}^{MT(an)}=\frac{e^{2}sh}{8{\xi }_{xy}^{2}(\varepsilon -\gamma
_{\varphi })}\sum_{n=0}^{\infty }\left( \frac{\gamma _{\varphi }+h(2n+1)+r/2%
}{\left\{ [(\gamma _{\varphi }+h(2n+1)][\gamma _{\varphi
}+h(2n+1)+r)]\right\} ^{1/2}}\right. - 
\]
\begin{equation}
-\left. \frac{\epsilon +h(2n+1)+r/2}{\left\{ [\epsilon +h(2n+1)][r+\epsilon
+h(2n+1)]\right\} ^{1/2}}\right) .  \label{sumcor}
\end{equation}

For the in-plane component of the fluctuation conductivity tensor the only
additional problem appears in the AL diagram, where the matrix elements of
the harmonic oscillator type, originating from the $B_{\parallel }$ $%
(q_{\parallel })$ blocks, have to be calculated. How to do this was
demonstrated in details in Section 4. The other contributions are
essentially analogous to their c-axis counterparts:

\begin{eqnarray}
\sigma _{xx}^{AL} &=&{\frac{{e^{2}}}{{4s}}}\sum_{n=0}^{\infty }(n+1)\left( {%
\frac{1}{\left\{ [\epsilon +h(2n+1)][r+\epsilon +h(2n+1)]\right\} ^{1/2}}}%
\right. -  \nonumber \\
&&{\frac{2}{\left\{ [\epsilon +h(2n+2)][r+\epsilon +h(2n+2)]\right\} ^{1/2}}}%
+  \label{l1} \\
&&\left. {\frac{1}{\left\{ [\epsilon +h(2n+3)][r+\epsilon +h(2n+3)]\right\}
^{1/2}}}\right) ,  \nonumber
\end{eqnarray}

\begin{equation}
\sigma _{xx}^{DOS}+\sigma _{xx}^{MT(reg)}=-{\frac{{e^{2}h(\kappa +\tilde{%
\kappa})}}{{2s}}}\sum_{n=0}^{1/h}{\frac{1}{\left\{ [\epsilon
+h(2n+1)][r+\epsilon +h(2n+1)]\right\} ^{1/2}}},  \label{12}
\end{equation}

and 
\begin{eqnarray}
\sigma _{xx}^{MT(an)} &=&{\frac{{e^{2}h}}{{4s(\epsilon -\gamma _{\varphi })}}%
}\sum_{n=0}^{\infty }\left( {\frac{1}{\left\{ [(\gamma _{\varphi
}+h(2n+1)][\gamma _{\varphi }+h(2n+1)+r)]\right\} ^{1/2}}}\right. - 
\nonumber \\
&&\left. {\frac{1}{\left\{ [\epsilon +h(2n+1)][r+\epsilon +h(2n+1)]\right\}
^{1/2}}}\right) .  \label{l3}
\end{eqnarray}
These results can in principle be already used for numerical evaluations and
fitting of the experimental data which was indeed successfully done in a
series of \ papers \cite{AHE,NVLBM96,WAHL98}.

The detailed comparison of the cited results with the experimental data \cite
{AHE,WAHL99}, especially in strong fields, raised the problem of
regularization of the DOS contribution. If in the absence of the magnetic
field its ultra-violet divergence was successfully cut off at $q\sim \xi
^{-1},$ in the case under consideration the cut off parameter depends on the
magnetic field and makes the fitting procedure ambiguous. The solution of
this problem was proposed in \cite{BV98}, where the authors calculated the
difference $\Delta \sigma _{zz}^{DOS}=$ $\sigma _{zz}^{DOS}(h,\epsilon
)-\sigma _{zz}^{DOS}(0,\epsilon )$ applying to formulas (\ref{doszz}) and (%
\ref{12}) the same trick which was already used in Section 2 for the
regularization of the free energy in magnetic field (Eq. (\ref{deltaF})).
The corresponding asymptotics for all out-of-plane fluctuation contributions
are presented in the following table:

\begin{tabular}{|l|l|l|l|}
\hline
& $h\ll \epsilon $ & $\epsilon \ll h\ll r\quad (3D)$ & $\max \{\epsilon
,r\}\ll h\quad (2D)$ \\ \hline
$\Delta \sigma _{zz}^{DOS}$ & $\frac{{e^{2}s\kappa }}{3\;2^{5}{\xi }_{xy}^{2}%
}\frac{r{(\epsilon +r/2)}}{{[\epsilon (\epsilon +r)]^{3/2}}}h^{2}$ & $0.428%
\frac{e^{2}s\kappa }{16{\xi }_{xy}^{2}}r\sqrt{\frac{h}{r}}$ & $\frac{%
e^{2}s\kappa }{8{\xi }_{xy}^{2}}r\ln \frac{\sqrt{h}}{\sqrt{\epsilon }+\sqrt{%
\epsilon +r}}$ \\ \hline
$\Delta \sigma _{zz}^{MT(reg)}$ & $\frac{{e^{2}s\tilde{\kappa}}}{{%
3\;2^{6}\xi }_{xy}^{2}}\frac{r^{2}}{{[\epsilon (\epsilon +r)]^{3/2}}}h^{2}$
& $0.428\frac{e^{2}s\tilde{\kappa}}{8{\xi }_{xy}^{2}}r\sqrt{\frac{h}{r}}$ & $%
-\sigma _{zz}^{MT(reg)}(0,\epsilon )-\frac{\pi ^{2}e^{2}s\tilde{\kappa}}{%
2^{8}{\xi }_{xy}^{2}}\frac{r^{2}}{h}$ \\ \hline
$-\Delta \sigma _{zz}^{AL}$ & $\frac{{e^{2}s}}{2^{8}{\xi }_{xy}^{2}}\frac{{%
r^{2}(\epsilon +r/2)}}{{[\epsilon (\epsilon +r)]^{5/2}}}h^{2}$ & $\sigma
_{zz}^{AL}(0,\epsilon )-\frac{3.24{e^{2}s}}{{\xi }_{xy}^{2}}\sqrt{\frac{r}{h}%
}$ & $\sigma _{zz}^{AL}(0,\epsilon )-\frac{7\zeta (3){e^{2}s}}{2^{9}{\xi }%
_{xy}^{2}}\frac{r^{2}}{h^{2}}$ \\ \hline
\begin{tabular}{c}
$-\Delta \sigma _{zz}^{MT(an)}$ \\ 
$\min \{\epsilon ,r\}\ll \gamma _{\varphi }$%
\end{tabular}
& $\frac{{e^{2}s}}{{3\;2^{7}\xi }_{xy}^{2}}\frac{r^{2}}{{[\epsilon (\epsilon
+r)]^{2}}}h^{2}$ & $\sigma _{zz}^{MT(an)}(0,\epsilon )-\frac{e^{2}s}{32{\xi }%
_{xy}^{2}}\sqrt{\frac{r}{\gamma _{\varphi }}}$ & $\sigma
_{zz}^{MT(an)}(0,\epsilon )-\frac{3\pi ^{2}{e^{2}s}}{2^{8}{\xi }_{xy}^{2}}%
\frac{{\max \{r,\gamma _{\varphi }\}}}{{h}}$ \\ \hline
\begin{tabular}{c}
$-\Delta \sigma _{zz}^{MT(an)}$ \\ 
$\gamma _{\varphi }\ll \min \{\epsilon ,r\}$%
\end{tabular}
& ${\frac{{e^{2}s}}{{3\;2^{7}\xi }_{xy}^{2}}}\frac{\sqrt{r}}{{\epsilon
\gamma _{\varphi }^{3/2}}}h^{2}$ & $\sigma _{zz}^{MT(an)}(0)-\frac{3.24e^{2}s%
}{64{\xi }_{xy}^{2}}\sqrt{\frac{r}{h}}$ & $\sigma _{zz}^{MT(an)}(0,\epsilon
)-\frac{3\pi ^{2}{e^{2}s}}{2^{8}{\xi }_{xy}^{2}}\frac{(r+\epsilon )}{h}$ \\ 
\hline
\end{tabular}

\begin{center}
Table 3
\end{center}

\begin{figure}[tbp]
\centerline {\epsfig {width=8cm,file=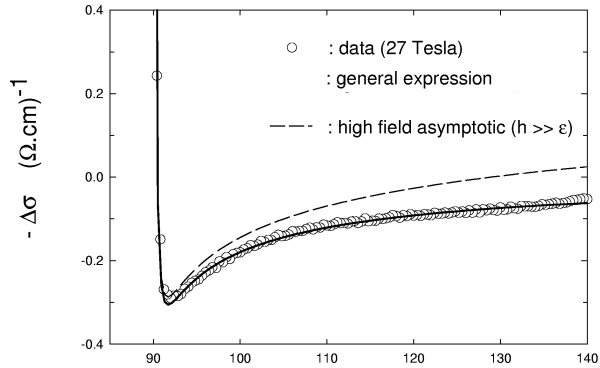,scale=.5}}
\caption{{}Magnetoconductivity versus temperature at 27 T for an underdoped
Bi-2212 single crystal. The solid line represents the theoretical
calculation. The symbols are the experimental magnetoconductivity $\Delta 
\protect\sigma _{zz}(B\shortparallel c\shortparallel I)$\protect\cite{Wahl00}
}
\label{Wahl2}
\end{figure}

The procedure described gives an excellent fitting up to very high fields 
\cite{Wahl00} which is shown in Fig . \ref{Wahl2}.

Let us start the analysis from the $2D$ case ($r\ll \epsilon $). One can see
that here the positive DOS contribution to magnetoconductivity turns out to
be dominant. It grows as $H^{2}$ up to $H_{c2}(\epsilon )$ and then crosses
to a slow logarithmic asymptote. At $H\sim H_{c2}(0)$ the value of $\Delta
\sigma _{zz}^{DOS}(h\sim 1,\epsilon )=-\sigma _{zz}^{DOS}(0,\epsilon )$
which means the total suppression of the fluctuation correction in such a
strong field. The regular part of the Maki-Thompson contribution does not
manifest itself in this case while the AL term can compete with the DOS one
in the immediate vicinity of $T_{c},$ where the small anisotropy factor $r$
can be compensated by the additional $\epsilon ^{3}$ in the denominator. The
anomalous MT contribution can contribute in the case of small pairbreaking
only, which is opposite to what is expected in HTS.

In the $3D$ case ($\epsilon \ll r$) the behavior of the magnetoconductivity
is more complex. In weak and intermediate fields the main, negative,
contribution to the magnetoconductivity occurs from the AL and MT terms. At $%
H\sim H_{c2}(\epsilon )(h\sim \epsilon )$ the paraconductivity is already
considerably suppressed by the magnetic field and the $h^{2}-$ dependence of
the magnetoconductivity changes through the $\sqrt{\frac{r}{h}}$ tendency to
the high field asymptote $-\sigma _{zz}^{(fl)}(0,\epsilon ).$ In this
intermediate region of fields ($\epsilon \ll h\ll r),$ side by side with the
decrease ($\sim \sqrt{\frac{r}{h}}$) of the main AL and MT contributions,
the growth of the still relatively small DOS term takes place. At the upper
limit of this region $(h\sim r)$ its positive contribution is of the same
order as the AL one and at high fields $(r\ll h\ll 1)$ the DOS contribution
determines the slow logarithmic decay of the fluctuation correction to the
conductivity which is completely suppressed only at $H\sim H_{c2}(0).$ The
regular part of the Maki- Thompson contribution is not of special importance
in the $3D$ case. It remains comparable with the DOS contribution in the
dirty case at fields $h\lesssim r$, but decreases rapidly $(\sim \frac{r}{h}%
) $ at strong fields ( $h\gtrsim r),$ in the only region where the robust $%
\Delta \sigma _{c}^{DOS}(h,\epsilon )\sim \ln \frac{h}{r}$ shows up
surviving up to $h\sim 1$.

\begin{figure}[tbp]
\epsfxsize=8cm
\centerline {\epsfbox{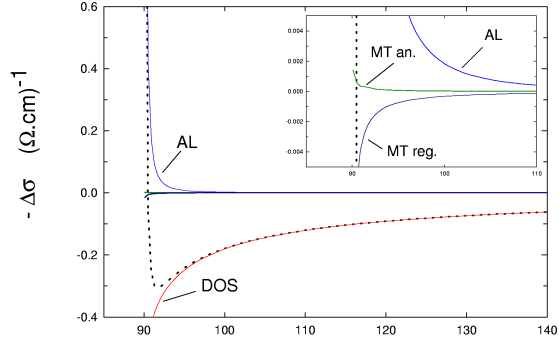}}
\caption{{}Decomposition of the calculation of total theoretical
magnetoconductivity for an underdoped Bi-2212 single crystal at 27 T. The
inset shows the regular and anomalous parts of \ the MT contribution which
are too small to be presented in the same scale as the AL and DOS
contributions \protect\cite{Wahl00}.}
\label{Wahl3}
\end{figure}

The temperature dependence of the different fluctuation contributions to the
magnetoconductivity calculated for an underdoped Bi-2212 single crystal at
the magnetic field 27 T is presented in Fig. \ref{Wahl3}.

The formulas for the in-plane magnetoconductivity are presented in the table:

\begin{tabular}{|l|l|l|}
\hline
& $h\ll \epsilon $ & 
\begin{tabular}{ll}
$\epsilon \ll h\ll r;$ & $\max \{\epsilon ,r\}\ll h$%
\end{tabular}
\\ \hline
$\Delta \sigma _{xx}^{AL}$ & $-\frac{{e^{2}}}{2^{8}{s}}\frac{{[8\epsilon
(\epsilon +r)+3r^{2}]}}{{[\epsilon (\epsilon +r)]^{5/2}}}h^{2}$ & 
\begin{tabular}{cc}
$-\sigma _{xx}^{AL}(0,\epsilon )+\frac{{e^{2}}}{{4s}}\frac{1}{\sqrt{2hr}};\;$
& $-\sigma _{xx}^{AL}(0,\epsilon )+{\frac{{e^{2}}}{{8s}}}\frac{1}{h}$%
\end{tabular}
\\ \hline
\begin{tabular}{c}
$\Delta \sigma _{xx}^{MT(an)}$ \\ 
$(\min \{\epsilon ,r\}\ll \gamma _{\varphi })$%
\end{tabular}
& $-{\frac{{e^{2}}}{3\;2^{5}{s}}\frac{{(\epsilon +r/2)}}{{[\epsilon
(\epsilon +r)]^{3/2}}}}h^{2}$ & $-\sigma _{xx}^{MT}(0,\epsilon )+\frac{e^{2}%
}{8s}\frac{1}{\gamma _{\varphi }}\ln \frac{\sqrt{\gamma _{\varphi }}}{\sqrt{%
2h}+\sqrt[=]{2h+r}}$ \\ \hline
\begin{tabular}{c}
$\Delta \sigma _{xx}^{MT(an)}$ \\ 
$(\gamma _{\varphi }\ll \min \{\epsilon ,r\})$%
\end{tabular}
& $-{\frac{{e^{2}}}{3\;2^{5}{s}}\frac{{1}}{\epsilon {\gamma ^{3/2}r^{1/2}}}}%
h^{2}$ & 
\begin{tabular}{cc}
$-\sigma _{xx}^{MT}(0,\epsilon )+{\frac{0.2{e^{2}}}{{s}}}\frac{1}{\sqrt{hr}}%
;\;$ & $-\sigma _{xx}^{MT}(0,\epsilon )+\frac{3\pi ^{2}e^{2}}{32s}\frac{1}{h}
$%
\end{tabular}
\\ \hline
\begin{tabular}{c}
$\Delta (\sigma _{xx}^{DOS}+$ \\ 
$\sigma _{xx}^{MT(reg)})$%
\end{tabular}
& $\frac{{e^{2}(\kappa +\tilde{\kappa})}}{3\;2^{7}s}\frac{{(\varepsilon +r/2)%
}}{{[\varepsilon (\varepsilon +r)]^{3/2}}}h^{2}$ & 
\begin{tabular}{cc}
$0.428\frac{e^{2}{(\kappa +\tilde{\kappa})}}{2^{6}s}\sqrt{\frac{h}{r};}\;$ & 
$\qquad \frac{e^{2}{(\kappa +\tilde{\kappa})}}{32s}\ln {\frac{\sqrt{h}}{(%
\sqrt{\epsilon }+\sqrt{\epsilon +r})}}$%
\end{tabular}
\\ \hline
\end{tabular}

\begin{center}
Table 4
\end{center}

Analyzing it one can see that in almost all regions the negative AL and MT
contributions govern the behavior of in-plane magnetoconductivity.
Nevertheless, similar to the c-axis case, the high field behavior is again
determined by the positive logarithmic $\Delta (\sigma _{xx}^{DOS}+\sigma
_{xx}^{MT(reg)})$ contribution, which is the only one to survive in strong
field. It is important to stress that the suppression of the DOS
contribution by a magnetic field takes place very slowly. Such robustness
with respect to the magnetic field is of the same physical origin as the
slow logarithmic dependence of the DOS-type corrections on temperature.

Another important problem which appears in the fitting of the resistive
transition shape in relatively strong fields with the fluctuation theory is
the much larger broadening of the transition than predicted by the
Abrikosov-Gorkov theory \cite{AG60}. Kim and Gray \cite{KG93} explained the
broadening of the c-axis peak with increasing magnetic field in terms of
Josephson coupling, describing a layered superconductor as a stack of
Josephson junctions. In Refs. \cite{IOT89,UD91} the self-consistent Hartree
approach was proposed for the extension of fluctuation theory beyond the
Gaussian approximation. It results in the considerable shift of $T_{c}(H)$
toward low temperatures with a corresponding broadening of the transition.
The renormalized reduced temperature $\tilde{\varepsilon}_{h}$ is determined
according to the self-consistent equation \cite{UD91}: 
\begin{equation}
\varepsilon _{h}=\tilde{\varepsilon}_{h}-\frac{1}{4}\ Gi_{\left( 2\right) }\
h\sum_{n=0}^{1/h}\frac{1}{[(\tilde{\varepsilon}_{h}+hn)(\tilde{\varepsilon}%
_{h}+h(n+1)+r)]^{1/2}}  \label{self}
\end{equation}
The authors of \cite{BLM97}, following the procedure proposed by Dorsey and
Ullah \cite{UD91}, modified the expressions (\ref{ALzz})- (\ref{l3}) by
account for (\ref{self}). In result they succeeded to fit quantitatively
both in-plane resistivity transition and the transverse resistivity peak for
BSCCO\ films strongly broadened by applied magnetic field.

\subsection{Fluctuations far from $T_{c}$ or in strong magnetic fields}

As was mentioned above the role of fluctuations is especially pronounced in
the vicinity of the critical temperature. Nevertheless for some phenomena
they can be still considerable far from the transition too. In these cases
the GL theory is certainly unapplicable since the short-wave and dynamical
fluctuation contributions have to be taken into account. It can be done in
the microscopic approach which we will demonstrate in several examples.

\subsubsection{Fluctuation magnetic susceptibility far from transition 
\protect\cite{AL73}.}

The given above qualitative estimations (\ref{3dkhi})-(\ref{anis}) for the
fluctuation diamagnetic susceptibility, based on the Langevin formula,
demonstrate that even at high temperatures $T\gg T_{c}$ it turns to be of
the order of $\chi _{P}$ for clean $3D$ superconductors and exceeds
noticeably this value for $2D$ systems. In order to develop the microscopic
theory \cite{M73,AL73,B74} let us start from the general expression for free
energy in the one-loop approximation:

\begin{equation}
F=T\sum_{\Omega _{k}}\mathrm{Tr}\{\ln [1-g\Pi (\Omega _{k},\mathbf{r,r}%
^{\prime })]\},  \label{oneloop}
\end{equation}
where $g$ is the effective interaction constant related with the transition
critical temperature by (\ref{gvst}). This approximation corresponds to the
ladder one (see (\ref{dyspro})) for the fluctuation propagator. The
polarization operator $\Pi (\Omega _{k},\mathbf{r,r}^{\prime })$ is
determined by expression (\ref{genprop}) but in the case of an applied
magnetic field the homogeneity of the system is lost and $\Pi (\Omega _{k},%
\mathbf{r,r}^{\prime })$ depends\ not on the space variable difference $%
\mathbf{r-}$ $\mathbf{r}^{\prime }$ but on each separately. Expanding $\Pi
(\Omega _{k},\mathbf{r,r}^{\prime })$ one can express the magnetic
susceptibility of a layered superconductor in a weak magnetic field
perpendicular to the layers in terms of the derivatives $\Pi _{x}=\frac{%
\partial }{\partial q_{x}}\Pi (\mathbf{q})$ \cite{AL73}:

\begin{equation}
\chi =-\frac{\partial ^{2}F}{\partial H^{2}}=-\frac{2}{3}e^{2}T\sum_{\Omega
_{k}}\int \frac{d^{3}\mathbf{q}}{(2\pi )^{3}}L^{3}\Pi _{x}(\Pi _{x}\Pi
_{yy}-\Pi _{y}\Pi _{xy}).
\end{equation}
The final expressions for the fluctuation diamagnetic susceptibility in the
clean and dirty cases for wide range of temperatures can be written as:

\begin{equation}
\chi _{fl}^{(3)}(T)=\frac{\chi _{P}}{3}\left\{ 
\begin{array}{c}
0.05(\ln ^{-2}(\omega _{D}/T_{c}))-\ln ^{-2}(T/T_{c})),\;\tau ^{-1}\ll T\ll
\omega _{D} \\ 
2\sqrt{T\tau }\ln ^{-1}(T/T_{c}),\;T_{c}\ll T\ll \tau ^{-1}
\end{array}
\right.  \label{khi3u}
\end{equation}

\begin{equation}
\chi _{fl}^{(2)}(T)=\frac{0.05}{p_{F}s}\chi _{P}\left( \frac{E_{F}}{T}%
\right) \frac{1}{\ln ^{3}(T/T_{c})}  \label{khi2u}
\end{equation}
Let us stress that these results are valid for the fluctuation diamagnetism
of a normal metal with $g>0$ too, if by $T_{c}$ one uses the formal value $%
T_{c}\sim E_{F}\exp (\frac{1}{\nu g}).$

\subsubsection{Fluctuation magnetoconductivity far from transition 
\protect\cite{L80}.}

Let us discuss the conductivity of the $2D$ electron system with \
impurities in a magnetic field at low temperatures.\ Even in the absence of
the field the effects of quantum interference of the non-interacting \
electrons in their scatterings on elastic impurities already results in the
appearance of a nontrivial temperature dependence of the resistance. This
result contradicts the statement of the classical theory of metals requiring
the saturation of the resistance at its residual value at low temperatures.
In a superconductor above the critical temperature this, so-called weak
localization (WL), effect is amplified by the Andreev reflection of
electrons on the fluctuation Cooper pair leading to appearance of the MT
correction to the conductivity. The characteristic feature of both the MT
and WL corrections is their extreme sensitivity to the dephasing time $\tau
_{\varphi }$ and to weak magnetic fields.

Beyond the GL region ($T\gtrsim T_{c}$) the MT correction is determined by
the same diagram 2 of Fig. \ref{conddia} but now the dynamic ($\Omega
_{k}\neq 0$) and short-wave-length ($q\sim \xi ^{-1}$) fluctuation modes
have to be taken into account. The corresponding calculations were performed
in \cite{AV80,L80} and the result can be written in the form:

\begin{equation}
\delta \sigma _{WL+MT}=\frac{e^{2}}{2\pi ^{2}}\left[ \alpha -\beta (T)\right]
Y\left( \Omega _{L}\tau \epsilon \right) ,  \label{larkMT}
\end{equation}
where we introduced the effective Larmour frequency for the diffusion motion 
$\Omega _{L}=$ $4\mathcal{D}eH$ with the diffusion coefficient $\mathcal{D}$ 
\footnote{%
Comparison of the expressions (\ref{dimlh}), (\ref{eigenmag}) and (\ref
{larkMT}) relates the Larmour frequency with the dimensionless field: $%
h=\Omega _{L}/2T_{c}$ introduced in section 2 and the diffusion coefficient
with the phenomenological GL constants $\mathcal{D}=1/m\alpha .$} and the
function

\begin{equation}
Y\left( x\right) =\ln x+\psi (\frac{1}{2}+\frac{1}{x})=\left\{ 
\begin{array}{c}
\frac{x^{2}}{24},\;x\ll 1 \\ 
\ln x,\;x\gg 1
\end{array}
\right. .
\end{equation}
The first term in this formula corresponds to the WL contribution ($\alpha
=1 $ if the spin-orbit interaction of the electrons with the impurities is
small while in the opposite limiting case $\alpha =-1/2$), the second
describes the MT contribution to magnetoconductivity. The function $\beta
\lbrack \ln (T/T_{c})]$ was introduced in Ref.\cite{L80}. At $T\rightarrow
T_{c}\;\beta (x)=1/x$ and (\ref{larkMT}) reduces to the already studied MT
correction in the vicinity of critical temperature. For $T\gg T_{c}\;\beta
(x)=1/x^{2}$ and the MT contribution gives a logarithmically small
correction to the WL\ result. Its zero-field value, being proportional to $\
\ln ^{-2}(T/T_{c}),$ decreases with the growth of the temperature faster
than both the AL contribution (in the dirty case $\delta \sigma _{AL}\sim
1/\ln (T/T_{c}))$ and the especially slow DOS contribution $\left( \delta
\sigma _{DOS}\sim \ln \ln (1/T_{c}\tau )-\ln \ln (T/T_{c})\right) $ (see
Ref. \cite{AV80,ARV83}).

It worth mentioning that for the region of temperatures $T\gg T_{c},$
analogous to Exp.(\ref{khi3u})- (\ref{khi2u}), the result (\ref{larkMT})$\;$%
can be applied both to superconducting and normal metals ($g>0$), if in
place of\ the critical temperature the formal value $T_{c}\sim E_{F}\exp (%
\frac{1}{\nu g})$ is undermined. The interplay of the localization and
fluctuation corrections was extensively studied (see, for example, \cite
{Ab85,GGZ83,Rz85,Mor87,BKF84})

\subsubsection{Fluctuations in magnetic fields near $H_{c2}(0)$ \protect\cite
{GL00}.}

As one can see from (\ref{l1})- (\ref{l3}), in the vicinity of the upper
critical field $H_{c2}(T)$ the fluctuation corrections diverge as $\epsilon
_{h}^{-1}$ for the $2D$ case and as $\epsilon _{h}^{-1/2}$ for the $3D$ case 
\footnote{$\epsilon _{h}$ is the renormalized by the magnetic field reduced
temperature $\epsilon _{h}=\epsilon +h$} (it is enough to keep just the
terms with $n=0$ in these formulas). This behavior is preserved in strong
magnetic fields too, but the coefficients undergo changes. A case of special
interest is $T\ll T_{c}$ (which means $H\rightarrow $ $H_{c2}(0)$) which
represents an example of a quantum phase transition \cite{GL00}. Microscopic
analysis of the magnetoconductivity permits us to study the effect of
fluctuations in magnetic fields of the order of $H_{c2}(0),$ where the GL
functional approach is inapplicable.

We restrict our consideration to the case of a dirty metal ($T\tau \ll 1$).
In this limit $|\tilde{\omega}_{n+\mu }-\tilde{\omega}_{-n}|\approx \tau
^{-1}$ and the Green function correlator (\ref{a2}) can be written in the
form

\begin{equation}
\mathcal{P}(\mathbf{q},\varepsilon _{1},\varepsilon _{2})=2\pi \nu \tau
^{2}\theta (-\varepsilon _{1}\varepsilon _{2})\left( \tau ^{-1}-|\varepsilon
_{1}-\varepsilon _{2}|-\widehat{\mathcal{D}}q^{2}\right) .  \label{lambdirty}
\end{equation}
Expressing $\Pi (\mathbf{q},\Omega _{k})$ in terms of $\mathcal{P}(\mathbf{q}%
,\varepsilon _{1},\varepsilon _{2})$ by means of (\ref{imppol}) and using
the definition of the critical temperature one can find an explicit formula
for the fluctuation propagator

\begin{eqnarray}
L^{-1}(\mathbf{q},\Omega _{k}) &=&g^{-1}-\Pi (\mathbf{q},\Omega _{k})=
\label{fardirpro} \\
&=&-\nu \left[ \ln \frac{T}{T_{c}}+\psi \left( \frac{1}{2}+\frac{|\Omega
_{k}|+\widehat{\mathcal{D}}\mathbf{q}^{2}}{4\pi T}\right) -\psi \left( \frac{%
1}{2}\right) \right] .  \nonumber
\end{eqnarray}
The prominent characteristic of this expression is that it is valid even
relatively far from the critical temperature (for temperatures $T\ll \min
\{\tau ^{-1},\omega _{D}\}$) and for $|\mathbf{q}|\ll l^{-1},|\Omega
_{k}|\ll \omega _{D}.$

One can rewrite this expression in a magnetic field applied along the c-axis
in the Landau representation by simply replacing $\left( \widehat{\mathcal{D}%
}\mathbf{q}^{2}\right) _{\Vert }$ $\Rightarrow $ $\Omega _{L}(n+1/2)$ \cite
{M73} :

\begin{eqnarray}
L_{n}^{-1}(q_{z},\Omega _{k}) &=&-\nu \left[ \ln \frac{T}{T_{c}}+\psi \left( 
\frac{1}{2}+\frac{|\Omega _{k}|}{4\pi T}\right. \right. + \\
&&+\left. \left. \frac{\Omega _{L}(n+1/2)+4\tau J^{2}\sin ^{2}\left(
q_{z}s/2\right) }{4\pi T}\right) -\psi \left( \frac{1}{2}\right) \right] . 
\nonumber
\end{eqnarray}

In the case of arbitrary temperatures and magnetic fields the expression for
the AL contribution to the conductivity takes the form:

\begin{eqnarray}
\sigma _{xx}^{AL}(\omega _{\nu }) &=&\nu e^{2}T\sum_{\Omega
_{k}}\sum_{n,m=0}^{\infty }B_{n,m}(\Omega _{k}+\omega _{\nu },\Omega
_{k})L_{m}(\Omega _{k})\times  \label{sigfield} \\
&&\times B_{m,n}(\Omega _{k},\Omega _{k}+\omega _{\nu })L_{n}(\Omega
_{k}+\omega _{\nu })  \nonumber
\end{eqnarray}
(we have restricted our consideration to the $2D$ case). The expression for $%
B_{n,m}(\Omega _{k},\omega _{\nu })$ can be rewritten as

\begin{eqnarray}
B_{n,m}(\Omega _{k}+\omega _{\nu },\Omega _{k}) &=&-\frac{2\pi \nu }{\sqrt{eH%
}}\tau ^{2}\mathcal{D}_{(2)}T\sum_{\varepsilon _{i}}\left[ \sqrt{n+1}\delta
_{m,n+1}+\sqrt{n}\delta _{m,n-1}\right] \times  \nonumber \\
&&\lambda _{n}(\varepsilon _{i}+\omega _{\nu },\Omega _{k}-\varepsilon
_{i})\lambda _{m}(\varepsilon _{i},\Omega _{k}-\varepsilon _{i})
\label{bfield}
\end{eqnarray}
with

\begin{equation}
\lambda _{m}(\varepsilon _{1},\varepsilon _{2})=\frac{1}{\tau }\frac{\Theta
\left( -\varepsilon _{1}\varepsilon _{2}\right) }{|\varepsilon
_{1}-\varepsilon _{2}|+\Omega _{L}(m+1/2)}.
\end{equation}

The critical field $H_{c2}(T)$ is determined by the equation $%
L_{0}^{-1}(q_{z}=0,\Omega _{k}=0)=0.$ This is why in the vicinity of $%
H_{c2}(T)$ the singular contribution to (\ref{sigfield}) originates only
from the terms with $L_{0}(0,\Omega _{k}).$ The frequency dependencies of
the functions $B_{n,m}(\Omega _{k}+\omega _{\nu },\Omega _{k})$ and $%
L_{1}(\Omega _{k})$ are weak although we cannot omit them to get
nonvanishing answer. It is enough to restrict ourselves to the linear
approximation in their frequency dependencies. If the temperature $T\ll
T_{c0}$ the sum over frequencies in (\ref{bfield}) can be approximated by an
integral. Transforming the boson frequency $\Omega _{k}$ summation to a
contour integration as was done above and making the analytic continuation
in the external frequency $\omega _{\nu }$ one can get an explicit
expression for the d.c. paraconductivity. In the same spirit the
contributions of all other diagrams from the Fig. \ref{conddia} which
contribute to fluctuation conductivity in the case under discussion are
calculated side by side with the AL one. The final answer can be presented
in the form: 
\begin{eqnarray}
\delta \sigma _{tot} &=&\frac{2e^{2}}{3\pi ^{2}}\left\{ -\ln \frac{\pi T_{c0}%
}{2\gamma T}+\frac{3\mathit{\gamma }_{E}T}{T_{c0}}\left( \frac{H_{c2}(T)}{%
H-H_{c2}(T)}\right) +\right.  \nonumber \\
&&+\psi \left[ \frac{T_{c0}}{2\mathit{\gamma }_{E}T}\left( \frac{H-H_{c2}(T)%
}{H_{c2}(T)}\right) \right] + \\
&&+\left. 4\left( \frac{T_{c0}}{2\mathit{\gamma }_{E}T}\frac{H_{c2}(T)}{%
H-H_{c2}(T)}\psi ^{\prime }\left[ \frac{T_{c0}}{2\mathit{\gamma }_{E}T}%
\left( \frac{H-H_{c2}(T)}{H_{c2}(T)}\right) \right] -1\right) \right\} , 
\nonumber
\end{eqnarray}
where $\mathit{\gamma }_{E}$ \ is the Euler constant. Let us consider some
limiting cases. If the temperature is relatively high $T/T_{c0}\gg \left(
H-H_{c2}(T)\right) /H_{c2}(T)$, we obtain the following formula for the
fluctuation conductivity: 
\begin{equation}
\delta \sigma =\frac{2\mathit{\gamma }_{E}e^{2}}{\pi ^{2}}\,\frac{T}{T_{c0}}%
\left( \frac{H_{c2}(T)}{H-H_{c2}(T)}\right) .  \label{lt}
\end{equation}
If $H<H_{c2}(0)$, we can introduce $T_{c}(H)$ and rewrite Eq.(\ref{lt}) in
the usual way 
\begin{equation}
\delta \sigma =\frac{3e^{2}}{2\mathit{\gamma }_{E}\pi ^{2}}\,\frac{T_{c0}}{%
T-T_{c}(H)}.  \label{lt2}
\end{equation}
If $H>H_{c2}(0)$, in the low-temperature limit $T/T_{c0}\ll \left(
H-H_{c2}(T)\right) /H_{c2}(T)$ we have 
\begin{equation}
\delta \sigma =-\frac{2e^{2}}{3\pi ^{2}}\,\ln \frac{H_{c2}(T)}{H-H_{c2}(T)}.
\label{st}
\end{equation}
One can see, that even at zero temperature a logarithmic singularity remains
and the corresponding correction is negative. It results from all three
fluctuation contributions, although the DOS one exceeds the others by
numerical factor. Let us recall that in the case of the c-axis conductivity
of a layered superconductor, or in granular superconductors above $T_{c},$
the DOS contribution exceeds the MT and AL ones parametrically \cite{BEL00}.

\subsection{The effect of fluctuations on the Hall conductivity\protect\cite
{AHL95}}

Let us start with a discussion of the physical meaning of the Hall
resistivity $\rho _{xy}$. In the case of only one type of carriers it
depends on their concentration $n$ and turns out to be independent of the
electron diffusion coefficient: $\rho _{xy}=H/\left( en\right) .$ The
fluctuation processes of the MT and DOS types contribute to the diffusion
coefficient, so their expected contribution to the Hall resistivity is zero.
For the Hall conductivity in a weak field one can write

\begin{equation}
\sigma _{xy}=\rho _{xy}\sigma _{xx}^{2}=\rho _{xy}\sigma _{xx}^{(n)2}+2\rho
_{xy}\sigma _{xx}^{(n)}\delta \sigma _{xx}=\sigma _{xy}^{(n)}\left( 1+2\frac{%
\delta \sigma _{xx}}{\sigma _{xx}}\right)
\end{equation}
so, evidently, the relative fluctuation correction to Hall conductivity is
twice as large as the fluctuation correction to the diagonal component. This
qualitative speculation is confirmed by the direct calculation of the MT
type diagram \cite{FET71}.

The AL process corresponds to an independent charge transfer which cannot be
reduced to a renormalization of the diffusion coefficient. It contributes
weakly to the Hall effect, and this contribution is related to the Cooper
pair particle-hole asymmetry. This effect was investigated in a set of
papers: \cite{FET71,APS71,IMOA79,VL90,UD91,AR92,AHL95}. Let us recall that
the proper general expression describing the paraconductivity contribution
to the Hall conductivity in the general case of arbitrary magnetic fields
and frequencies (in the TDGL theory limits) was already carried out above in
the phenomenological approach (see Eq. (\ref{condgeneral})). The microscopic
consideration of this value can be done in the spirit of the calculation of $%
\sigma _{xx}^{AL}$ (see (\ref{analicon})) and after the analytical
continuation results in

\begin{eqnarray}
\sigma _{xy}^{AL} &=&\left( \frac{2h\nu (0)}{\pi }\right)
^{2}\sum_{n=0}^{\infty }\left( n+1\right) \int_{-\pi /s}^{\pi /s}\frac{dk_{z}%
}{2\pi }\int_{-\infty }^{\infty }\coth \frac{z}{2T}dz\times  \nonumber \\
&&\times \left[ \mbox{Im}L_{n}^{R}(z)\frac{\partial }{\partial z}\mbox{Re}%
L_{n+1}^{R}(z)-\mbox{Im}L_{n+1}^{R}(z)\frac{\partial }{\partial z}\mbox{Re}%
L_{n}^{R}(z)\right] .  \label{hallge}
\end{eqnarray}
where dimensionless magnetic field $h$\ was introduced by (\ref{dimlh}). The
phenomenological expression (\ref{condgeneral}) can be obtained from this
formula by carrying out the frequency integration in the same way as was
done in the calculation of (\ref{analicon}) (the essential region of
integration is $z\ll T$).

One can see from (\ref{hallge}) that if $\mbox{Im}L_{n}^{R}(-z)=-\mbox{Im}%
L_{n}^{R}(z)$ and $\mbox{Re}L_{n}^{R}(-z)=\mbox{Re}L_{n}^{R}(z)$ the Hall
conductivity is equal to zero, or, in terms of the phenomenological
parametrization, the reality of $\gamma _{GL}$\ results in a zero Hall
effect. Physically it is possible to say that this zero is the direct
consequence of electron-hole symmetry. However, from the formula (\ref
{genprop}) one can see that an energy dependence of the density of states or
the electron interaction constant $g$ immediately results in the appearance
of an imaginary part of $\gamma _{GL}$. In the weak interaction approximation

\begin{equation}
\mbox{Im}\gamma _{GL}=\frac{\nu (0)}{2}\left( \frac{\partial \ln T_{c}}{%
\partial E}\right) _{E=E_{F}}.  \label{imgamma}
\end{equation}
Usually this value is small in comparison with $\mbox{Re}\gamma _{GL}$ by a
ratio of the order of $T_{c}/E_{F}.$ Taking into account the terms of the
order of $\ \mbox{Im}\gamma _{GL}$ in (\ref{hallge}) and using the explicit
form of the fluctuation propagator for layered superconductor (\ref{layerpro}%
) one can find

\begin{eqnarray}
\sigma _{xy}^{AL} &=&e^{2}T\frac{\mbox{Im}\gamma _{GL}}{2\nu (0)}h \\
&&\times \int_{-\pi /s}^{\pi /s}\frac{dk_{z}}{2\pi }\frac{1}{\left[ \epsilon
+r\sin ^{2}\left( k_{z}s/2\right) )\right] ^{2}}F(\frac{\epsilon +r\sin
^{2}\left( k_{z}s/2\right) )}{2h}),  \nonumber
\end{eqnarray}
where

\begin{equation}
F(x)=4x^{2}\left[ \psi (x)+x\psi ^{\prime }(x)-1-\psi (\frac{1}{2}+x)\right]
..
\end{equation}
For $H\rightarrow 0$ the expression for the fluctuation Hall
paraconductivity takes the form

\begin{equation}
\sigma _{xy}^{AL}=\frac{e^{2}T}{6s}\left( \frac{\mbox{Im}\gamma _{GL}}{\nu
(0)}\right) h\frac{\epsilon +r/2}{\left[ \epsilon \left( \epsilon +r\right) %
\right] ^{3/2}}=\frac{e^{3}\Phi _{0}T}{6\pi s}\left( \frac{\mbox{Im}\gamma
_{GL}}{\nu (0)}\right) \frac{\epsilon +r/2}{\left[ \epsilon \left( \epsilon
+r\right) \right] ^{3/2}}h.  \label{ALhall}
\end{equation}
One can see that in the $2D$ case the temperature dependence of the AL
fluctuation correction to the\ Hall conductivity

\[
\sigma _{xy}^{AL}\approx \frac{e^{3}\Phi _{0}}{12\pi s}\left( \frac{T_{c}}{%
E_{F}}\right) \frac{h}{\epsilon ^{2}} 
\]
turns out to be more singular than the MT one.

\subsection{Fluctuations in the ultra-clean case \protect\cite{LSV00}}

When dealing with the superconductor electrodynamics in the fluctuation
regime, it is necessary to remember that in the vicinity of the critical
temperature the role of the effective size of a fluctuation Cooper pair is
played by the GL coherence length $\xi _{GL}(T)=\xi _{0}/\sqrt{\epsilon }$.
So, as was already mentioning above, the case of a pure enough
superconductor with electron mean free path $\ell \gg \xi _{0}$ has to be
formally subdivided into the clean ($\xi _{0}\ll \ell \ll \xi _{GL}(T)$) and
ultra-clean ($\xi _{GL}(T)\ll \ell $) limits. The nontrivial cancellation of
the contributions, previously divergent in $T\tau $ (see, for example, (\ref
{DOSint})), will be shown in this Section. This results in a reduction of
the total fluctuation correction in the ultra-clean case to the AL term
only. We will base on \cite{LSV00} restricting our consideration to the case
of a 2D electron system.

In terms of the parameter $T\tau ,$ used in the theory of disordered alloys,
three different domains of the metal purity can be distinguished: $T\tau \ll
1$ (dirty case), $1\ll T\tau \ll 1/\sqrt{\epsilon }$ (clean case) and $1/%
\sqrt{\epsilon }\ll T\tau $ (ultra-clean case of non-local electrodynamics).
The latter case was rarely discussed in the literature \cite{BM90,RV94,AG95}
in spite of the fact that it becomes of primary importance for metals of
very modest purity, let us say, with $T\tau \approx 10$. Really, in this
case the condition $T\tau \geq 1/\sqrt{\epsilon }$ , which in terms of the
reduced temperature is read as $10^{-2}\leq \epsilon \ll 1,$ practically
covers all the experimentally accessible range of temperatures for the
fluctuation conductivity measurements. As regards the usually considered
local clean case ($1\ll T\tau \ll 1/\sqrt{\epsilon }$) for the chosen value $%
T\tau \approx 10,$ it would not have any range of applicability. Indeed, the
equivalent condition for the allowed temperature interval is $\epsilon \ll
(T\tau )^{-2},$ and it almost contradicts the $2D$ thermodynamical
Ginzburg-Levanyuk criterion for the mean field approximation applicability ($%
Gi_{(2)}=\frac{T_{c}}{E_{F}}\ll \epsilon $). Moreover, as we will show
below, for transport coefficients the higher order corrections become
comparable with the mean field results much before they are important for
thermodynamical quantities, namely at $\epsilon \sim \sqrt{Gi_{(2)}}$ \cite
{LO73,VD83}. So in practice one can speak about the dirty and the non-local
ultra-clean limits only.

As we saw above the 2D AL contribution turns out to be completely
independent of the electron mean free path $\ell $ \cite{AL68} . The
anomalous Maki-Thompson contribution, being induced by the pairing on the
Brownian diffusive trajectories \cite{VBML99}, naturally depends on $T\tau ,$
but in an indirect way. It turns out to be $\tau -$independent up to $T\tau
\sim 1/\sqrt{\epsilon },$ (see (\ref{d25_1})) and diverges as $T\tau \ln
(T\tau )$ for $T\tau \gg 1/\sqrt{\epsilon }$ \cite{BM90,RV94}. The analogous
problem takes place in the case of the DOS contribution: its standard
diagrammatic technique calculations lead to a negative correction (\ref
{DOSint}) \cite{ILVY93} \ evidently strongly divergent when $T\tau
\rightarrow \infty .$ In the derivation of all these results the local form
of the fluctuation propagator and Cooperons were used. This is why the
direct extension of their validity for $T\tau \gg 1/\sqrt{\epsilon }%
\rightarrow \infty $ \ is incorrect.

One can notice\ \cite{LSV00} that at the upper limit of the clean case, when 
$T\tau \sim 1/\sqrt{\epsilon },$ both the DOS and anomalous MT (\ref{d25_1})
contributions turn out to be of the same order of magnitude but of opposite
signs. So one can suspect that in the case of a correct procedure of
impurity averaging in the ultra-clean case the large negative DOS
contribution can be cancelled with the positive anomalous MT one. In the
case of a $2D$ electron spectrum the Cooperon can be calculated exactly for
the case of an arbitrary electron mean free path:

\begin{equation}
\lambda (\mathbf{q},\varepsilon _{1},\varepsilon _{2})=\left( 1-{\frac{%
\Theta (-\varepsilon _{1}\varepsilon _{2})}{{\tau \sqrt{(\widetilde{%
\varepsilon }_{1}-\widetilde{\varepsilon }_{2})^{2}+v_{F}^{2}q^{2}}}}}%
\right) ^{-1}.  \label{a14}
\end{equation}
One can see that this expression can be reduced to (\ref{lamb}) in the case
of $v_{F}q\ll |\widetilde{\varepsilon }_{1}-\widetilde{\varepsilon }_{2}|$.
\ Let us stress that this result was carried out without any expansion over
the Cooper pair center of mass momentum $\mathbf{q}$ and is valid in the $2D$
case for an arbitrary $\ell q.$

The fluctuation propagator in the $2D$ case of an arbitrary mean free path
can be written as \cite{LSV00} 
\begin{eqnarray}
-[\nu L(\mathbf{q},\Omega _{k})]^{-1} &=&\ln \frac{T}{T_{c}}%
+\sum_{n=0}^{\infty }\left\{ \frac{1}{n+1/2}\right.  \label{genpro} \\
&&\ \left. -\frac{1}{\sqrt{\left( n+\frac{1}{2}+\frac{\Omega _{k}}{4\pi T}+%
\frac{1}{4\pi T\tau }\right) ^{2}+\frac{v_{F}^{2}\mathbf{q}^{2}}{16\pi
^{2}T^{2}}}-\frac{1}{4\pi T\tau }}\right\} .  \nonumber
\end{eqnarray}
Near $T_{c}$ $\ln \frac{T}{T_{c}}\approx \epsilon $ and in the local limit,
when only small momenta $\ell q\ll 1$ are involved in the final
integrations, the Exp. (\ref{genpro}) can be expanded in $v_{F}q/\max
\{T,\tau ^{-1}\}$ and reduces to the appropriate local expression.

Let us demonstrate the specifics of the non-local calculations for the
example of the Maki-Thompson contribution. We restrict our consideration to
the vicinity of the critical temperature, where the static approximation is
valid. Using the non-local expressions for the Cooperon and the propagator
one can find after integration over electronic momentum: 
\begin{eqnarray}
Q^{(MT)}\left( \omega _{\upsilon }\right) &=&-4\pi \nu
v_{F}^{2}e^{2}T^{2}\sum_{\varepsilon _{n}}\int \frac{d^{2}\mathbf{q}}{(2\pi
)^{2}}L(\mathbf{q},0)\times  \label{MT1} \\
&&\times \left[ \mathcal{M}\left( \stackrel{\sim }{\epsilon }_{n},\stackrel{%
\sim }{\epsilon }_{n+\nu },\mathbf{q}\right) +\mathcal{M}\left( \stackrel{%
\sim }{\epsilon }_{n+\nu },\stackrel{\sim }{\epsilon }_{n},\mathbf{q}\right) %
\right] ,  \nonumber
\end{eqnarray}
where 
\[
\mathcal{M}\left( \alpha ,\beta ,\mathbf{q}\right) =\frac{\ R_{q}(2\alpha 
\mathbf{)}R_{q}(\alpha +\beta \mathbf{)}-\Theta (\alpha \beta )R_{q}(2\alpha 
\mathbf{)}R_{q}(2\beta \mathbf{)}}{(\beta -\alpha )^{2}\left( R_{q}(2\alpha 
\mathbf{)}-\frac{1}{\tau }\right) \left( R_{q}(2\beta \mathbf{)}-\frac{1}{%
\tau }\right) R_{q}(\alpha +\beta \mathbf{)}} 
\]
and $R_{q}\mathbf{(}x\mathbf{)}=\sqrt{x^{2}+v_{F}^{2}\mathbf{q}^{2}}$. The
analogous consideration of the DOS diagrams 2 and 4 which are the leading
ones in the clean case \cite{ILVY93} results in similar expressions.

One can see that, after analytical continuation over the external frequency $%
\omega _{\nu }\rightarrow -i\omega $ and the consequent tending $\omega
\rightarrow 0,$ each of the DOS or MT type diagrams is written in the form
of a Laurent series of the type $C_{-2}(T\tau )^{2}+C_{-1}(T\tau
)+C_{0}+C_{1}(T\tau )^{-1}+...$ and is divergent at $T\tau \rightarrow
\infty $ in accordance with (\ref{DOSint}). Nevertheless the expansion in a
Laurent series of the sum of these non-local diagrams leads to the exact
cancellation of all divergent contributions. The leading order of the sum of
the MT and DOS contributions in the limit of $T\tau \gg 1$ turns out to be
proportional to $(T\tau )^{-1}$ only and disappears in the ultra-clean
limit. So the correct accounting for non-local scattering processes in the
ultra-clean limit results in a total quantum correction negligible in
comparison with the AL contribution. Nevertheless, its formal independence
on impurities concentration (see (\ref{ALmicro})) was re-examined for
ultra-clean case too in \cite{AHL95} and there it was demonstrated that this
statement is valid in a rigorous sense only in the case of direct current
and absence of a magnetic field. Let us recall that the normal Drude
conductivity in the ultra-clean case takes the form

\begin{equation}
\ \sigma _{\pm }\left( \omega \right) =\sigma _{xx}\pm i\sigma _{xy}=\frac{%
e^{2}n\tau /m}{1-i(\omega \mp \omega _{c})\tau },
\end{equation}
where $\omega _{c}$ is the cyclotron frequency. When $\tau \rightarrow
\infty $ \ the real part of the conductivity vanishes. The analysis of the
AL diagram in the ultra-clean case demonstrates that each of the Green
functions blocks $B$ acquires the same denominator. As a result the
expression for the fluctuation conductivity contains the same Drude like
pole but of second order

\begin{equation}
\sigma _{\pm }^{AL}(\omega )=\frac{\sigma _{xx}^{AL(l)}\pm i\sigma
_{xy}^{AL(l)}}{(1-i(\omega \mp \omega _{c})\tau )^{2}}  \label{Dr}
\end{equation}
($\sigma _{\alpha \beta }^{AL(l)}$ is the component of the paraconductivity
tensor calculated above in the local limit (\ref{ALmicro}-(\ref{ALhall})).
The origin of this pole can be recognized by means of the following
speculation. The electric field does not interact directly with the
fluctuation Cooper pairs, but it produces the effect by interaction with the
quasiparticles forming these pairs only. The characteristic time of the
change of a quasiparticle state is of the order of $\tau $. Consequently the
single-particle Drude type conductivity in an a.c. field has a first order
pole, while in the AL paraconductivity it is of second order \cite{AHL95}.
In spite of this difference one can see that the AL conductivity, like the
Drude one, vanishes at $\omega \neq 0,\tau \rightarrow \infty $ because in
the absence of impurities the interaction of the electrons does not produce
any effective force acting on the superconducting fluctuations, while the
d.c. paraconductivity conserves its usual $\tau -$independent form. It is
impossible to distinguish the motion of the electron liquid from the
condensate motion in current experiments without additional scattering.

The non-local form of the Cooperon and fluctuation propagator have to be
taken into account not only for the ultra-clean case but in every problem
where relatively large bosonic momenta are involved: the consideration of
dynamical and short wavelength fluctuations beyond the vicinity of critical
temperature, the effect of relatively strong magnetic fields on fluctuations
etc. Recently such an approach was developed in a number of studies \cite
{RV94,LSV00,ERS99,Ax99}.

\subsection{The effect of fluctuations on the one-electron density of states
and on tunneling measurements}

\subsubsection{ Density of states \protect\cite{CCRV90}.}

The appearance of non-equilibrium Cooper pairing above $T_{c}$ leads to a
redistribution of the one-electron states around the Fermi level. A
semi-phenomenological study of the fluctuation effects on the density of
states (DOS) of a dirty superconducting material was first carried out while
analyzing the tunneling experiments of granular $Al$ in the fluctuation
regime just above $T_{c}$ \cite{CA68}. The second metallic electrode was in
the superconducting regime and its well developed gap gave a bias voltage
around which a structure, associated with the superconducting fluctuations
of $Al$, appeared. The measured DOS energy dependence has a dip at the Fermi
level \footnote{%
Here we refer the energy $E$ to the Fermi level, where we assume $E=0$.},
reaches its normal value at some energy $E_{0}(T),$ show a maximum at an
energy value equal to several times $E_{0}$, finally decreases towards its
normal value at higher energies. The characteristic energy $E_{0}$ was found
to be of the order of the inverse of the GL relaxation time $\tau _{GL}$
introduced above.

The presence of a depression at $E=0$ and of a peak at $E\sim (1/{\ \tau
_{GL}})$ in the DOS above $T_{c}$ are precursor effects of the appearance of
the superconducting gap in the quasiparticle spectrum at temperatures below $%
T_{c}$. The microscopic calculation of the fluctuation contribution to the
one-electron DOS can be carried out within the diagrammatic technique \cite
{ARW70,CCRV90}.

\begin{figure}[tbp]
\epsfxsize=4in
\centerline {\epsfbox{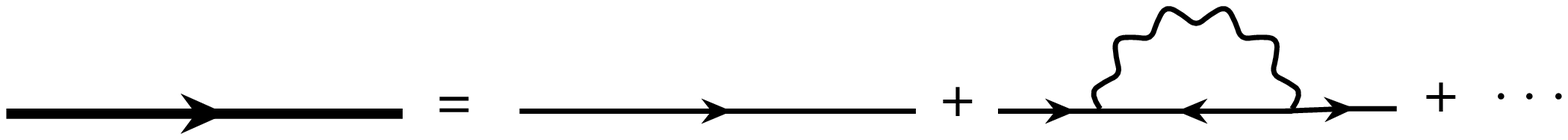}}
\caption{{}The one-electron Green function with the first order fluctuation
correction.}
\label{dendia}
\end{figure}

Let us start from the discussion of a clean superconductor. As is well known
the one-electron DOS is determined by the imaginary part of the retarded
Green function integrated over momentum. This definition permits\ us to
express the appropriate fluctuation correction in terms of the fluctuation
propagator:

\begin{equation}
\delta \nu ^{(c)}(E,\epsilon )=-\frac{1}{\pi }\mbox{Im}\int \frac{d^{D}%
\mathbf{p}}{(2\pi )^{D}}\delta G^{R}(\mathbf{p},E)=-\frac{1}{\pi }\mbox{Im}%
R^{R}(E)
\end{equation}
where $R^{R}(E)$ is the retarded analytical continuation of the expression
corresponding to the diagram of Fig. \ref{dendia}:

\begin{equation}
R(\varepsilon _{n})=T\sum_{\Omega _{k}}\int \frac{d^{D}\mathbf{q}}{(2\pi
)^{D}}L(\mathbf{q},\Omega _{k})\int \frac{d^{D}\mathbf{p}}{(2\pi )^{D}}G^{2}(%
\mathbf{p},\varepsilon _{n})G(\mathbf{q-p},\Omega _{k}-\varepsilon _{n}).
\end{equation}

The result of the integration of the last expression depends strongly of the
electron spectrum dimensionality: for the two important cases of isotropic $%
3D$ and $2D$ electron spectra one finds \cite{CCRV90}

\begin{eqnarray}
\frac{\delta \nu _{(3)}^{(c)}(E,\epsilon )}{\nu _{(3)}(0)} &=&-\frac{\left(
4\pi \right) ^{3/2}}{7\zeta (3)}\sqrt{Gi_{(3,c)}}\mbox{Re}\frac{\sqrt{T_{c}}%
}{\sqrt{\tau _{GL}^{-1}-2iE+\varkappa _{3}^{2}T_{c}}}\times  \label{rho3} \\
&&\left\{ \frac{1}{\tau _{GL}^{-1}-iE+\tau _{GL}^{-1/2}\left[ \tau
_{GL}^{-1}-2iE+\varkappa _{3}^{2}T_{c}\right] ^{1/2}}\right\} ,  \nonumber
\end{eqnarray}

\begin{eqnarray}
\frac{\delta \nu _{(2)}^{(c)}(E,\epsilon )}{\nu _{(2)}(0)} &=&-\frac{\left(
4\pi \right) ^{2}}{7\zeta (3)}Gi_{(2,c)}\frac{T_{c}^{2}}{\left[
E^{2}+\varkappa _{2}^{2}T_{c}\tau _{GL}^{-1}\right] }\times  \label{nu2} \\
&&\left\{ 1-\frac{E}{\sqrt{E^{2}+\varkappa _{2}^{2}T_{c}\tau _{GL}^{-1}}}\ln 
\frac{E+\sqrt{E^{2}+\varkappa _{2}^{2}T_{c}\tau _{GL}^{-1}}}{\varkappa \sqrt{%
T_{c}\tau _{GL}}}\right\} ,  \nonumber
\end{eqnarray}
where $\varkappa _{D}=\pi \sqrt{\pi D/7\zeta (3)}.$

In a dirty superconductor the calculations may be carried out in a similar
way with the only difference that the impurity renormalization of the Cooper
vertices has to be taken into account \cite{ARW70}. The value of the
fluctuation dip at the Fermi level can be written in the form:

\begin{equation}
\frac{\delta \nu ^{(d)}(0)}{\nu (0)}\sim -\left\{ 
\begin{array}{ll}
\sqrt{Gi_{(3,d)}}\epsilon ^{-3/2}, & D=3 \\ 
Gi_{(2,d)}\epsilon {^{-2}}, & D=2
\end{array}
\right. .  \label{dirtyds}
\end{equation}
At large energies $E\gg \tau _{GL}^{-1}$ the DOS recovers its normal value,
according to the same laws (\ref{dirtyds}) but with the substitution $%
\epsilon \rightarrow E/T_{c}$. It is interesting that the critical exponents
of the fluctuation correction of the DOS change when moving from a dirty to
a clean superconductor \cite{CCRV90}: the analysis of (\ref{rho3})-(\ref{nu2}%
) gives

\begin{equation}
\frac{\delta \nu ^{(c)}(0)}{\nu (0)}\sim -\left\{ 
\begin{array}{ll}
\sqrt{Gi_{(3,c)}}\epsilon ^{-1/2}, & D=3 \\ 
Gi_{(2,c)}\epsilon ^{-1}, & D=2
\end{array}
\right. .  \label{cleands}
\end{equation}
\begin{figure}[tbp]
\epsfxsize=3.5in
\centerline {\epsfbox{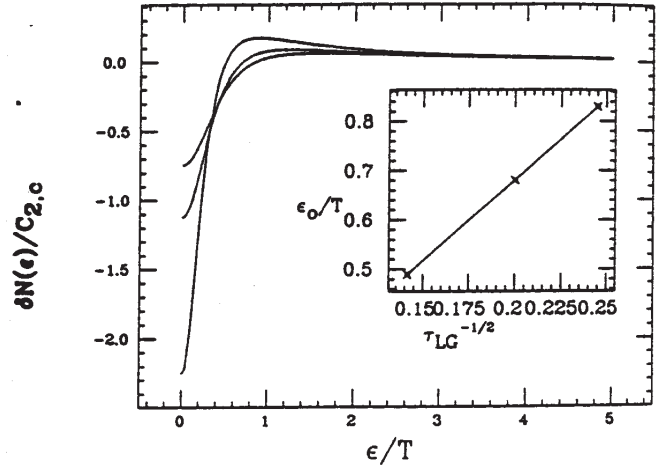}}
\caption{{}The theoretical curve of the energy dependence for the normalised
correction to the single-particle density of states vs energy for a clean
two-dimensional superconductor above $T_{c}.$ }
\end{figure}

Another important respect in which the character of the DOS renormalization
differs strongly for the clean and dirty cases is the energy scale at which
this renormalization occurs. In the dirty case this energy turns out to be 
\cite{ARW70} $E_{0}^{(d)}\sim T-T_{c}\sim \tau _{GL}^{-1},$ while in the
clean case $E_{0}^{(c)}\sim \sqrt{T_{c}(T-T_{c})}$ \cite{CCRV90}. To
understand this important difference one has to study the character of the
electron motion in both cases \cite{CCRV90}. The relevant energy scale in
the dirty case is the inverse of the time necessary for the electron to
diffuse over a distance equal to the coherence length $\xi (T)$. This energy
scale coincides with the inverse relaxation time: $t_{\xi }^{-1}=\mathcal{D}%
\xi ^{-2}(T)\sim \tau _{GL}^{-1}\sim T-T_{c}.$ In the clean case, the
ballistic motion of the electrons gives rise to a different characteristic
energy scale $t_{\xi }^{-1}\sim v_{F}\xi ^{-1}(T)\sim (T_{c}\tau
_{GL}^{-1})^{1/2}\sim \sqrt{T_{c}(T-T_{c})}.$

One can check that the integration of (\ref{rho3})-(\ref{nu2}) over all
positive energies gives zero: 
\begin{equation}
\int_{0}^{\infty }\delta \nu (E)dE=0  \label{N}
\end{equation}
This ``sum rule'' is a consequence of a conservation law: the number of
quasiparticles is determined by the number of cells in the crystal and
cannot be changed by the interaction. So the only effect which can be
produced by the inter-electron interaction is a redistribution of the energy
levels near the Fermi energy. The sum rule (\ref{N}) plays an important role
in the understanding of the manifestation of the fluctuation DOS
renormalization in the observable phenomena. As we will see in the next
Section the singularity in the tunneling current (at zero voltage), due to
the density of states renormalization, turns out to be much weaker than that
in the DOS itself ($\ln \epsilon $ instead of $\epsilon ^{-1}$ or $\epsilon
^{-2}$, see (\ref{dirtyds})-(\ref{cleands})). A similar smearing of the DOS
singularity occurs in the opening of the pseudo-gap in the c-axis optical
conductivity, in the NMR relaxation rate etc. These features are due to the
fact that we must always form the convolution of the DOS with some slowly
varying function: for example, a difference of Fermi functions in the case
of the tunnel current. The sum rule then leads to an almost perfect
cancellation of the main singularity at low energies. The main non-zero
contribution then comes from the high energy region where the DOS correction
has its `tail'. Another important consequence of the conservation law (\ref
{N}) is the considerable increase of the characteristic energy scale of the
fluctuation pseudo-gap opening with respect to $E_{0}$: this is $eV_{0}=\pi
T $ for tunneling and $\omega \sim \tau ^{-1}$ for the c-axis optical
conductivity.

\subsubsection{The effect of fluctuations on the tunnel current \protect\cite
{VD83}.}

It is quite evident that the renormalization of the density of states near
the Fermi level, even of only one of the electrodes, will lead to the
appearance of anomalies in the voltage-current characteristics of a tunnel
junction. The quasiparticle current flowing through it may be written as a
convolution of the densities of states with the difference of the electron
Fermi distributions in each electrode (L and R):

\begin{eqnarray}
I_{qp} &=&\frac{1}{eR_{n}\nu _{L}(0)\nu _{R}(0)}\times  \label{ambar} \\
&&\int_{-\infty }^{\infty }\left( \tanh \frac{E+eV}{2T}-\tanh \frac{E}{2T}%
\right) \nu _{L}(E)\nu _{R}(E+eV)dE,  \nonumber
\end{eqnarray}
where $R_{n}$ is the Ohmic resistance per unit area and $\nu _{L}(0)$, $\nu
_{R}(0)$ are the densities of states at the Fermi levels in each of
electrodes in the absence of interaction. One can see that for low
temperatures and voltages the expression in parenthesis is a sharp function
of energy near the Fermi level. Nevertheless, depending on the properties of
the DOS functions, the convolution (\ref{ambar}) may exhibit different
properties. If the energy scale of the DOS correction is much larger than $T$
, the expression in parenthesis in (\ref{ambar}) acts as a delta-function
and the zero-bias anomaly in the tunnel conductivity strictly reproduces the
anomaly of the density of states around the Fermi level:

\begin{equation}
\frac{\delta G(V)}{G_{n}(0)}=\frac{\delta \nu (eV)}{\nu (0)},  \label{cond}
\end{equation}
where $G(V)$ is the differential tunnel conductance and $G_{n}(0)$ is the
background value of the Ohmic conductance supposed to be bias independent, $%
\delta G(V)=G(V)-G_{n}(0)$. This situation, for instance, occurs in a
junction with one amorphous electrode \cite{AA79a}, where the dynamically
screened Coulomb interaction is strongly retarded, which leads to a
considerable suppression of the density of states in the vicinity of the
Fermi level, within $\tau ^{-1}\gg T.$

It is worth stressing that the proportionality between the tunneling current
and the electron DOS of the electrodes is widely accepted as an axiom, but
generally speaking this is not always so. As one can see from the previous
subsection, the opposite situation occurs in the case of the DOS
renormalization due to the electron-electron interaction in the Cooper
channel: in this case the DOS correction varies strongly already in the
scale of $E_{0}\sim E_{\ker }\ll T$ and the convolution in (\ref{ambar})
with the DOS (\ref{nu2}) has to be carried out without the simplifying
approximations assumed to obtain (\ref{cond}). We will show that the
fluctuation induced pseudo-gap like structure in the tunnel conductance
differs drastically from the anomaly of the density of states (\ref{nu2}),
both in its temperature singularity near $T_{c}$ and in the energy range of\
its manifestation.

Let us first discuss the effect of the fluctuation suppression of the
density of states on the properties of a tunnel junction between a normal
metal and a superconductor above $T_{c}$. The effect under discussion turns
out to be most pronounced in the case of thin superconducting films ($d\ll
\xi (T)$) and layered superconductors like HTS cuprates. In order to derive
the explicit expression for the fluctuation contribution to the differential
conductance of a tunnel junction with one thin film electrode close to its $%
T_{c}$ we differentiate (\ref{ambar}) with respect to voltage, and
substitute the DOS correction given by (\ref{nu2}). This results in (see 
\cite{VD83}): 
\begin{eqnarray}
\frac{\delta G_{fl}(V,\epsilon )}{G_{n}(0)} &=&\frac{1}{2T}\int_{-\infty
}^{\infty }\frac{dE}{\cosh ^{2}\left( \frac{E+eV}{2T}\right) }\delta \nu
^{(2)}(E,\epsilon )\sim  \label{flcon1} \\
&\sim &Gi_{(2)}\ln \left( \frac{2}{\sqrt{\epsilon }+\sqrt{\epsilon +r}}%
\right) \mbox{Re}\psi ^{\prime \prime }\left( \frac{1}{2}-\frac{ieV}{2\pi T}%
\right) .  \nonumber
\end{eqnarray}

\begin{figure}[tbp]
\centerline {\epsfig {file=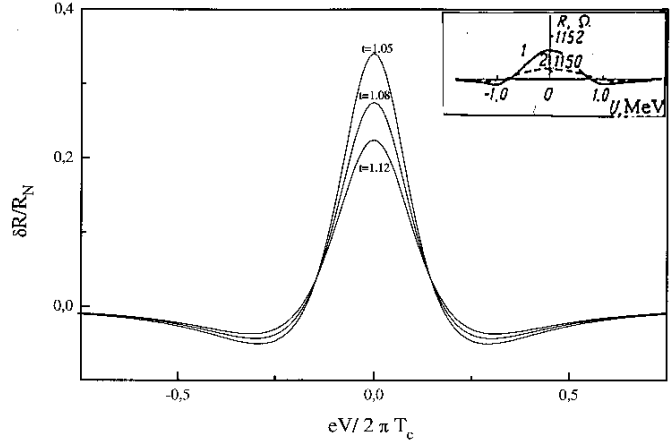, width=9cm}}
\caption{The theoretical prediction for the fluctuation-induced zero-bias
anomaly in tunnel-junction resistance as a function of voltage for reduced
temperatures $\protect\epsilon =0.05$ (top curve), $\protect\epsilon =0.08$
(middle curve) and $\protect\epsilon =0.12$ (bottom curve). The insert shows
the experimentally observed differential resistance as a function of voltage
in an Al-I-Sn junction just above the transition temperature}
\label{difconduc}
\end{figure}

It is important to emphasize several nontrivial features of the result
obtained. First, the sharp decrease ($\epsilon ^{-2(1)}$) of the density of
electron states in the immediate vicinity of the Fermi level generated by
fluctuations surprisingly results in a much more moderate growth of the
tunnel resistance at zero voltage ($\ln 1/\epsilon $). Second, in spite of
the manifestation of the DOS renormalization at the characteristic scales $%
E_{0}^{(d)}\sim T-T_{c}$ or $E_{0}^{(cl)}\sim \sqrt{T_{c}(T-T_{c})}$, the
energy scale of the anomaly developed in the $I-V$ characteristic is much
larger: $eV=\pi T\gg $ $E_{0}$ (see Fig. \ref{difconduc}).

In the inset of Fig. \ref{difconduc} the result of measurements of the
differential resistance of the tunnel junction $Al-I-Sn$ at temperatures
slightly above the critical temperature of $Sn$ electrode is presented. This
experiment was done \cite{Khachat} with the purpose of checking the theory
proposed \cite{VD83}. The non-linear differential resistance was precisely
measured at low voltages which permitted the observation of the fine
structure of the zero-bias anomaly. The reader can compare the shape of the
measured fluctuation part of the differential resistance (the inset in Fig. 
\ref{difconduc}) with the theoretical prediction. It is worth mentioning
that the experimentally measured positions of the minima are $eV\approx \pm
3T_{c}$, while the theoretical prediction following from (\ref{flcon1}) is $%
eV=\pm \pi T_{c}$. Recently similar results on an aluminium film with two
regions of different superconducting transition temperatures were reported 
\cite{PIP95}. The observations of the pseudogap anomalies in tunneling
experiments at temperatures above $T_{c}$ obtained by a variety of
experimental techniques were reported in \cite{tao97,MSW97,SKN97,RRG97,CCV99}%
..

We will now consider the case of a symmetric junction between two
superconducting electrodes at temperatures above $T_{c}$. In this case,
evidently, the correction (\ref{flcon1}) has to be multiplied by a factor of
''two'' because of the possibility of fluctuation pairing in both
electrodes. Furthermore, in view of the extraordinarily weak ($\sim \ln
1/\epsilon $) temperature dependence of the first order correction,
different types of high order corrections may manifest themselves on the
energy scale $eV\sim T-T_{c}$ or $\sqrt{T_{c}(T-T_{c})}$. Among them are the
familiar AL and MT corrections which take place in the first order of $Gi$
but in the second order of the barrier transparency. Another type of higher
order correction appears in the first order of barrier transparency but in
the second of fluctuation strength ($\sim Gi^{2}$) \cite{VD83}. Such
corrections are generated by the interaction of fluctuations through the
barrier and they can be evaluated directly from (\ref{ambar}) applied to a
symmetric junction . The second order correction in $Gi$ can be written as 
\cite{VD83}:

\begin{equation}
\delta G_{fl}^{(2)}\left( 0,\epsilon \right) \sim \int_{-\infty }^{\infty }%
\frac{dE}{\cosh ^{2}{\ }\left( \frac{E}{2T}\right) }\left[ \delta \nu
^{(2)}(E,\epsilon )\right] ^{2}\sim \frac{Gi_{\left( 2\right) }^{2}}{%
\epsilon ^{3}}  \label{ds2ord}
\end{equation}
This nonlinear fluctuation correction turns out to be small by $Gi^{2}$ but
its strong singularity in temperature and opposite sign with respect to $%
\delta G_{fl}^{(1)}$ make it interesting. Apparently it leads to the
appearance of a sharp maximum at zero voltage in $G(V)$ with a
characteristic width $eV\sim T-T_{c}$ in the immediate vicinity of $T_{c}$
(one can call this peak as the hyperfine structure). This result was
confirmed in \cite{Rei93} but to our knowledge such corrections were never
observed in tunneling experiments.

One can see that $\delta G_{fl}^{\left( 1\right) }$ and $\delta G_{fl}^{(2)}$%
\ become of the same order at $\epsilon _{cr}^{\ast }\sim \sqrt[3]{Gi}$,
i.e. the critical region where nonlinear fluctuations effects become
important in the problem under consideration starts much before the
thermodynamical criterion $\epsilon _{cr}\sim Gi.$ In the next Section we
will discuss this early manifestation of nonlinear fluctuation effects in
transport phenomena.

\subsection{Nonlinear fluctuation effects \protect\cite{LO01}}

As we\ have already seen in the temperature region $Gi\ll \epsilon \ll 1$
the thermodynamic fluctuations of the order parameter $\Psi $ can be
considered to be Gaussian. Nevertheless the example of the previous Section
demonstrates that in transport phenomena nonlinear effects, related with the
interaction of fluctuations (higher order corrections) can manifest
themselves much earlier. It has been found in paper \cite{LO73}, that
nonlinear fluctuation phenomena restrict the Gaussian region in the
fluctuation conductivity of a superconducting film to a new temperature
scale: $\sqrt{Gi_{\left( 2d\right) }}\ll \epsilon \ll 1$ (see also \cite
{Pat71,KK72,VD83,Rei92,MT89}). In this Section we obtain expressions for the
conductivity in the temperature region $Gi_{\left( 2d\right) }<\epsilon <%
\sqrt{Gi_{\left( 2d\right) }}$, where both the perturbation theory works
well and the nonlinear fluctuation effects are important.

Let us start from the correlator (\ref{dirtypro}) which can be expressed by
means of the $Gi_{\left( 2d\right) }$ number: 
\begin{equation}
\langle \Psi _{\mathbf{k}}^{\ast }\Psi _{\mathbf{k}}\rangle ={\frac{T}{\nu }}%
{\frac{1}{\epsilon +{\frac{\pi \mathcal{D}}{8T}}\mathbf{k}^{2}}}={\frac{%
32\pi ^{3}}{7\zeta (3)}Gi_{\left( 2d\right) }\frac{T^{2}}{{k^{2}+{\frac{%
8T\epsilon }{\pi \mathcal{D}}}}}}.  \label{corgi}
\end{equation}
The long-wave-length fluctuations with $k^{2}<k_{\mathrm{min}%
}^{2}=8T\epsilon /\pi \mathcal{D}$ can be considered as a local condensate.
They lead to the formation of the pseudogap

\begin{equation}
\Delta _{pg}=\left[ \int_{k^{2}\lesssim k_{\mathrm{min}}^{2}}\frac{d^{2}k}{%
\left( 2\pi \right) ^{2}}\langle \Psi _{\mathbf{k}}^{\ast }\Psi _{\mathbf{k}%
}\rangle \right] ^{1/2}\simeq {T}\sqrt{Gi_{\left( 2d\right) }}.  \label{pg}
\end{equation}
in the single-particle spectrum of excitations.

Not very close to the transition ($\epsilon >\sqrt{Gi_{\left( 2d\right) }}$)
only excitations with energies $E>\Delta _{pg}$ are important. The pseudogap
does not play any role for them. Thus, in this region of temperatures it is
sufficient to consider fluctuations in the linear approximation only (see 
\cite{AL68,M68,T70}). \ However, in the temperature region $\epsilon <\sqrt{%
Gi_{\left( 2d\right) }}$ the nonlinear fluctuation contribution of the
excitations with energies $E<\Delta _{pg}$ becomes essential.

To take into account the spatial dependence of the order parameter we will
use the results obtained in \cite{LO71}. It was shown there that the spatial
variations of $\Delta _{pg}$ act on single-particle excitations in the same
way as magnetic impurities do (the analogy between the effect of
fluctuations and magnetic impurities was observed in many papers, see for
example, \cite{VDS88}). In this case, the total pairbreaking rate $\Gamma $
can be written as a sum of the pairbreaking rate due to the magnetic
impurities and the fluctuation term. Thus, the self-consistent equation for $%
\Gamma $ can be written in the following form \cite{LO71}: 
\begin{equation}
\Gamma =\int {\frac{d^{2}k}{(2\pi )^{2}}}\,\frac{\langle \Psi _{k}^{\ast
}\Psi _{k}\rangle }{E+{\frac{1}{2}}\mathcal{D}k^{2}+\Gamma }+{\frac{1}{\tau
_{s}}}.  \label{Gamma}
\end{equation}

In the region $E\lesssim \Gamma $, $\Gamma \gg T\epsilon $ we obtain from
Eqs.(\ref{Gamma}), (\ref{corgi}): 
\begin{equation}
\Gamma \sim T\left( Gi_{\left( 2d\right) }\,\right) ^{1/2}\simeq \Delta
_{pg},  \label{Gam2}
\end{equation}
which coincides with the results obtained in \cite{Pat71,BCAW85}.

Let us note, that the pair-breaking rate $\Gamma $ was found to be of the
order of the pseudogap $\Delta _{pg}$. Thus, a wide maximum appears in the
density of states at $E\sim \Delta _{pg}$. As we already saw (\ref{d25_1}),
in purely $2D$ case the Maki-Thompson correction to the conductivity
saturates for $T\epsilon <\Gamma $ (where $\Gamma =8T\gamma _{\varphi }/\pi $%
) and takes the form \cite{LO01}: 
\begin{equation}
{\frac{\delta \sigma ^{MT}}{\sigma _{n}}}\sim {\frac{T}{\Gamma }}Gi_{\left(
2d\right) }\ln {\frac{\pi \Gamma }{8T\epsilon }}.  \label{MT}
\end{equation}
As it can be seen from Eqs.(\ref{Gam2}) such a saturation takes place when $%
\epsilon <\sqrt{Gi_{\left( 2d\right) }}$. Similar results have been obtained
in \cite{Pat71,KK72,Rei92}, with slightly different numerical coefficients%
\footnote{%
Note, that the numerical coefficient in Eq.(\ref{MT}) depends on the
definition of $Gi_{\left( 2d\right) }$ and how the summation of higher order
diagrams is made.}. However, its exact value is not very important since in
the region $T\epsilon <\Gamma $ the Maki-Thompson correction is less
singular than the Aslamazov-Larkin one and can be neglected. The latter does
not saturate when $T$ tends to $T_{c}$ but becomes more and more singular.

In the presence of the pseudogap if there is no equilibrium, the fluctuating
Cooper pair lifetime increases with respect to the GL one: $\tau _{fl}=a\tau
_{GL}$ ($a>1$). Recall, that analogous changes in the coefficient $a$ in the
TDGL equations appear below the transition temperature (see e.g. \cite
{Sch66,GE68,SS75,LO75,LO86}). The growth of the coefficient $a$ and,
consequently, the increase of the fluctuation lifetime is because the
quasiparticles require more time to attain thermal equilibrium (the
corresponding time we denote as $\tau _{e}$). A rough estimate gives $a\sim
\Delta _{pg}\tau _{e}$. In the case of weak energy relaxation, $\tau _{e}$
has to be determined from the diffusion equation taking account of the
pseudogap (see \cite{LO75,LO86,LO77}). Note, that in this complicated case
the coefficient $a$ becomes a non-local operator. Rough estimates give the
following value for the thermal equilibrium transition time $\tau _{e}\sim (%
\mathcal{D}k_{\mathrm{min}}^{2})^{-1}\sim (T\epsilon )^{-1}$. Taking into
account Eq.(\ref{pg}) we obtain from (\ref{sigmaint}) for the
paraconductivity contribution in the discussed limit of the weak energy
relaxation \cite{LO01} : 
\begin{equation}
{\frac{\delta \sigma }{\sigma _{n}}}\sim {\frac{Gi_{\left( 2d\right) }^{3/2}%
}{\epsilon ^{2}}}.  \label{sig1}
\end{equation}

Let us discuss now the role of the energy relaxation processes,
characterized by a quasiparticle lifetime $\tau _{\varepsilon }$. Nonelastic
electron scattering off phonons and other possible collective excitations
can decrease $\tau _{\varepsilon }$ significantly. These processes together
with additional pairbreaking processes (due to magnetic impurities or a
magnetic field) lead to a decrease of the nonlinear effects. In view of
these processes, one can write the following interpolation formula for the
non-linear fluctuation conductivity \cite{LO01}: 
\begin{equation}
{\frac{\delta \sigma }{\sigma _{n}}}\sim {\frac{Gi_{\left( 2d\right) }^{3/2}%
}{\epsilon \left( \epsilon +{\frac{1}{T\tau _{\varepsilon }}}\right) \left(
1+{\frac{\Gamma }{T\sqrt{Gi_{\left( 2d\right) }}}}\right) }}.  \label{inter}
\end{equation}

Note that Eqs.(\ref{sig1}-\ref{inter}) are valid only if the parameters $%
\Gamma $ and $\tau _{\varepsilon }$ are such that the correction to
conductivity $\delta \sigma $ is larger than the usual Aslamazov-Larkin
correction Eq.(\ref{ALmicro}). \ If $\Gamma >T$, $T\tau _{\varepsilon }<%
\sqrt{Gi_{\left( 2d\right) }}$ or if $T^{2}\tau _{\varepsilon }/\Gamma
<Gi_{\left( 2d\right) }$, than nonlinear effects are negligible and the
usual result (\ref{ALmicro}) is valid for all $\epsilon >Gi_{\left(
2d\right) }$.

We see that the paraconductivity can exceed the value of the normal
conductivity $\sigma _{n}$ in the region $Gi_{\left( 2d\right) }<\epsilon
<Gi_{\left( 2d\right) }^{3/4}$. Let us recall, that in this region
corrections to all the thermodynamic coefficients are still small and the
linear theory is well applicable.

\subsection{The effect of fluctuation on the optical conductivity 
\protect\cite{FV96}}

The optical conductivity of a layered superconductor can be expressed by the
same analytically continued electromagnetic response operator $Q_{\alpha
\beta }^{(R)}(\omega )$ (see Exp.(\ref{sigmaQ})) but in contrast to the d.c.
conductivity case, calculated without the assumption $\omega \rightarrow 0$.
Let us recall that the paraconductivity tensor in an a.c. field was already
studied in Section 4 in the framework of the TDGL\ equation \cite{SchH} and
the most interesting asymptotics \ for our discussion (\ref{sigomx})-(\ref
{sigomz}), valid for $\omega \ll T$ in the $2D$ regime, were calculated
there. The microscopic calculation of the AL diagram \cite{AV80} shows that
in the vicinity of $T_{c}$ and for $\omega \ll T$ the leading singular
contribution to the response operator $Q_{\alpha \beta }^{AL\,(R)}$ arises
from the fluctuation propagators rather than from the $B_{\alpha }$ blocks,
which confirms the TDGL results. Nevertheless the DOS and MT corrections can
be calculated only by the microscopic method, as was done in \cite{AV80,FV96}%
..

Let us note that the external frequency $\omega _{\nu }$ enters in the
expression for the DOS contribution to $Q_{\alpha \beta }(\omega )$ only by
means of the Green's function $G(\mathbf{p},\omega _{n+\nu })$ and it is not
involved in $q$ integration. So, near $T_{c}$, even in the case of an
arbitrary external frequency, we can restrict consideration to the static
limit, taking into account only the propagator frequency $\Omega _{k}=0,$
and to get \cite{FV96}: 
\[
\mbox{Re}\sigma _{\alpha \beta }^{DOS}(\omega )=-{\frac{e^{2}}{2\pi s}}\hat{%
\kappa}\left( \omega ,T,\tau \right) A_{\alpha \beta }\ln \left[ \frac{2}{%
\sqrt{\epsilon +r}+\sqrt{\epsilon }}\right] , 
\]
where the anisotropy tensor $A_{\alpha \beta }$ was introduced in (\ref{d15}%
). Let us stress that, in contrast to the AL frequency dependent
contribution, this result has been found with only the assumption $\epsilon
\ll 1$, so it is valid for any frequency, and impurity concentration. The
function $\hat{\kappa}\left( \omega ,T,\tau \right) $ was calculated in \cite
{FV96} exactly but we present here only its asymptotics for the clean and
dirty cases: 
\[
\hat{\kappa}_{d}\left( \omega ,T\ll \tau ^{-1}\right) ={\ \frac{8}{\pi }}%
\left\{ 
\begin{array}{c}
\frac{7\zeta (3)}{2\pi ^{2}},\;\mbox{ }\omega \ll T\ll \tau ^{-1} \\ 
\left( \frac{T}{\omega }\right) ^{2},\;\mbox{ }T\ll \omega \ll \tau ^{-1} \\ 
-\frac{\pi T^{2}}{\omega ^{3}\tau },\;T\ll \tau ^{-1}\ll \omega
\end{array}
\right. , 
\]
\[
\hat{\kappa}_{cl}\left( \omega ,T\gg \tau ^{-1}\right) ={\ \ \frac{\pi ^{3}}{%
28\zeta (3)}}\left\{ 
\begin{array}{c}
\left( T\tau \right) ^{2},\;\mbox{ }\omega \ll \tau ^{-1}\ll T \\ 
\left( \frac{T}{\omega }\right) ^{2},\;\mbox{ }\tau ^{-1}\ll \omega \ll T \\ 
-4\left( \frac{T}{\omega }\right) ^{3},\;\tau ^{-1}\ll T\ll \omega
\end{array}
\right. . 
\]
\ 

The general expression for the MT\ contribution is too cumbersome, so we
restrict ourselves here to the important $2D$ overdamped regime ($r\ll
\epsilon \leq \gamma _{\varphi }$): 
\[
\sigma _{zz}^{MT(an)(2D)}(\omega )=\frac{e^{2}s}{2^{7}\eta _{(2)}}\frac{r^{2}%
}{\gamma _{\varphi }\epsilon }\left\{ 
\begin{array}{c}
1,\;\mbox{ }\tilde{\omega}\ll \tau _{\varphi }^{-1} \\ 
\left( \frac{8T_{c}\gamma _{\varphi }}{\pi \omega }\right) ^{2},\;\mbox{ }%
\tilde{\omega}\gg \tau _{\varphi }^{-1}
\end{array}
\right. 
\]
\[
\sigma _{xx}^{MT(an)(2D)}(\omega )=\frac{e^{2}}{8s}\left\{ 
\begin{array}{c}
\frac{1}{\gamma _{\varphi }}\ln \frac{\gamma _{\varphi }}{\epsilon },\;\mbox{
}\omega \ll \tau _{\varphi }^{-1} \\ 
\left( \frac{8T_{c}\gamma _{\varphi }}{\pi \omega }\right) ^{2},\;\mbox{ }%
\omega \gg \tau _{\varphi }^{-1}
\end{array}
\right. . 
\]

Let us discuss the results obtained. Because of the large number of
parameters entering the expressions we restrict our consideration to the
most interesting $c$-axis component of the fluctuation conductivity tensor
in the $2D$ region (above the Lawrence-Doniach crossover temperature).

The AL contribution describes the fluctuation condensate response to the
applied electromagnetic field. The current associated with it can be treated
as the precursor phenomenon of the screening currents in the superconducting
phase. As was demonstrated above the characteristic \ ''binding energy '' of
fluctuation Cooper pair is of the order of $T-T_{c}$, so it is not
surprising that the AL contribution decreases when the electromagnetic field
frequency exceeds this value. Indeed $\omega ^{AL}\sim T-T_{c}$ is the only
relevant scale for $\sigma ^{AL}$: its frequency dependence does not contain 
$T,\tau _{\varphi }$ and $\tau $. The independence from the latter is due to
the fact that elastic impurities do not present obstacles for the motion of
Cooper pairs. The interaction of the electromagnetic wave with the
fluctuation Cooper pairs resembles, in some way, the anomalous skin-effect
where the reflection is determined by the interaction with the free electron
system.

The anomalous MT contribution also is due to fluctuation Cooper pairs, but
this time they are formed by electrons moving along self-intersecting
trajectories. Being the contribution related with the Cooper pair electric
charge transfer it does not depend on the elastic scattering time but it
turns out to be extremely sensitive to the phase-breaking mechanisms. So two
characteristic scales turn out to be relevant in its frequency dependence: $%
T-T_{c}$ and $\tau _{\varphi }^{-1}$. In the case of HTS, where $\tau
_{\varphi }^{-1}$ has been estimated as at least $0.1T_{c},$ for
temperatures up to $5\div 10K$ above $T_{c}$ the MT contribution is
overdamped, it is determined by the value of $\tau _{\varphi }$ and is
almost temperature independent.

The DOS contribution to $\mbox{Re}\sigma (\omega )$ is quite different from
those above. In the wide range of frequencies $\omega \ll \tau ^{-1}$ the
lack of electron states at the Fermi level leads to the opposite sign effect
in comparison with the AL and MT contributions: $\mbox{Re}\sigma
^{DOS}(\omega )$ turns out to be negative and this means an increase of the
surface impedance, or, in other words, decrease of the reflectance.
Nevertheless, the applied electromagnetic field affects the electron
distribution and at very high frequencies $\omega \sim \tau ^{-1}$ the DOS
contribution changes its sign. It is interesting that the DOS contribution,
as a one-electron effect, depends on the impurity scattering in a similar
manner to the normal Drude conductivity. The decrease of $\mbox{Re}\sigma
^{DOS}(\omega )$ starts at frequencies $\omega \sim \min \{T,\tau ^{-1}\}$
which for HTS are much higher than $T-T_{c}$ and $\tau _{\varphi }^{-1}$.

\begin{figure}[tbp]
\centerline{\epsfig {file=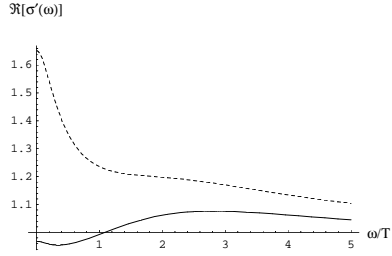,width=9cm}}
\caption{The theoretical dependence \protect\cite{FV96} of the real part of
the conductivity, normalized by the Drude normal conductivity, on $\protect%
\omega /T$, $\Re \left[ \protect\sigma ^{\prime }(\protect\omega )\right] =%
\mathrm{Re}\left[ \protect\sigma (\protect\omega )\right] /\protect\sigma ^{%
\mathrm{n}}$. The dashed line refers to the $ab$-plane component of the
conductivity tensor whose Drude normal conductivity is $\protect\sigma %
_{\parallel }^{\mathrm{n}}=N(0)e^{2}\protect\tau v_{F}^{2}$. The solid line
refers to the $c$-axis component whose Drude normal conductivity is $\protect%
\sigma _{\perp }^{\mathrm{n}}=\protect\sigma _{\parallel }^{\mathrm{n}%
}J^{2}s^{2}/v_{F}^{2}$. In this plot we have put $T\protect\tau
=0.3,E_{F}/T=50,r=0.01,\protect\epsilon =0.04,T\protect\tau _{\protect\varphi
}=4.$}
\label{FedericiPRB.2}
\end{figure}

The $\omega $-dependence of $\mbox{Re}\sigma _{zz}^{tot}$ with the most
natural choice of parameters ($T_{c}r\ll T_{c}\epsilon \leq \tau _{\varphi
}^{-1}\ll \min \{T,\tau ^{-1}\}$) is presented in Fig. \ref{FedericiPRB.2}.

Let us discuss it referring to a strongly anisotropic layered
superconductor. The positive AL and MT contributions to\ $\sigma _{zz}^{tot}$
, being suppressed by the square of the interlayer transparency, are small
in magnitude and they vary in the low frequency region $\omega \sim \min
\{T-T_{c},\tau _{\varphi }^{-1}\}.$ The DOS contribution is proportional to
the first order of transparency and remains in this region almost
invariable. With a further increase of frequency $\min \{T-T_{c},\tau
_{\varphi }^{-1}\}\lesssim \omega $ the AL and MT contributions decay; $%
\mbox{Re}\sigma _{\perp }$ remains negative up to $\omega \sim \min \{T,\tau
^{-1}\}$, then it changes its sign at $\omega \sim \tau ^{-1},$ reaches
maximum and rapidly decreases. The following high frequency behavior is
governed by the Drude law. So one can see that the characteristic
pseudo-gap-like behavior in the frequency dependence of the c-axis optical
conductivity takes place: a transparency window appears in the range $\omega
\in \lbrack T-T_{c},\tau ^{-1}]$.

In the case of the $ab$-plane optical conductivity the two first positive
contributions are not suppressed by the interlayer transparency, and exceed
considerably the negative DOS contribution in a wide range of frequencies.
Any pseudo-gap like behavior is therefore unlikely in $\sigma
_{xx}^{tot}(\omega )$: the reflectivity will be of the metallic kind.

\subsection{Thermoelectric power above the superconducting transition 
\protect\cite{RS94,VLF97}}

Thermoelectric effects are difficult both to calculate and to measure if
compared with electrical transport properties. At the heart of the problem
lies the fact that the thermoelectric coefficients in metals are the small
resultant of two opposing currents which almost completely cancel. In
calculating the thermoelectric power one finds that the electrons above the
Fermi level carry a heat current that is nearly the negative of that carried
by the electrons below $E_{F}$. In the model of a monovalent metal in which
band structure and scattering probabilities are symmetric about $E_{F}$,
this cancellation would be exact; in a real metal a small asymmetry
survives. Because of their compensated nature, thermoelectric effects are
very sensitive to the characteristics of the electronic spectrum, presence
of impurities and peculiarities of scattering mechanisms. The inclusion of
many-body effects, such as electron-phonon renormalization, multi-phonon
scattering, drag effect, adds even more complexity to the problem of
calculating the thermoelectric power. Among such effects, there is also the
influence of thermodynamical fluctuations on the thermoelectric transport in
a superconductor above the critical temperature. This problem has been
attracting the attention of theoreticians for more than twenty years, since
the paper of Maki \cite{M74} appeared, where the logarithmically divergent
AL contribution was predicted for the two-dimensional case. So the AL term
turns out to be less singular compared with the corresponding correction to
conductivity.

In every case where the main AL and MT fluctuation corrections are
suppressed for some reason, the contribution connected with fluctuation
renormalization of the one-electron density of states (DOS) can become
important. The analogous situation also occurs in the case of the
thermoelectric coefficient \cite{VLR92,VLF97}. Although the DOS term has the
same temperature dependence as the AL contribution \cite{M74,RS94}, it turns
out to be the leading fluctuation contribution in both the clean and dirty
cases, due to its specific dependence on the electron mean free path.

We introduce the thermoelectric coefficient $\vartheta $ in the framework of
linear response theory as 
\[
\vartheta =\frac{1}{T}\lim_{\omega \rightarrow 0}\frac{\mbox{Im}[%
Q^{(eh)R}(\omega )]}{\omega } 
\]
where $Q^{(eh)R}(\omega )$ is the Fourier representation of the retarded
correlation function of electric $J^{e}$ and heat $J^{h}$ current operators
in Heisenberg representation: 
\[
Q^{(eh)R}(X-X^{\prime })=-\Theta (t-t^{\prime })\langle \langle \left[
J^{h}(X),J^{e}(X^{\prime })\right] \rangle \rangle . 
\]
Here $X=(\mathbf{r},t)$ and $\langle \langle \cdots \rangle \rangle $
represents both thermodynamical averaging and averaging over random impurity
positions. The correlation function $Q^{(eh)R}$ in the diagrammatic
technique is represented by a bubble with two exact electron Green's
functions and two external field vertices, the first, $e\mathbf{v}$,
associated with the electric current operator and the second, $\frac{i}{2}%
(\varepsilon _{n}+\varepsilon _{n+\nu })\mathbf{v}$, associated with the
heat current operator ($\varepsilon _{n}$ is fermionic Matsubara frequency) 
\cite{VL90}. The first order fluctuation corrections to $Q^{(eh)}(\omega
_{\nu })$ are represented by the same diagrams as for conductivity (see Fig.%
\ref{conddia}).

The first diagram describes the AL contribution to thermoelectric
coefficient and was calculated in \cite{M74,RS94} with the electron-hole
asymmetry factor taken into account in the fluctuation propagator. Diagrams
2-4 represent the Maki-Thompson contribution, neither anomalous nor regular
parts of these diagrams contribute to $\vartheta $ in any order of
electron-hole asymmetry \cite{VL90,RS94}. The contribution from diagrams
5-10 describes the correction to $\vartheta $ due to fluctuation
renormalization of the one-electron density of states. Evaluating it in the
same way as (\ref{d21}) but with one heat current vertex one obtains a
vanishing result if electron-hole asymmetry is not taken into account. The
first possible source of this factor is contained in the fluctuation
propagator; it was used in \cite{RS94} for the AL diagram but for the DOS
contribution this correction results in non-singular contributions to $%
\vartheta $ only and can be neglected. Another source of electron-hole
asymmetry is connected with expansion of energy-dependent functions in
powers of $\xi /E_{F}$ near the Fermi level: 
\begin{equation}
\nu (\xi )\mathbf{v}^{2}(\xi )=\nu (0)\mathbf{v}^{2}(0)+\xi \left[ \frac{%
\partial (\nu (\xi )\mathbf{v}^{2}(\xi ))}{\partial \xi }\right] _{\xi =0}.
\label{exp}
\end{equation}
Only the second term in Eq. (\ref{exp}) contributes to the thermoelectric
coefficient. Performing the integration over $\xi $ , summations over
fermionic frequencies and analytical continuation of the result obtained we
find that the contribution to the thermoelectric coefficient associated with
the DOS renormalization takes the form 
\begin{equation}
\vartheta _{\mathrm{2D}}^{\mathrm{DOS}}=\frac{1}{4\pi ^{2}}\frac{eT_{c}}{\nu
(0)\mathbf{v}_{F}^{2}}\left[ \frac{\partial (\nu \mathbf{v}^{2})}{\partial
\xi }\right] _{\xi =0}\kappa ^{\ast }(T\tau )\ln [\frac{2}{\sqrt{\epsilon }+%
\sqrt{\epsilon +r}}],  \label{final1}
\end{equation}
where 
\begin{eqnarray}
\kappa ^{\ast }(T\tau ) &=&-{\frac{1+{\frac{\pi }{8T\tau }}}{T\tau \left[
\psi \left( {\frac{1}{2}}+{\frac{1}{4\pi T\tau }}\right) -\psi \left( {\frac{%
1}{2}}\right) -{\frac{1}{4\pi T\tau }}\psi ^{\prime }\left( {\frac{1}{2}}%
\right) \right] }} \\
&=&\left\{ 
\begin{array}{c}
\frac{8\pi ^{2}}{7\zeta (3)}T\tau \approx 9,4T\tau \;\;T\tau \gg 1 \\ 
\left( T\tau \right) ^{-1}\;\;T\tau \ll 1
\end{array}
\right. .  \nonumber
\end{eqnarray}

Summing Eq. ({\ref{final1}) with the AL contribution \cite{RS94} one can
find the total correction to the thermoelectric coefficient in the case of a
2D superconducting film of thickness $d$ : 
\[
\frac{\vartheta ^{DOS}+\vartheta ^{AL}}{\vartheta _{0}}=-0.17\frac{1}{%
E_{F}\tau }\frac{1}{p_{F}d}\ln \left( \frac{T_{c}}{T-T_{c}}\right) \left[
\kappa ^{\ast }(T_{c}\tau )+5.3\ln \frac{\omega _{D}}{T_{c}}\right] , 
\]
Assuming $\ln (\omega _{D}/T_{c})\approx 2$ one finds that the DOS
contribution dominates the AL one for any value of impurity concentration: $%
\kappa ^{\ast }$ has a minimum at $T\tau \approx 0.3$ and even at this point
the DOS term is twice as large. In both limiting cases $T\tau \ll 1$ and $%
T\tau \gg 1$ this difference strongly increases.}

In practice, although the Seebeck coefficient $S=-\vartheta /\sigma $ is
probably the easiest to measure among the thermal transport coefficients,
the comparison between experiment and theory is complicated by the fact that 
$S$ cannot be calculated directly; it is rather a composite quantity of the
electrical conductivity and thermoelectric coefficient. As both $\vartheta $
and $\sigma $ have corrections due to superconducting fluctuations, the
total correction to the Seebeck coefficient is given by 
\begin{equation}
\Delta S=S_{0}\left( \frac{\Delta \vartheta }{\vartheta _{0}}-\frac{\Delta
\sigma }{\sigma _{0}}\right)  \label{S}
\end{equation}
We see that the fluctuations result in a decrease of the absolute value of
the overall Seebeck coefficient as the temperature approaches $T_{c}$.

The situation is complicated additionally in HTS materials, where the
temperature behavior of the background value of the thermoelectric power
remains unknown. This does not permit to extract precisely from the
experimental data the fluctuation part $\Delta \vartheta $ to compare it
with the theoretical prediction. Nevertheless the very sharp maximum in the
Seebeck coefficient experimentally observed in a few papers \cite
{H90,ZSY92,KB97} seems to be unrelated to the fluctuation effects. This
conclusion is supported by recent analysis of the temperature dependence of
the thermoelectric coefficient close to the transition in Refs. \cite{MVF94}.

\subsection{The effect of fluctuations on NMR characteristics \protect\cite
{RV94}}

\subsubsection{Preliminaries.}

In this Section we discuss\ the contribution of superconducting fluctuations
to the spin susceptibility and the NMR relaxation rate. For both these
effects the interplay of different fluctuation contributions is unusual with
respect to the case of the conductivity. Like in the case of the optical
conductivity, the fluctuation contributions to the spin susceptibility and
the NMR relaxation rate can manifest themselves as the opening of a
pseudogap already in the normal phase, a phenomenon which is characteristic
to HTS compounds.

We begin with the dynamic spin susceptibility $\chi _{\pm }^{(R)}(\mathbf{k}%
,\omega )=\chi _{\pm }(\mathbf{k},i\omega _{\nu }\rightarrow \omega +i0^{+})$
where 
\begin{equation}
\chi _{\pm }(\mathbf{k},\omega _{\nu })=\int_{0}^{1/T}d\tau e^{i\omega _{\nu
}\tau }\langle \langle \hat{T}_{\tau }\left( \hat{S}_{+}(\mathbf{k},\tau )%
\hat{S}_{-}(-\mathbf{k},0)\right) \rangle \rangle .
\end{equation}
Here $\hat{S}_{\pm }$ are the spin raising and lowering operators, $\hat{T}%
_{\tau }$ is the time ordering operator, and the brackets denote thermal and
impurity averaging in the usual way. The uniform, static spin susceptibility
is given by ${\chi _{s}}=\chi _{\pm }^{(R)}(\mathbf{k}\rightarrow 0,\omega
=0)$ while the dynamic NMR relaxation rate is given by 
\begin{equation}
{\frac{1}{{T_{1}T}}}=\lim_{\omega \rightarrow 0}{\frac{A}{\omega }}{\int \,}%
\frac{{d}^{3}{\mathbf{k}\,}}{\left( 2\pi \right) ^{3}}{\mbox{Im}\chi _{\pm
}^{(R)}(\mathbf{k},\omega )}
\end{equation}
where $A$ is a positive constant involving the gyromagnetic ratio.

For non-interacting electrons $\chi _{\pm }^{0}(\mathbf{k},\omega _{\nu })$
is determined by the usual loop diagram. Simple calculations lead to the
well known results for $T\ll E_{F}$: $\chi _{s}^{0}=\nu $ (Pauli
susceptibility) and $\left( 1/T_{1}T\right) ^{0}=A\pi \nu ^{2}$ (Korringa
relaxation). We will present the fluctuation contributions in a
dimensionless form by normalizing to the above results.

To leading order in $Gi$ the fluctuation contributions to $\chi _{\pm }$ can
be discussed with the help of the same diagrams drawn for the conductivity
in Fig. \ref{conddia}. It is important to note that the role of the external
vertices (electron interaction with the external field) is now played by the 
$\hat{S}_{\pm }(\mathbf{k},\tau )$ operators. This means that the two
fermion lines attached to the external vertex must have opposite spin labels
(up and down). Consequently, the Aslamazov-Larkin diagram for $\chi _{\pm }$
does not exist since one cannot consistently assign a spin label to the
central fermion for spin-singlet pairing. The next set of diagrams to
consider is the Maki-Thompson contribution. While the MT diagrams for $\chi
_{\pm }$ appear to be identical to those for the conductivity, there is an
important difference in topology which arises from their spin structure. It
is easy to see, by drawing the fluctuation propagator explicitly as a ladder
of attractive interaction lines, that the MT diagram is a non-planar graph
with a single fermion loop. In contrast the MT graph for the conductivity is
planar and has two fermion loops. The number of loops, in accordance with
the rules of diagrammatic technique \cite{AGD}, affects the sign of the
contribution.

The diagrams 5 and 6 represent the effect of fluctuations on the
single-particle self energy, leading to a decrease in the DOS. The DOS
diagrams 7 and 8 include impurity vertex corrections (note that these have
only a single impurity scattering line as additional impurity scattering in
the form of a ladder has a vanishing effect). Finally 9 and 10 are the DOS
diagrams with the Cooperon impurity corrections.

\subsubsection{Spin Susceptibility\protect\cite{MA77,RV94}.}

We note that, when the external frequency and momentum can be set to zero at
the outset, as is the case for ${\chi _{s}}$, there is no anomalous MT piece
(which as we shall see below is the most singular contribution to $1/T_{1}$%
). The MT diagram 2 then yields a result which is identical to the sum of
the DOS diagrams 5 and 6.

In the clean limit ($T_{c}\tau \gg 1)$ the fluctuation contribution is given
by $\chi _{s}^{\mathrm{fl}}={\chi _{s}}_{2}+{\chi _{s}}_{5}+{\chi _{s}}_{6}$%
; all other diagrams turn out to be negligible. In the dirty case ($%
T_{c}\tau \ll 1$), the DOS diagrams 5 and 6, together with the regular part
of the MT diagram (2), yield the same result as in the clean limit (of the
order $\mathcal{O}(T_{c}/E_{F})$). One can see, that this contribution is
negligible in comparison with the expected dominant one for the dirty case
of the order $\mathcal{O}(1/E_{F}\tau )$. A thorough study of all diagrams
shows that the important graphs in the dirty case are those with the
Cooperon impurity corrections MT 3 and 4, and the DOS ones 9 and 10. This is
the unique example known to us where the Cooperons, which play a central
role in the weak localization theory, give the leading order result in the
study of superconducting fluctuations. Diagrams 3 and 4 give one half of the
final result given below; diagrams 9 and 10 provide the other half. The
total fluctuation susceptibility is $\chi _{s}^{\mathrm{fl}}={\chi _{s}}_{3}+%
{\chi _{s}}_{4}+{\chi _{s}}_{9}+{\chi _{s}}_{10}$. Interesting, that in both
the clean and dirty cases $\chi _{s}^{\mathrm{fl}}/{\chi _{s}}^{(0)}$ can be
expressed by the same formula if one expresses the coefficient in terms of
the GL number $Gi_{\left( 2\right) }$ (\ref{Gi2micro}): 
\begin{equation}
{\frac{\chi _{s}^{\mathrm{fl}}}{{\chi _{s}}^{(0)}}}=-2Gi_{\left( 2\right)
}\ln \left( \frac{2}{\sqrt{\epsilon }+\sqrt{\epsilon +r}}\right) .
\label{dirty.chi}
\end{equation}

It is tempting to explain the negative sign of the fluctuation contribution
to the spin susceptibility in Eq.~(\ref{dirty.chi}) as arising from a
suppression of the DOS at the Fermi level. But one must keep in mind that
only the contribution of diagrams 5 and 6 can strictly be interpreted in
this manner; the MT graphs and the coherent impurity scattering described by
the Cooperons do not permit such a simple interpretation.

\subsubsection{Relaxation Rate\protect\cite{MA76,KF89,H90,RV94}.}

The calculation of the fluctuation contribution to $1/T_{1}$ requires rather
more care than ${\chi _{s}}$ because of the subtleties of analytic
continuation. Let us define the local susceptibility

\[
K(\omega _{\nu })=\int (d\mathbf{k})\chi _{+-}(\mathbf{k},\omega _{\nu }). 
\]

In order to write down the fluctuation contribution to $1/T_{1}$ for the
case of an arbitrary impurity concentration including the ultra-clean case
let us start from the anomalous MT contribution and evaluate it using the
standard contour integration techniques 
\begin{equation}
\lim_{\omega \rightarrow 0}{\frac{1}{\omega }}\mbox{Im}K^{{(}an{)}R}(\omega
)=-{\frac{\pi \nu ^{2}}{8}}\int (d\mathbf{q})L(\mathbf{q},0)\mathcal{K}(%
\mathbf{q}),  \label{eqn.a}
\end{equation}
\begin{eqnarray}
\mathcal{K}(\mathbf{q}) &=&2\tau \int_{-\infty }^{\infty }{\frac{dz}{\cosh
^{2}(z/4T\tau )}}{\frac{1}{\left( \sqrt{l^{2}q^{2}-(z-i)^{2}}-1\right) }%
\times }  \nonumber \\
&&\frac{1}{\left( \sqrt{l^{2}q^{2}-(z+i)^{2}}-1\right) }.  \label{eqn.b}
\end{eqnarray}
We have used the impurity vertices in the general form (\ref{a14}). The
first simple limiting case for (\ref{eqn.b}) is $lq\ll 1$, when the square
roots in the denominator can be expanded and $\mathcal{K}(\mathbf{q})=2\pi /%
\mathbf{D}q^{2}$. As we already know from Section 8.4 this corresponds to
the usual local approximation and covers the domain $T_{c}\tau \ll 1/\sqrt{%
\epsilon }$. Introducing the pair breaking rate $\gamma _{\varphi }$ as an
infrared cut off one can find: 
\begin{equation}
{\frac{\delta \left( 1/T_{1}\right) ^{MT(an)}}{\left( 1/T_{1}\right) ^{0}}}=%
\frac{28\zeta (3)}{\pi ^{4}}Gi_{\left( 2,d\right) }{\frac{1}{\epsilon
-\gamma _{\varphi }}}\ln (\epsilon /\gamma _{\varphi }).  \label{mt.dirty}
\end{equation}
The other limiting case is the ``ultra-clean limit'' when the characteristic 
$q$-values satisfy $lq\gg 1$. This is obtained when $T\tau \gg 1/\sqrt{%
\epsilon }\gg 1$. From (\ref{eqn.b}) we then find $\mathcal{K}(\mathbf{q}%
)=4\ln (lq)/vq$, which leads to 
\begin{equation}
{\frac{\delta (1/T_{1})^{MT(an)}}{(1/T_{1})^{0}}}=\frac{\pi ^{3}}{\sqrt{%
14\zeta (3)}}Gi_{\left( 2,cl\right) }{\frac{1}{\sqrt{\varepsilon }}}\ln
(T\tau \sqrt{\epsilon }).
\end{equation}

We note that in all cases the anomalous MT contribution leads to an \textit{%
enhancement} of the NMR relaxation rate over the normal state Korringa
value. In particular, the superconducting fluctuations above $T_{c}$ have
the \textit{opposite} sign to the effect for $T\ll T_{c}$ (where $1/T_{1}$
drops exponentially with $T$). One might argue that the enhancement of $%
1/T_{1}$ is a precursor to the coherence peak just below $T_{c}$. Although
the physics of the Hebel-Slichter peak (pile-up of the DOS just above gap
edge and coherence factors) appears to be quite different from that embodied
in the MT process, we note that both effects are suppressed by strong
inelastic scattering.

We now discuss the DOS and the regular MT contributions which are important
when strong dephasing suppresses the anomalous MT contribution discussed
above. The local susceptibility arising from diagrams 5 and 6 can be easily
evaluated. The other remaining contribution is from the regular part of the
MT diagram. It can be shown that this regular contribution is exactly one
half of the total DOS contribution from diagrams 5 and 6. All other diagrams
either vanish (as is the case for graphs 7 and 8) or contribute at higher
order in $1/E_{F}\tau $ (this applies to the graphs with the Cooperon
corrections). The final results can be presented in a unique way for the
clean (but not ultra-clean) and dirty cases by means of the ${Gi}_{\left(
2\right) }$ number : 
\begin{equation}
{\frac{\delta (1/T_{1})^{DOS}}{(1/T_{1})^{0}}}=-12Gi_{\left( 2\right) }\ln
\left( {{\frac{2}{{\ }\sqrt{\epsilon }+\sqrt{\epsilon +r}}}}\right) .
\end{equation}

The negative DOS contribution to the NMR relaxation rate is evident from the
Korringa formula and it sign seems very natural while the sign of the
positive Maki-Thompson contribution can generate a questions about its
physical origin. Let us consider a self-intersecting trajectory and the
motion of the electron along it with fixed spin orientation (let us say
''spin up''). If, after passing a full turn, the electron interacts with the
nucleus and changes its spin state and momentum to the opposite value it can
pass again along the previous trajectory moving in the opposite direction .
Interaction of the electron with itself on the previous stage of the motion
is possible due to the retarded character of the Cooper interaction and such
a pairing process, in contrast to the AL one, turns out to be an effective
mechanism for relaxation near $T_{c}$. This purely quantum process opens a
new mechanism of spin relaxation, and so contributes positively to the
relaxation rate $1/T_{1}$.

In the case of the nuclear magnetic relaxation rate calculations, the
electron interaction causing nuclear spin flip is considered. If one would
try to imagine an AL process of this type he would be in trouble, because
the electron-nuclei scattering with spin-flip evidently transforms the
initial singlet state of the fluctuation Cooper pair in a triplet-one, which
is forbidden in the scheme discussed. So the formally discovered absence of
the AL contribution to the relaxation rate is evident enough.

It is worth mentioning that the cancellation of the MT and DOS contributions
to conductivity found in Section 8.4 is crucial for the fluctuation
contributions to the NMR relaxation rate. In fact, the MT and DOS
contributions here have the same structure as in the conductivity while the
AL contribution is absent. So the full fluctuation correction to the NMR
relaxation rate in clean superconductor\ simply disappears.

\subsubsection{Discussion.}

The main results of this Section, valid for $\epsilon \ll 1$, can be
summarized as follows:

(1) Fluctuations lead to a suppression of the spin susceptibility ${\chi _{s}%
}$ , due to the combined effect of the reduction of the single particle
density of states arising from the self energy contributions, and of the
regular part of the MT process.

(2) ``Cooperon'' impurity interference terms, involving impurity ladders in
the particle-particle channel, are crucial for the ${\chi_s}$ suppression in
the dirty limit.

(3) The processes which dominate the results in (1) and (2) above have
usually been ignored in fluctuation calculations (conductivity, $1/T_{1}$,
etc.). The spin susceptibility is unusual in that the AL and the anomalous
MT terms, which usually dominate, are absent.

(4) For weak pair-breaking ($1/\tau _{\varphi }\ll T_{c}$), an enhancement
of $1/T_{1}T$ , coming from the positive anomalous MT term, takes place \cite
{KF89,RV94}.

(5) Strong dephasing suppresses the anomalous MT contribution, and $1/T_{1}$
is then dominated by the less singular DOS and the regular MT terms. Being
negative, these contributions lead to a suppression of spectral weight and a
decrease in $1/T_{1}$.

An intensive controversy took place in recent years in relation to the
magnetic field dependence of the fluctuation contribution to $1/T_{1}$. The
situation here resembles much the situation with the magnetoconductivity: a
positive MT contribution is suppressed by the magnetic field while the
magnetic field dependent part of the DOS contribution increases with the
growth of the field. But in contrast to the magnetoconductivity, which can
be measured extremely precisely, the NMR relaxation rate measurements are
much more sophisticated. The result of this delicate competition, depending
on many parameters ($r,\gamma _{\varphi },\tau ,$), was found in HTS
materials to be qualitatively different in experiments of various groups.
The absence of a strong positive AL contribution, possible d-pairing,
killing the MT contribution \cite{KF89}, small magnitude of the sum of MT
and DOS effects even in the case of s-pairing, lack of the precise values of 
$r,\gamma _{\varphi },\tau ,$ leading to contradictive theoretical
predictions\cite{RV94,CLRV96,ERS99,MRV00}, the dispersion in the quality of
samples and experimental methods were the reason of this discussion \cite
{CLRV96,Mitr99,B95,Gor99,CRLRV00,Zheng99}.

\section{Conclusions}

Several comments should be made in conclusion. As was mentioned in the
Introduction the first ''fluctuation boom'' took place at the end of 60's -
beginning of 70's, just after the discovery of the fluctuation smearing of
the superconducting transition and formulation of the microscopic theory of
fluctuations. The discovery of HTS reanimated this interest and, in order to
account for the specifics of these layered structures with high critical
temperatures, low charge carrier concentration and other particularities,
considerable progress in studies of fluctuation phenomena was achieved (see
for instance the conference proceedings \cite{AV97,BDPW98} and the extensive
review article \cite{VBML99}). As it is recognized now the optimally or
overdoped phases of HTS compounds present an example of a ''bad'' Fermi
liquid. The accounting for superconducting fluctuations is identical to
including of the electron-electron interaction beyond the Fermi-liquid
approximation. As a result a lot of anomalies of the normal state properties
of such HTS compounds can be explained. The situation was found to be much
more sophisticated in underdoped phases where the quasiparticle approach,
which from we have started this Chapter, fails.

In the fluctuation theory discussed above, as in modern statistical physics
in general, two methods have been used mainly: they are the diagrammatic
technique and the method of functional (continual) integration over the
order parameter. Each of them as we have seen, has its own advantages and
disadvantages and in different parts of this review we used the former or
the latter.

The years of the fluctuation boom coincided with the maximum development of
the diagrammatic methods of many body theory in Condensed Matter Theory.
This methods turns out to be extremely powerful: any physical problem, after
its clear formulation and writing down the Hamiltonian, can be reduced to
the summation of some classes of diagrams. The diagrammatic technique is
especially comfortable for problems containing some small parameter. In the
theory of superconducting fluctuations such a small parameter exists: as we
have seen, it is the Ginzburg-Levanyuk number $Gi_{(D)}$ which is expressed
as some powers of the ratio $\max \{T_{c},\tau ^{-1}\}/E_{F}.$ This is why
superconducting fluctuations led to the appearance of the small corrections
to different physical values in a wide range of temperatures, and due to
this smallness these corrections can be evaluated quantitatively. On the
other hand their specific dependence on nearness to the critical temperature 
$T-T_{c}$ permits to separate them in experiment from other effects.

In those cases when fluctuations are small it is possible to restrict their
summation to the ladder approximation only. The diagrammatic technique
permits in a unique way to describe the quantum and classical fluctuations,
the thermodynamical and transport effects

In the description of thermodynamic fluctuations the method of functional
integration turned out to be simpler. The ladder approximation in the
diagrammatic approach is equivalent to the Gaussian approximation in
functional integration. The method of functional integration turns out to be
more effective too in the case of strong fluctuations, for instance, in the
immediate vicinity of the phase transition. The final equations of the
renormalization group carried out by means of functional integrations turn
out to be equivalent to the result of the summation of the parquet diagrams
series. Nevertheless the former derivation is much more simple.

There is one another reason why we have tried to use both methods and even
to carry out some results in both ways. In its explosive development of the
last decades physics became an ''oral science'' . In the process of such
direct communication near a blackboard it is difficult to write and to read
some cumbersome formulas. The language of diagrams is much more
comprehensive: by drawing them the speaker demonstrates that this one is
small and that one has to be taken into account for this and that reasons
clear for the experienced listener. The success of the diagrammatic
technique in some sense is similar to the success of geometry in Ancient
Greece, where the science was ''oral'' too.

This advantage of the diagrammatic technique transforms into its
disadvantage when there is no direct communication between the speaker and
listener. It is difficult to learn the diagrammatic technique by a textbook
on your own, when no one helps you to find the necessary insight on a
complex graph. May be because of similar reasons geometry disappeared in
Middle Ages when direct communications between scientists was minimal while
the ''written'' algebra had continued to develop. Operating with Osvald
Spengler ''prosymbols'' we can say that the diagrammatic technique belongs
more to the Ancient Greece culture style with its ''finite body'' prosymbol,
while functional integration, side by side with the Vikings travels to
unknown lands and Leibnitz analysis of infinitesimals, is an evident modern
contribution to West-European culture with its ''infinite space'' prosymbol.

That is why, suspecting that the modern physics in the near future can fall
down to a ''New Middle Ages'' period, we have carried out some results by
means of functional integration instead of the diagrammatic technique.

\section{\protect\bigskip Acknowledgments}

In the first place we would like to express our deep gratitude to R.S.
Thompson and T. Mishonov, who were the first readers of the manuscript and
made a lot of valuable comments. We are grateful \ to our colleagues and
friends G. Balestrino, A. Buzdin, F. Federici, V. Galitski, D. Geshkenbein,
A. Koshelev, D. Livanov, Yu.N. Ovchinnikov, A. Rigamonti, G. Savona,
collaboration and discussions with whom helped us in writing this work.
A.A.Varlamov acknowledges the financial support of COFIN-MURST 2000 and the
Scientific Exchange Programme of the University of Minnesota. A considerable
part of this work was \ written during the visits in their frameworks. A.I.
Larkin acknowledges the financial support of the NSF Grant No. DRM-9812340.

\end{document}